\documentclass[11pt,a4paper]{article}
\usepackage{jheppub}
\usepackage[caption=false]{subfig}
\usepackage{grffile}

\newcommand{\beq}{\begin{eqnarray}}
\newcommand{\eeq}{\end{eqnarray}}
\newcommand{\non}{\nonumber\\}

\newcommand{\p}{\partial}

\def\arccoth{\qopname\relax o{arccoth}}
\newcommand{\bphi}{\boldsymbol{\phi}}
\newcommand{\bvarphi}{\boldsymbol{\varphi}}
\newcommand{\bdelta}{\boldsymbol{\delta}}
\newcommand{\bDelta}{\boldsymbol{\Delta}}
\newcommand{\bJ}{\mathbf{J}}
\newcommand{\dc}{\delta{\mkern-1mu}c}
\newcommand{\dphi}{\delta{\mkern-1mu}\phi}
\newcommand{\df}{\delta{\mkern-2.5mu}f}
\newcommand{\dlambda}{\delta{\mkern-2.5mu}\lambda}
\newcommand{\bdphi}{\bdelta{\mkern-1mu}\bphi}
\newcommand{\dtheta}{\delta{\mkern-1mu}\theta}
\newcommand{\Og}{{\rm O}}
\newcommand{\Lag}{\mathcal{L}}
\renewcommand{\d}{{\mathrm{d}}}
\renewcommand{\i}{\mathrm{i}}
\renewcommand{\div}{{\qopname\relax o{div}}}

\title{Near-BPS baby Skyrmions} 

\author{Sven Bjarke Gudnason$^1$,}
\affiliation{$^1$Institute of Contemporary Mathematics, School of
  Mathematics and Statistics, Henan University, Kaifeng, Henan 475004,
  P.~R.~China}
\emailAdd{gudnason(at)henu.edu.cn}
\author{Marco Barsanti$^{2,3}$ and}
\emailAdd{marco.barsanti@phd.unipi.it}
\author{Stefano Bolognesi$^{2,3}$}
\affiliation{$^2$Department of Physics ``E. Fermi'', University of Pisa
  Largo Pontecorvo, 3, Ed. C, 56127 Pisa, Italy}
\affiliation{$^3$INFN, Sezione di Pisa, Largo Pontecorvo, 3, Ed. C,
  56127 Pisa, Italy}
\emailAdd{stefanobolo(at)gmail.com}

\abstract{
We consider the baby-Skyrme model in the regime close to the so-called
restricted baby-Skyrme model, which is a BPS model with area-preserving 
diffeomorphism invariance.
The perturbation takes the form of the standard kinetic Dirichlet term 
with a small coefficient $\epsilon$.  
Classical solutions of this model, to leading order in $\epsilon$,
are called restricted harmonic maps.
In the BPS limit ($\epsilon\to0$) of the model with the potential
being the standard pion-mass term, the solution with unit topological
charge is a compacton. 
Using analytical and numerical arguments we obtain solutions to
the problem for topological sectors greater than one.
We develop a perturbative scheme in $\epsilon$ with which we can
calculate the corrections to the BPS mass.
The leading order ($\mathcal{O}(\epsilon^1)$) corrections show that
the baby Skyrmion with topological charge two is energetically
preferred. 
The binding energy requires us to go to the third order in $\epsilon$
to capture the relevant terms in perturbation theory, however, the
binding energy contributes to the total energy at order $\epsilon^2$. 
We find that the baby Skyrmions -- in the near-BPS regime -- are
compactons of topological charge two, that touch each other on their
periphery at a single point and with orientations in the attractive
channel. 
}

\begin{document}
\maketitle

\section{Introduction}

Nuclear binding energies are relatively small compared to the total
nuclear mass, i.e.~roughly of the order $1/100$ and slightly less than
that for the light nuclei.
A challenge for solitonic nuclear models, such as the Skyrme model, is
to reproduce this result since they usually overestimate the binding
energies.
This sparked the interest, which has grown in the recent years, in
looking for BPS solitonic models and small perturbations around them.  
In BPS models the energy is proportional to the topological charge, in
this case the baryon number, so there is no binding energy between the
nuclei. A small perturbation of the BPS model would give, presumably,
a small binding energy.
These ``near-BPS'' models have small binding energies already at the
classical level. One model of this kind is the 
so-called BPS Skyrme model which consists of a Lagrangian with a
potential plus a sextic term for the
pion fields, which is also the squared topological charge density
\cite{Adam:2010fg,Bonenfant:2010ab,Adam:2010ds}. The ordinary
Dirichlet (two-derivative kinetic) term and the Skyrme 
(four-derivative) term are considered as small perturbations. The BPS
Skyrme model has the nice phenomenological feature of having an
infinite-dimensional moduli space for static solutions consisting of
all volume-preserving diffeomorphism maps. This resonates well with the
liquid drop model of an incompressible fluid so successful in describing
nuclear matter. Various theoretical and phenomenological studies have
been done regarding the near-BPS Skyrme model
\cite{Bonenfant:2012kt,Adam:2013wya,Adam:2014dqa,Adam:2015lpa,Adam:2015lra}
but a fundamental question still remains unanswered: When the non-BPS
perturbations are gradually switched off, the solution in an arbitrary
baryonic sector flows to a particular BPS solution, one of many
in the infinite dimensional moduli space, but which one?  Clearly,
it is the infinite number of zero modes -- the very same feature which
makes the model attractive -- that makes the question hard to answer,
both analytically and numerically. Attempts to answer this question by
direct numerical study have not given a clear definite answer
\cite{Gillard:2015eia}. It was suggested, in modeling the nuclear
stars, that the answer could be the axially symmetric multi-Skyrmion
\cite{Adam:2014dqa}, but this, for reasons that we will explain,
cannot be the true minimum of the energy.  It is like searching for a
needle in a haystack; clearly we first need a good and well-motivated
analytical guess and then we can embark on the numerical study. In
this paper we will provide, with the use of a toy model, an analytical
guess, at least for a certain class of potentials.

The baby-Skyrme model, as the name suggests, is a toy model for the
Skyrme model in one spacetime dimension less, i.e.~$2+1$ instead of 
$3+1$, and with one target-space dimension less, $S^2$ instead of
$S^3$ \cite{Leese:1989gi,Piette:1994jt,Piette:1994ug,Piette:1994mh}.
In this sense it has been used in the past as a test bed for various
ideas and conjectures about the more difficult model. Among the
various similarities between the two models, both of them possess a
BPS restricted model.\footnote{We will use the names BPS baby-Skyrme
  model and restricted baby-Skyrme model indistinguishably in this
  paper, as often done in the literature.}
For the baby-Skyrme model, the restricted version contains only a
four-derivative term (the square of the topological charge) and a
potential. This BPS model has area-preserving diffeomorphism 
invariance for the static solutions and it was actually discovered
much earlier than the related model in $3+1$
dimensions \cite{Gisiger:1996vb}. 
The soliton structure depends very much on the type of potential that
is chosen, and in particular on the potential behavior close to the
minimum. In this paper we will concentrate on the type of potentials
that are quadratic at the minimum and for which the soliton solution
of the BPS model is a ``compacton''
\cite{Adam:2009px,Adam:2010jr,Speight:2010sy}. 
A compacton owe its name to the fact that all nontrivial behavior is
contained in a compact region of space, outside of which the field is
exactly at the minimum of the potential.
The topic of baby Skyrmions is interesting in its own right; 
particularly, much attention has been paid recently to the
applications in magnetic materials, see ref.~\cite{Fert2017} for a
review and
refs.~\cite{Barton-Singer:2018dlh,Schroers:2019hhe,Ross:2020hsw} for
some recent theoretical work.  
So what we describe in this work may hopefully be applied also in
these cases, with opportune modifications. 

The BPS property of a soliton model is often related to the existence
of a supersymmetric extension, but this is not the case for the
BPS-Skyrme model in 3+1 dimensions, since its target space is not
K\"ahler \cite{Zumino:1979et}.
In fact, the first attempt at supersymmetrizing the Skyrme model
yielded a supersymmetric model which was in fact more similar to the
baby Skyrme model \cite{Bergshoeff:1984wb,Freyhult:2003zb} in that the
target space was effectively $\mathbb{C}P^1\sim S^2$, although these
extensions contain extra terms in the Lagrangian -- in particular four
time derivatives, which are absent in the Skyrme model and the baby
Skyrme model.
The supersymmetric extension of exactly the baby Skyrme model was
constructed later as an $\mathcal{N}=1$ supersymmetric model (having
two supercharges) \cite{Adam:2011hj,Bolognesi:2014ova}. 
This model, however, suffered from the fact that the limit in which it
would become the restricted baby-Skyrme model does not exist, which
was otherwise expected from the BPS property of the purely bosonic
version of the baby-Skyrme model.
It turns out that extended supersymmetry, namely $\mathcal{N}=2$
supersymmetry (with four supercharges), is necessary for the
construction of the BPS baby Skyrme model; a further perhaps
surprising fact is that supersymmetry automatically eliminates the
kinetic term, indeed yielding the restricted baby-Skyrme model as its
bosonic sector \cite{Adam:2013awa,Bolognesi:2014ova}.
The baby Skyrmion solutions (compactons) turn out to preserve only a
quarter of supersymmetry (namely one supercharge)
\cite{Nitta:2014pwa,Nitta:2015uba}. 
A successful attempt at constructing a supersymmetric version of the
normal Skyrme term in 3+1 dimensions was made in
ref.~\cite{Gudnason:2015ryh}, by enlarging the 
target space from SU(2) to SL($2,\mathbb{C}$) and this model, like its 
lower-dimensional cousin, also automatically eliminates the kinetic
term.
This supersymmetric version of the normal Skyrme model (without the
kinetic term) does contain soliton solutions, but they are not BPS
\cite{Gudnason:2016iex}.

Deformations of BPS models by small perturbations have been discussed
in various cases, see for example
refs.~\cite{Bolognesi:2013nja,Bolognesi:2014ova}.
The general feature is that the moduli space of solutions is lifted by
some effective potential and the solutions, when the perturbation is
small, lie close to the minima of said potential.
In the case of BPS models with volume preserving diffeomorphism
symmetry, the moduli space is infinite dimensional and thus the problem
becomes considerably more difficult.
A rigorous mathematical definition of the problem has been put forward
in ref.~\cite{Speight:2014fqa}.
In particular, if the perturbation is the kinetic Dirichlet term,
solutions have been coined ``restricted harmonic''. Harmonic because
they minimize the Dirichlet term, restricted because the minimization
is constrained to a sub-space of the functional space, namely the
infinite-dimensional moduli space of solutions of the BPS equations.
In this paper, we will show that the concept of restricted harmonicity is
still not enough to solve the problem in the case of compactons.
In fact, for compactons, the restricted harmonic maps are still an
infinite-dimensional subspace of the moduli space. Finding the correct 
solution thus requires to go to even higher order in the
perturbation. This in particular affects the way the binding energy is
going to zero as the perturbation is switched off. 

In this paper, we focus on the baby-Skyrme model with a standard pion
mass term and the perturbation is only the Dirichlet (kinetic) term
with coefficient $\epsilon$. When $\epsilon$ is switched off, the
restricted baby-Skyrme model is recovered.
We address the near BPS behavior of the solutions in the model with
topological charges $Q=1,2$ and $4$ by large brute-force numerical
simulations and by developing a framework for perturbation theory
around the BPS solution.
We find that the correction to the energy (mass) of the axially
symmetric compactons, to leading order in $\epsilon$, comes from the
kinetic (Dirichlet) term and remains inside the compacton domain.
This is specific to the case of the compactons.
This leading-order correction thus cannot teach us anything about the
binding energy of the solutions in the near-BPS regime.
A main result of this paper, in the part of the perturbation theory,
is that the binding energy does not appear at any particular order, if
we strictly truncate the correction to a specific polynomial order.
Nevertheless, we can calculate the binding energy using a linearized
perturbation and it shows that the contribution to the binding energy
is in some sense all-order, which we could choose to interpret as a
``nonperturbative effect''. This happens because the perturbation that
gives rise to the binding energy takes the form
$\exp(-r/\sqrt{\epsilon})$ and thus does not have any nonvanishing
Taylor coefficients upon expansion.
For the choice of potential we work with in this paper, we find that
the near-BPS solutions take the form of charge-2 baby Skyrmions placed
very close to each other so that their periphery just touch.

The paper is organized as follows. In sec.~\ref{due} we give a
brief review of the baby-Skyrme model and of the solutions in the BPS
limit. In sec.~\ref{tre} we develop a systematic framework for
calculating the perturbations to the BPS soliton, which describes the
solutions in the regime near the restricted BPS model with the
Dirichlet term being a small perturbation. In order to calculate the
binding energy, we find that it is necessary to go to the third order
or the next-to-next-to-leading order in perturbation theory.
We test the perturbative scheme on axially symmetric compactons, for
which we have a comparison with the exact numerical solutions. 
In sec.~\ref{quattro} we present the results of the
full numerical PDE computations for topological charges $Q=1,2,4$ and
calculate the total energy to be used for comparison with the results
in sec.~\ref{cinque}, where we calculate the binding energies in
perturbation theory, by solving a linear PDE.
In sec.~\ref{sei} we contemplate what the solutions with large
topological charges look like. Finally we conclude with a discussion
in sec.~\ref{sette}.
This paper studies primarily the standard pion mass term, but the
cases of other potentials are relegated to appendix
\ref{app:potentials}.

\section{The model}\label{due}

The model is based on the BPS baby-Skyrme model with a non-BPS
deformation which is manifested as the kinetic term with coefficient
$\epsilon$
\beq
\mathcal{L}[\bphi] =
\epsilon\mathcal{L}_2[\bphi]
+ \mathcal{L}_4[\bphi]
- V(\bphi)
+ \frac12\lambda(\bphi\cdot\bphi - 1),
\label{eq:L}
\eeq
where the kinetic term and the Skyrme terms, respectively, are given by
\begin{align}
\mathcal{L}_2[\bphi] &= -\frac12 (\p_{\mu}\bphi\cdot\p^{\mu}\bphi),\\
\mathcal{L}_4[\bphi] &= -\frac14 (\p_{\mu}\bphi\cdot\p^{\mu}\bphi) (\p_{\nu}\bphi\cdot\p^{\nu}\bphi)
+\frac14 (\p_{\mu}\bphi\cdot\p_{\nu}\bphi) (\p^{\mu}\bphi\cdot\p^{\nu}\bphi),
\end{align}
and $\bphi=(\phi^1,\phi^2,\phi^3)$ is a real 3-vector on which 
the nonlinear sigma model constraint $\bphi\cdot\bphi=1$ is imposed by
the Lagrange multiplier, $\lambda$, and the metric signature is here
taken to be the mostly positive one. 
The symmetry of the Lagrangian for $V=0$ is $\tilde{G}=\Og(3)$.
In the vacuum, this symmetry is spontaneously broken
to $\tilde{H}\simeq\Og(2)$, which gives rise to the target space 
$\tilde G/\tilde H \simeq\Og(3)/\Og(2)\simeq S^2$.

If we choose the potential of the form
\beq
\label{genpot}
V(\bphi) = \frac1pm^2(1-\phi^3)^p,
\eeq
with $\frac12\leq p<2$, then the soliton solution, called a
baby Skyrmion, has support on a compact domain in the limit of
$\epsilon\to 0$ and is then dubbed a compacton \cite{Adam:2010jr}.
In this limit the model is then called the BPS baby-Skyrme model or
restricted baby-Skyrme model.

In this paper, we will use the pion mass term
\beq
\label{masspot}
V(\bphi) = m^2(1-\phi^3),
\eeq
which turns the baby Skyrmion into a compacton in the limit
$\epsilon\to 0$. 

The topological degree of the baby Skyrmion or compacton is given by
\beq
Q = \frac{1}{4\pi}\int\d^2x\; \mathcal{Q}_{12}
  = -\frac{1}{4\pi}\int\d^2x\; \bphi\cdot\p_1\bphi\times\p_2\bphi,
\eeq
which also counts the number of baby Skyrmions in $\mathbb{R}^2$. 

For the analytic calculations, it will prove convenient to use the
following parametrization
\beq
\bphi =
\frac{1}{1+|\omega|^2}
\begin{pmatrix}
  \omega + \bar{\omega}\\
  -\i(\omega - \bar{\omega})\\
  1-|\omega|^2
\end{pmatrix},
\label{eq:omegacoords}
\eeq
in terms of which the nonlinear sigma model constraint
$\bphi\cdot\bphi=1$ is manifest and we can thus write the Lagrangian
components as 
\begin{align}
  \Lag_2[\omega,\bar{\omega}]
  &= -2\frac{\p_\mu\omega\p^\mu\bar{\omega}}{(1+|\omega|^2)^2},\\
  \Lag_4[\omega,\bar{\omega}]
  &= -2\frac{(\p_\mu\omega\p^\mu\bar{\omega})(\p_\nu\omega\p^\nu\bar{\omega})-(\p_\mu\omega\p_\nu\bar{\omega})(\p^\mu\omega\p^\nu\bar{\omega})}{(1+|\omega|^2)^4},
\end{align}
and the potential as
\beq
V(\omega,\bar\omega) = 2m^2 \frac{|\omega|^2}{1+|\omega|^2}.
\eeq
In this parametrization, the topological charge reads
\beq
Q = -\frac{\i}{2\pi}\int \d^2x\;
\frac{\epsilon^{ij}\p_i\omega\p_j\bar\omega}{(1+|\omega|^2)^2}.
\eeq
We will keep both formalisms in this paper, as $\omega$ is useful for
analytical calculations and $\bphi$ is useful for the numerical
analysis.

\subsection{BPS solutions}

The model contains a BPS submodel, which is given by setting
$\epsilon=0$ in the Lagrangian \eqref{eq:L}:
\beq
\Lag^{\rm BPS}[\bphi] = \Lag_4[\bphi] - V(\bphi)
  + \lambda(\bphi\cdot\bphi - 1).
\label{eq:LBPS}
\eeq
In order to find the BPS equation and corresponding solutions for each
topological sector, $Q$, it proves convenient to rewrite the static
energy density (ignoring the Lagrange multiplier term)
\begin{align}
\mathcal{E}^{\rm BPS}[\bphi] &=
\frac14(\p_i\bphi\cdot\p_i\bphi)(\p_j\bphi\cdot\p_j\bphi)
-\frac14(\p_i\bphi\cdot\p_j\bphi)(\p_i\bphi\cdot\p_j\bphi)
+ m^2(1-\phi^3)\non
&= \frac14(\p_i\bphi\times\p_j\bphi)\cdot(\p_i\bphi\times\p_j\bphi)
+ m^2(1-\phi^3)\non
&= \frac14(\bphi\cdot\p_i\bphi\times\p_j\bphi)(\bphi\cdot\p_i\bphi\times\p_j\bphi)
+ m^2(1-\phi^3)\non
&= \frac12(\bphi\cdot\p_1\bphi\times\p_2\bphi)(\bphi\cdot\p_1\bphi\times\p_2\bphi)
+ m^2(1-\phi^3)\non
&= \frac12\left[(\bphi\cdot\p_1\bphi\times\p_2\bphi) \mp m\sqrt{2(1-\phi^3)}\right]^2
\pm m\sqrt{2(1-\phi^3)}(\bphi\cdot\p_1\bphi\times\p_2\bphi),
\label{eq:Bogomolny_completion}
\end{align}
where we have performed a Bogomol'nyi completion in the last step.
The last line of the above equation contains the square of a real
expression (the BPS equation) and a boundary term, which we will show
shortly is proportional to the topological charge.
Setting the expression in the square equal to zero yields the BPS
equation, which thus reads:
\beq
\mathcal{Q}_{12}
= -\bphi\cdot\p_1\bphi\times\p_2\bphi
= \mp m\sqrt{2(1-\phi^3)},
\label{eq:BPSeq}
\eeq
and when satisfied, the total energy is proportional to the
topological degree of the baby Skyrmion, $Q$. 

Changing parametrization of $\bphi$ to stereographic coordinates
\eqref{eq:omegacoords}, the BPS equation now reads 
\beq
\frac{\p_1\omega\p_2\bar{\omega} - \p_2\omega\p_1\bar{\omega}}{(1+|\omega|^2)^2}
=\frac{\p_r\omega\p_\theta\bar{\omega} - \p_\theta\omega\p_r\bar{\omega}}{r(1+|\omega|^2)^2}
= \mp\i m\sqrt{\frac{|\omega|^2}{1+|\omega|^2}}.
\eeq
Inserting the Ansatz $\omega=e^{\i N\theta}\zeta(r)$, we get
\beq
\frac{\p_r\zeta}{r}
  = \pm \frac{m}{2N}(1+\zeta^2)^{\frac32},
\label{eq:BPSeq_zeta}
\eeq
which we can write as
\beq
\int\frac{\d\zeta}{(1+\zeta^2)^{\frac32}}
= \frac{\zeta}{\sqrt{1+\zeta^2}}
= \pm\frac{m}{2N} \int \d{r}\; r
= \pm\frac{m}{4N} r^2 \pm\kappa
= \pm(\xi^2 + \kappa).
\eeq
Choosing the lower sign, we get
\beq
\frac{\zeta}{\sqrt{1+\zeta^2}} = -\xi^2-\kappa,
\eeq
which with $\kappa=-\xi_0^2$ can be written as
\beq
\zeta
= \frac{\xi_0^2-\xi^2}{\sqrt{(1-\xi_0^2+\xi^2)(1+\xi_0^2-\xi^2)}}.
\eeq
If we set the integration constant $\xi_0=1$, we can move the
coordinate singularity to $\xi=0$, obtaining the solution
\beq \label{Backsol}
\zeta
= \frac{1-\xi^2}{\xi\sqrt{2-\xi^2}}, \qquad
\xi\in[0,1],
\eeq
where $\xi\equiv\frac{r}{R}$ and the compacton radius is
\beq
R = \sqrt{\frac{4N}{m}}.
\label{eq:radius}
\eeq

It will prove useful to calculate the BPS mass which is simply the
boundary term of eq.~\eqref{eq:Bogomolny_completion}: 
\begin{align}
M^{\rm BPS} &= \int \d^2x\; \mathcal{E}^{\rm BPS}[\bphi] \non
&= \pm \i4m \int \d^2x\;\frac{|\omega|}{(1+|\omega|^2)^{\frac52}}
  \epsilon^{ij}\p_i\omega\p_j\bar\omega \non
&= \pm 16\pi m N \int \d r\;
  \frac{\zeta^2}{(1+\zeta^2)^{\frac52}}\p_r\zeta \non
&= - 16\pi m N \int_\infty^0 \d\zeta\;
   \frac{\zeta^2}{(1+\zeta^2)^{\frac52}} \non
&= \frac{16\pi m N}{3}.
\end{align}
On the third line above, we see that the entire expression is indeed a
boundary term, as promised, and it can be written as an integral over
the target space coordinate $\zeta$ (fourth line).
In the fourth line we have chosen the lower sign, corresponding to the
boundary conditions $\zeta(0)\to\infty$ and $\zeta(R)=0$. 

Note that the topological charge of this axially symmetric
configuration is
\beq
Q = -2N\int \d r\;\frac{\zeta\p_r\zeta}{(1+\zeta^2)^2} = N.
\eeq
For axially symmetric baby Skyrmions, we will use $N$ (which is equal
to $Q$) to denote the topological charge, whereas for more complicated
configurations, $Q$ is the total topological charge.
For instance, later it will be useful to consider a $Q=4$ baby
Skyrmion that is composed by two $N=2$ axially symmetric solutions.

\subsection{Energy bound}

In the previous section, we have shown that the energy is bounded from
below by the Bogomol'nyi type bound for the BPS sector $\epsilon=0$,
\beq
E_{4+0} \geq M^{\rm BPS} = \frac{16\pi m |Q|}{3}.
\eeq
In addition, it can also be shown that there is a bound in the energy
for the first term in the Lagrangian \eqref{eq:L},
\begin{align}
E_{2} &=
  2\epsilon\int \d^2x\;
  \frac{\p_i\omega\p_i\bar\omega}{(1+|\omega|^2)^2} \non
&= 4\epsilon\int \d^2x\;
  \frac{|\p\omega|^2 + |\bar\p\omega|^2}{(1+|\omega|^2)^2} \non
&\geq 4\epsilon \left|\int \d^2x\;
  \frac{|\p\omega|^2 - |\bar\p\omega|^2}{(1+|\omega|^2)^2}\right| \non
&\geq 2\epsilon \left|-\i\int \d^2x\;
  \frac{\epsilon^{ij}\p_i\omega\p_j\bar\omega}{(1+|\omega|^2)^2}\right| \non
&\geq 4\pi\epsilon |Q| = M^{\rm lump}.
\end{align}
The total energy in the model \eqref{eq:L} is thus bounded from below by
\begin{align}
  E_{2+4+0} &\geq M^{\rm BPS} + M^{\rm lump} \non
  &\geq 16\pi|Q|\left(\frac{\epsilon}{4} + \frac{m}{3}\right).
\end{align}
This bound is, however, only satisfied in the two limits:
$\epsilon\to 0$ (BPS baby-Skyrmion limit) and $\epsilon\to\infty$
(BPS lump limit) \cite{Bolognesi:2014ova}.

\section{Perturbation in \texorpdfstring{$\epsilon$}{epsilon}}
\label{tre}

We will now consider making a perturbation in $\epsilon$ around a
background solution.
That is, the background, $\bvarphi$, is a BPS compacton, which
minimizes the BPS submodel \eqref{eq:LBPS}:
\beq
\delta\Lag^{\rm BPS}[\bvarphi]=0.
\eeq
For reference, it will be useful to write the explicit form of the
charge-$N$ axially symmetric compacton solution \eqref{Backsol} in the
vector coordinate $\bvarphi$:
\beq
\bvarphi =
\begin{pmatrix}
  \sin f(r)\cos(N\theta - \alpha)\\
  \sin f(r)\sin(N\theta - \alpha)\\
  \cos f(r)
\end{pmatrix},\qquad
f(r) = \arccos\left(-1 + \frac{4r^2}{R^2} - \frac{2r^4}{R^4}\right),
\eeq
with $R$ the compacton radius \eqref{eq:radius} and
$x+\i y=r e^{\i\theta}$ the standard polar coordinates in
$\mathbb{R}^2$.

We will consider the corrections to the energy order by order in the
following sections.

\subsection{Leading-order correction}
\label{tre.uno}

The leading order correction, proportional to $\epsilon$, comes from
inserting the background solution into the kinetic term. However,
since $\bvarphi$ depends on the moduli parameters of the BPS sector,
not all the possible background solutions are equivalent choices at
this order. Recently, it has been conjectured that the right choice
among all the possible maps is given by the minimizer of $E_2$ on the
moduli space of the BPS solutions
\cite{Speight:2014fqa,Bolognesi:2014ova}.
A map respecting this request is said to be \emph{restricted harmonic}
and, in the case of a single compacton, it is identified by the
axially symmetric solution \eqref{Backsol}. The complete proof of this 
statement is given in the next section.

The first order correction in terms of $\epsilon$ is therefore
\begin{align}\label{1order}
\epsilon M^{\rm LO}(N) &= -\epsilon\int \d^2x\; \Lag_2[\bvarphi] \non
&= 4\pi\epsilon\int \d r\;
  \frac{r\zeta_r^2 + \frac{N^2}{r}\zeta^2}{(1+\zeta^2)^2} \non
&= 4\pi\epsilon\int \d\xi\;
  \frac{\xi\zeta_\xi^2 + \frac{N^2}{\xi}\zeta^2}{(1+\zeta^2)^2} \non
&= 4\pi\epsilon\left(2\log 2 + \frac{7N^2}{24}\right).
\end{align}
The parenthesis on the last line takes the value $1.678$ for $N=1$,
which is about $5/3$ of the energy bound for the kinetic term.

To this order, the energy reads
\begin{align}
\label{0+1order}
  E(\epsilon,N) &= M^{\rm BPS}(N) + \epsilon M^{\rm LO}(N) \non
  &= \frac{16\pi m N}{3}
    + 8\pi\epsilon\log 2
    +\frac{7\pi\epsilon N^2}{6}.
\end{align}
We can now consider the energy per $N$ as a function of $N$.
In particular, this function has a minimum
\beq
\frac{\d}{\d N}\left(\frac{E(\epsilon,N)}{N}\right) =
  -\frac{8\pi\epsilon\log 2}{N^2}
  +\frac{7\pi\epsilon}{6} = 0,
\eeq
which is
\beq
N_\star = 4\sqrt{\frac{3\log 2}{7}} \simeq 2.180.
\label{eq:Nstar}
\eeq
This means that for an axially symmetric configuration with
topological charge $N=1,2$, the leading-order energy (i.e.~to
$\mathcal{O}(\epsilon)$) per charge, $Q=N$, decreases as a function of
$N$.
This implies that an axially symmetric 2-Skyrmion has lower energy
than 2 well separated 1-Skyrmions.\footnote{ This happens also for
  large values of $\epsilon=\mathcal{O}(1)$, where the charge $Q$
  solutions are composed by two $N=2$ solutions as end caps,
  sandwiching $(Q-4)/2$ stretched $N=2$ solutions for even $Q$ and
  $(Q-5)/2$ stretched $N=2$ solutions with an extra squashed $N=1$
  solution \cite{Foster:2009vk}. In ref.~\cite{Foster:2009vk} they
  have been coined baby-Skyrmion chains. }

This, in particular, means that for parametrically small $\epsilon$,
the lowest energy configuration with topological charge $Q=2M$ will
consist of $M$ 2-Skyrmions at a separation distance that is not
determined at this order in the calculation.

\subsection{Restricted harmonic maps}\label{tre.due}

In the previous section, the notion of restricted harmonic maps
\cite{Speight:2014fqa} was briefly discussed.
In what follows, we will formally define such a map that is used as
the zeroth-order background of a field expansion in a near-BPS
model.
After the general definition, we focus on the near-BPS baby-Skyrme
model, giving a review of the criterion obtained in
ref.~\cite{Speight:2014fqa} to identify the correct restricted
harmonic map.
The axially symmetric solution used in eq.~\eqref{1order} is verified
to respect this criterium.
In the last part of this section, we further verify that even two
axially symmetric compactons sitting side-by-side with a random
orientation are restricted harmonic.
The last result will be useful in sec.~\ref{cinque} in which we
discuss the interaction between two near-BPS baby Skyrmions and
calculate the binding energy. 

Given a set of scalar fields $\phi^a$ mapping $d+1$ dimensional
Minkowski spacetime $(\mathcal{M},\eta)$ with flat metric $\eta$ to
the target space $(\mathcal{N},h)$ with $\mathcal{N}=S^2$ being the
2-sphere with metric $h$, a near-BPS Lagrangian $\Lag$ can be written
as  
\beq
\Lag=\Lag^{\rm BPS}+\epsilon\Lag_2,
\label{eq:Lat}
\eeq
where $\Lag^{\rm BPS}$ describes the pure BPS sector and $\epsilon$ is
a small parameter multiplying the Dirichlet term 
\begin{equation}
\Lag_2=-\frac{1}{2}h_{ab}\p_\mu\phi^a\p^\mu\phi^b,
\end{equation}
with $h_{ab}$ representing the metric of the target manifold
$\mathcal{N}$.

Taking into account only the BPS sector, we denote by
$\varphi^a(x,\lambda)$ the generic static solution of the
model that depends on the moduli parameters $\lambda$.
Explicitly, we consider a BPS system for which $\lambda$ consists of
all the possible orbits in the group of volume-preserving
diffeomorphisms.

In the limit of $\epsilon\to 0$, the total static energy $E$ of the
system \eqref{eq:Lat}, at the lowest order in $\epsilon$, takes the form
\begin{equation}\label{en}
  E = E^{\rm BPS}[\varphi^a(x,\lambda^*)]
  + \epsilon E_2[\varphi^a(x,\lambda^*)],
\end{equation}
where $\lambda^*\subset \lambda$ is the subset of $\lambda$ that
minimizes the Dirichlet energy $E_2$ 
\begin{equation}
E_2=\frac{1}{2}\int_{\mathcal{M}} \d^dx\; h_{ab}\p_i\phi^a\p_i\phi^b.
\end{equation}
The map $\varphi^a(x,\lambda^*)$ that locally minimizes $E_2$
within the domain of the group of volume-preserving diffeomorphisms is 
said to be \emph{restricted harmonic}.

A possible proof for the expression \eqref{en} can be performed
assuming a Taylor expansion for the field $\phi$ in terms of
$\epsilon$,
\begin{equation}
\phi=\phi^{(0)}+\epsilon\phi^{(1)}+\epsilon^2\phi^{(2)}+\cdots
\end{equation}
All orders of $\phi$ can be solved by minimizing the total action $S$
order by order.
It is important to note that a perturbative expansion in terms of
$\epsilon$ is not always possible for the exact static solution
$\phi$, since the dependence on such a parameter could be
nonanalytic.
Although the leading order energy (i.e.~to order
$\mathcal{O}(\epsilon)$) does take the form of eq.~\eqref{en}, it
turns out that higher-order contributions (of order
$\mathcal{O}(\epsilon^2)$ and beyond) include a nonanalytic
contribution to the energy which captures the binding energy, see
the next section.  
We therefore assume this hypothesis here only to give a simple
justification of the expression \eqref{en}; for a more complete
treatment of the topic, see
refs.~\cite{Speight:2014fqa,Bolognesi:2014ova}.

Since we are dealing only with static configurations, we consider the
static energy $E$ to be minimized instead of the action $S$ and we
analogously write a Taylor series for the energy
\beq
E = E^{(0)} + \epsilon E^{(1)} + \epsilon^2 E^{(2)} + \cdots
\eeq
Hence, the zeroth-order of the static energy is given by 
\begin{equation}
  E^{(0)}=-\int_{\mathcal{M}} \d^dx \; \Lag^{\rm BPS}\big(\phi^{(0)}\big),
  \label{eq:En-zero}
\end{equation}
whose static solution is by definition
\begin{equation}
  \phi^{(0)}=\varphi^a(x,\lambda).
  \label{BPSs}
\end{equation}
The first-order in the expansion reads
\begin{align}
  \epsilon E^{(1)} &= -\int_{\mathcal{M}}
  \d^dx\;\left[\epsilon\phi^{(1)}\left(\frac{\p\Lag^{\rm
        BPS}}{\p\phi}\bigg\lvert_{\phi^{(0)}} -
    \p_i\left(\frac{\p\Lag^{\rm BPS}}{\p\p_i\phi}\right)\bigg\lvert_{\phi^{(0)}}\right)
    +\epsilon\Lag_2\big(\phi^{(0)}\big)\right]\non
  &= -\int_{\mathcal{M}} \d^dx\;\epsilon\Lag_2\big(\phi^{(0)}\big),\label{S1}
\end{align}
where the first term in the parenthesis on the first line vanishes due
to the equations of motion. 
Since the functional $E^{(1)}$ contains only the background solution,
the minimum of $E^{(1)}$ must be sought not among all the field
configurations but only within the moduli space of the BPS solution
\eqref{BPSs}.
Therefore, following the principle of least energy (action)
\begin{equation}
  \delta E^{(1)}=0\quad\Rightarrow\quad
  \frac{\p E_2[\varphi^a(x,\lambda)]}{\p\lambda}=0,
\label{dS1}
\end{equation}
whose solution $\varphi^a(x,\lambda^*)$ is by definition the
restricted harmonic map discussed above.
We clarify that for an infinite moduli space, such as the group of
volume-preserving diffeomorphism, the derivative with respect to
$\lambda$ does not have a mathematically well-defined meaning and we
use it here only to simplify the notation.
Summing the zeroth-order \eqref{eq:En-zero} and the first
order \eqref{S1} of the energy calculated with the solution
$\varphi^a(x,\lambda^*)$ we obtain the expression \eqref{en}, as we
wanted to prove. 

After the formal definition, we need some practical tools to calculate
the restricted harmonic maps of a given system. A mathematical
criterion that is able to identify whether a map is restricted
harmonic (or not) has been proposed in ref.~\cite{Speight:2014fqa}.
In the following, we review the theorem given there, which we use for
the near-BPS baby-Skyrme model in the case of a single and two
compactons. 

Given a smooth map $\phi$ from the manifold $\mathcal{M}$ with metric
$g=g_{ij}\,\d x^i\otimes\d x^j$ to the manifold $\mathcal{N}$ with metric
$h=h_{ab}\,\d\phi^a\otimes\d\phi^b$, the Dirichlet energy is defined as  
\begin{equation}
E_2=\int_{\mathcal{M}} \frac{1}{2}\d^dx\; h_{ab}g^{ij}\p_i\phi^a\p_j\phi^b,
\end{equation}
where $d$ is the number of dimensions of the space $\mathcal{M}$.
Using the map $\phi$ we can construct the pull-back $\phi^*h$ of the
metric $h$ to $\mathcal{M}$  
\begin{equation}
\phi^*h = h_{ab}\frac{\p\phi^a}{\p x^i}\frac{\p\phi^b}{\p x^j}\d x^i\otimes\d x^j,
\end{equation}
which is necessary for what follows.
Among all the maps $\phi$ with finite Dirichlet energy connected by a
volume-preserving diffeomorphisms, a map $\tilde{\phi}$ is restricted
harmonic if and only if the one-form $\div\,\tilde{\phi^*}h$ on
$\mathcal{M}$ is exact \cite{Speight:2014fqa}.
It is useful to recall that the divergence of a symmetric
$(0,2)$ tensor $\omega=\omega_{ij}\d x^i\otimes\d x^j$ on
$\mathcal{M}$ it is nothing but 
\begin{equation}
\div\,\omega=D^i\omega_{ij}\d
x^j=g^{ik}\big(\p_k\omega_{ij}-\Gamma^l_{ki}\omega_{lj}-\Gamma^l_{kj}\omega_{il}\big)\d x^j,
\end{equation}
with the connection $\Gamma$ (Christoffel symbols) defined as 
\begin{equation}
\Gamma^i_{jk}=\frac{1}{2}g^{il}\left(\frac{\p g_{lj}}{\p x^k}+\frac{\p g_{lk}}{\p x^j}-\frac{\p g_{jk}}{\p x^l}\right).
\end{equation}
Using this theorem we verify that the axially symmetric solution used
in eq.~\eqref{1order} is restricted harmonic.
Using polar coordinates on $\mathcal{M}=\mathbb{R}^2$ and the vector
notation $\phi^a$ for the field, with the constraint $\phi^a\phi^a=1$,
we write a generic $Q=N$ axially symmetric compacton in the form 
\begin{equation}\label{N1s}
  \phi^a=
  \begin{pmatrix}
    \sin f(r)\cos (N\theta-\alpha)\\
    \sin f(r)\sin (N\theta-\alpha)\\
    \cos f(r)
  \end{pmatrix},
\end{equation}
where the function $f$ depends only on the radial coordinate and
$\alpha$ is the orientation phase \cite{Gisiger:1996vb}.
With this choice, the metric $h$ reduces to the standard euclidean
metric and the pull-back $\phi^*h$ of $h$ can be written as
\begin{equation}\label{eq:pul1} 
\phi^*h=\frac{\p\phi^a}{\p \tilde{x}^i}\frac{\p\phi^a}{\p
  \tilde{x}^j}\d\tilde{x}^i\otimes \d\tilde{x}^j=(f')^2\d r^2
  +N^2(\sin f)^2\d\theta^2,
\end{equation}
with $\d\tilde{x}^i=(\d r,\d\theta)$.
Note that there is no longer any dependence on the orientation phase
$\alpha$.
Taking the divergence of this tensor yields
\begin{align}
  \div\,\phi^*h&=\,\tilde{D}^i\tilde{\p}_i\phi^a\tilde{\p}_j\phi^a\d\tilde{x}^j\non
  &=\tilde{g}^{ik}\big(\tilde{\p}_k(\tilde{\p}_i\phi^a\tilde{\p}_j\phi^a)-\tilde{\Gamma}^l_{ki}\tilde{\p}_l\phi^a\tilde{\p}_j\phi^a-\tilde{\Gamma}^l_{kj}\tilde{\p}_i\phi^a\tilde{\p}_l\phi^a\big)\d\tilde{x}^j\non
  &=\left(2f' f''+\frac{(f')^2}{r}-\frac{N^2\sin^2f}{r^3}\right)\d r\non
  &\equiv\omega_i\d\tilde{x}^i,
\end{align}
resulting in a one-form whose exactness we must prove.
Here, it is useful to recall that according to Poincar\'e's lemma all
closed forms on a contractible manifold are exact.
Hence, to complete the proof we must verify that
$\d\big(\div\,\phi^*h\big)=0$, with $\d$ being the exterior
derivative. 
Explicitly,
\begin{align}
  \d\big(\div\,\phi^*h\big)=\d\omega&=(\tilde{\p}_i\omega_j-\tilde{\p}_j\omega_i)\d\tilde{x}^i\wedge\d\tilde{x}^j\non
  &=2(\p_r\omega_\theta-\p_\theta\omega_r)\d r\wedge\d\theta\non
  &=0,\label{exp1}
\end{align}
which gives the necessary and sufficient condition for eq.~\eqref{N1s}
to be restricted harmonic. 

The same proof can be extended to the case of two axially symmetric
compactons sitting side-by-side without overlap.
A solution for two separated compactons with $Q=N+N$ and random
orientations can be written as 
\begin{equation}\label{N11}
\phi^a=\begin{cases}
\phi^a_1(x+x_0,y)\quad &x<0,\\
\phi^a_2(x-x_0,y)\quad &x\geq0,\\
\end{cases}
\end{equation}
where $\phi^a_1,\, \phi_2^a$ are axially symmetric with respect to the
points $(-x_0,0)$ and $(x_0,0)$ respectively, i.e., 
\beq
\phi^a_1=
\begin{pmatrix}
  \sin f(r_+)\cos(N\theta_+-\alpha)\\
  \sin f(r_+)\sin(N\theta_+-\alpha)\\
  \cos f(r_+)
\end{pmatrix},\qquad
\phi^a_2=
\begin{pmatrix}
  \sin f(r_-)\cos(N\theta_--\beta)\\
  \sin f(r_-)\sin(N\theta_--\beta)\\
  \cos f(r_-)
\end{pmatrix},
\eeq
with
\begin{align}
  r_+&=\sqrt{(x+x_0)^2+y^2},\qquad&
  r_-&=\sqrt{(x-x_0)^2+y^2},\non
  \theta_+&=\arctan\left(\frac{y}{x+x_0}\right),\qquad&
  \theta_-&=\arctan\left(\frac{y}{x-x_0}\right),
\end{align}
and $\alpha,\, \beta$ are two independent phases.
Note that in order to have two separated compactons we require
$|x_0|> R$, where $R$ is the compacton radius \eqref{eq:radius}.

Following the same procedure as adopted above, we calculate the
pull-back of the metric $h$ induced by $\phi$: 
\begin{equation}\label{pul}
\phi^*h=\frac{\p\phi^a}{\p x^i}\frac{\p\phi^a}{\p x^j}\d x^i\otimes\d
x^j=\frac{\p\phi_1^a}{\p x^i}\frac{\p\phi_1^a}{\p x^j}\d x^i\otimes\d
x^j+\frac{\p\phi_2^a}{\p x^i}\frac{\p\phi_2^a}{\p x^j}\d x^i\otimes\d x^j,
\end{equation}
where $\d x^i=(\d x,\d y)$ are the Cartesian coordinates in
$\mathcal{M}=\mathbb{R}^2$.
In the last equality of eq.~\eqref{pul}, we used the propriety of
the compacton that the fields are constant outside of its radius so 
that
\begin{align}
&\p_i\phi^a(x,y)=\p_i\phi_1^a(x+x_0,y)+\p_i\phi_2^a(x-x_0,y)\non
&\p_i\phi_1^a(x+x_0,y)\p_j\phi_2^a(x-x_0,y)=0,\quad \forall\quad i,j.
  \label{pro}
\end{align}
At this point it is useful to manipulate the expression \eqref{pul}
by a series of change of variables in order to write $\phi^*h$ as a
sum of symmetric $(0,2)$ tensors of the form \eqref{eq:pul1}.
Hence,
\begin{align}
  \phi^*h&=\frac{\p\phi_1^a}{\p x^i}\frac{\p\phi_1^a}{\p x^j}\d x^i\otimes \d x^j+\frac{\p\phi_2^a}{\p x^i}\frac{\p\phi_2^a}{\p x^j}\d x^i\otimes \d x^j\non
  &=\frac{\p\phi_1^a}{\p x_+^i}\frac{\p\phi_1^a}{\p x_+^j}\d x_+^i\otimes \d x_+^j+\frac{\p\phi_2^a}{\p x_-^i}\frac{\p\phi_2^a}{\p x_-^j}\d x_-^i\otimes \d x_-^j\non
  &=\frac{\p\phi_1^a}{\p \tilde{x}_+^i}\frac{\p\phi_1^a}{\p \tilde{x}_+^j}\d\tilde{x}_+^i\otimes \d\tilde{x}_+^j+\frac{\p\phi_2^a}{\p \tilde{x}_-^i}\frac{\p\phi_2^a}{\p \tilde{x}_-^j}\d\tilde{x}_-^i\otimes \d\tilde{x}_-^j,
\end{align}
where $x_\pm^i=(x_\pm,y_\pm)$ are defined by
\beq
x_\pm = x\pm x_0, \qquad
y_\pm = y,
\eeq
while $\tilde{x}_\pm^i=(r_\pm,\theta_\pm)$ are the polar coordinates
\beq
r_\pm = \sqrt{x_\pm^2+y_\pm^2},\qquad
\theta_\pm = \arctan\left(\frac{y_\pm}{x_\pm}\right).
\eeq
Defining for simplicity
\begin{equation}\label{pul2}
  \phi^*h\equiv a_{ij}^+\d\tilde{x}_+^i\otimes \d\tilde{x}_+^j
  +a^-_{ij}\d\tilde{x}_-^i\otimes \d\tilde{x}_-^j,
\end{equation}
the final result for the two tensors is 
\begin{equation}
  a_{ij}^\pm \d\tilde{x}_\pm^i\otimes \d\tilde{x}_\pm^j=
  \big(f'(r_\pm)\big)^2\d r_\pm^2
  +N^2\sin^2f(r_\pm)\d\theta_\pm^2,
\end{equation}
which takes the same form as that of eq.~\eqref{eq:pul1}, but with
translated coordinates.

Applying the divergence to eq.~\eqref{pul2} and using the linearity 
\begin{align}
  \div\,\phi^*h =
  \div\,(a^++a^-)=\div\,a^++\div\,a^-&=\tilde{D}^ia_{ij}^+\d\tilde{x}^j_++\tilde{D}^ia_{ij}^-\d\tilde{x}^j_-\non
  &\equiv\omega^+_j\d\tilde{x}_+^j+\omega^-_j\d\tilde{x}_-^j,\label{1f}
\end{align}
with 
\beq
\omega^\pm_j\d\tilde{x}_\pm^j = \left(2f'(r_\pm)f''(r_\pm)
+\frac{\big(f'(r_\pm)\big)^2}{r_\pm}
-N^2\frac{\sin^2f(r_\pm)}{r_\pm^3}\right)\d r_\pm,
\eeq
we obtain a one-form whose exactness we must verify.
Using Poincar\'e's lemma again as well as the linearity of the
exterior derivative, we get 
\begin{align}
  \d(\div\,\phi^*h)
  &= \d\big(\omega^+_j\d\tilde{x}_+^j+\omega^-_j\d\tilde{x}_-^j\big)
  =\sum_{\pm}\d\big(\omega^+_jd\tilde{x}_+^j\big)\non
  &=\sum_{\pm}(\tilde{\p}_i\omega^\pm_j-\tilde{\p}_j\omega^\pm_i)\d\tilde{x}_\pm^i\wedge
  \d\tilde{x}_\pm^j \non
  &=2\sum_{\pm}\left(\p_{r_\pm}\omega_\theta^\pm-\p_{\theta_\pm}\omega_r^\pm\right)\d r_\pm\wedge \d\theta_\pm =0,\label{exp}
\end{align}
which finally proves that the map \eqref{N11} is restricted harmonic.

The results of this section show that both an axially symmetric
compacton with random orientation and two nonoverlapping
axially-symmetric compactons, with random orientations are restricted
harmonic maps.
In this paper, we therefore use the first solution as the
zeroth-order background for a single near-BPS baby Skyrmion, as already
anticipated in sec.~\ref{tre.uno}, and the second solution for two
near-BPS baby Skyrmions.

With this last result, we are now able to calculate the energy of two
near-BPS baby Skyrmions with topological charge $Q=N+N$ at the leading
order in $\epsilon$, as already performed for the single case $Q=N$ in
eq.~\eqref{0+1order}.
Using the restricted harmonic map \eqref{N11} and the properties
\eqref{pro}, we get
\begin{align}
E^{\textrm{side-by-side}}(\epsilon,N\oplus N) &= M^{\rm BPS}(N\oplus N) + \epsilon M^{\rm LO}(N\oplus N) \non
&= 2M^{\rm BPS}(N) + 2\epsilon M^{\rm LO}(N),
\end{align}
where $N\oplus N$ means an axially symmetric charge-$N$ compacton
side-by-side of another axially symmetric charge-$N$ compacton. 
At this order, the energy of the $Q=N+N$ configuration is therefore
simply the sum of the energies of the two components with no
information about the interaction between two near-BPS baby
Skyrmions. An investigation of the next-to-leading order (NLO) is
required to calculate the binding energy of such a configuration.

\subsection{NLO and \texorpdfstring{N$^2$LO}{N2LO} corrections}

We will now consider the next-to-leading order (NLO) and
next-to-next-to-leading order (N$^2$LO) corrections to the
energy, which corresponds to taking into account the corrections of
order $\mathcal{O}(\epsilon^2)$ and $\mathcal{O}(\epsilon^3)$,
respectively.  
To this end, we will perform a linear perturbation of the model
\beq
\bphi = \bvarphi + \bdphi,
\eeq
where $\bdphi=(\dphi^1,\dphi^2,\dphi^3)$ is a small perturbation about
the background solution $\bvarphi=(\varphi^1,\varphi^2,\varphi^3)$.

A brief spoiler of the results in this section, is that we
need to go to the third order in $\epsilon$ to include the terms
that will give rise to a nontrivial behavior of the fluctuations
outside of the compacton domain(s).
It will turn out to give, however, a nonanalytic contribution in
$\epsilon$.
Therefore, we will write down all the terms up to order
$\mathcal{O}(\epsilon^3)$, but we will solve the equation of motion
for the fluctuation to linearized order, which does indeed capture
this nonanalytic contribution to the energy.
A key step will be to find the correct form of the fluctuations that 
does not take the fields out of the $\Og(3)$ group.

For the NLO and N$^2$LO corrections, we need to calculate the
variation up to third order (in the fields) of the Lagrangian
\eqref{eq:L}: 
\begin{align}
\Lag^{\rm perturb}[\bvarphi,\bdphi] &=
\left.\frac{\p\Lag}{\p\phi^a}\right|\dphi^a
+\left.\frac12\frac{\p^2\Lag}{\p\phi^a\p\phi^b}\right|\dphi^a\dphi^b
+\left.\frac16\frac{\p^3\Lag}{\p\phi^a\p\phi^b\p\phi^c}\right|\dphi^a\dphi^b\dphi^c
\non
&\phantom{=\ }
+\left.\frac{\p\Lag}{\p\p_\mu\phi^a}\right|\p_\mu\dphi^a 
+\left.\frac12\frac{\p^2\Lag}{\p\p_\mu\phi^a\p\p_\nu\phi^b}\right|\p_\mu\dphi^a\p_\nu\dphi^b
\non
&\phantom{=\ }
+\left.\frac16\frac{\p^3\Lag}{\p\p_\mu\phi^a\p\p_\nu\phi^b\p\p_\rho\phi^c}\right|\p_\mu\dphi^a\p_\nu\dphi^b\p_\rho\dphi^c\non
&= \frac{\lambda_0 + \dlambda}{2}\bdphi^2
+\dlambda\bvarphi\cdot\bdphi
+\left.\epsilon\frac{\p\Lag_2}{\p\p_\mu\phi^a}\right|\p_\mu\dphi^a \non
&\phantom{=\ }
+\left.\frac12\frac{\p^2\Lag}{\p\p_\mu\phi^a\p\p_\nu\phi^b}\right|\p_\mu\dphi^a\p_\nu\dphi^b
+\left.\frac16\frac{\p^3\Lag}{\p\p_\mu\phi^a\p\p_\nu\phi^b\p\p_\rho\phi^c}\right|\p_\mu\dphi^a\p_\nu\dphi^b\p_\rho\dphi^c\non
&= \frac{\lambda_0 + \dlambda}{2}\bdphi^2
+\dlambda\bvarphi\cdot\bdphi
-\epsilon J_a^\mu\p_\mu\dphi^a 
-\frac12 V_{a b}^{\mu\nu}\p_\mu\dphi^a\p_\nu\dphi^b\non
&\phantom{=\ }
-\frac16\Gamma_{a b c}^{\mu\nu\rho}\p_\mu\dphi^a\p_\nu\dphi^b\p_\rho\dphi^c,
\label{eq:Lperturb}
\end{align}
where the $|$ denotes that the expression (to the left of the bar) is
evaluated on the background by setting $\bphi=\bvarphi$ and we have
defined 
\begin{equation}
J_a^\mu \equiv \left.-\frac{\p\Lag_2}{\p\p_\mu\phi^a}\right|
= \p^\mu\varphi^a, \qquad
V_{a b}^{\mu\nu} \equiv
\left.-\frac{\p^2\Lag}{\p\p_\mu\phi^a\p\p_\nu\phi^b}\right|, \qquad
\Gamma_{a b c}^{\mu\nu\rho} \equiv
\left.-\frac{\p^3\Lag}{\p\p_\mu\phi^a\p\p_\nu\phi^b\p\p_\rho\phi^c}\right|.
\end{equation}
We have furthermore replaced the Lagrange multiplier constant
$\lambda$ with an expansion $\lambda\to\lambda_0+\dlambda$,
where it is understood that $\lambda_0$ is the Lagrange multiplier
that solves the sigma-model constraint for the background BPS
solution.
$\dlambda$ is then a Lagrange multiplier that ensures that
the perturbation does not bring the total field
$\bphi=\bvarphi+\bdphi$ out of the O(3) group, i.e.~it should still
preserve the unit length constraint
$\bphi\cdot\bphi=1+\mathcal{O}(\epsilon^4)$.
More precisely, if we vary the above Lagrangian density with respect
to $\dlambda$ we obtain
\beq
\left(\frac12\bdphi^2 + \bvarphi\cdot\bdphi\right) = 0,
\label{eq:lambda1_constraint}
\eeq
which is solved by \cite{Piette:1994ug}
\beq
\bdphi = \bDelta\times\bvarphi
+ \frac12\bDelta\times(\bDelta\times\bvarphi),
\label{eq:Delta_form}
\eeq
which satisfies the constraint \eqref{eq:lambda1_constraint} up to
$\mathcal{O}(\bDelta^4)$, which we shall verify is of order
$\mathcal{O}(\epsilon^4)$ and hence will give a contribution to the
energy of order $\mathcal{O}(\epsilon^5)$.
Notice that the above form automatically restricts $\bDelta$ to be
orthogonal to the background solution $\bvarphi$ to leading order.

The Hessian of the Lagrangian density with respect to the derivatives
of the fields, $V_{a b}^{\mu\nu}$, is given by  
\begin{align}
V_{a b}^{\mu\nu} &\equiv V_{0 a b}^{\mu\nu} + \epsilon V_{1 a b}^{\mu\nu}, \non
V_{0 a b}^{\mu\nu} &\equiv (\p_\rho\bvarphi\cdot\p^\rho\bvarphi)\eta^{\mu\nu}\delta^{a b}
+2\p^\mu\varphi^a\p^\nu\varphi^b
-\p^\mu\bvarphi\cdot\p^\nu\bvarphi\delta^{a b}
-\p_\rho\varphi^a\p^\rho\varphi^b\eta^{\mu\nu} 
-\p^\mu\varphi^b\p^\nu\varphi^a, \non
V_{1 a b}^{\mu\nu} &\equiv \eta^{\mu\nu}\delta^{a b},
\end{align}
the derivative of the Hessian is
\begin{align}
\Gamma_{a b c}^{\mu\nu\rho} &=
\eta^{\mu\nu}\big(2\p^\rho\varphi^c\delta^{a b}
-\p^\rho\varphi^a\delta^{b c}
-\p^\rho\varphi^b\delta^{c a}\big)
+\eta^{\nu\rho}\big(2\p^\mu\varphi^a\delta^{b c}
-\p^\mu\varphi^b\delta^{c a}
-\p^\mu\varphi^c\delta^{a b}\big) \non
&\phantom{=\ }
+\eta^{\mu\rho}\big(2\p^\nu\varphi^b\delta^{c a}
-\p^\nu\varphi^c\delta^{a b}
-\p^\nu\varphi^a\delta^{b c}\big),
\end{align}
and the Lagrange multiplier of the background, $\lambda_0$, reads
\begin{align}
\lambda_0 &=
-(\bvarphi\cdot\p^2\bvarphi)
  (\p_\mu\bvarphi\cdot\p^\mu\bvarphi)
+(\bvarphi\cdot\p_\mu\p_\nu\bvarphi)
  (\p^\mu\bvarphi\cdot\p^\nu\bvarphi)
-m^2\varphi^3.
\end{align}

Substituting the form of the variation \eqref{eq:Delta_form} into the
perturbation Lagrangian density \eqref{eq:Lperturb} yields
\begin{align}
\Lag^{\rm perturb}[\bvarphi,\bDelta] &=
\Lag_2^{\rm perturb}[\bvarphi,\bDelta]
+\Lag_3^{\rm perturb}[\bvarphi,\bDelta]
+\mathcal{O}(\epsilon^4),\\
\Lag_2^{\rm perturb}[\bvarphi,\bDelta] &=
\frac{\lambda_0}{2}
\left(\bDelta^2 - (\bDelta\cdot\bvarphi)^2\right)
-\epsilon\bJ^\mu\cdot\p_\mu\bDelta\times\bvarphi
-\epsilon\bJ^\mu\cdot\bDelta\times\p_\mu\bvarphi\non
&\phantom{=\ }
-\frac12V_{0ab}^{\mu\nu}\varepsilon^{acd}\p_\mu(\Delta^c\varphi^d)\varepsilon^{bef}\p_\nu(\Delta^e\varphi^f),\\
\Lag_3^{\rm perturb}[\bvarphi,\bDelta] &=
-\frac\epsilon2(\bvarphi\cdot\bDelta)(\bJ_\mu\cdot\p^\mu\bDelta)
-\frac\epsilon2(\bJ_\mu\cdot\bDelta)(\p^\mu\bvarphi\cdot\bDelta)
-\frac\epsilon2(\bJ_\mu\cdot\bDelta)(\bvarphi\cdot\p^\mu\bDelta)\non
&\phantom{=\ }
+\frac\epsilon2(\bJ_\mu\cdot\p^\mu\bvarphi)\bDelta^2
+\epsilon(\bJ_\mu\cdot\bvarphi)(\bDelta\cdot\p^\mu\bDelta)
\non
&\phantom{=\ }
-\frac12V_{0ab}^{\mu\nu}\varepsilon^{acd}\p_\mu(\Delta^c\varphi^d)\p_\nu\left(\Delta^b(\bDelta\cdot\bvarphi)- \bDelta^2\varphi^b\right) \non
&\phantom{=\ }
-\frac16\Gamma_{a b c}^{\mu\nu\rho}
  \varepsilon^{a d e}\p_\mu(\Delta^d\varphi^e)
  \varepsilon^{b f g}\p_\nu(\Delta^f\varphi^g)
  \varepsilon^{c h i}\p_\rho(\Delta^h\varphi^i)\non
&\phantom{=\ }
-\frac\epsilon2V_{1ab}^{\mu\nu}\varepsilon^{acd}\p_\mu(\Delta^c\varphi^d)\varepsilon^{bef}\p_\nu(\Delta^e\varphi^f),
\label{eq:L3perturb}
\end{align}
which is the complete Lagrangian up to third order in $\epsilon$,
(i.e.~discarding terms of $\mathcal{O}(\epsilon^4)$).
Notice that the term with $\dlambda$ is first needed at the fifth order in
$\epsilon$ and can thus safely be ignored -- its job was to produce
the form \eqref{eq:Delta_form}.

The reason for keeping terms up to third order in $\epsilon$, as we
shall see, is to retain the last term in eq.~\eqref{eq:L3perturb},
which will enable the fluctuation to propagate outside of the compacton
and hence give rise to binding energy.
In order to calculate the energy consistently, we have thus kept all
other terms up to third order in $\epsilon$.
However, for calculating $\bDelta$ we will linearize its equation of
motion
\begin{align}
&-\p_\mu\left[V_{c b}^{\mu\nu}\varepsilon^{a c d}\varepsilon^{b e f}\p_\nu(\Delta^e\varphi^f)\right]\varphi^d
+\lambda_0\Delta^a
-\lambda_0(\bvarphi\cdot\bDelta)\varphi^a
+\frac\epsilon2(\bvarphi\cdot\bDelta)\p_\mu J_a^\mu
+\frac\epsilon2(\p_\mu\bJ^\mu\cdot\bDelta)\varphi^a \non
&\qquad
-\epsilon(\p_\mu\bJ^\mu\cdot\bvarphi)\Delta^a
= -\epsilon\varepsilon^{a b c}\varphi^b\p_\mu J_c^\mu,
\label{eq:Delta_perturb}
\end{align}
where we have used that $V_{c b}^{\mu\nu}$ is symmetric under the
simultaneous exchange of $\mu\leftrightarrow\nu$ and
$c\leftrightarrow b$.

Let us now consider what happens outside of the compacton.
Since $\varphi^a\to\delta^{a3}$ is at its vacuum and all derivatives
of the background field vanish identically, most terms including the
source term (right-hand side of eq.~\eqref{eq:Delta_perturb}) switch
off.
In fact, the only remaining terms outside of the compacton background
are $\epsilon V_{1ab}^{\mu\nu}\subset V_{ab}^{\mu\nu}$ and the
$\lambda_0$ terms in eq.~\eqref{eq:Delta_perturb}.
Thus the linearized equation of motion of the perturbation, $\bDelta$,
reduces outside of the compacton background to
\beq
\epsilon\left(\p^2\Delta^a - \p^2\Delta^3\delta^{3a}\right)
-m^2\left(\Delta^a - \Delta^3\delta^{3a}\right) = 0.
\label{eq:outsidecompacton}
\eeq
It is interesting to see that the kinetic term here (outside the
compacton) is given by $\epsilon V_{1ab}^{\mu\nu}$ which is third
order in $\epsilon$ (i.e.~$\mathcal{O}(\epsilon^3)$).
Thus, had we only kept terms up to second order in $\epsilon$ in the
Lagrangian \eqref{eq:Lperturb}, the equation of motion would have been
a constraint setting $\bDelta=0$.
If $\bDelta=0$ outside the compacton, then there is no information
propagating between two compactons and hence there is no binding
energy. 

A natural question then arises: At which order in $\epsilon$ is the
binding energy of baby Skyrmions captured?
If we do not restrict to the linearized equation of motion for the
fluctuation, it is clear that the above conclusion about the equation
outside of the compacton does not change (although we do not prove
this).
It is instructive to look at the solution for the fluctuation outside
of the compacton as the solution to eq.~\eqref{eq:outsidecompacton},
which reads
\beq
\bDelta =
\begin{pmatrix}
  c_1\\
  c_2\\
  0
\end{pmatrix}
\exp\left(-\frac{m r}{\sqrt{\epsilon}}\right),
\label{eq:tail}
\eeq
where $c_{1,2}$ are constants. 
This solution illustrates perfectly the problem of describing the
binding energy at a specific order.
In fact, although the tail does not vanish for a nonvanishing
$\epsilon$, any order in $\epsilon$ vanishes due to the exponential.
In some sense, this solution is all-order in $\epsilon$ or
``nonperturbative''.
Importantly, this shows that the expansion around the BPS compacton
background contains contributions nonanalytic in $\epsilon$.

A word of caution is that although we include this ``nonperturbative''
effect of the all-order tail for the linear perturbation, we solve a
linearized equation for the perturbation which only captures every
effect to second order in $\epsilon$ and some effects -- like the
crucial tail \eqref{eq:tail} -- at third order in $\epsilon$.
Nevertheless, we calculate the full energy to third order in
$\epsilon$ using this solution to the linearized equation of motion.

Writing out eq.~\eqref{eq:Delta_perturb} explicitly in the static
case, we get 
\begin{equation}
X \Delta_{ii}^a
+X^{ab} \Delta_{ii}^b
+X_{ij}^{ab} \Delta_{ij}^b
+X_i^{ab} \Delta_{i}^b
+\Lambda\Delta^a
+\Lambda^{ab}\Delta^b
= -\epsilon\varepsilon^{a b c}\varphi^b\varphi_{ii}^c,
\label{eq:Xlineq}
\end{equation}
where we have defined
\begin{align}
  X &\equiv \epsilon,\\
  X^{a b} &\equiv
  -\epsilon\varphi^a\varphi^b
  +\varphi_j^a\varphi_j^b,\\
  X_{ij}^{ab} &\equiv -\varphi_i^a\varphi_j^b,\\
  X_i^{a b} &\equiv
  -2\epsilon\varphi_i^a\varphi^b
  -\varphi_{ij}^a\varphi_j^b
  +\varphi_{jj}^a\varphi_i^b
  -2\varphi_i^a\varphi_{jj}^b
  +2\varphi_j^a\varphi_{ij}^b
  +(\bvarphi_j\cdot\bvarphi_j)(\varphi^a\varphi_i^b-\varphi_i^a\varphi^b)\non
  &\phantom{=\ }
  -(\bvarphi_i\cdot\bvarphi_j)(\varphi^a\varphi_j^b-\varphi_j^a\varphi^b),\\
  \Lambda &\equiv -m^2\varphi^3,\\
  \Lambda^{a b} &\equiv
  -\frac\epsilon2\varphi_{ii}^a\varphi^b
  +\frac\epsilon2\varphi^a\varphi_{ii}^b
  +m^2\varphi^3\varphi^a\varphi^b
  -(\bvarphi_{ij}\cdot\bvarphi_j)\varphi_i^a\varphi^b
  +(\bvarphi_{ii}\cdot\bvarphi_j)\varphi_j^a\varphi^b
  +(\bvarphi_i\cdot\bvarphi_j)\varphi_{ij}^a\varphi^b \non
  &\phantom{=\ }
  -(\bvarphi_j\cdot\bvarphi_j)\varphi_{ii}^a\varphi^b
  +(\bvarphi_i\cdot\bvarphi_j)^2\varphi^a\varphi^b
  -(\bvarphi_i\cdot\bvarphi_i)^2\varphi^a\varphi^b.
  \label{eq:Xtensors}  
\end{align}

The energy density of the perturbation can be written as
\begin{align}
\mathcal{E}^{\rm perturb}[\bvarphi,\bDelta] &=
\mathcal{E}_2^{\rm perturb}[\bvarphi,\bDelta]
+\mathcal{E}_3^{\rm perturb}[\bvarphi,\bDelta],\label{eq:Eperturb}
\end{align}
with the second order (NLO) in $\epsilon$:
\begin{align}
\mathcal{E}_2^{\rm perturb}[\bvarphi,\bDelta] &=
\epsilon\bvarphi_i\cdot\bDelta_i\times\bvarphi
+\frac{m^2}{2}\varphi^3\big(\bDelta^2 - (\bDelta\cdot\bvarphi)^2\big)\non
&\phantom{=\ }
+\frac12(\bvarphi_j\cdot\bvarphi_j)\left[
  (\bvarphi_i\cdot\bvarphi_i)(\bvarphi\cdot\bDelta)^2
  -(\bvarphi_i\cdot\bDelta)^2
  -2(\bvarphi\cdot\bDelta)(\bvarphi_i\cdot\bDelta_i)
\right]\non
&\phantom{=\ }  
-\frac12(\bvarphi_i\cdot\bvarphi_j)\left[
  (\bvarphi_i\cdot\bvarphi_j)(\bvarphi\cdot\bDelta)^2
  -(\bvarphi_i\cdot\bDelta)(\bvarphi_j\cdot\bDelta)
  -2(\bvarphi\cdot\bDelta)(\bvarphi_i\cdot\bDelta_j)
\right]\non
&\phantom{=\ }  
+\frac12(\bvarphi_i\cdot\bDelta_i)^2
+\frac12(\bvarphi_i\cdot\bDelta_j)^2
-(\bvarphi_i\cdot\bDelta_j)(\bvarphi_j\cdot\bDelta_i),
\end{align}
the third order (NNLO) in $\epsilon$
\begin{align}
\mathcal{E}_3^{\rm perturb}[\bvarphi,\bDelta] &=
\mathcal{E}_{3,{\rm quad}}^{\rm perturb}[\bvarphi,\bDelta]
+\mathcal{E}_{3,{\rm cubic}}^{\rm perturb}[\bvarphi,\bDelta],\\
\mathcal{E}_{3,{\rm quad}}^{\rm perturb}[\bvarphi,\bDelta] &=
\frac\epsilon2\bDelta_i\cdot\bDelta_i
-\frac\epsilon2(\bvarphi\cdot\bDelta_i)^2
-\frac\epsilon2(\bvarphi\cdot\bDelta)(\bvarphi_i\cdot\bDelta_i)
+\frac\epsilon2(\bvarphi_i\cdot\bDelta)(\bvarphi\cdot\bDelta_i),
\end{align}
\begin{align}
\mathcal{E}_{3,{\rm cubic}}^{\rm perturb}[\bvarphi,\bDelta] &=
-\frac13(\bvarphi_i\cdot\bDelta_i)\bvarphi_j\cdot\bDelta\times\bDelta_j
-\frac13(\bvarphi_i\cdot\bDelta_j)\bvarphi_i\cdot\bDelta\times\bDelta_j\non
&\phantom{=\ }
+\frac23(\bvarphi_i\cdot\bDelta_j)\bvarphi_j\cdot\bDelta\times\bDelta_i
-(\bvarphi_i\cdot\bDelta_j)\bvarphi\cdot\bDelta_i\times\bDelta_j\non
&\phantom{=\ }
+\frac13(\bvarphi_i\cdot\bvarphi_i)(\bvarphi\cdot\bDelta)\bvarphi_j\cdot\bDelta\times\bDelta_j
-\frac13(\bvarphi_i\cdot\bvarphi_j)(\bvarphi\cdot\bDelta)\bvarphi_i\cdot\bDelta\times\bDelta_j\non
&\phantom{=\ }
-\frac13(\bvarphi_i\cdot\bDelta)(\bvarphi_i\cdot\bDelta)\bvarphi\cdot\bvarphi_j\times\bDelta_j
+\frac13(\bvarphi_i\cdot\bDelta)(\bvarphi_j\cdot\bDelta)\bvarphi\cdot\bvarphi_i\times\bDelta_j\non
&\phantom{=\ }
+\frac56(\bvarphi_i\cdot\bvarphi_i)(\bDelta\cdot\bDelta)\bvarphi\cdot\bvarphi_j\times\bDelta_j
-\frac56(\bvarphi_i\cdot\bvarphi_j)(\bDelta\cdot\bDelta)\bvarphi\cdot\bvarphi_i\times\bDelta_j\non
&\phantom{=\ }
-\frac16(\bvarphi_i\cdot\bvarphi_i)(\bDelta\cdot\bDelta_j)\bvarphi\cdot\bDelta\times\bvarphi_j
+\frac16(\bvarphi_i\cdot\bvarphi_j)(\bDelta\cdot\bDelta_j)\bvarphi\cdot\bDelta\times\bvarphi_i\non
&\phantom{=\ }
+\frac12(\bvarphi_i\cdot\bDelta_j)(\bvarphi\cdot\bDelta)\bDelta\cdot\bvarphi_i\times\bvarphi_j
+\frac16(\bvarphi_i\cdot\bDelta_j)(\bvarphi\cdot\bDelta)\bvarphi\cdot\bvarphi_i\times\bDelta_j\non
&\phantom{=\ }
-\frac56(\bvarphi_i\cdot\bDelta_i)(\bvarphi\cdot\bDelta)\bvarphi\cdot\bvarphi_j\times\bDelta_j
+\frac23(\bvarphi_i\cdot\bDelta_j)(\bvarphi\cdot\bDelta)\bvarphi\cdot\bvarphi_j\times\bDelta_i\non
&\phantom{=\ }
+\frac16(\bvarphi_i\cdot\bDelta_j)(\bvarphi\cdot\bDelta_j)\bvarphi\cdot\bDelta\times\bvarphi_i
+\frac16(\bvarphi_i\cdot\bDelta_i)(\bvarphi\cdot\bDelta_j)\bvarphi\cdot\bDelta\times\bvarphi_j\non
&\phantom{=\ }
-\frac13(\bvarphi_i\cdot\bDelta_j)(\bvarphi\cdot\bDelta_i)\bvarphi\cdot\bDelta\times\bvarphi_j
-\frac12(\bvarphi_i\cdot\bvarphi_i)(\bvarphi_j\cdot\bDelta)\bvarphi\cdot\bDelta\times\bDelta_j\non
&\phantom{=\ }
+\frac12(\bvarphi_i\cdot\bvarphi_j)(\bvarphi_i\cdot\bDelta)\bvarphi\cdot\bDelta\times\bDelta_j
+\frac16(\bvarphi_i\cdot\bDelta)(\bvarphi_i\cdot\bDelta_j)\bvarphi\cdot\bDelta\times\bvarphi_j\non
&\phantom{=\ }
+\frac16(\bvarphi_i\cdot\bDelta)(\bvarphi_j\cdot\bDelta_j)\bvarphi\cdot\bDelta\times\bvarphi_i
-\frac13(\bvarphi_i\cdot\bDelta)(\bvarphi_j\cdot\bDelta_i)\bvarphi\cdot\bDelta\times\bvarphi_j,
\end{align}
where the former expression contains only quadratic terms in $\bDelta$
which contribute to the linear equation of motion and the latter
gathers the cubic terms.

Notice that the entire perturbation energy vanishes manifestly for
$\bDelta\propto\bvarphi$. Due to the nonlinearity in the form of $\bdphi$
in eq.~\eqref{eq:Delta_form}, a mixing can occur if both a transverse
$\bDelta_{\bot}$ and a longitudinal perturbation $\dc(x)\bvarphi$ is
turned on at the same time. Inserting
$\bDelta=\dc\bvarphi+\bDelta_{\bot}$ into quadratic part of the
perturbation Lagrangian yields 
\beq
\Lag_{2}^{\rm perturb} + \Lag_{3,{\rm quad}}^{\rm perturb}
\supset \frac\epsilon2\left[
  \dc\,\p_\mu\bvarphi\cdot\p^\mu\bDelta_{\bot}
  + \p_\mu\dc\,\p^\mu\bvarphi\cdot\bDelta_{\bot}\right].
\eeq
The solution for $\dc$ -- which is sourceless -- with vanishing
boundary condition at $r=0$ and $r\to\infty$ (as measured from the
compacton origin) yields $\dc=0$.
Because of the absence of a quadratic term for $\dc$, variation
neither gives an auxiliary equation (algebraic) nor a dynamic equation
for $\dc$.
We checked that the quadratic terms in $\dc$ do not appear in the
cubic part of the Lagrangian either and hence can first appear at
order $\mathcal{O}(\epsilon^4)$, which we shall not consider in this
paper\footnote{We are working with the perturbation of the form
  \eqref{eq:Delta_form} for which
  $\bphi\cdot\bphi=1+\mathcal{O}(\epsilon^4)$, so this relation should
  be improved to higher order before making conclusions about the
  terms present at order $\mathcal{O}(\epsilon^4)$. Then we expect in
  such case the quadratic terms of $\dc$ would not even appear at
  order $\mathcal{O}(\epsilon^4)$, but that is far beyond the scope of
  this paper. }.
We shall thus only consider $\bDelta=\bDelta_{\bot}$ which is
transverse to the background solution $\bvarphi$, i.e.~for which it
holds $\bvarphi\cdot\bDelta_{\bot}=0$.

At this point, it will prove useful to specialize to the case of the
background BPS solution for $\bvarphi$, the transverse perturbations
for $\bDelta=\bDelta_{\bot}$ and switch to polar coordinates in
$\mathbb{R}^2$, for which the perturbation energy reads
\beq
\mathcal{E}^{\rm perturb}[f,\df,\dtheta] =
\mathcal{E}_2^{\rm perturb}[f,\df,\dtheta]
+\mathcal{E}_{3,{\rm quad}}^{\rm perturb}[f,\df,\dtheta]
+\mathcal{E}_{3,{\rm cubic}}^{\rm perturb}[f,\df,\dtheta],
\label{eq:Eperturb_df_dtheta}
\eeq
with
\begin{align}
&\mathcal{E}_2^{\rm perturb}[f,\df,\dtheta] =
\frac{\epsilon N^2}{2r^2}\sin(2f)\df
+\epsilon f_r\df_r 
+\frac{N^2}{2r^2}\sin^2(f)\df_r^2
+\frac{f_r^2}{2r^2}\dtheta_\theta^2\non
&\phantom{=\ }
+\frac12\left(m^2\cos f + \frac{N^2}{r^2}\cos(2f)f_r^2\right)\df^2
+\frac12\left(m^2\cos f - \frac{N^2}{r^2}\sin^2(f)f_r^2\right)\dtheta^2\non
&\phantom{=\ }
+\frac{N}{r^2}\sin(f)f_r
  \left[2\df_r\dtheta_\theta - \df_\theta\dtheta_r\right]
+\frac{N^2}{2r^2}\sin(2f)f_r
  \left[2\df_r\df + \dtheta_r\dtheta\right]
+\frac{N}{r^2}\cos(f)f_r^2\df\dtheta_\theta,
\end{align}
for the NLO terms,
\begin{align}
\mathcal{E}_{3,{\rm quad}}^{\rm perturb}[f,\df,\dtheta] &=
\frac\epsilon2\left(\df_r^2+\frac{\df_\theta^2}{r^2}\right)
+\frac{\epsilon N^2}{2r^2}\cos(2f)\df^2
+\frac\epsilon2\left(\dtheta_r^2 + \frac{\dtheta_\theta^2}{r^2}\right)\non
&\phantom{=\ }
-\frac\epsilon2\left(f_r^2 - \frac{N^2}{r^2}\cos^2f\right)\dtheta^2
-\frac{\epsilon N}{r^2}\cos(f)\left(\df_\theta\dtheta - \df\dtheta_\theta\right),
\end{align}
for the NNLO terms quadratic in $\bDelta$ and
\begin{align}
&\mathcal{E}_{3,{\rm cubic}}^{\rm perturb}[f,\df,\dtheta] =
\frac{1}{2r^2}\left(
  N\sin(f)\df_r
  +f_r\dtheta_\theta
  +N\cos(f)f_r\df
  \right)\times\non&
\left[
  2N\cos(f)\df_r\df
  +2\df_r\dtheta_\theta
  +2N\cos(f)\dtheta_r\dtheta
  -2\df_\theta\dtheta_r
  -N\sin(f)f_r\df^2
  -N\sin(f)f_r\dtheta^2
\right],
\end{align}
for the NNLO terms cubic in $\bDelta$, and we have defined the
transverse $\bDelta$ perturbations 
\beq
\bDelta_\bot \equiv
\begin{pmatrix}
  -\sin N\theta\\
  \cos N\theta\\
  0
\end{pmatrix}\df -
\begin{pmatrix}
  \cos f\cos N\theta\\
  \cos f\sin N\theta\\
  -\sin f
\end{pmatrix}\dtheta,
\label{eq:Delta_transverse}
\eeq
and the background BPS compacton solution is described by the radial
profile 
\beq
f = f^{\rm BPS} = \arccos\left(-1 + \frac{4r^2}{R^2} - \frac{2r^4}{R^4}\right),
\eeq
with $R$ being the compacton radius of eq.~\eqref{eq:radius}.

The corresponding equations of motion can be written as
\begin{align}
  X_{rr}
  \begin{pmatrix}
    \df_{rr}\\
    \dtheta_{rr}
  \end{pmatrix}
  +\frac{1}{r}X_r
  \begin{pmatrix}
    \df_r\\
    \dtheta_r
  \end{pmatrix}
  +\frac{1}{r^2}X_{\theta\theta}
  \begin{pmatrix}
    \df_{\theta\theta}\\
    \dtheta_{\theta\theta}
  \end{pmatrix}
  +\frac{1}{r}X_\theta
  \begin{pmatrix}
    \df_\theta\\
    \dtheta_\theta
  \end{pmatrix}
  +\frac{1}{r}X_{r\theta}
  \begin{pmatrix}
    \df_{r\theta}\\
    \dtheta_{r\theta}
  \end{pmatrix}\non
  +\Lambda
  \begin{pmatrix}
    \df\\
    \dtheta
  \end{pmatrix}
  =
  \begin{pmatrix}
    -\epsilon\left(f_{rr} + \frac{1}{r}f_r - \frac{N^2}{2r^2}\sin(2f)\right)\\
    0
  \end{pmatrix},
  \label{eq:Xpde_polar}
\end{align}
where we have defined the matrices
\begin{align}
  X_{rr} &=
  \begin{pmatrix}
    \epsilon + \frac{N^2}{r^2}\sin^2f & 0\\
    0 & \epsilon
  \end{pmatrix},\\
  X_r &=
  \begin{pmatrix}
    \epsilon - \frac{N^2}{r^2}\sin^2f + \frac{N^2}{r}\sin(2f)f_r & 0\\
    0 & \epsilon
  \end{pmatrix},\\
  X_{\theta\theta} &=
  \begin{pmatrix}
    \epsilon & 0\\
    0 & \epsilon + f_r^2
  \end{pmatrix},
\end{align}
\begin{align}
  X_\theta &=
  \begin{pmatrix}
    0 & X_\theta^{\dtheta}\\
    X_\theta^{\df} & 0
  \end{pmatrix},\\
  X_\theta^{\df} &= \frac{2\epsilon N}{r}\cos f
    -\frac{N}{r}\sin(f)f_{rr}
    +\frac{N}{r^2}\sin(f)f_r,\\
  X_\theta^{\dtheta} &= -\frac{2\epsilon N}{r}\cos f
    +\frac{2N}{r}\sin(f)f_{rr}
    +\frac{N}{r}\cos(f)f_r^2
    -\frac{2N}{r^2}\sin(f)f_r,\\
  X_{r\theta} &= \frac{N}{r}\sin(f)f_r
  \begin{pmatrix}
    0 & 1\\
    1 & 0
  \end{pmatrix},
\end{align}
\begin{align}
  \Lambda &=
  \begin{pmatrix}
    \Lambda^{\df} & 0\\
    0 & \Lambda^{\dtheta}
  \end{pmatrix},\\
  \Lambda^{\df} &=
  -\frac{\epsilon N^2}{r^2}\cos(2f)
  +\frac{N^2}{r^2}\sin(2f)f_{rr}
  +\frac{N^2}{r^2}\cos(2f)f_r^2
  -\frac{N^2}{r^3}\sin(2f)f_r
  -m^2\cos f,\\
  \Lambda^{\dtheta} &=
  \epsilon f_r^2
  -\frac{\epsilon N^2}{r^2}\cos^2(f)
  +\frac{N^2}{2r^2}\sin(2f)f_{rr}
  +\frac{N^2}{r^2}\cos^2(f)f_r^2
  -\frac{N^2}{2r^3}\sin(2f)f_r
  -m^2\cos f.
\end{align}

First notice that there is only a source term (right-hand side of
eq.~\eqref{eq:Xpde_polar}) for the upper equation, viz.~for $\df$. 
Notice then that the off-diagonal components reside only in the
matrices $X_\theta$, $X_{r\theta}$ and hence couple the second
equation to $\df_\theta$ and $\df_{r\theta}$.
The source for the $\dtheta$ perturbation is thus the coupling to the
nontrivial $\theta$ behavior in $\df$. 
It is now clear that, if we restrict to axially symmetric
perturbations such that $\df=\df(r)$, the second equation decouples
and is sourceless; hence it is satisfied by the trivial solution
$\dtheta=0$, everywhere.

The perturbation governing axially symmetric baby Skyrmions is thus
simply described by the equation
\begin{align}
\left(\epsilon + \frac{N^2}{r^2}\sin^2f\right) \df_{rr}
+\frac1r\left(\epsilon - \frac{N^2}{r^2}\sin^2f + \frac{N^2}{r}\sin(2f)f_r\right) \df_r
+\Lambda^{\df}\df \qquad\non
= -\epsilon\left(f_{rr} + \frac{1}{r}f_r - \frac{N^2}{2r^2}\sin2f\right),
\label{eq:axially_sym_EOM_df}
\end{align}
and the corresponding energy for the axially symmetric perturbation is
\begin{align}
\mathcal{E}_2^{\rm perturb}[f,\df] &=
\frac{\epsilon N^2}{2r^2}\sin(2f)\df
+\epsilon f_r\df_r  
+\frac{N^2}{2r^2}\sin^2(f)\df_r^2\non
&\phantom{=\ }  
+\frac12\left(m^2\cos f + \frac{N^2}{r^2}\cos(2f)f_r^2\right)\df^2
+\frac{N^2}{r^2}\sin(2f)f_r\df_r\df,
\label{eq:E2axial}
\end{align}
for the NLO terms,
\begin{align}
\mathcal{E}_{3,{\rm quad}}^{\rm perturb}[f,\df] &=
\frac\epsilon2\df_r^2
+\frac{\epsilon N^2}{2r^2}\cos(2f)\df^2,
\label{eq:E3axialquad}
\end{align}
for the NNLO terms quadratic in $\df$ and
\begin{align}
&\mathcal{E}_{3,{\rm cubic}}^{\rm perturb}[f,\df] =
\frac{N^2}{2r^2}\sin(2f)\df_r^2\df
+\frac{N^2}{r^2}\left(1 - \frac32\sin^2f\right)f_r\df_r\df^2
-\frac{N^2}{4r^2}\sin(2f)f_r^2\df^3,
\label{eq:E3axialcubic}
\end{align}
for the NNLO terms cubic in $\df$.

\subsection{Axially symmetric solutions}

In order to verify our perturbative scheme, we start with axially
symmetric baby Skyrmions.
It will prove useful to interpret the perturbation $\df$ as follows
\begin{align}
\bphi &= \bvarphi + \bdphi \non
&= \bvarphi + \bDelta_\bot\times\bvarphi + \frac12\bDelta_\bot\times(\bDelta_\bot\times\bvarphi)\non
&= 
\begin{pmatrix}
  \sin f\cos N\theta\\
  \sin f\sin N\theta\\
  \cos f
\end{pmatrix}
+
\begin{pmatrix}
  \cos f\cos N\theta\\
  \cos f\sin N\theta\\
  -\sin f
\end{pmatrix}
\df
+\frac12
\begin{pmatrix}
  -\sin f\cos N\theta\\
  -\sin f\sin N\theta\\
  -\cos f
\end{pmatrix}
\df^2\non
&\simeq
\begin{pmatrix}
  \sin(f+\df)\cos N\theta\\
  \sin(f+\df)\sin N\theta\\
  \cos(f+\df)
\end{pmatrix}
+ \mathcal{O}(\df^3),
\end{align}
where we have used eq.~\eqref{eq:Delta_transverse} and set
$\dtheta=0$.
It is now clear that $\df$ indeed is an additive correction to the BPS
background profile function $f$. 
  
The naive attempt is to solve eq.~\eqref{eq:axially_sym_EOM_df} with
boundary conditions
\beq
\df(0)=0, \qquad
\df(\infty)=0. \label{eq:df_normal_BC}
\eeq
Had the soliton background solution been smooth, it would probably
have worked out.
However, the first derivative of the background solution is
discontinuous at $r=R$ with the compacton radius $R$ given by
eq.~\eqref{eq:radius}. 
Since the full soliton solution is smooth, the perturbation should
counteract this cusp (jump in the first derivative), which we can
write as a new condition at $r=R$:
\beq
\lim_{r\to R^-} \p_r\phi^3 - \lim_{r\to R^+}\p_r\phi^3 = 0,
\eeq
which with the perturbation $\df$ of eq.~\eqref{eq:Delta_transverse}
yields
\beq
\p_r\df(R^-) - \p_r\df(R^+) = \frac{4}{R}.
\label{eq:BCjump}
\eeq
We will refer to this extra condition in addition to the conditions
\eqref{eq:df_normal_BC} as the ``new boundary conditions.''

\begin{figure}[!htp]
  \begin{center}
    \mbox{\includegraphics[width=0.49\linewidth]{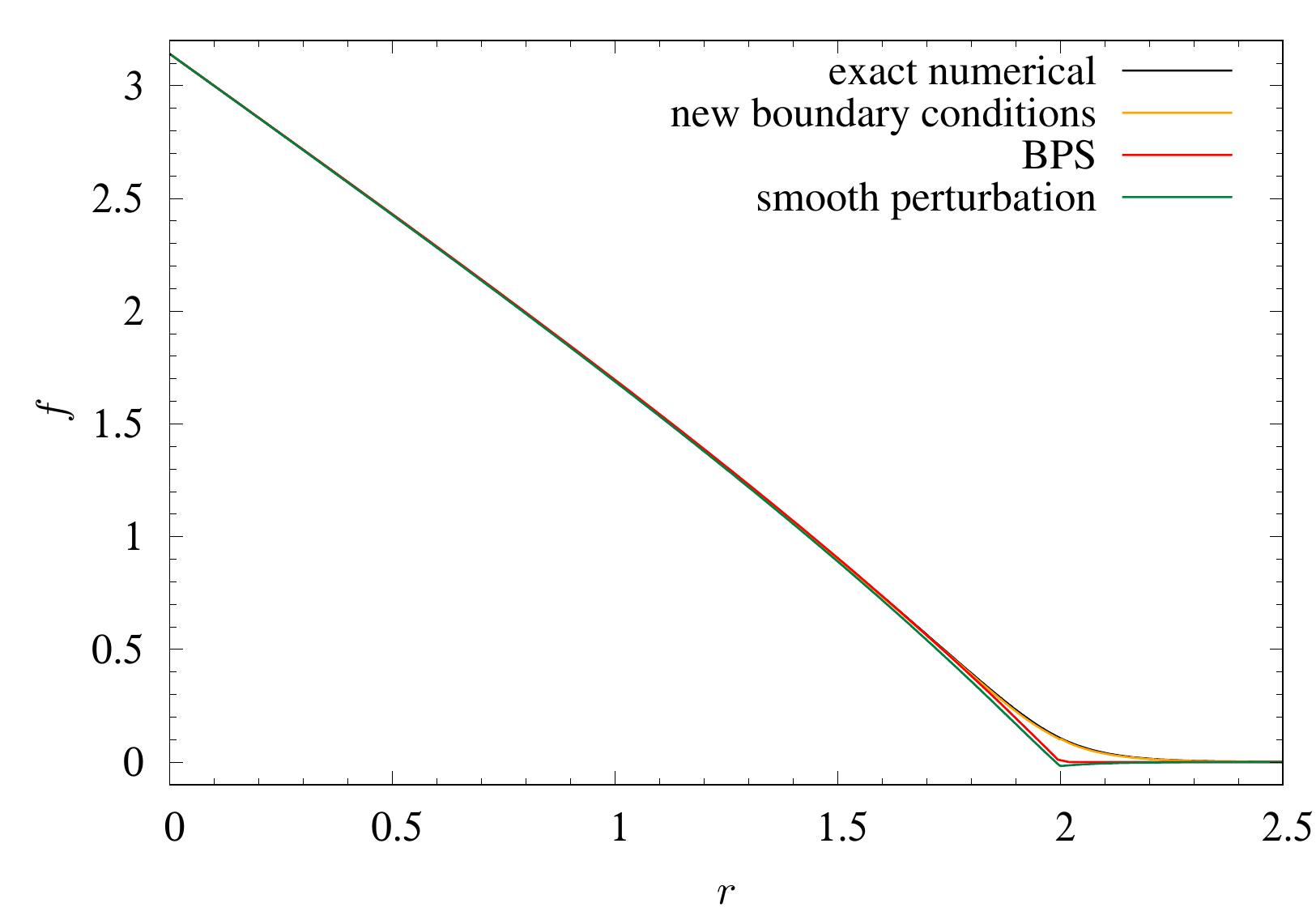}
      \includegraphics[width=0.49\linewidth]{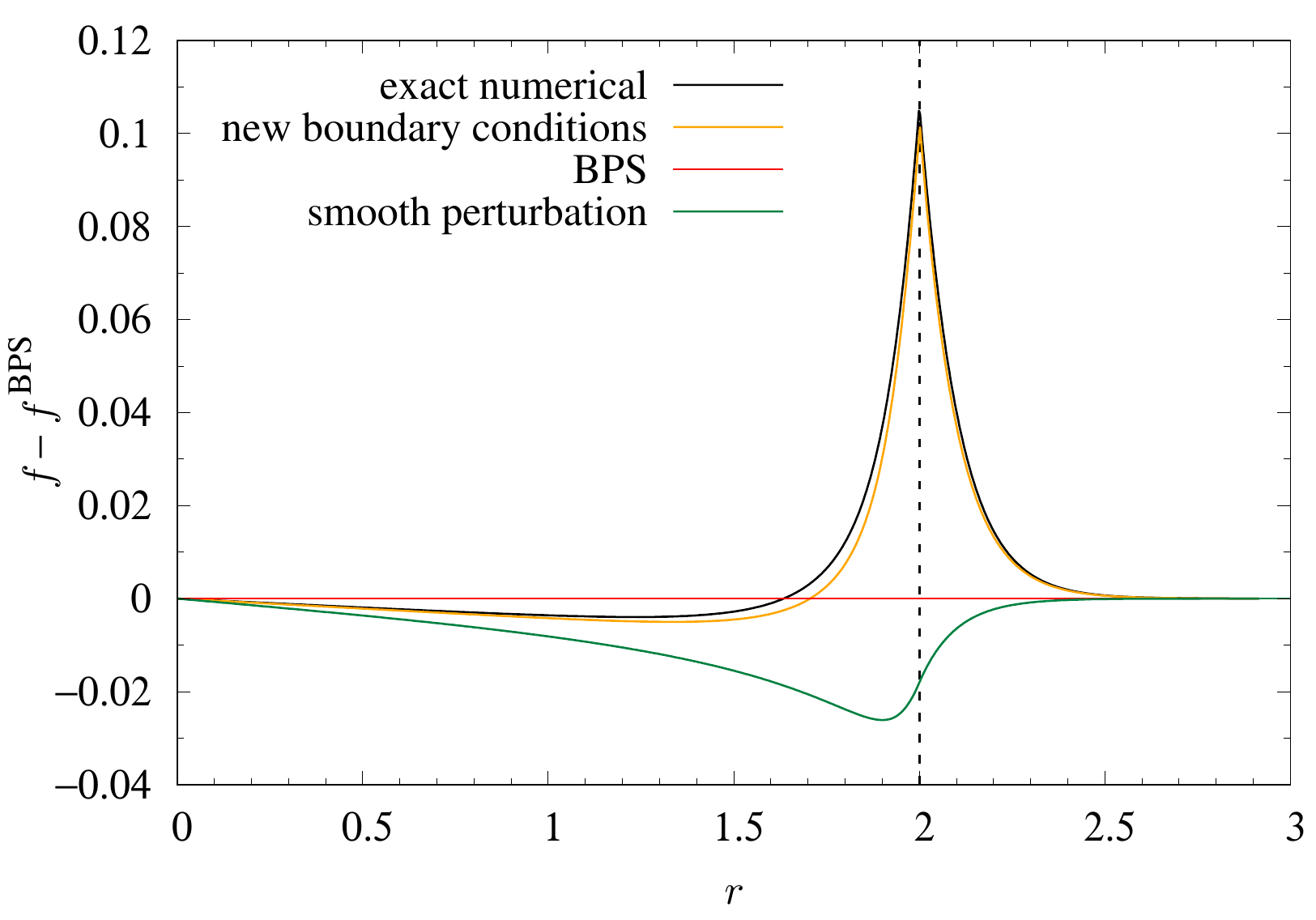}}
    \caption{The profile function $f=\arccos(\phi^3)$ for $N=1$ and
      $\epsilon=0.01$ with $m=1$. The BPS background profile function
      is shown as a red solid curve.
      The smooth perturbation (dark-green curve) does not capture the
      kink at the compacton radius $r=R=2$. The perturbation subject
      to the new boundary condition \eqref{eq:BCjump} (orange) on the
      other hand lies very close to the exact numerical solution
      (black).
    } 
    \label{fig:perturb7profile}
  \end{center}
\end{figure}
In fig.~\ref{fig:perturb7profile} is shown the profile function
$f=\arccos(\phi^3)$ for the $N=1$ axially symmetric baby Skyrmion with
$\epsilon=0.01$ obtained by different methods.
Since the BPS solution is a good starting point, we subtract off the
BPS profile to better see the differences in the panel on the
right-hand side. 
The naive implementation with a smooth perturbation is shown as the
dark-green curve and does not resemble an improvement with respect to
the BPS background solution compared to the exact solution
(numerical). 
The perturbation subject to the new boundary condition
\eqref{eq:BCjump} (orange curve) on the other hand gives an incredible
improvement over the BPS solution and the discrepancy with respect to
the exact solution (numerical) is really tiny and is expected to be
due to nonlinearities (since we have linearized its equation of motion
\eqref{eq:axially_sym_EOM_df}).

\begin{figure}[!ht]
  \begin{center}
    \mbox{\subfloat{\includegraphics[width=0.49\linewidth]{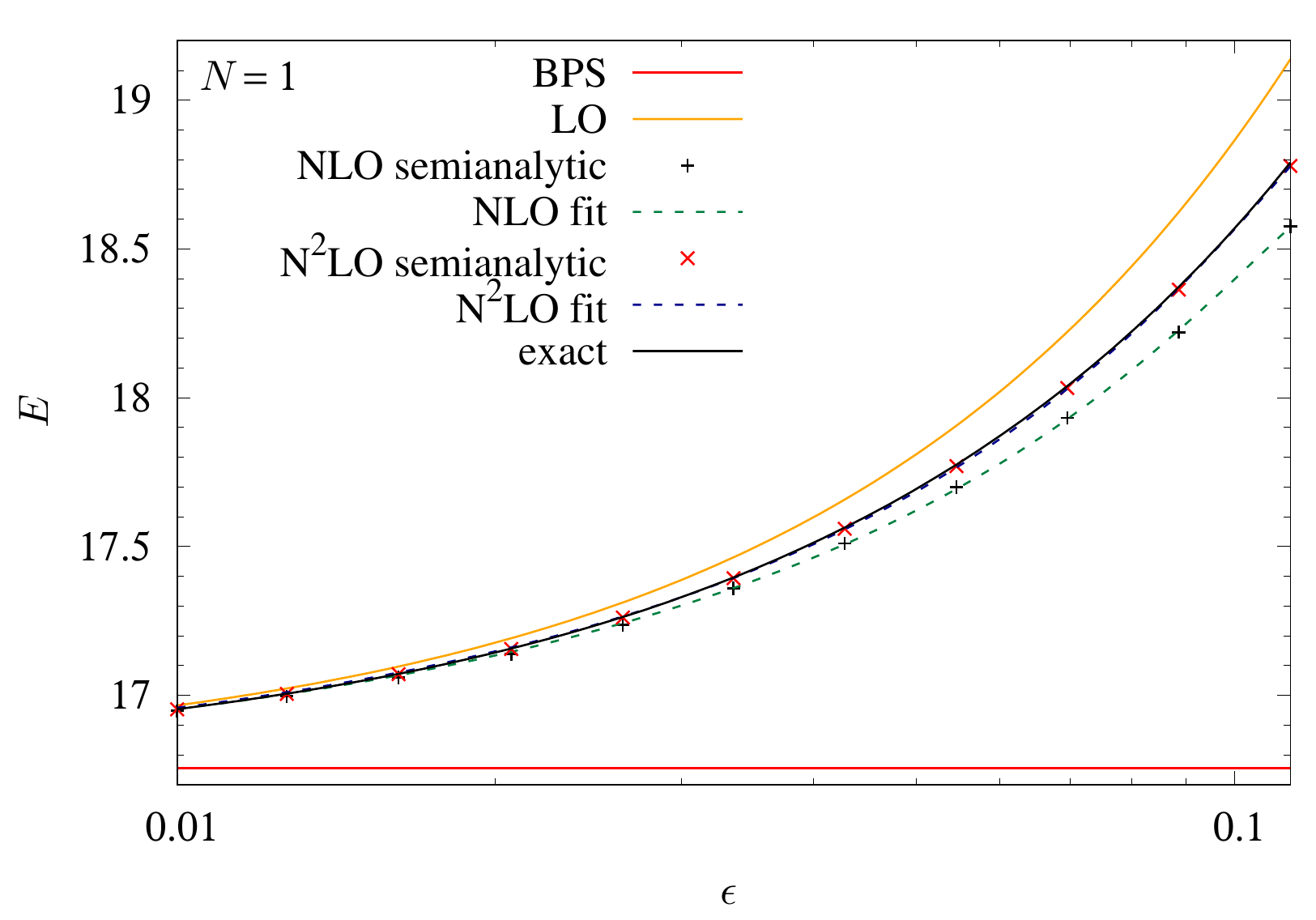}}
      \subfloat{\includegraphics[width=0.49\linewidth]{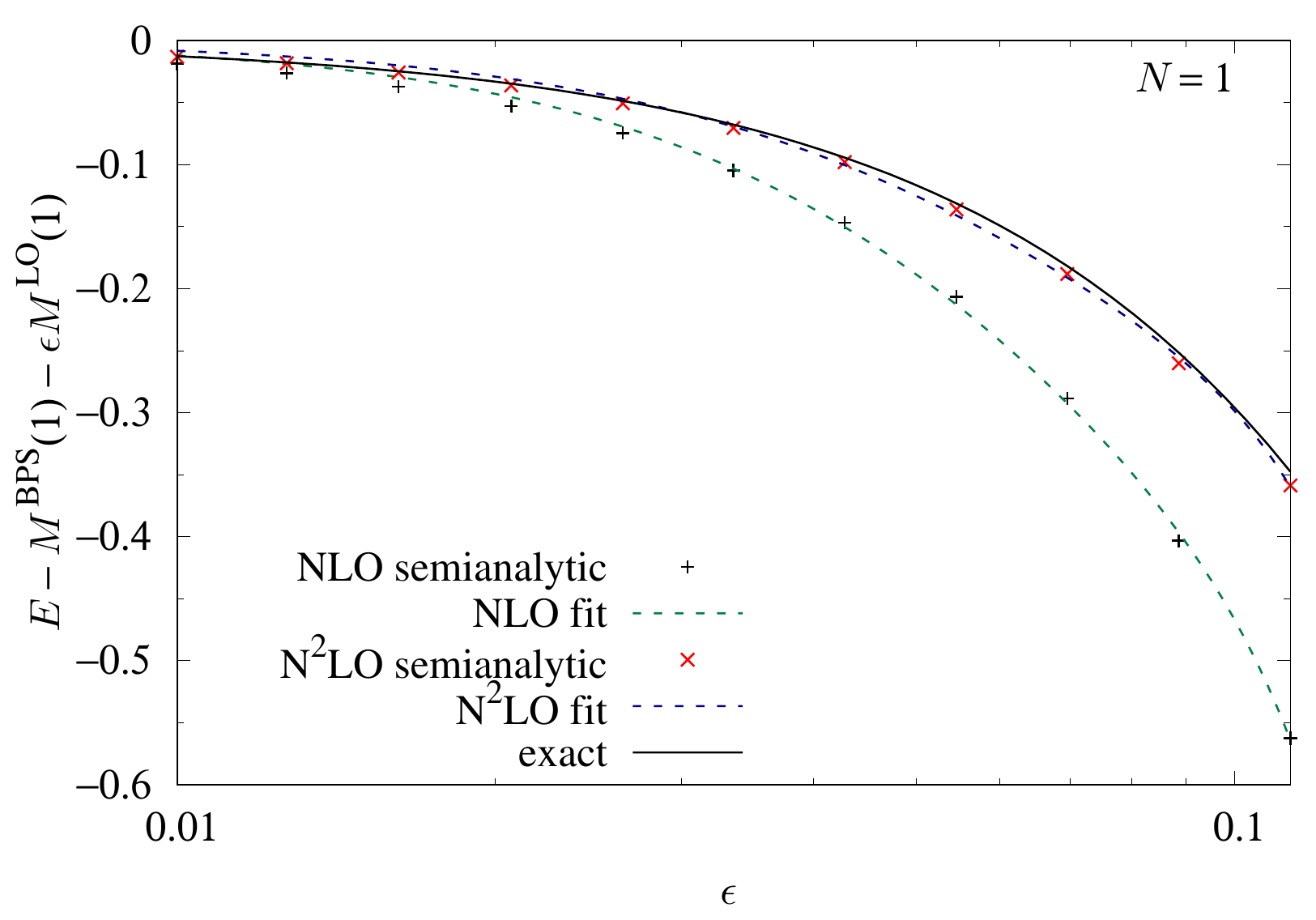}}}
    \mbox{\subfloat{\includegraphics[width=0.49\linewidth]{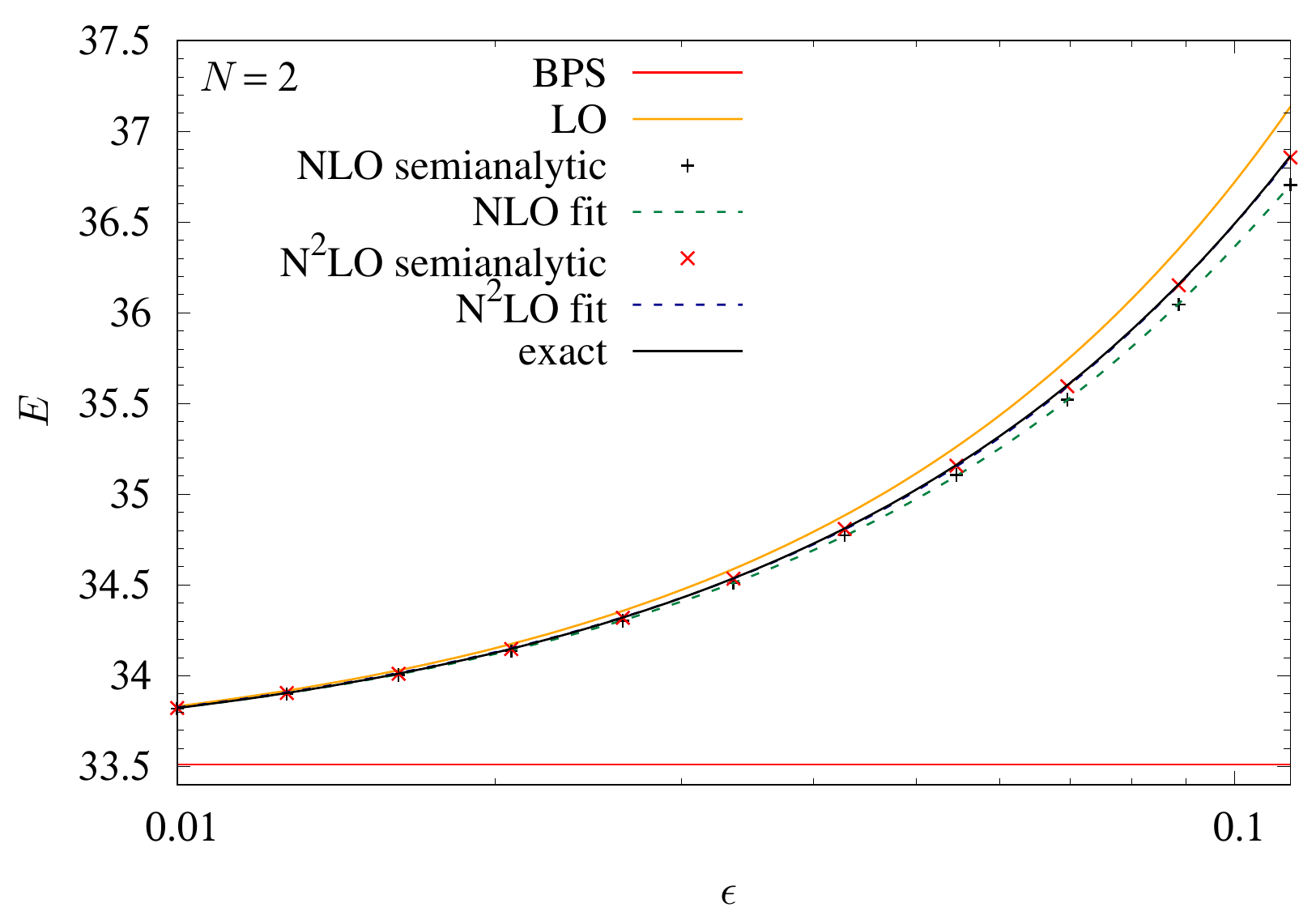}}
      \subfloat{\includegraphics[width=0.49\linewidth]{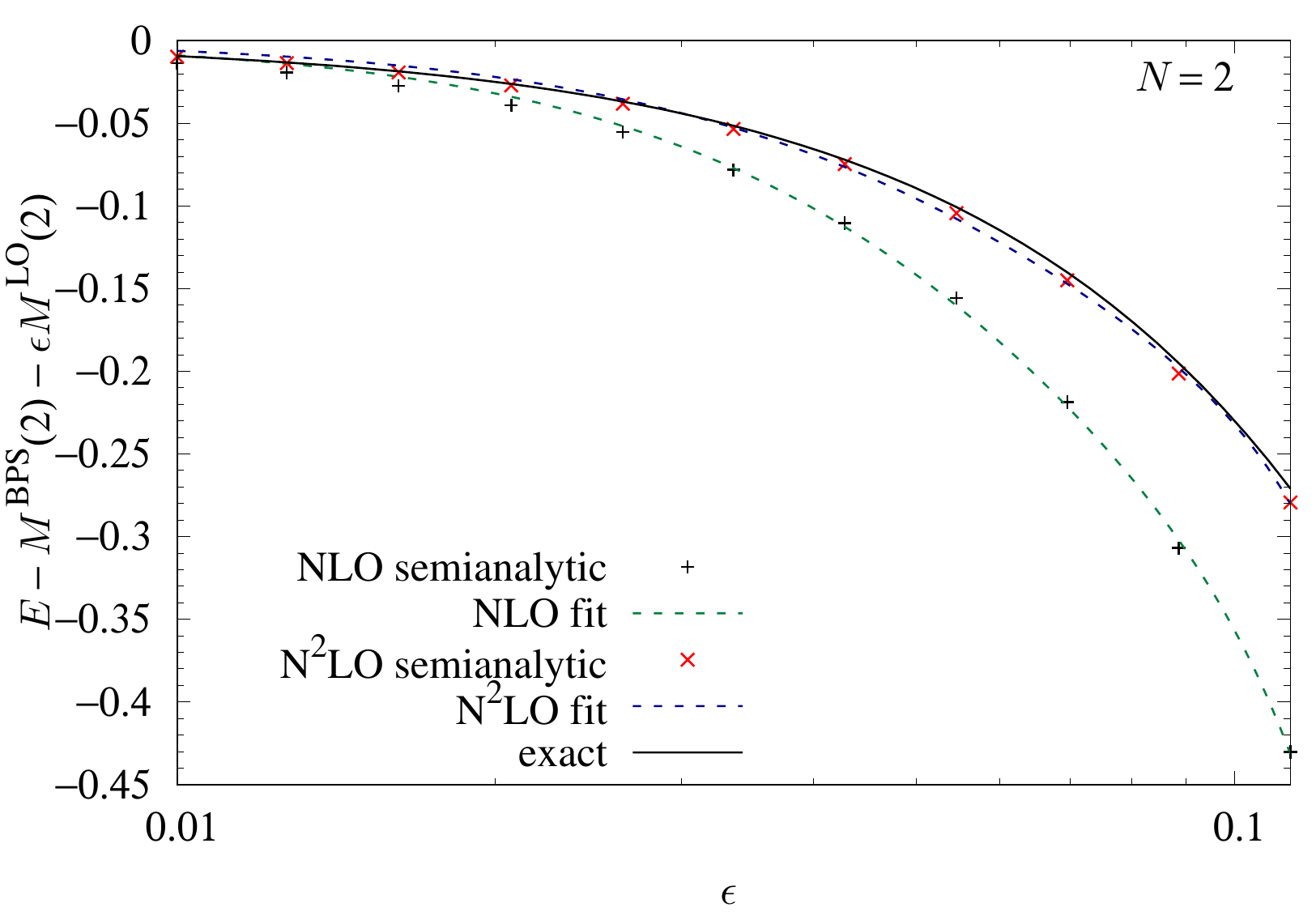}}}
    \mbox{\subfloat{\includegraphics[width=0.49\linewidth]{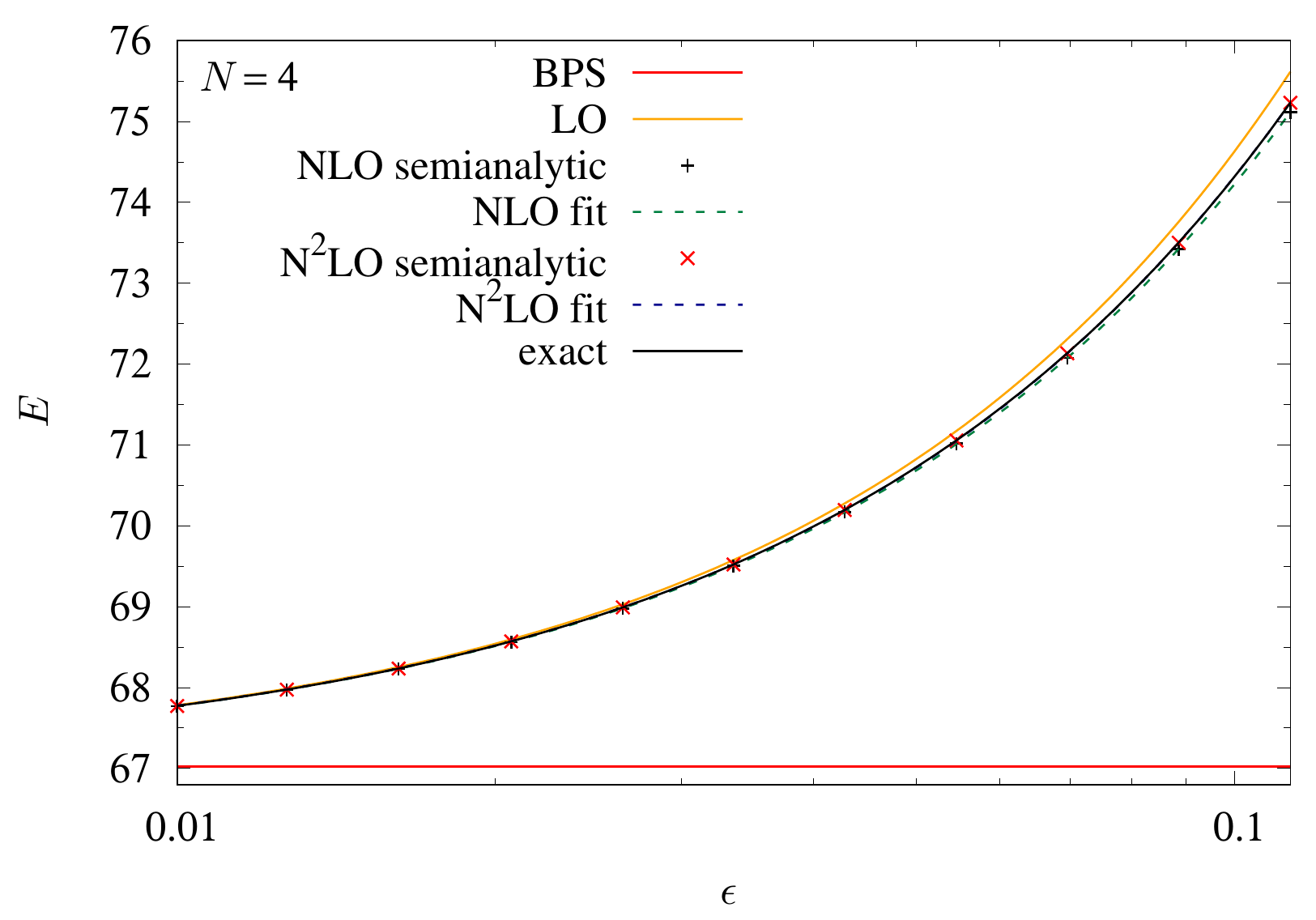}}
      \subfloat{\includegraphics[width=0.49\linewidth]{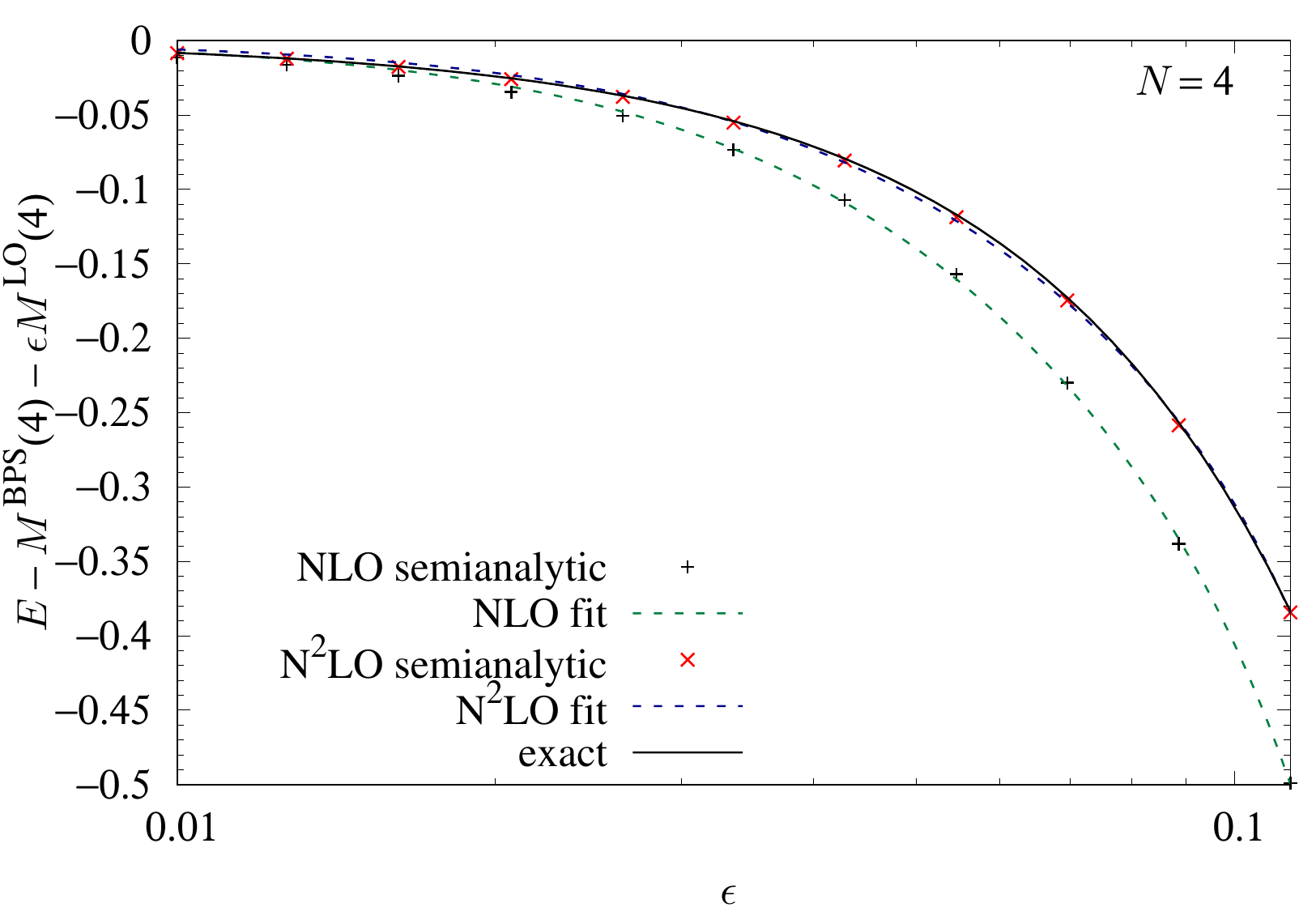}}}
    \caption{The mass of the baby Skyrmions in the perturbative scheme
    as a function of $\epsilon$ (on a logarithmic scale): The red line
    is the BPS bound, the orange curve is the leading order (LO)
    correction, the black pluses are the corrections calculated at
    next-to-leading order (NLO) using the linear perturbation, the
    green-dashed curve is a fit to the latter points, the red crosses
    are the corrections calculated at next-to-next-to-leading order
    (N$^2$LO) using the same linear perturbation, the dark-blue curve
    is a fit to the latter points and the black curve is the exact
    energy calculated using the full (nonlinear) equations of
    motion. The left columns show the total energy while the right
    columns show the energy relative to the LO correction. }
    \label{fig:perturb_energy_axial}
  \end{center}
\end{figure}

We are now ready to calculate the perturbative corrections to the
energy of the $N=1,2,4$ axially symmetric baby Skyrmions and the
results are shown in fig.~\ref{fig:perturb_energy_axial}.
The corrections are calculated using the linear perturbation, which is
a solution to equation \eqref{eq:axially_sym_EOM_df} subject to
the new boundary condition \eqref{eq:BCjump}.
The NLO energy is calculated using the energy to order $\epsilon^2$
using eq.~\eqref{eq:E2axial}, whereas the N$^2$LO energy is calculated
as the sum of eqs.~\eqref{eq:E3axialquad} and
\eqref{eq:E3axialcubic}.
We can see that the perturbative LO corrections overshoot the energy
for every $N$ and the NLO corrections correct this overshooting too
strongly, yielding an undershoot.
Including finally the N$^2$LO corrections, the result is extremely
close to the exact numerical one.
Notice by careful inspection of the figure, that the red crosses
actually match the exact result (black curves) better than the fit 
(dark-blue curves).

\begin{figure}[!ht]
  \begin{center}
    \includegraphics[width=0.5\linewidth]{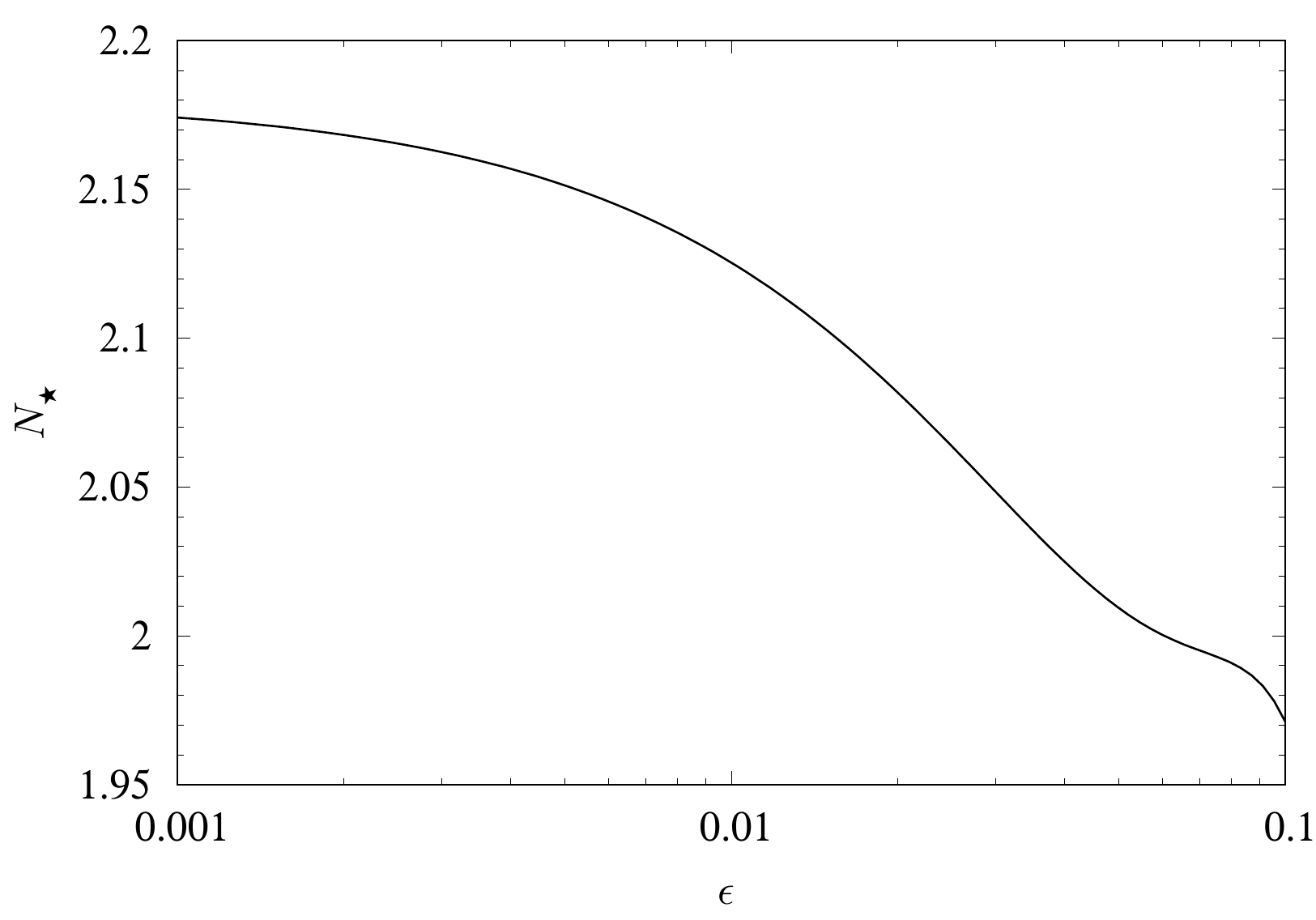}
    \caption{The critical value, $N_\star$, of the topological charge
      of the axially symmetric baby Skyrmion, as a function of
      $\epsilon$. $N_\star$ corresponds to the smallest energy per
      topological charge and hence the globally stable axially
      symmetric configuration.}
    \label{fig:Nstar}
  \end{center}
\end{figure}

The fit to the N$^2$LO data (for $m=1$) gives
\begin{align}
E(\epsilon,N) &=
M^{\rm BPS}(N)
+ \epsilon M^{\rm LO}(N)
+ \epsilon^2 M^{\rm NLO}(N)
+ \epsilon^3 M^{{\rm N}^2{\rm LO}}(N)
+ \epsilon^4 M^{{\rm residual},4}(N)\non
&= \frac{16\pi N}{3}
+ \epsilon\left(8\pi\log 2 + \frac{7\pi N^2}{6}\right)
+ \epsilon^2\left(-128.9 + 43.50 N - 6.869 N^2\right) \non
&\phantom{=\ }
+ \epsilon^3\left(1458 - 450.9 N + 57.20 N^2\right)
+ \epsilon^4\left(-6044 + 1884 N - 238.4 N^2\right).
\label{eq:E_N2LO_compacton_fit}
\end{align}
This higher-order result in turn gives a correction to $N_\star$ of
eq.~\eqref{eq:Nstar}:
\beq
N_\star = 2.180 - 6.023\epsilon + 54.60\epsilon^2
  + 55.52\epsilon^3 - 2083\epsilon^4 + \mathcal{O}(\epsilon^5).
\eeq
Although we write the expansion to fourth order in $\epsilon$, it
should not be trusted beyond the third order. The higher orders
simply represent the all-order contributions from the tail that are
important for capturing the binding energies.
$N_\star$ is shown in fig.~\ref{fig:Nstar}.
As can be seen from the figure, taking into account higher-order
corrections in $\epsilon$ decreases $N_\star$. However, the $N=2$ baby
Skyrmion remains the energetically stable axially symmetric solution.
The drop in $N_\star$ after $\epsilon\gtrsim 0.1$ is possibly just an
artifact of the fit and should not be taken too seriously at this
stage.

\section{Numerical calculations}
\label{quattro}

In this section, we present numerical solutions to the full equations
of motion. 
This is a computationally extremely expensive work and can be carried
out only for the 2-dimensional case unless very clever adaptive
methods are being used. 
The numerical calculations performed here are carried out with a
fourth-order 5-point stencil finite-difference method using the
arrested Newton flow \cite{Gudnason:2020arj} on a square grid of sizes
up to $\sim 3226^2\sim 10^7$ with lattice spacing down to
$\sim 0.0037$.  
Our numerical accuracy is about $10^{-6}$ or better.
Such expensive grids would take a very long time on conventional CPU
clusters, so our code is written in CUDA C and is executed on a GPU
cluster.
In order to avoid warp divergence, the code is made such that the bulk
of the lattice is launched as one kernel without any conditionals and
the edges are launched as two different additional kernels
corresponding to the vertical and horizontal edges of the lattice
(also without any conditionals). 

In this section we explore the numerical solutions of baby Skyrmions
that do not possess axial symmetry, however, we also check the axially
symmetric configurations both for ensuring that they exist in the
given range of parameter space and for comparing our numerical
accuracy to that obtained using ODEs for the axially symmetric system. 

Our aim here is to confirm whether the energetically most favorable
configuration -- in the case of small but finite $0<\epsilon\ll 1$ --
consists of $N=2$ baby Skyrmions sitting next to each other at a
finite or vanishing distance between them.
Such a scenario is supported by the notion of restricted harmonic
maps.

\newcommand{\fwidth}{0.235\linewidth}
\begin{figure}[!t]
  \begin{center}
  \mbox{
    \rotatebox{90}{\quad$\epsilon=1$}
    \includegraphics[width=\fwidth]{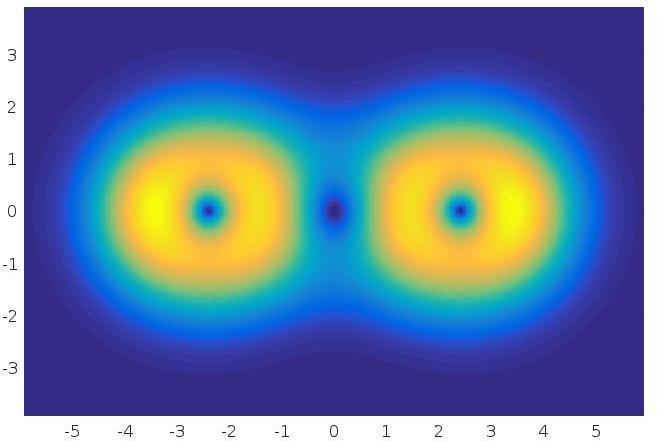}
    \includegraphics[width=\fwidth]{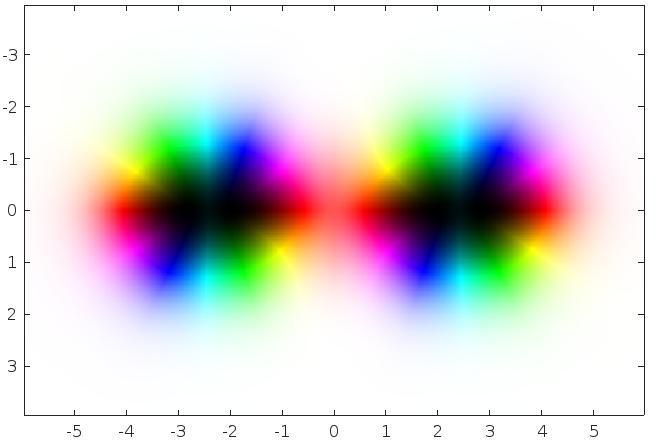}
    \includegraphics[width=\fwidth]{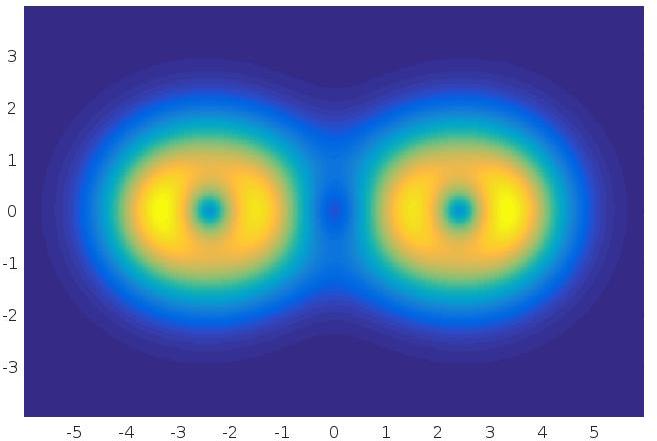}
    \includegraphics[width=\fwidth]{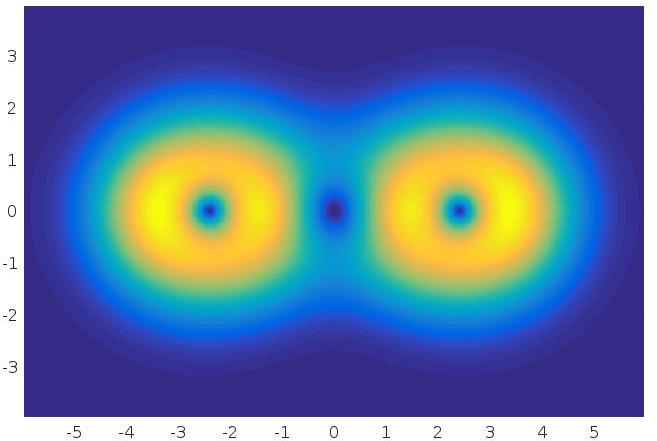}}
  \mbox{
    \rotatebox{90}{\quad$\epsilon=0.616$}
    \includegraphics[width=\fwidth]{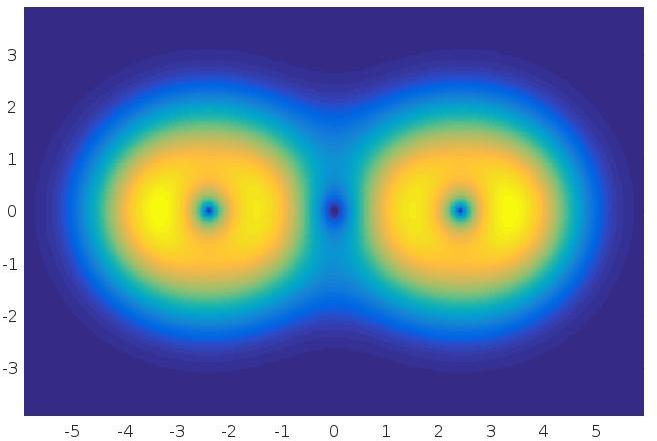}
    \includegraphics[width=\fwidth]{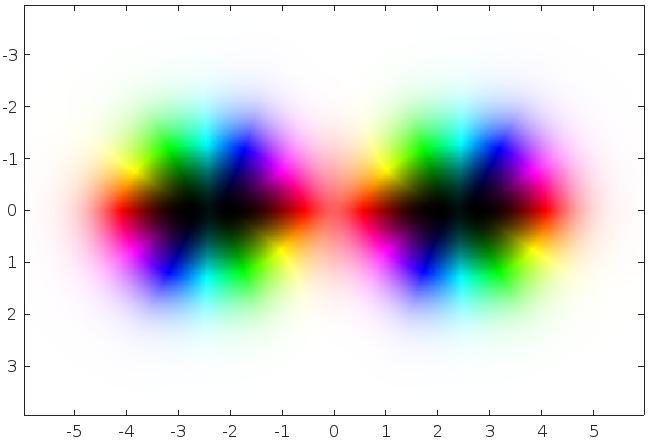}
    \includegraphics[width=\fwidth]{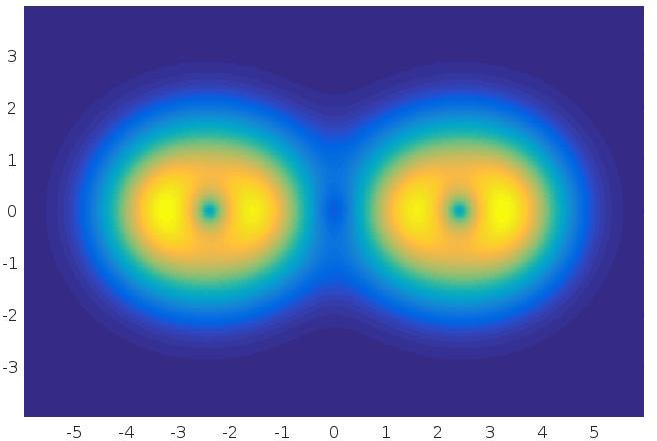}
    \includegraphics[width=\fwidth]{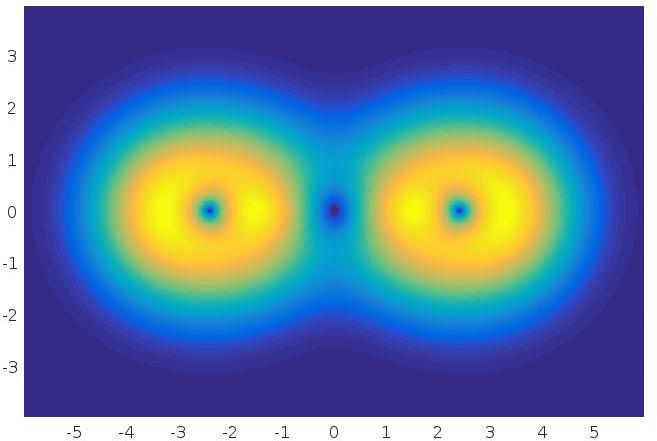}}
  \mbox{
    \rotatebox{90}{\quad$\epsilon=0.379$}
    \includegraphics[width=\fwidth]{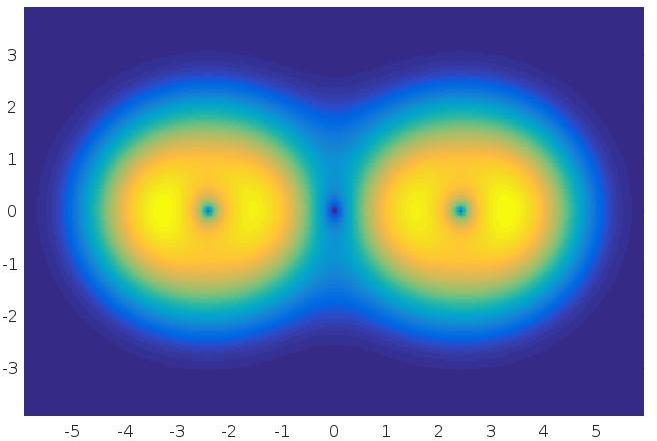}
    \includegraphics[width=\fwidth]{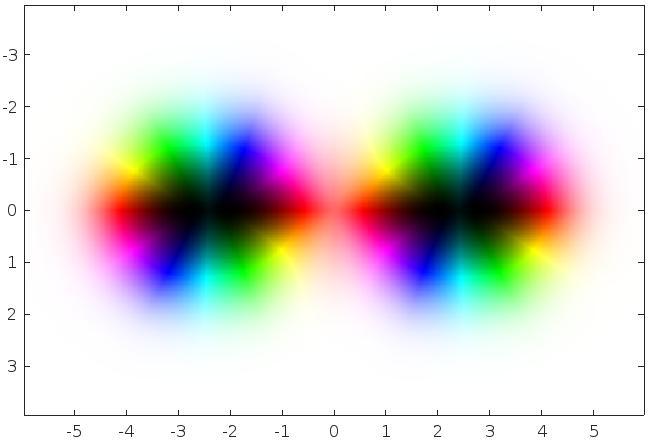}
    \includegraphics[width=\fwidth]{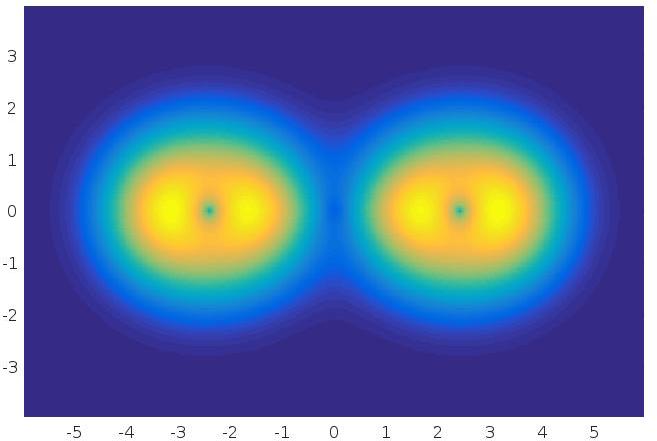}
    \includegraphics[width=\fwidth]{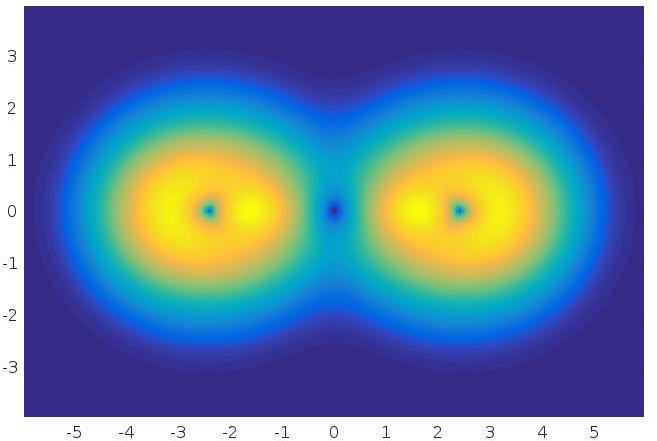}}
  \mbox{
    \rotatebox{90}{\quad$\epsilon=0.183$}
    \includegraphics[width=\fwidth]{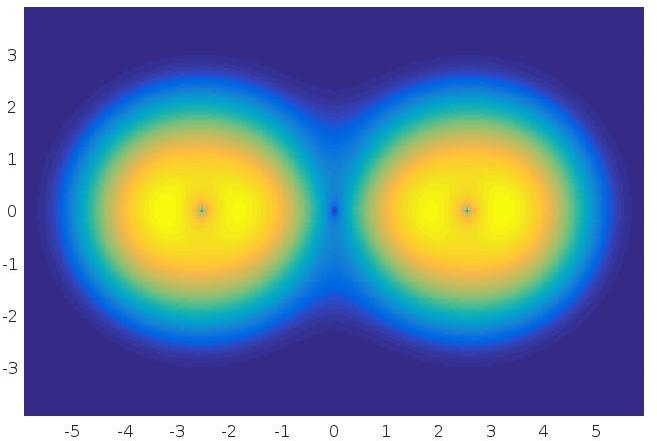}
    \includegraphics[width=\fwidth]{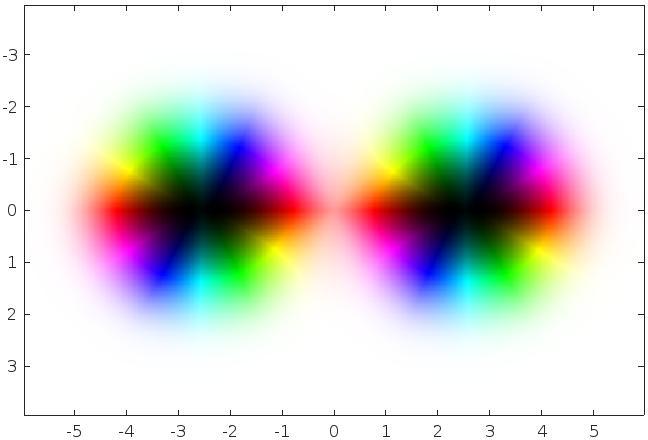}
    \includegraphics[width=\fwidth]{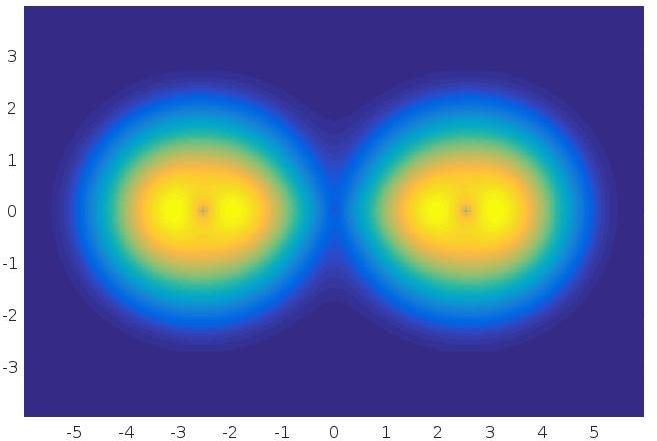}
    \includegraphics[width=\fwidth]{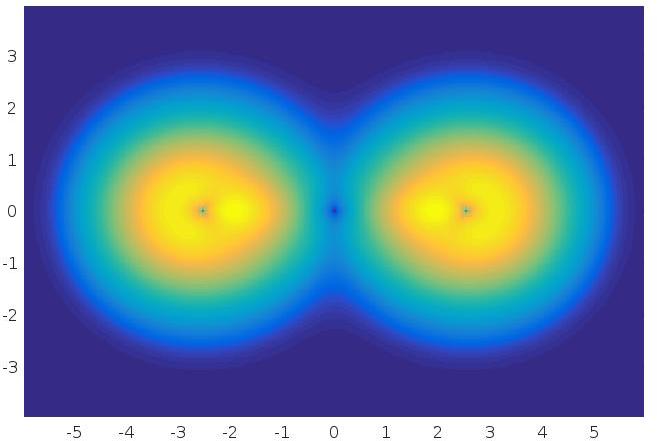}}
  \mbox{
    \rotatebox{90}{\quad$\epsilon=0.0886$}
    \includegraphics[width=\fwidth]{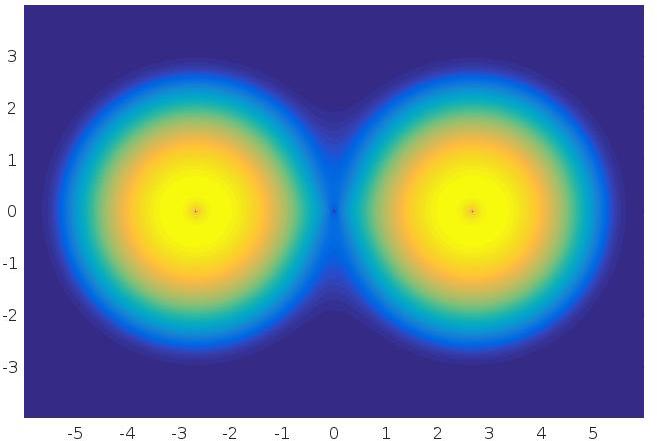}
    \includegraphics[width=\fwidth]{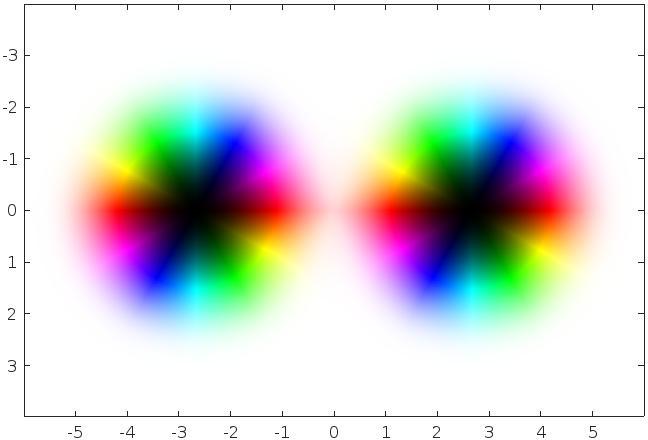}
    \includegraphics[width=\fwidth]{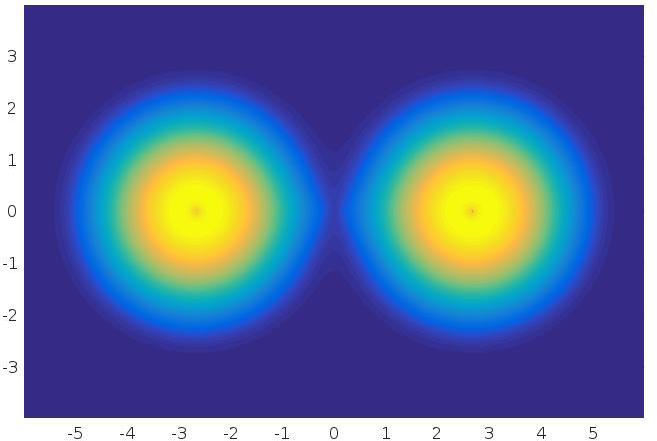}
    \includegraphics[width=\fwidth]{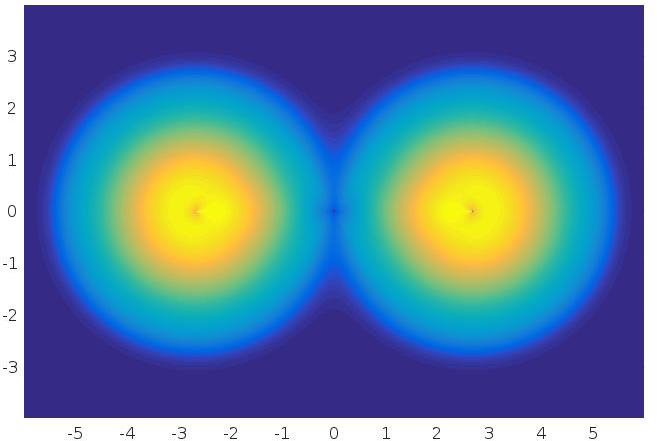}}
  \mbox{
    \rotatebox{90}{\quad$\epsilon=0.0428$}
    \includegraphics[width=\fwidth]{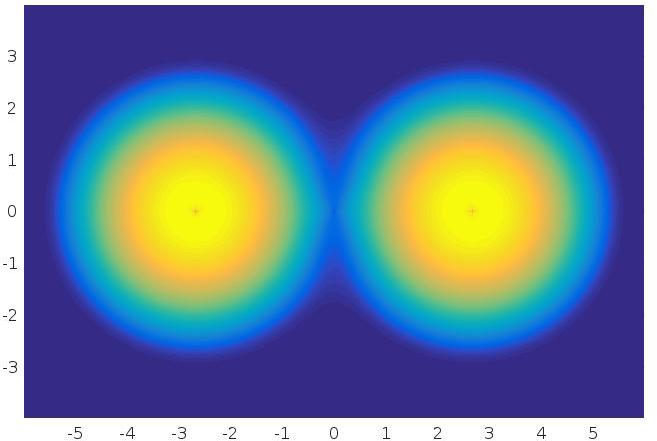}
    \includegraphics[width=\fwidth]{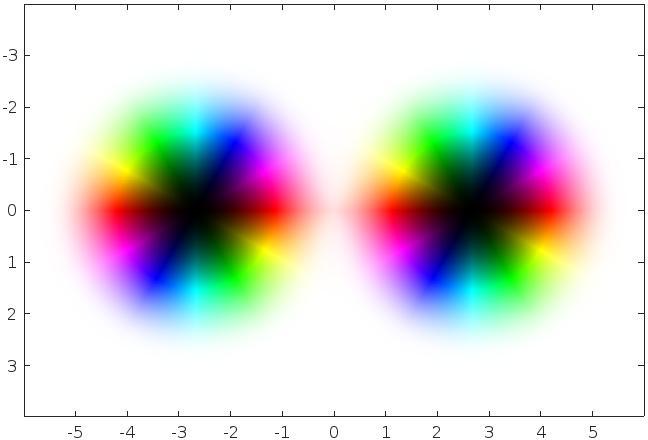}
    \includegraphics[width=\fwidth]{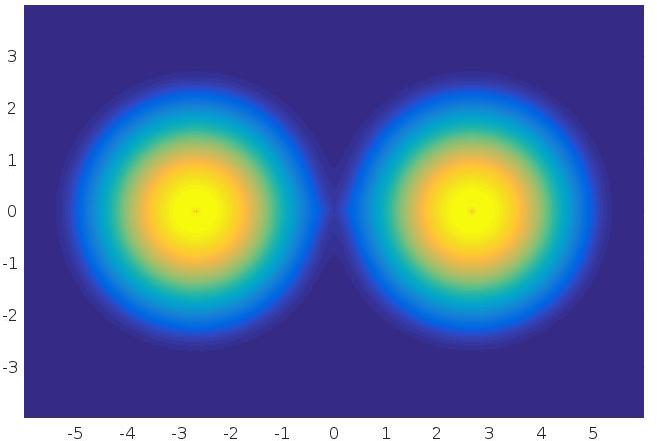}
    \includegraphics[width=\fwidth]{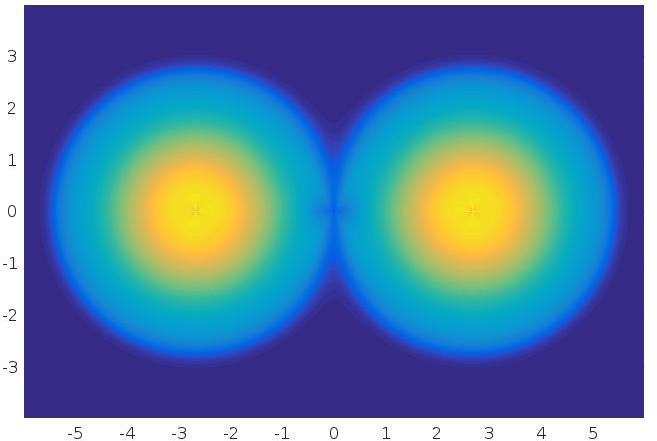}}
  \mbox{
    \rotatebox{90}{\quad$\epsilon=0.0207$}
    \includegraphics[width=\fwidth]{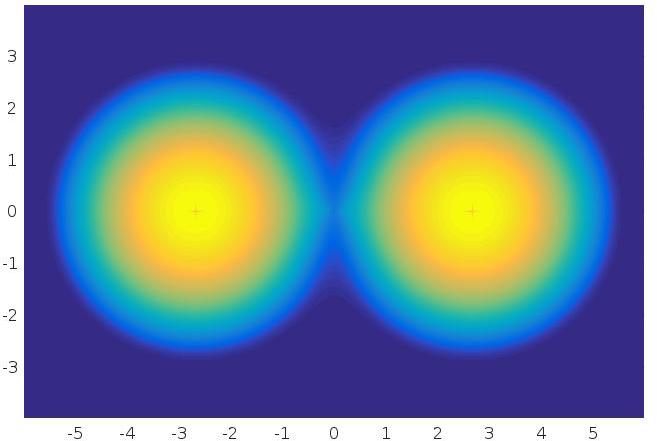}
    \includegraphics[width=\fwidth]{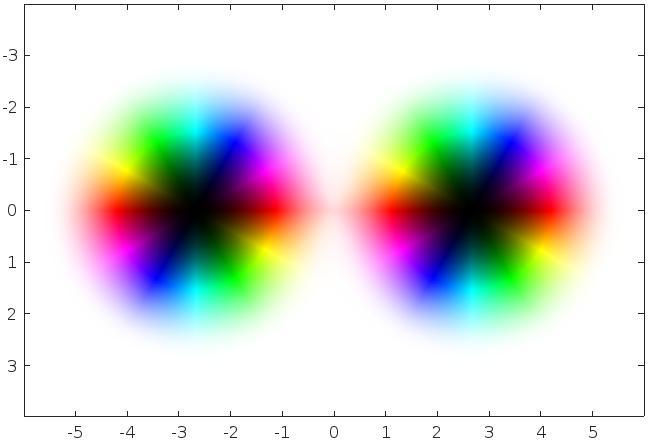}
    \includegraphics[width=\fwidth]{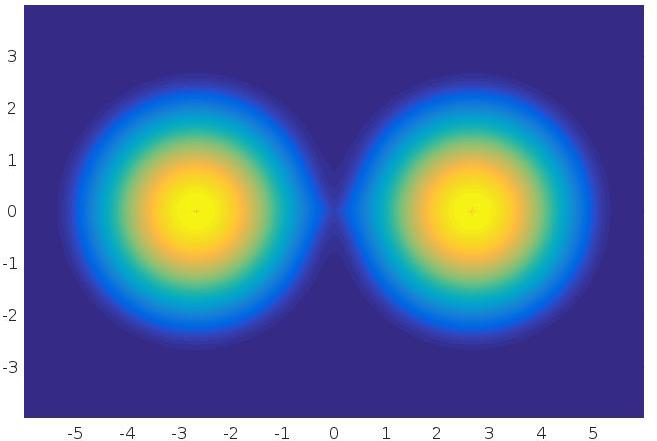}
    \includegraphics[width=\fwidth]{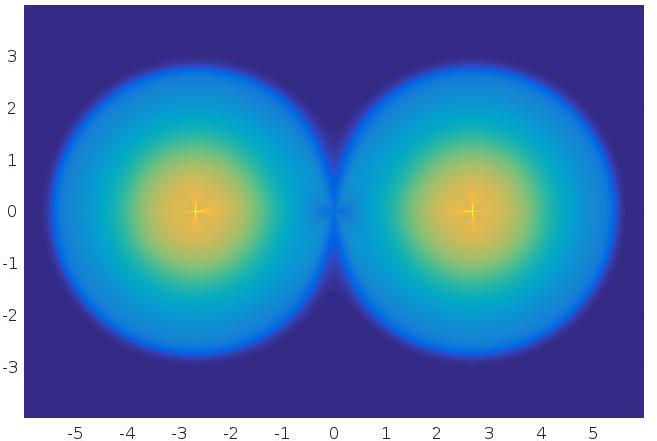}}
  \caption{The $Q=2+2$ baby Skyrmion solution as a function of
    $\epsilon$, which is the stable $Q=4$ solution.
  The columns display the topological charge density, the field
  orientation using the color scheme described in the text, the total
  energy and finally the kinetic term $-\epsilon\Lag_2$.
  We note that the two stretched $N=2$ baby Skyrmions become almost
  perfectly round in the last row (for $\epsilon=0.0207$).
  }
  \label{fig:N=2+2}
  \end{center}
\end{figure}

\begin{figure}[!t]
  \begin{center}
  \mbox{
    \rotatebox{90}{\quad$\epsilon=1$}
    \includegraphics[width=\fwidth]{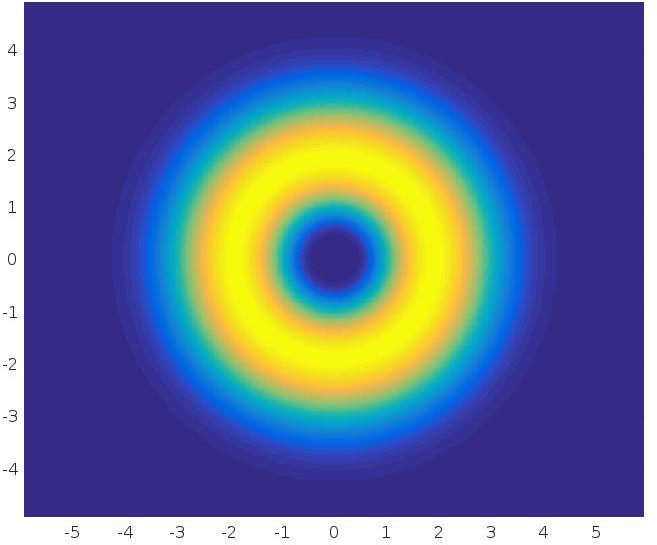}
    \includegraphics[width=\fwidth]{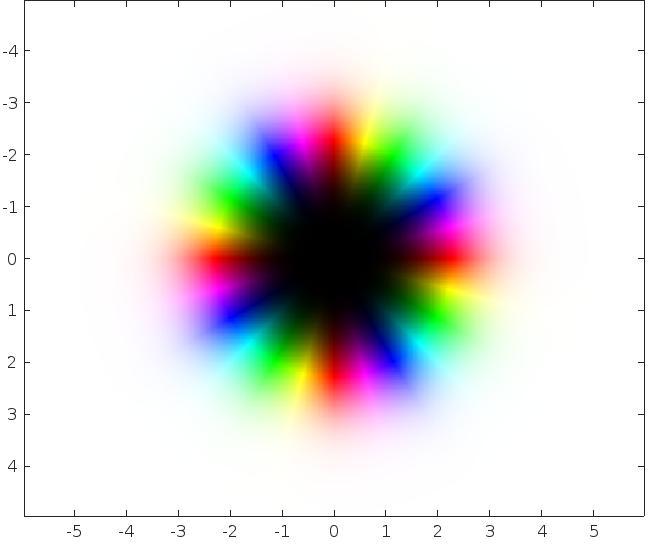}
    \includegraphics[width=\fwidth]{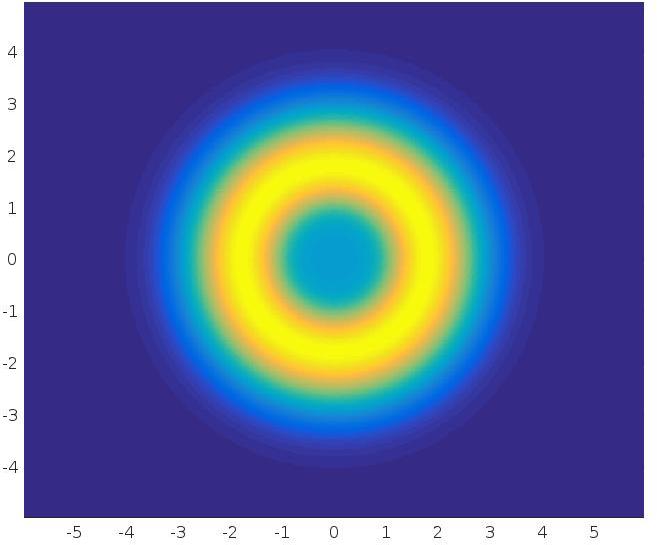}
    \includegraphics[width=\fwidth]{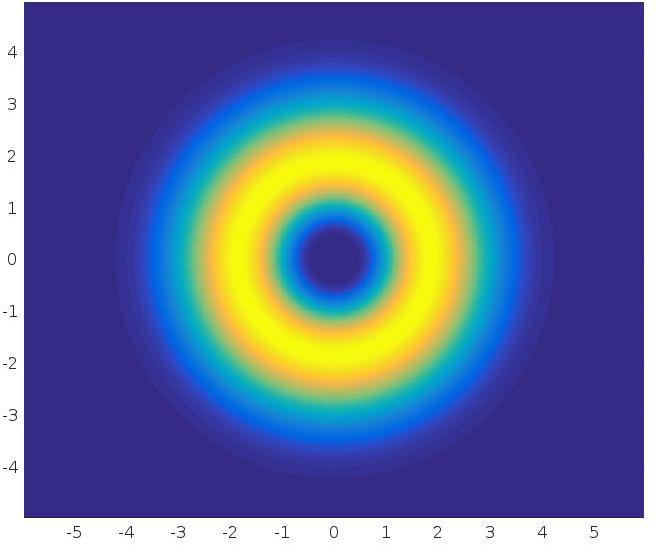}}
  \mbox{
    \rotatebox{90}{\quad$\epsilon=0.616$}
    \includegraphics[width=\fwidth]{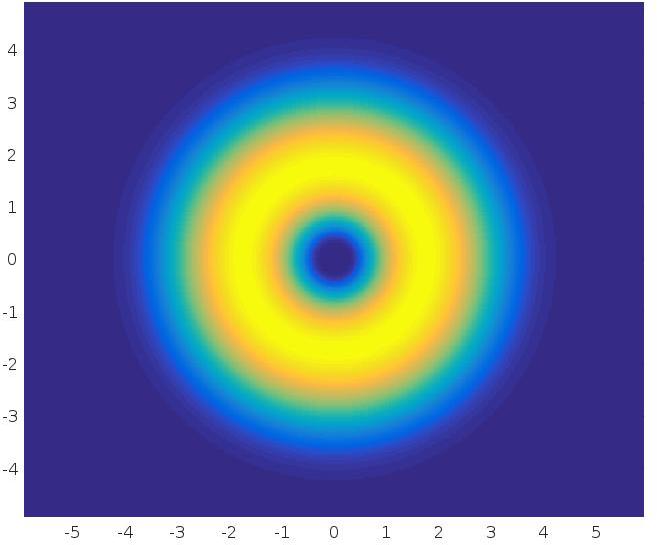}
    \includegraphics[width=\fwidth]{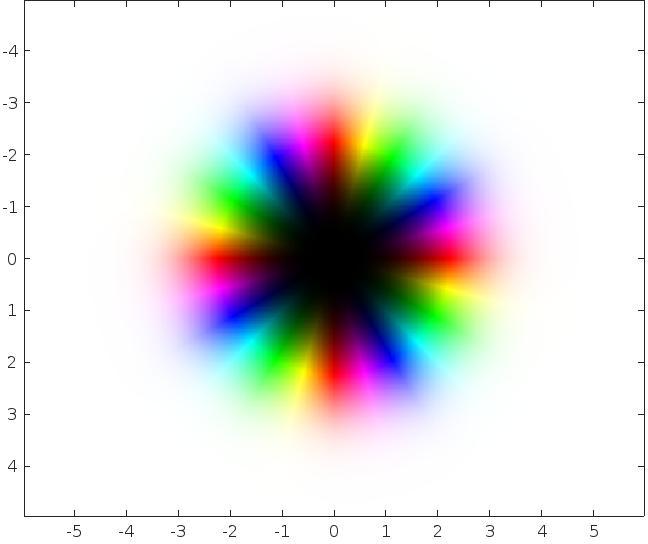}
    \includegraphics[width=\fwidth]{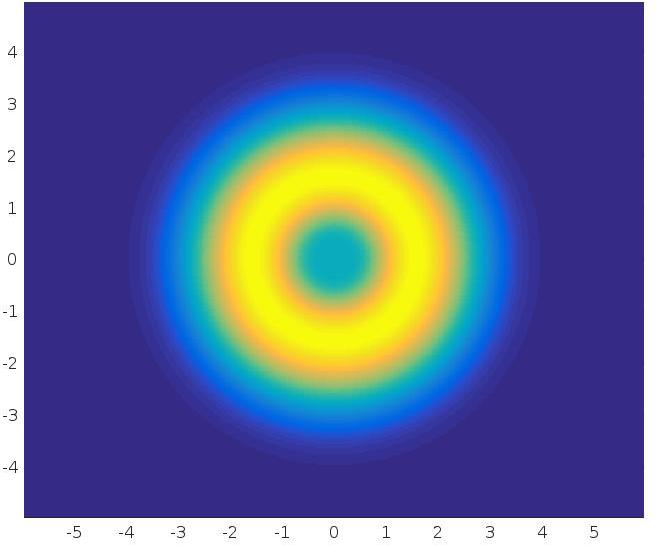}
    \includegraphics[width=\fwidth]{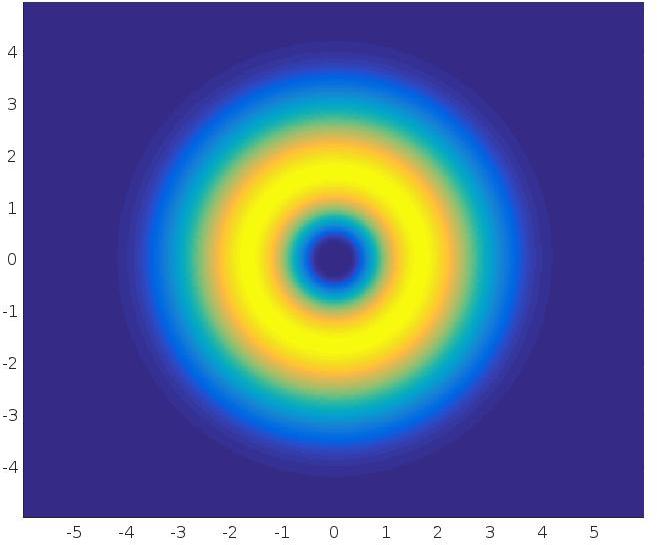}}
  \mbox{
    \rotatebox{90}{\quad$\epsilon=0.379$}
    \includegraphics[width=\fwidth]{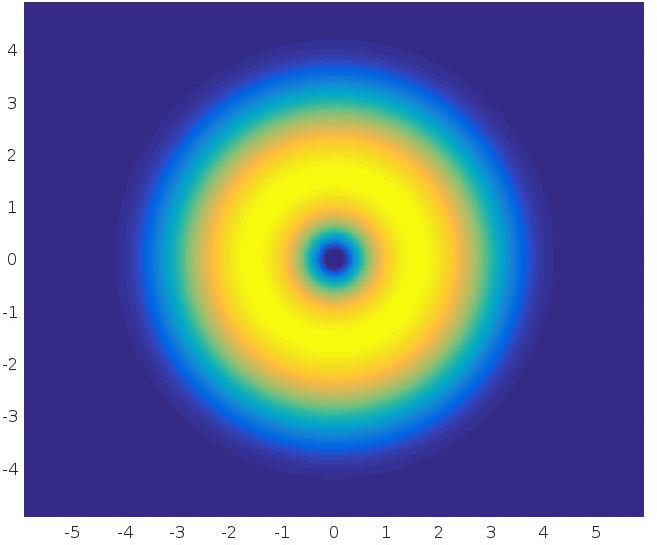}
    \includegraphics[width=\fwidth]{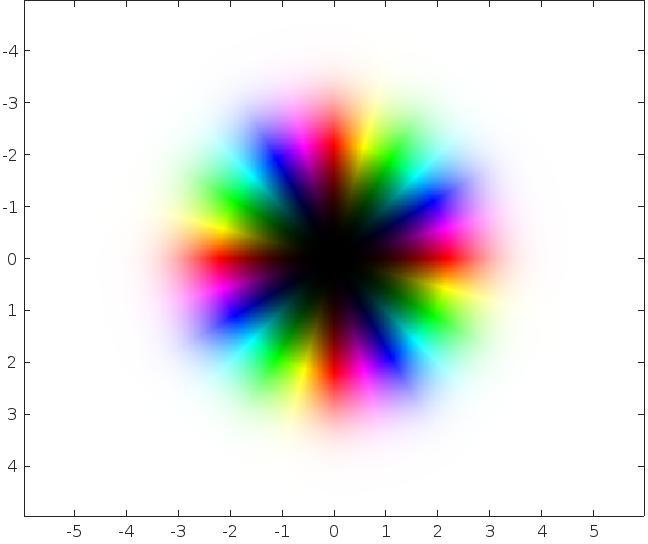}
    \includegraphics[width=\fwidth]{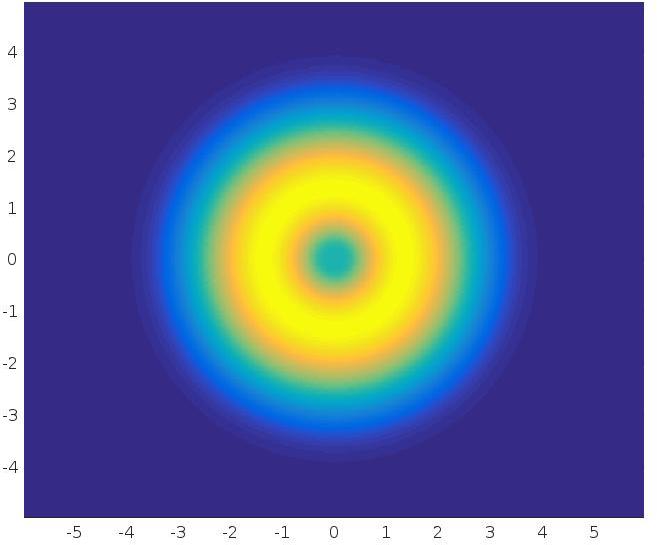}
    \includegraphics[width=\fwidth]{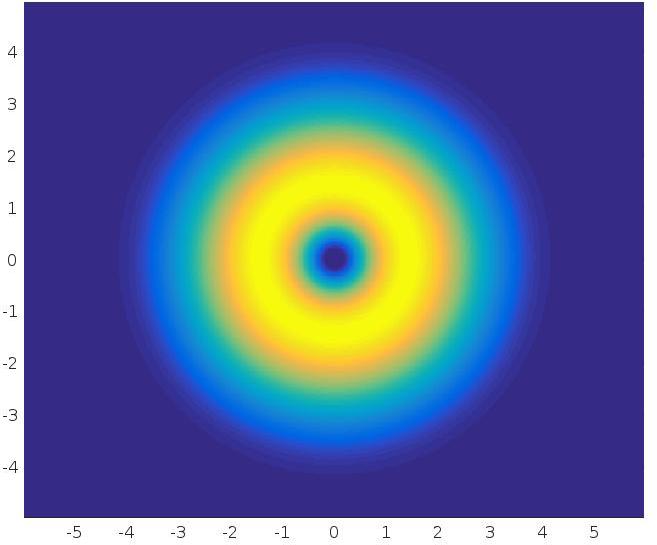}}
  \mbox{
    \rotatebox{90}{\quad$\epsilon=0.183$}
    \includegraphics[width=\fwidth]{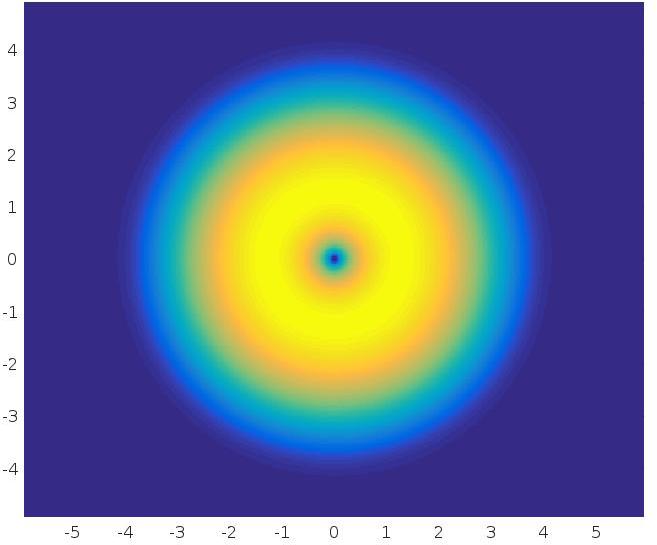}
    \includegraphics[width=\fwidth]{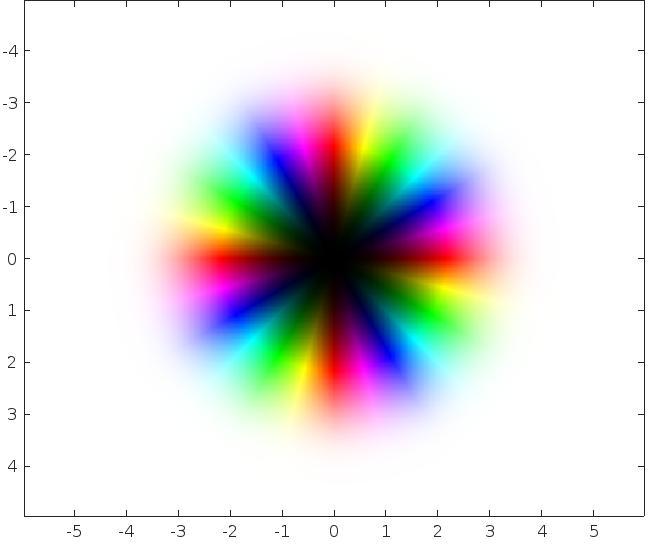}
    \includegraphics[width=\fwidth]{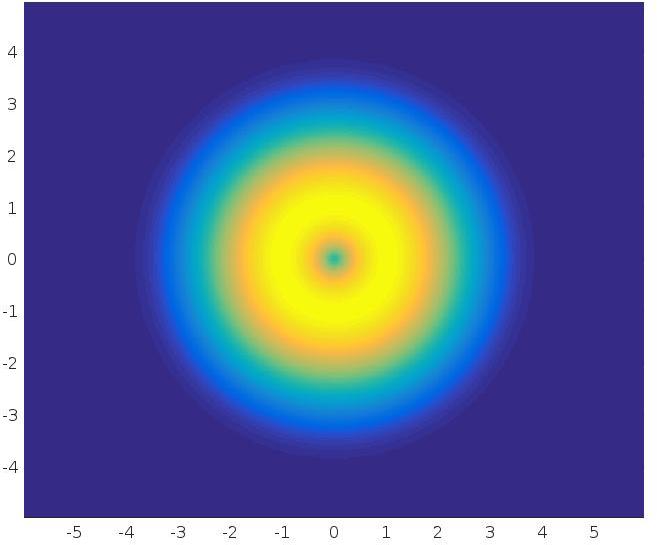}
    \includegraphics[width=\fwidth]{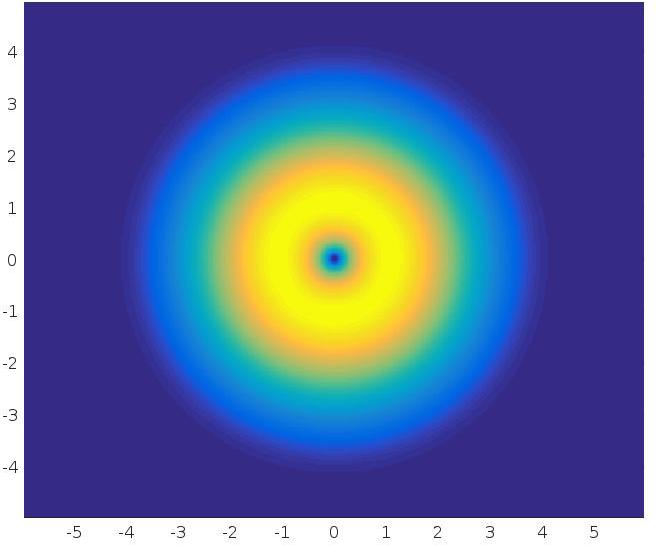}}
  \mbox{
    \rotatebox{90}{\quad$\epsilon=0.144$}
    \includegraphics[width=\fwidth]{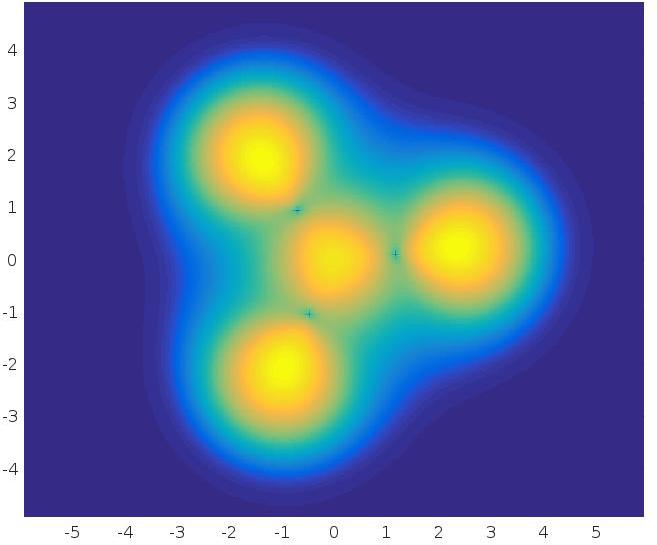}
    \includegraphics[width=\fwidth]{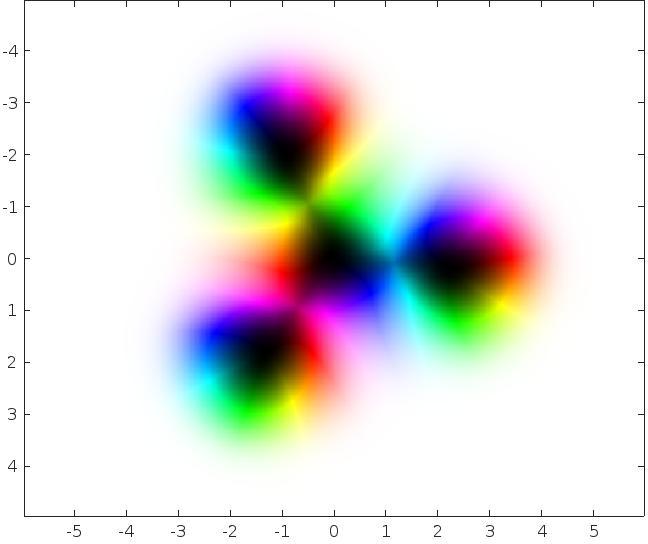}
    \includegraphics[width=\fwidth]{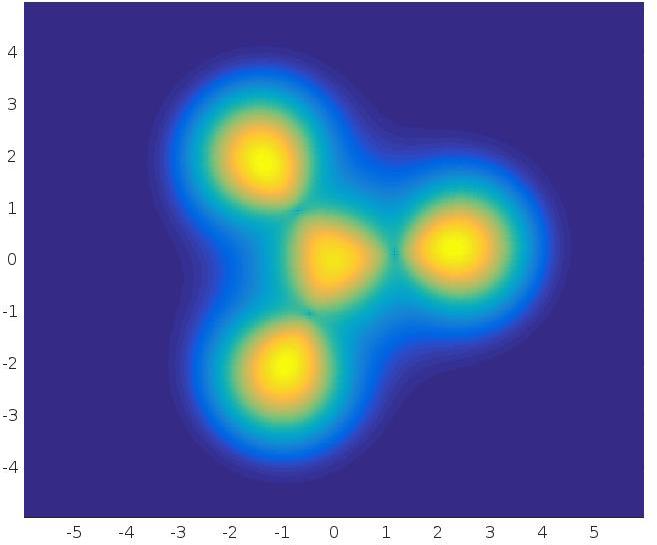}
    \includegraphics[width=\fwidth]{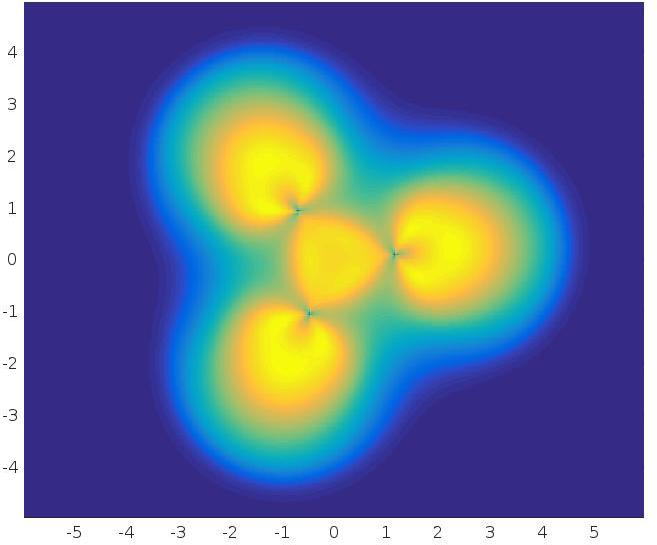}}
  \caption{The $N=4$ baby Skyrmion solution as a function of
    $\epsilon$.
  The columns display the topological charge density, the field
  orientation using the color scheme described in the text, the total
  energy and finally the kinetic term $-\epsilon\Lag_2$.
  The last row (i.e.~for $\epsilon=0.144$) shows that the solution
  has decayed into a lower-energy state with tetrahedral symmetry
  instead of axial symmetry. 
  }
  \label{fig:N=4}
  \end{center}
\end{figure}

\begin{figure}[!t]
  \begin{center}
  \mbox{
    \rotatebox{90}{\quad$\epsilon=1$}
    \includegraphics[width=\fwidth]{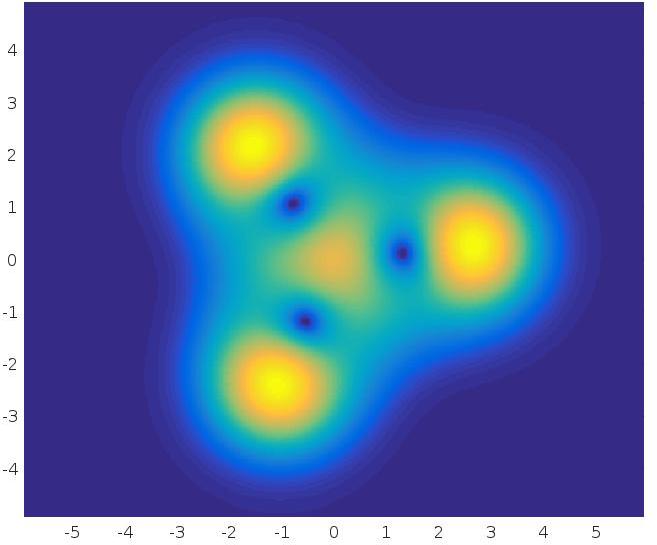}
    \includegraphics[width=\fwidth]{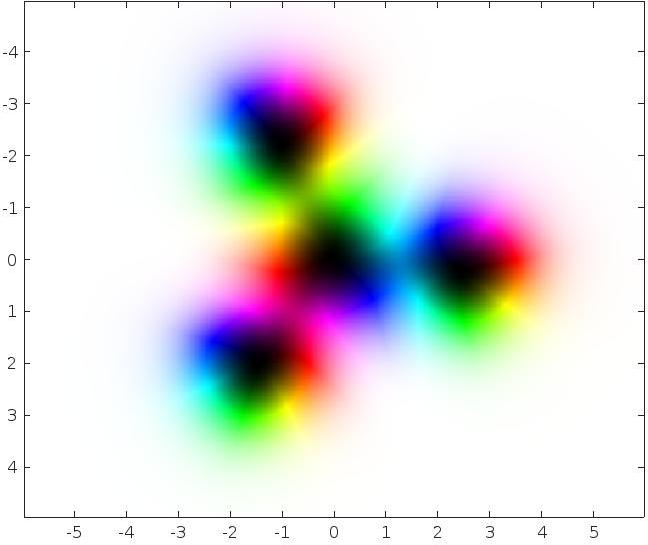}
    \includegraphics[width=\fwidth]{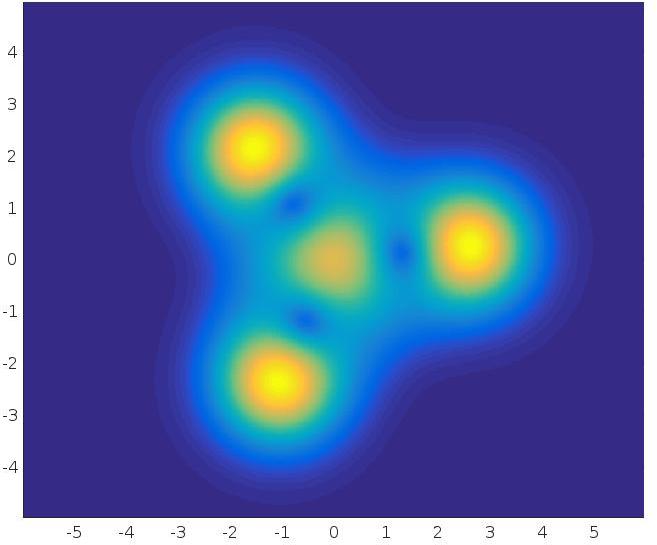}
    \includegraphics[width=\fwidth]{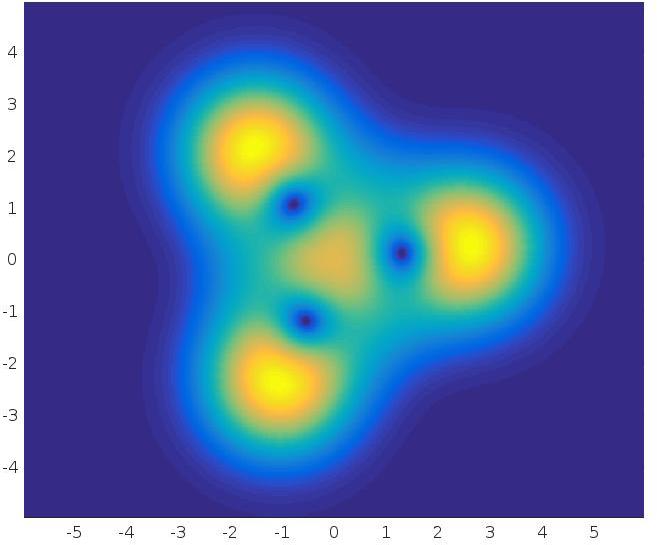}}
  \mbox{
    \rotatebox{90}{\quad$\epsilon=0.379$}
    \includegraphics[width=\fwidth]{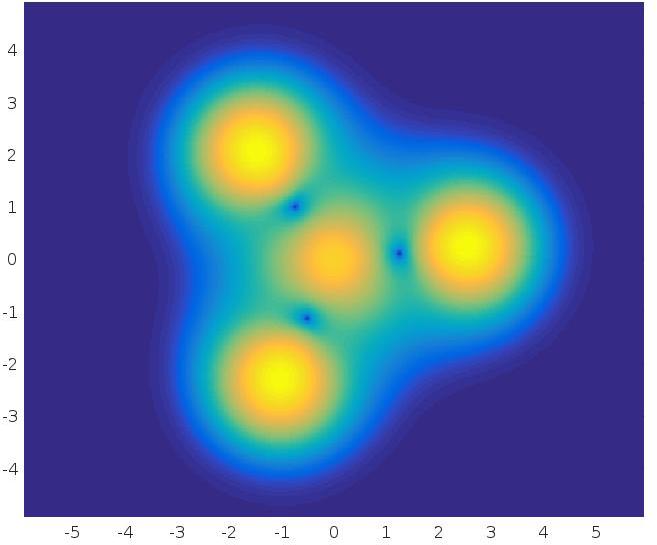}
    \includegraphics[width=\fwidth]{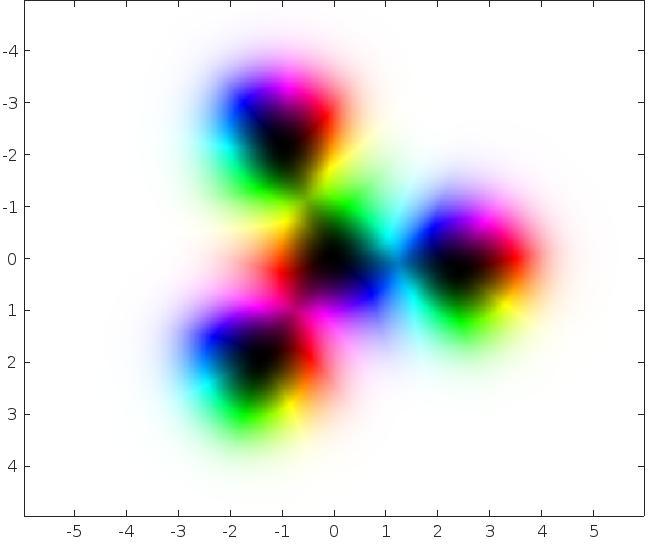}
    \includegraphics[width=\fwidth]{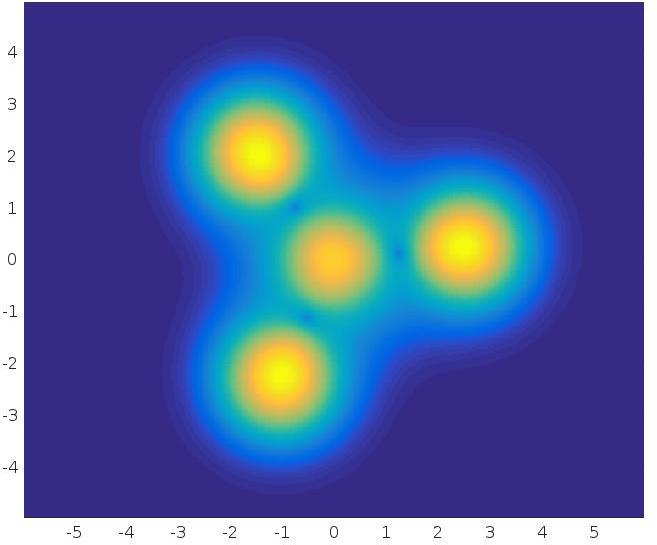}
    \includegraphics[width=\fwidth]{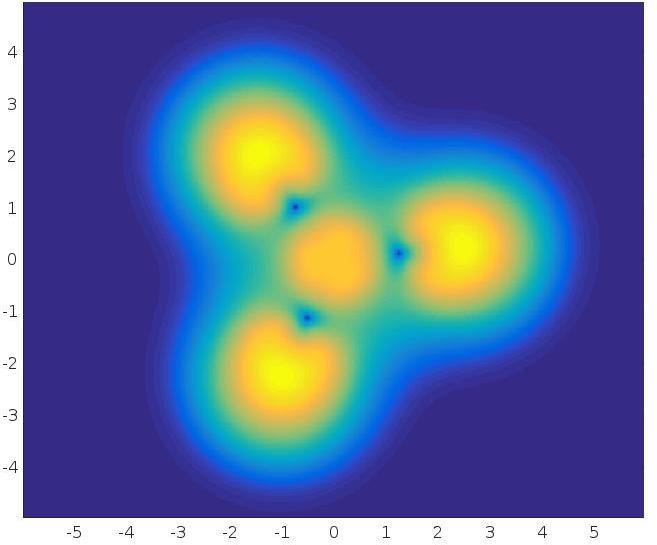}}
  \mbox{
    \rotatebox{90}{\quad$\epsilon=0.183$}
    \includegraphics[width=\fwidth]{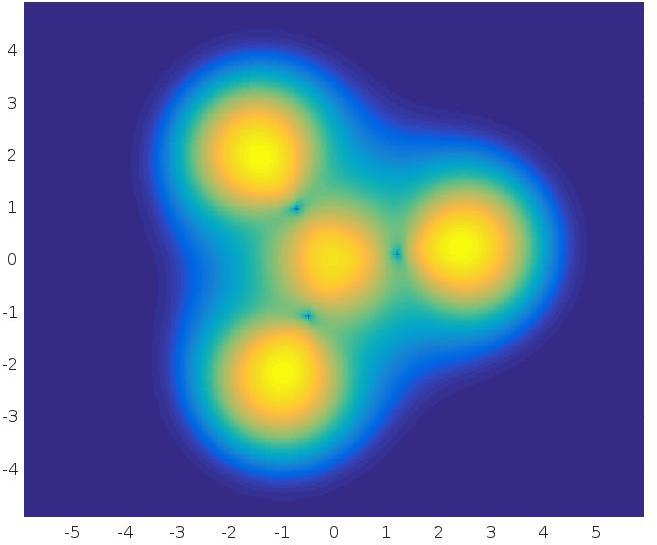}
    \includegraphics[width=\fwidth]{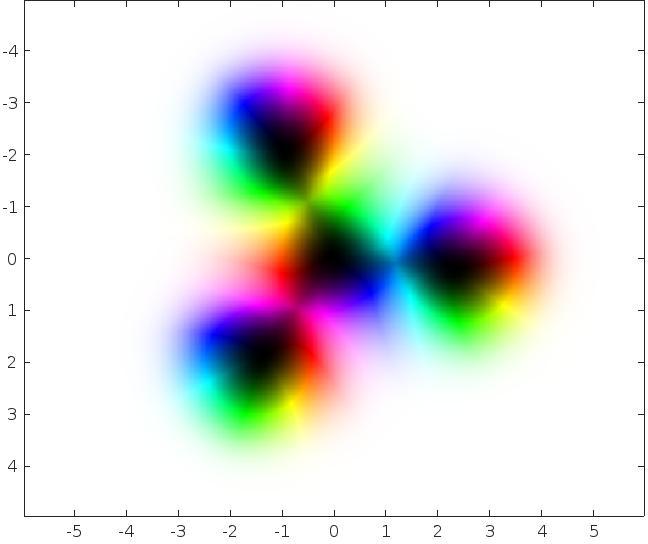}
    \includegraphics[width=\fwidth]{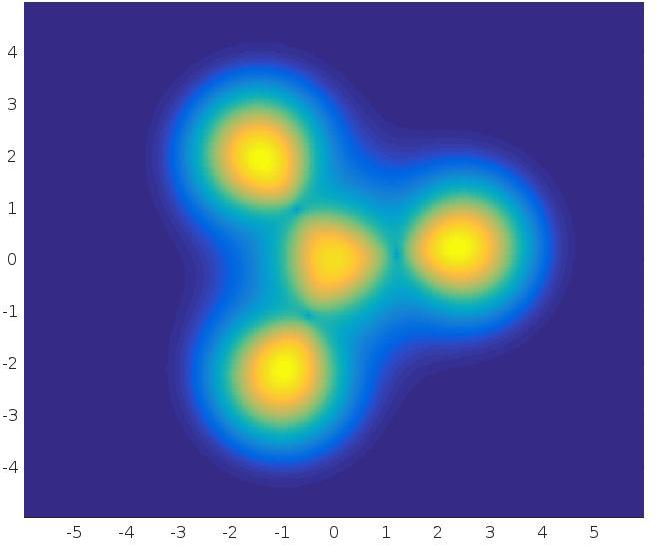}
    \includegraphics[width=\fwidth]{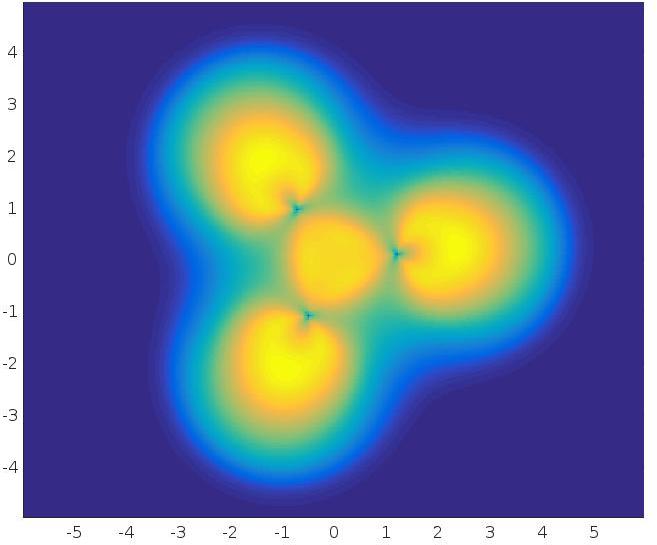}}
  \mbox{
    \rotatebox{90}{\quad$\epsilon=0.0886$}
    \includegraphics[width=\fwidth]{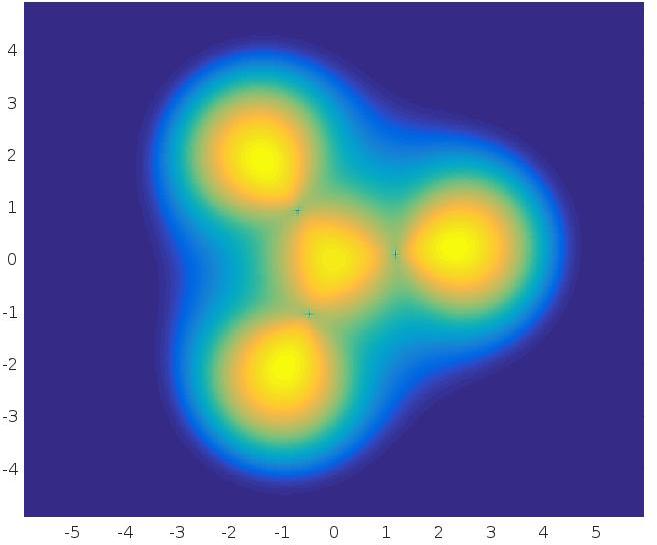}
    \includegraphics[width=\fwidth]{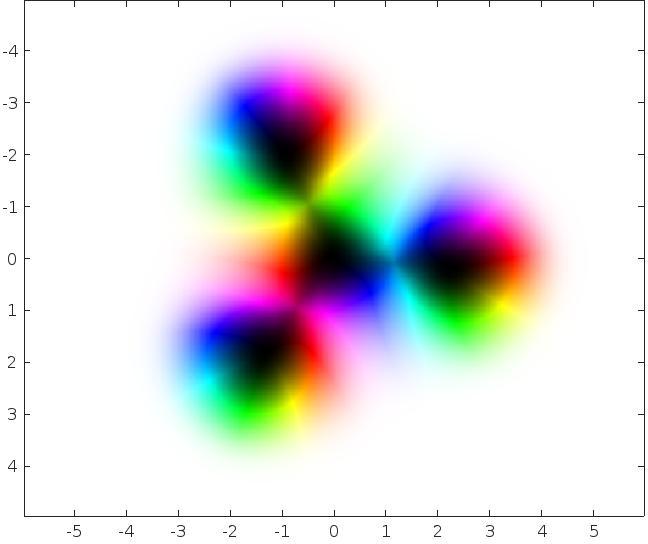}
    \includegraphics[width=\fwidth]{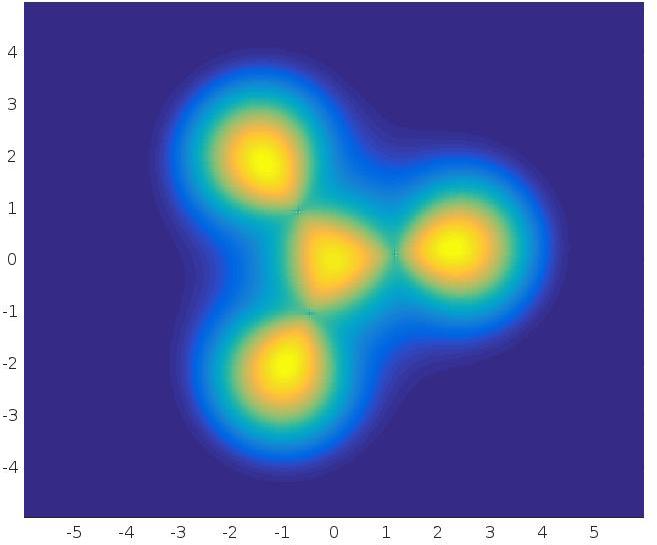}
    \includegraphics[width=\fwidth]{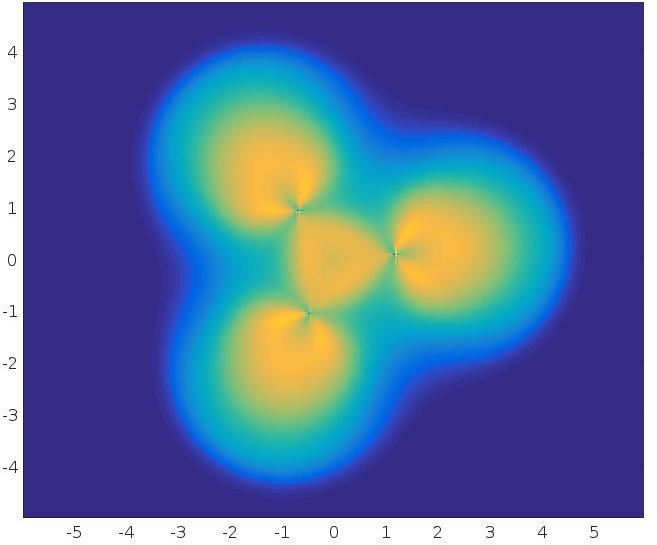}}
  \mbox{
    \rotatebox{90}{\quad$\epsilon=0.0428$}
    \includegraphics[width=\fwidth]{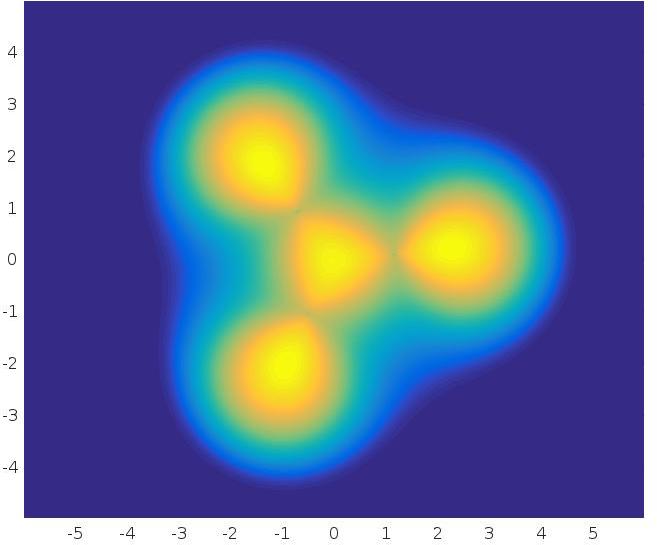}
    \includegraphics[width=\fwidth]{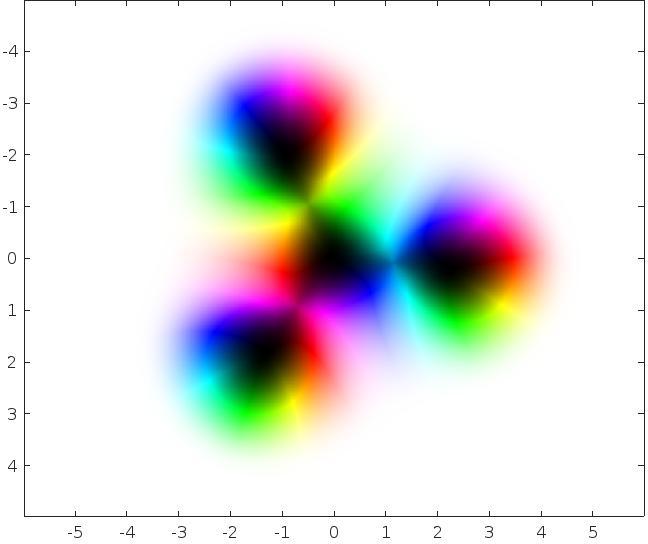}
    \includegraphics[width=\fwidth]{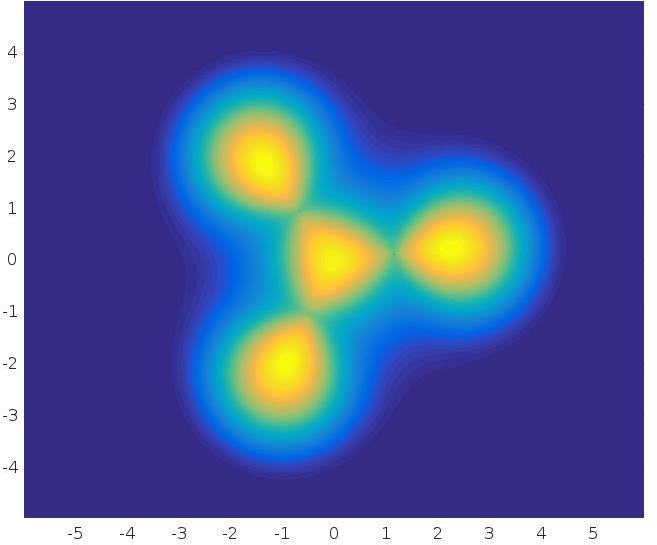}
    \includegraphics[width=\fwidth]{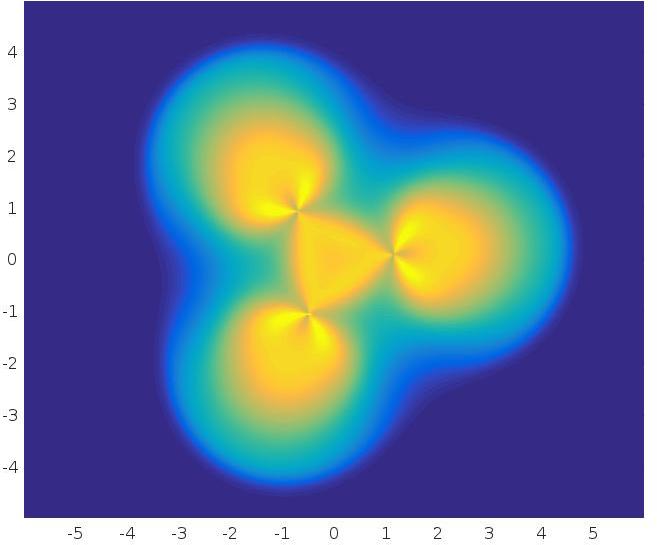}}
  \mbox{
    \rotatebox{90}{\quad$\epsilon=0.0207$}
    \includegraphics[width=\fwidth]{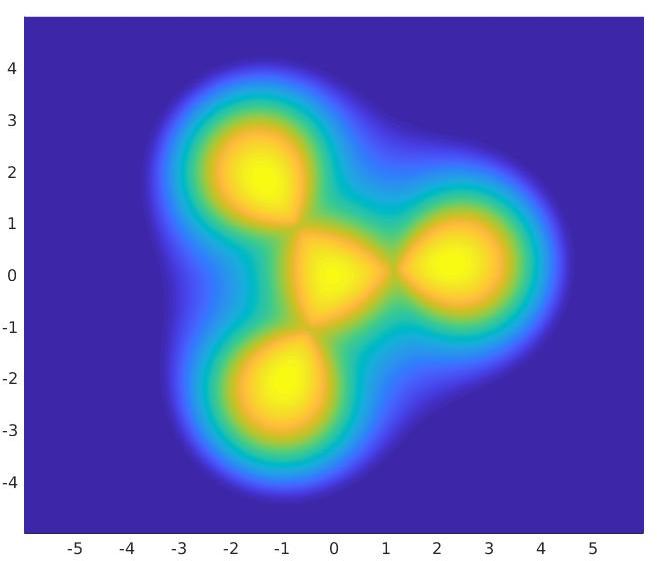}
    \includegraphics[width=\fwidth]{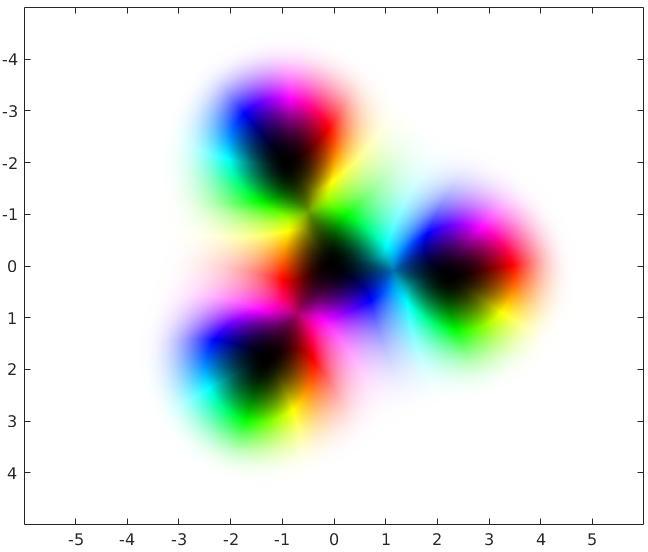}
    \includegraphics[width=\fwidth]{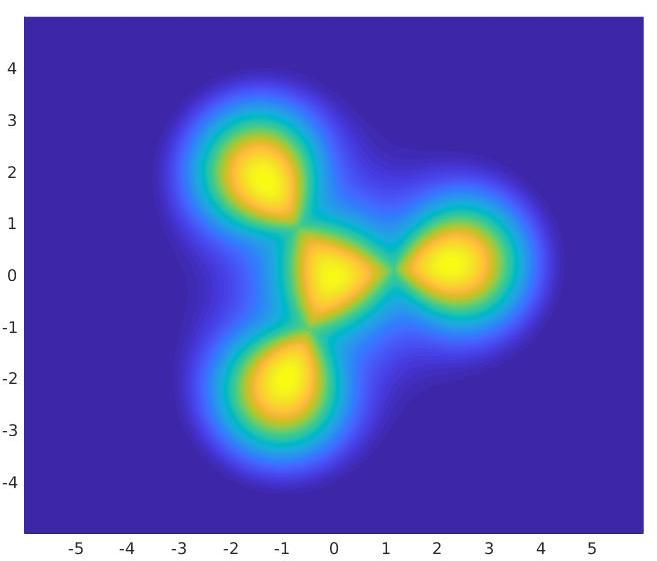}
    \includegraphics[width=\fwidth]{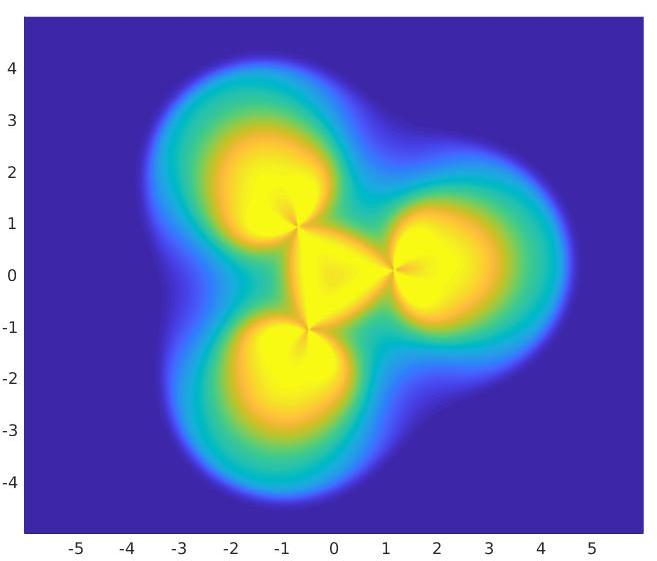}}
  \caption{The $Q=1+1+1+1$ tetrahedral baby-Skyrmion solution as a
    function of $\epsilon$.
  The columns display the topological charge density, the field
  orientation using the color scheme described in the text, the total
  energy and finally the kinetic term $-\epsilon\Lag_2$.
  We note that the tetrahedrally symmetric baby Skyrmion exists also
  for large $\epsilon$ (i.e.~$\epsilon=1$).
  }
  \label{fig:N=1+1+1+1_triangle}
  \end{center}
\end{figure}

\begin{figure}[!t]
  \begin{center}
  \mbox{
    \rotatebox{90}{$\epsilon=0.183$}
    \includegraphics[width=\fwidth]{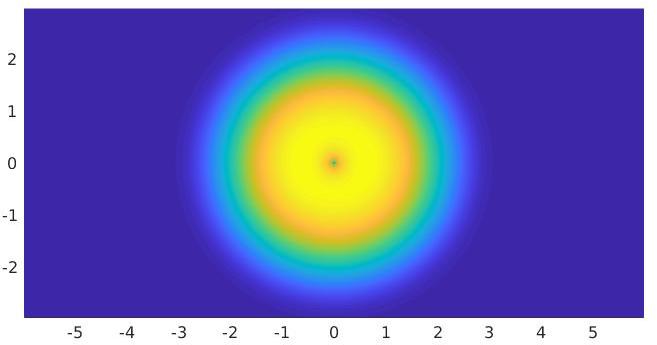}
    \includegraphics[width=\fwidth]{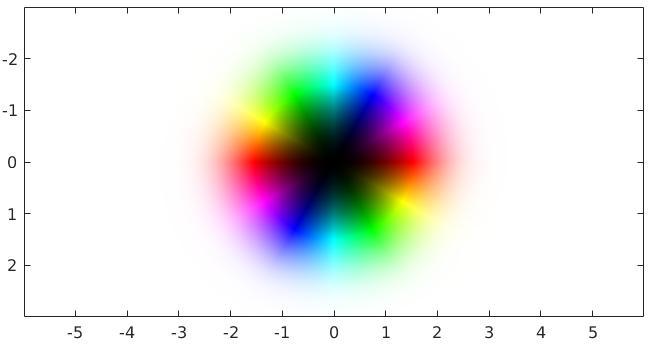}
    \includegraphics[width=\fwidth]{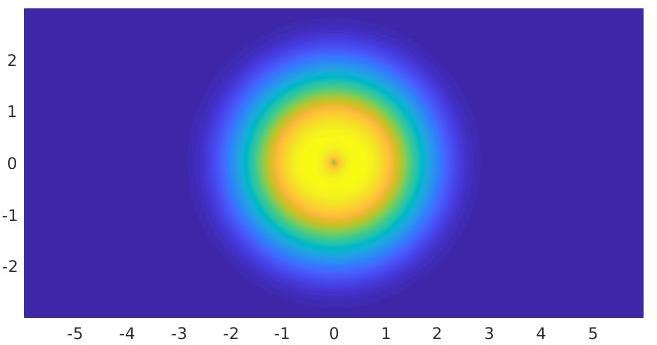}
    \includegraphics[width=\fwidth]{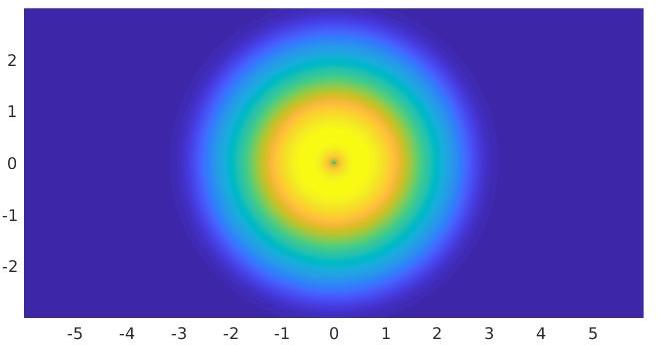}}
  \mbox{
    \rotatebox{90}{$\epsilon=0.144$}
    \includegraphics[width=\fwidth]{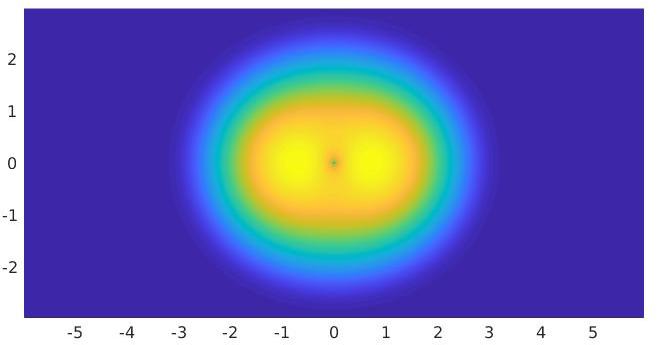}
    \includegraphics[width=\fwidth]{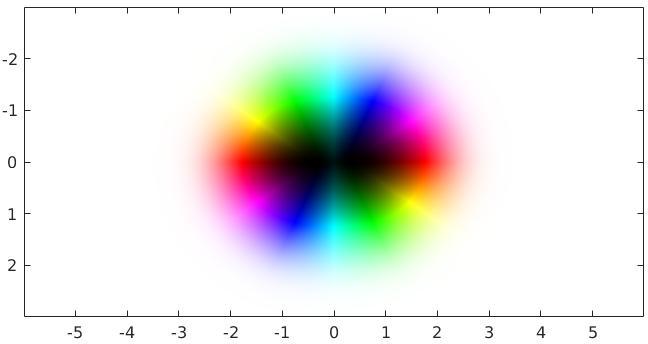}
    \includegraphics[width=\fwidth]{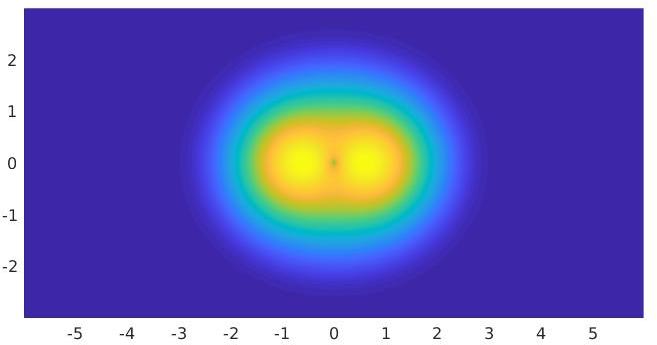}
    \includegraphics[width=\fwidth]{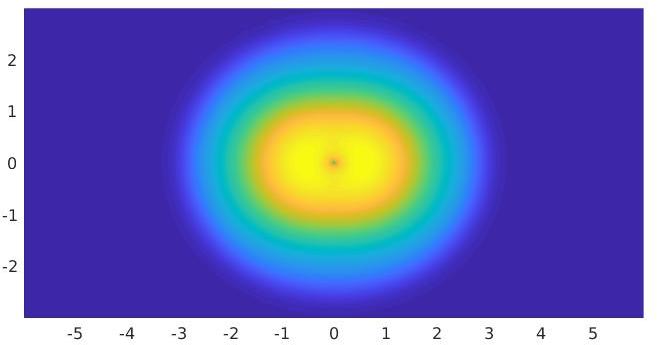}}
  \mbox{
    \rotatebox{90}{$\epsilon=0.113$}
    \includegraphics[width=\fwidth]{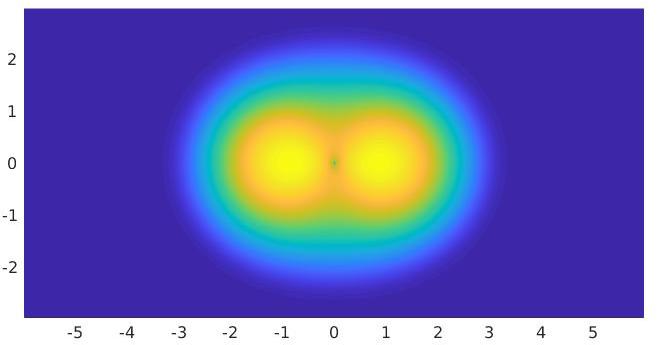}
    \includegraphics[width=\fwidth]{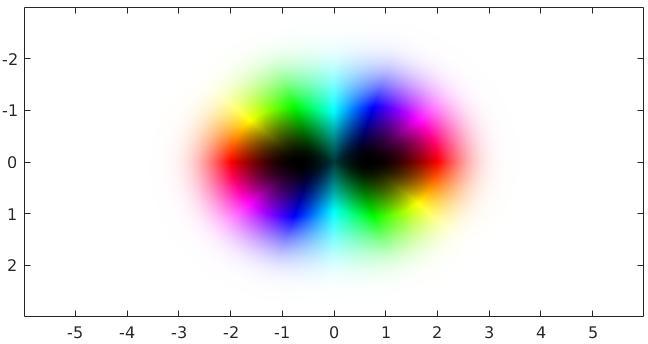}
    \includegraphics[width=\fwidth]{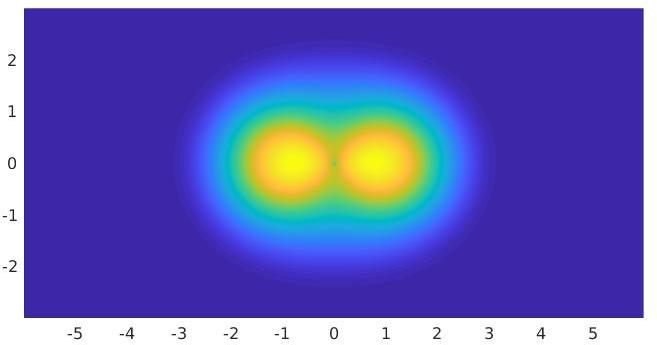}
    \includegraphics[width=\fwidth]{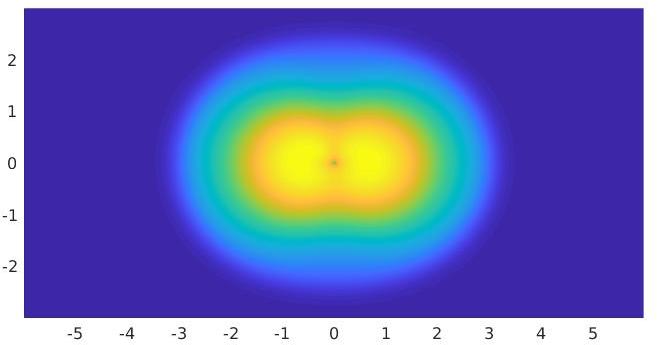}}
  \mbox{
    \rotatebox{90}{$\epsilon=0.0886$}
    \includegraphics[width=\fwidth]{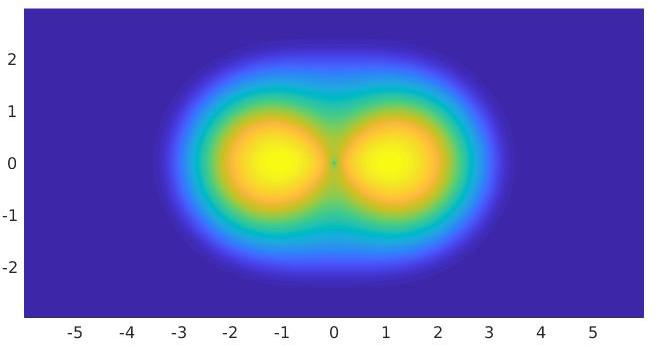}
    \includegraphics[width=\fwidth]{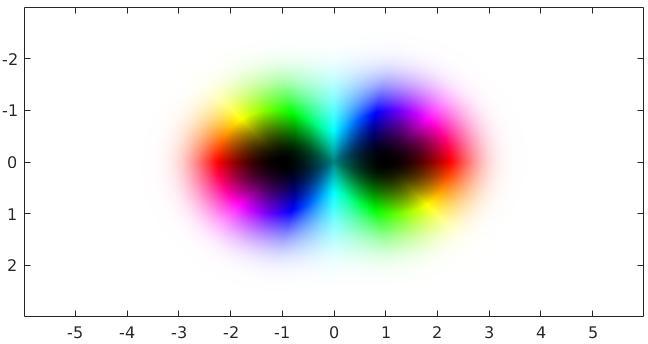}
    \includegraphics[width=\fwidth]{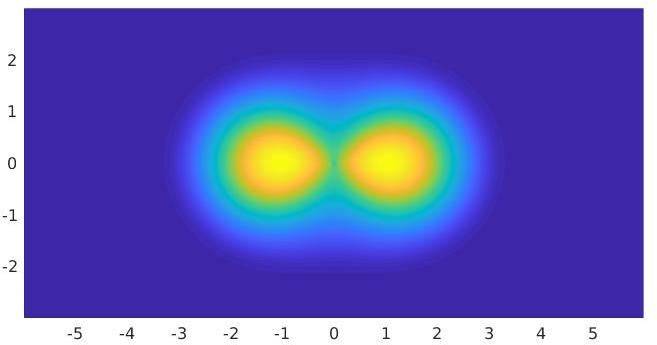}
    \includegraphics[width=\fwidth]{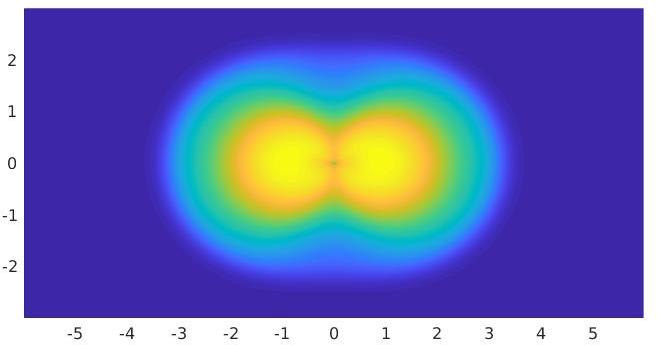}}
  \mbox{
    \rotatebox{90}{$\epsilon=0.0695$}
    \includegraphics[width=\fwidth]{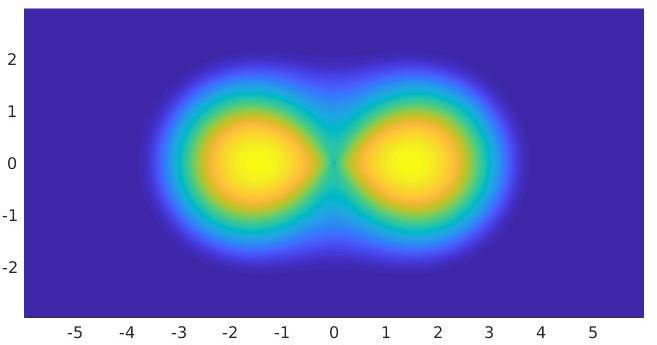}
    \includegraphics[width=\fwidth]{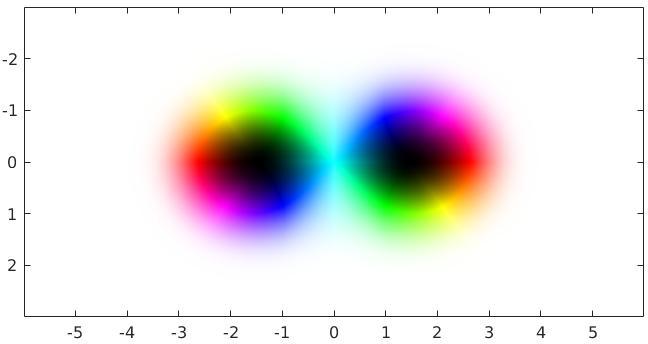}
    \includegraphics[width=\fwidth]{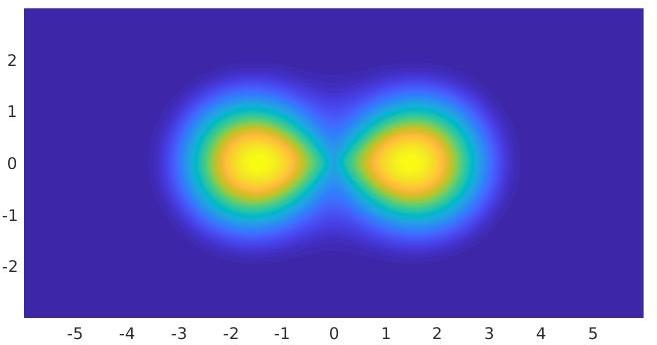}
    \includegraphics[width=\fwidth]{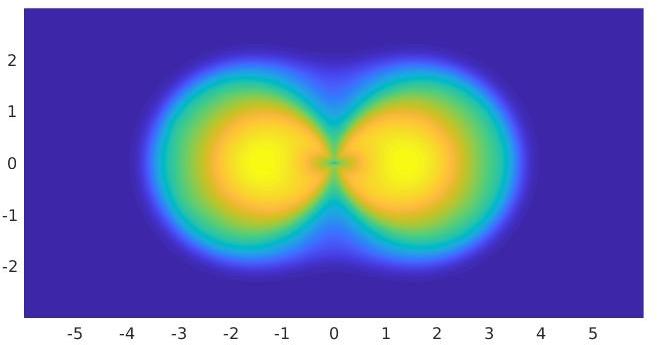}}
  \mbox{
    \rotatebox{90}{$\epsilon=0.0546$}
    \includegraphics[width=\fwidth]{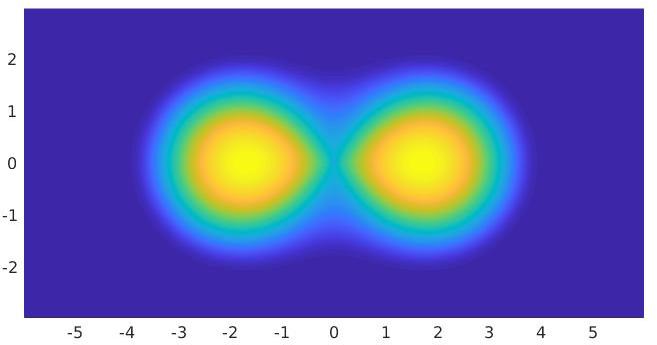}
    \includegraphics[width=\fwidth]{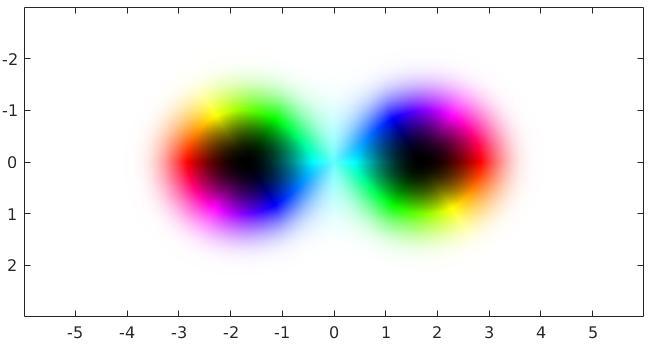}
    \includegraphics[width=\fwidth]{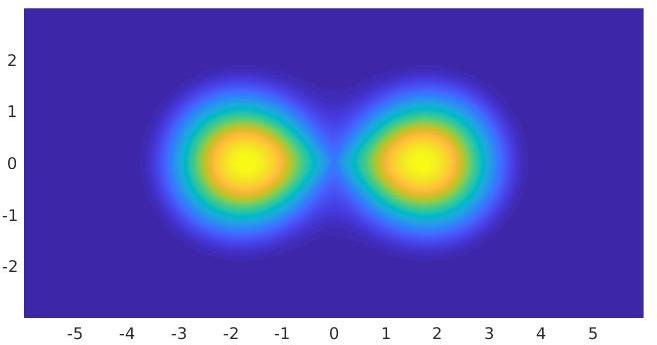}
    \includegraphics[width=\fwidth]{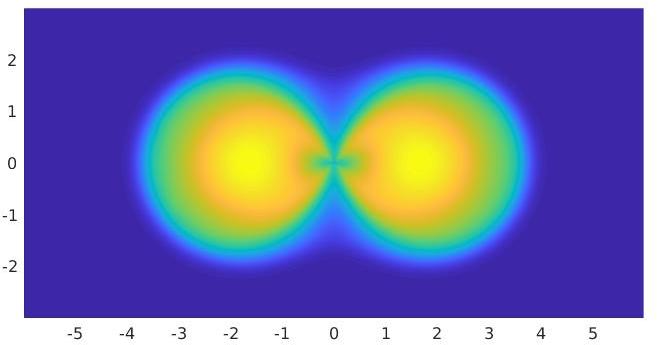}}
  \mbox{
    \rotatebox{90}{$\epsilon=0.0428$}
    \includegraphics[width=\fwidth]{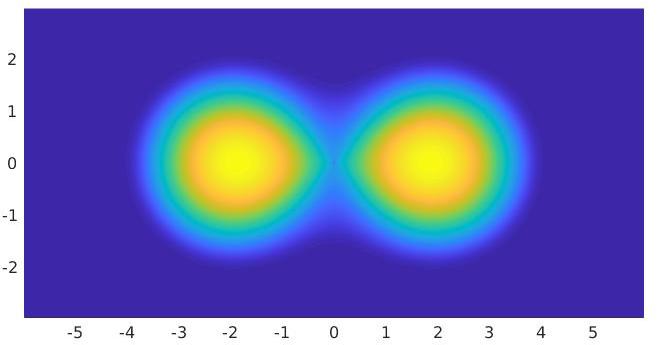}
    \includegraphics[width=\fwidth]{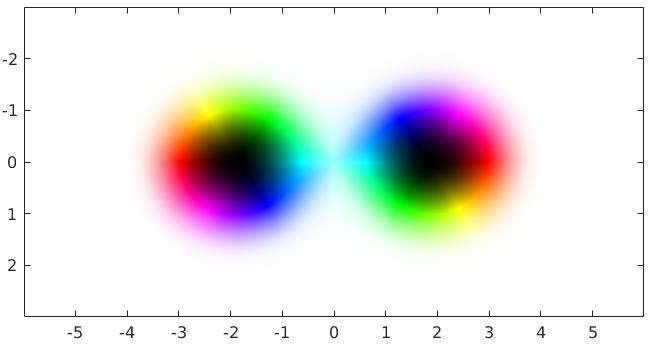}
    \includegraphics[width=\fwidth]{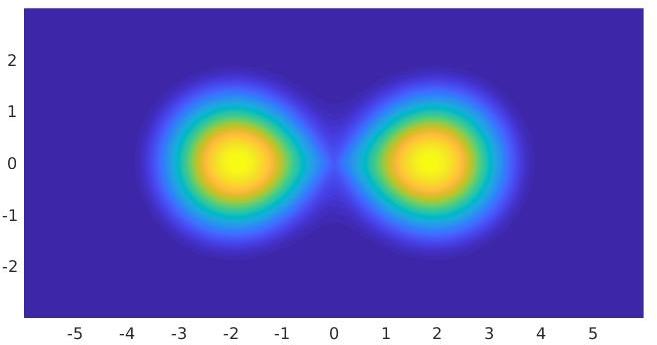}
    \includegraphics[width=\fwidth]{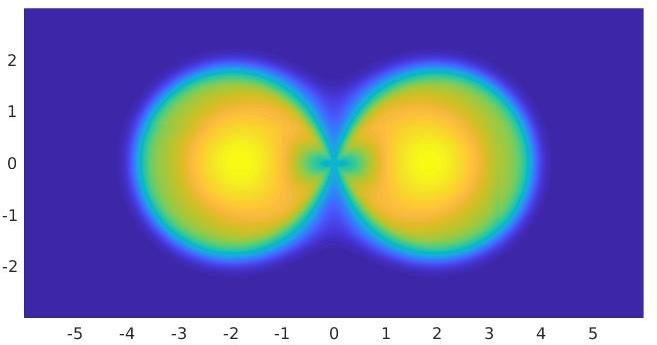}}
  \mbox{
    \rotatebox{90}{$\epsilon=0.0207$}
    \includegraphics[width=\fwidth]{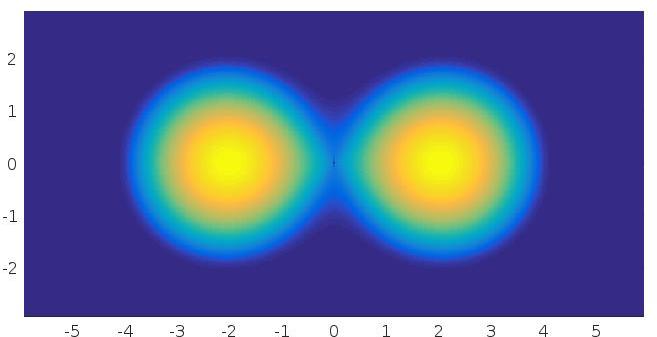}
    \includegraphics[width=\fwidth]{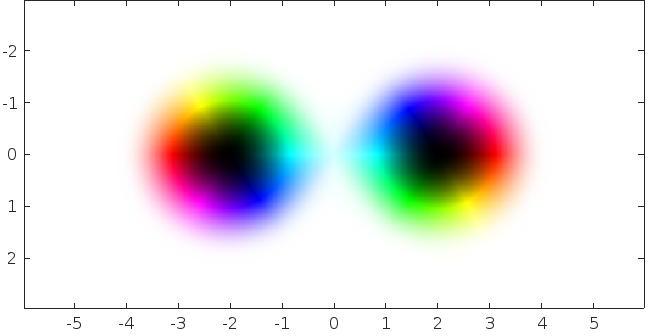}
    \includegraphics[width=\fwidth]{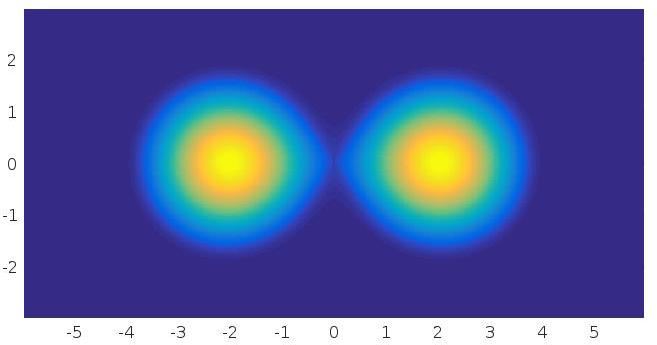}
    \includegraphics[width=\fwidth]{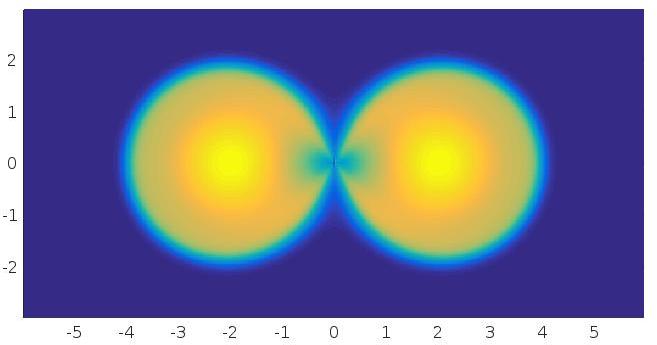}}
  \caption{The $Q=1+1$ baby-Skyrmion solution as a function of
    $\epsilon$.
  The columns display the topological charge density, the field
  orientation using the color scheme described in the text, the total
  energy and finally the kinetic term $-\epsilon\Lag_2$.
}
  \label{fig:N=1+1}
  \end{center}
\end{figure}

\begin{figure}[!t]
  \begin{center}
  \mbox{
    \rotatebox{90}{\quad$\epsilon=1$}
    \includegraphics[width=\fwidth]{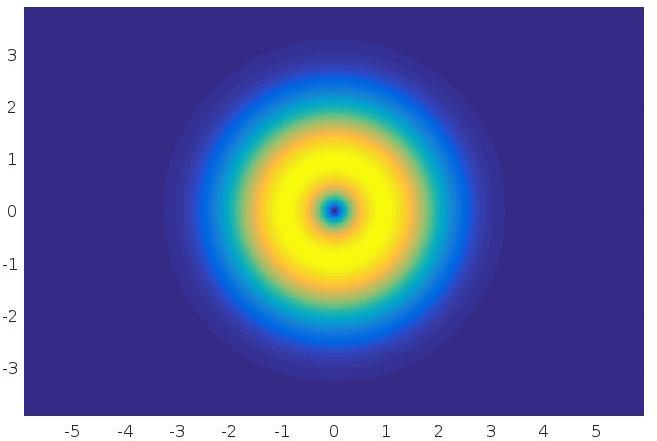}
    \includegraphics[width=\fwidth]{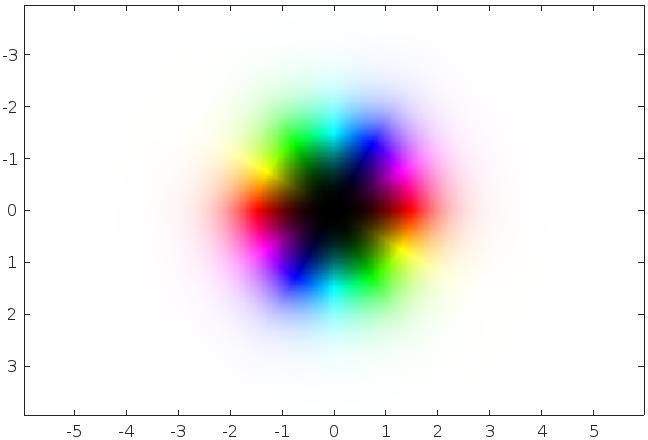}
    \includegraphics[width=\fwidth]{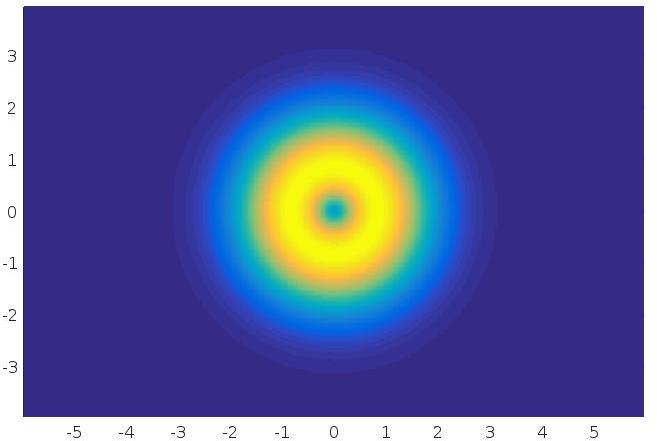}
    \includegraphics[width=\fwidth]{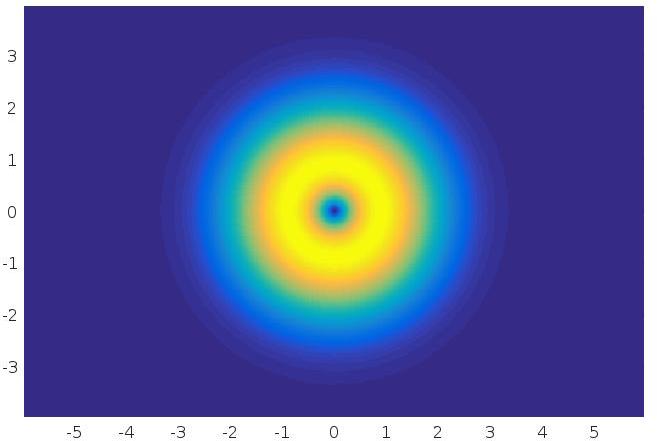}}
  \mbox{
    \rotatebox{90}{\quad$\epsilon=0.616$}
    \includegraphics[width=\fwidth]{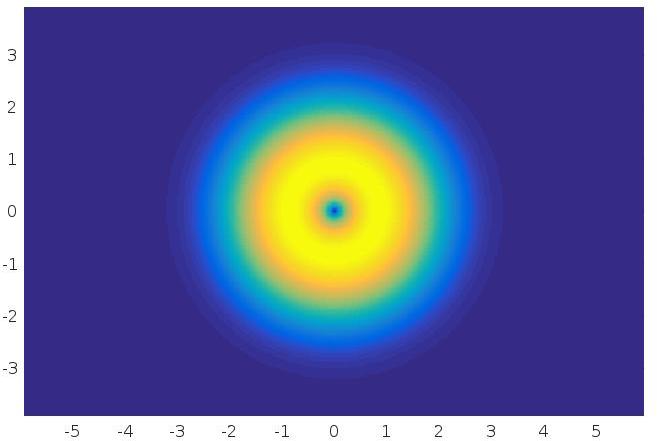}
    \includegraphics[width=\fwidth]{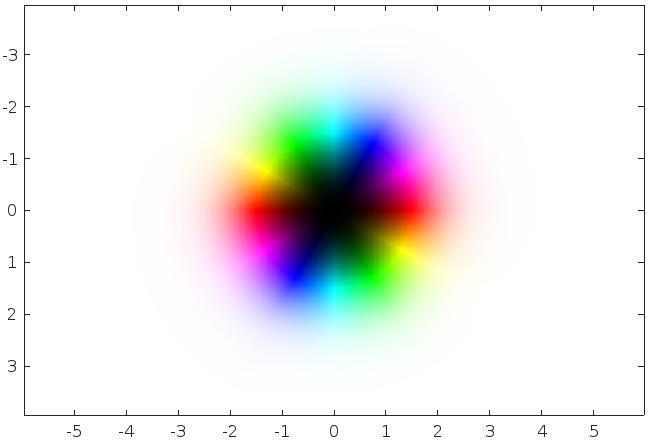}
    \includegraphics[width=\fwidth]{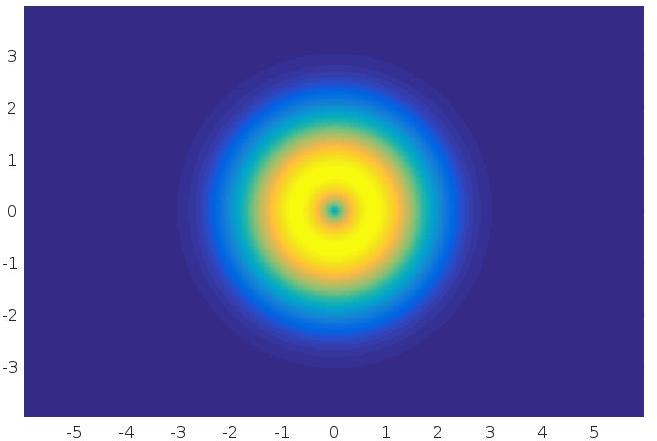}
    \includegraphics[width=\fwidth]{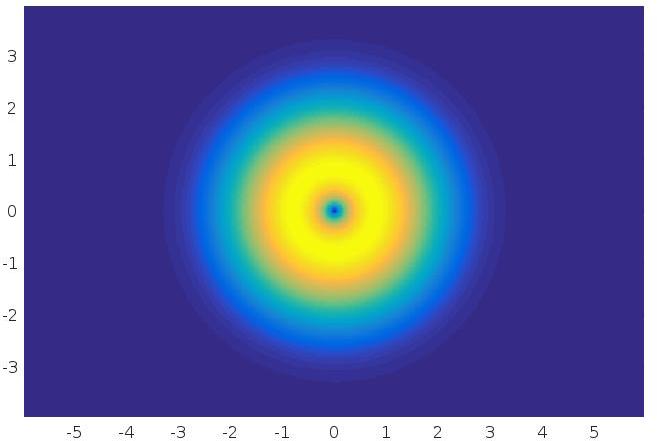}}
  \mbox{
    \rotatebox{90}{\quad$\epsilon=0.379$}
    \includegraphics[width=\fwidth]{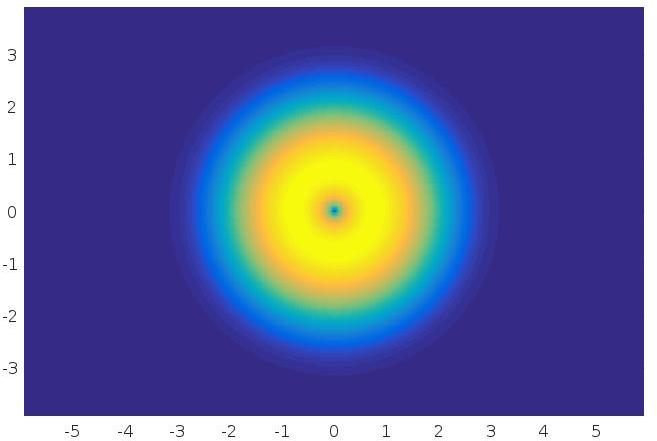}
    \includegraphics[width=\fwidth]{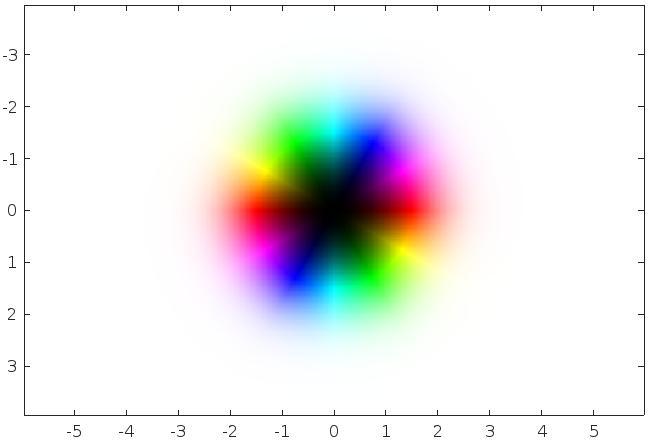}
    \includegraphics[width=\fwidth]{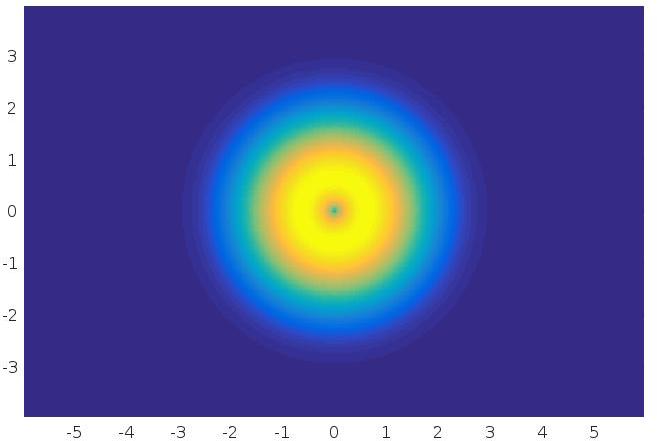}
    \includegraphics[width=\fwidth]{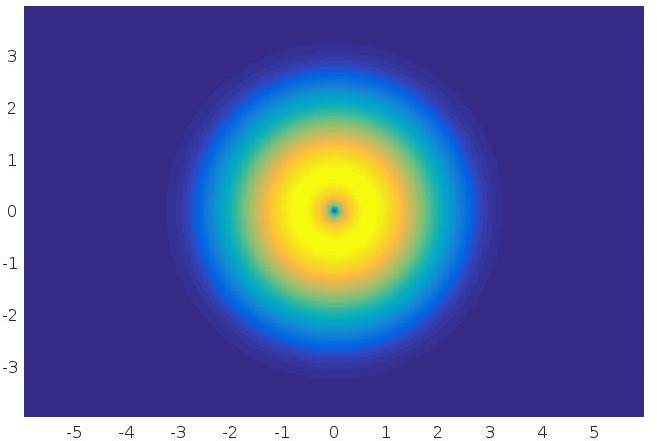}}
  \mbox{
    \rotatebox{90}{\quad$\epsilon=0.183$}
    \includegraphics[width=\fwidth]{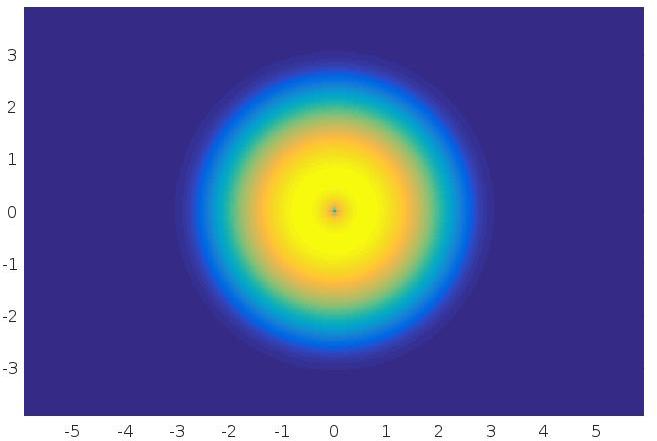}
    \includegraphics[width=\fwidth]{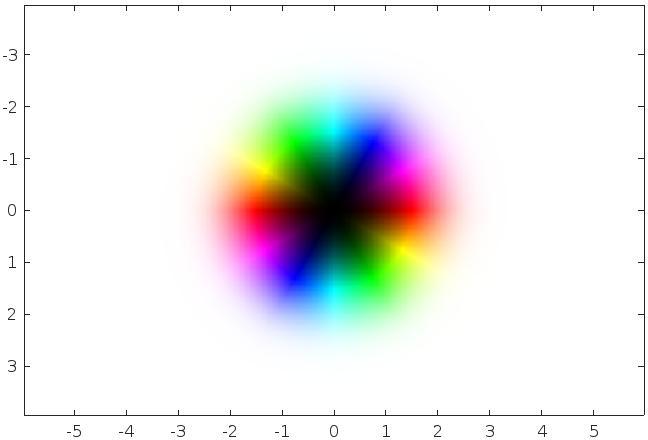}
    \includegraphics[width=\fwidth]{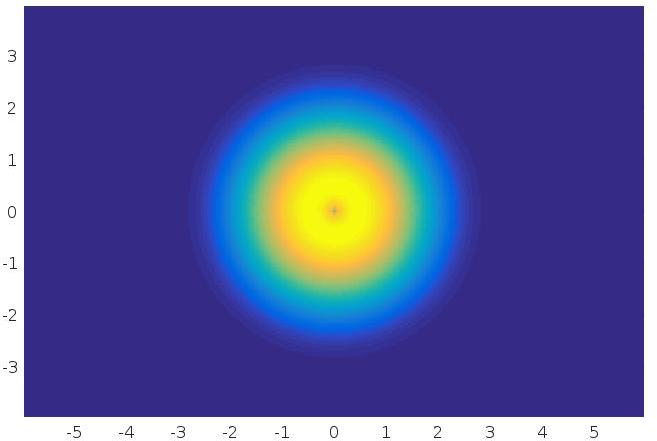}
    \includegraphics[width=\fwidth]{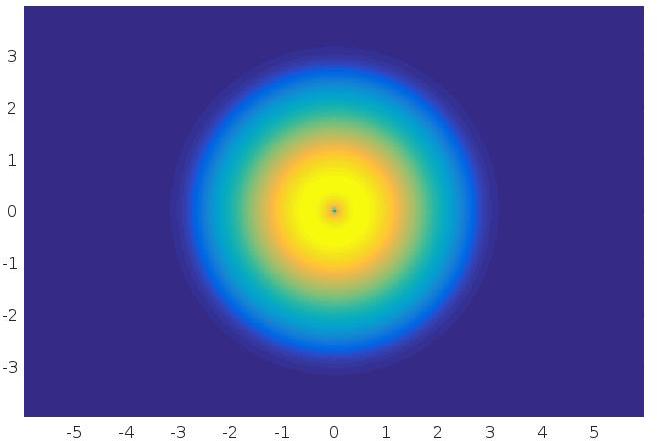}}
  \mbox{
    \rotatebox{90}{\quad$\epsilon=0.0886$}
    \includegraphics[width=\fwidth]{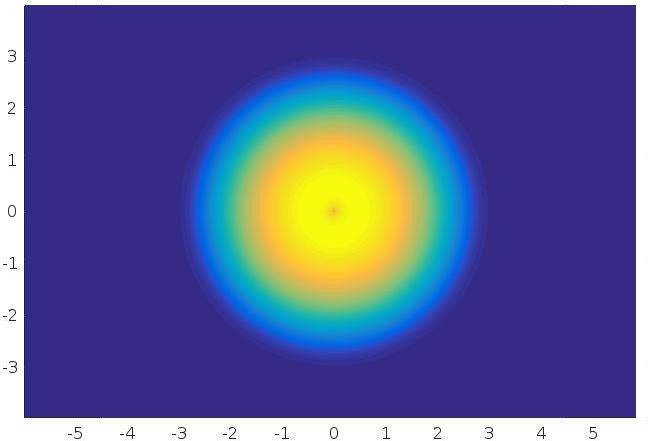}
    \includegraphics[width=\fwidth]{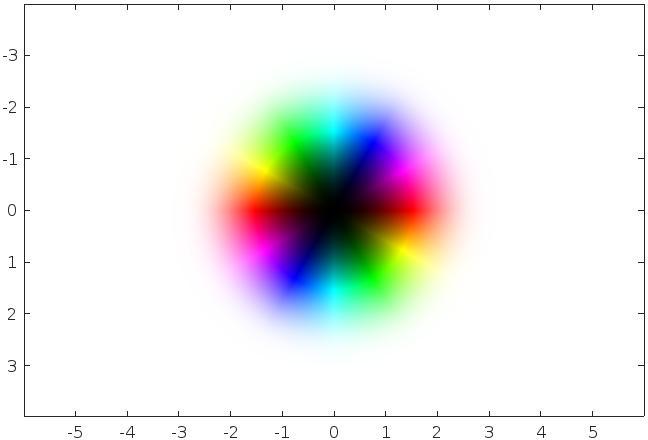}
    \includegraphics[width=\fwidth]{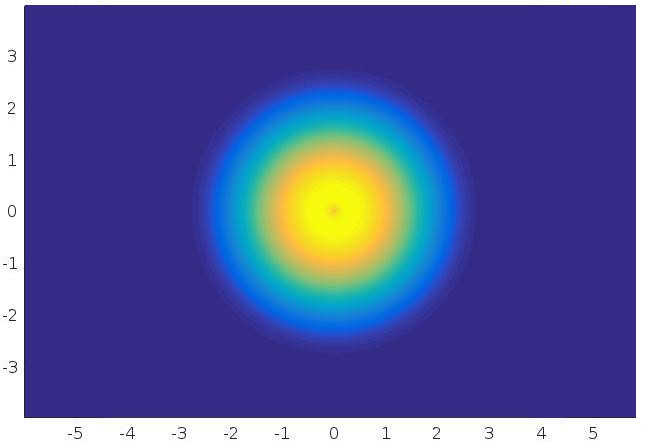}
    \includegraphics[width=\fwidth]{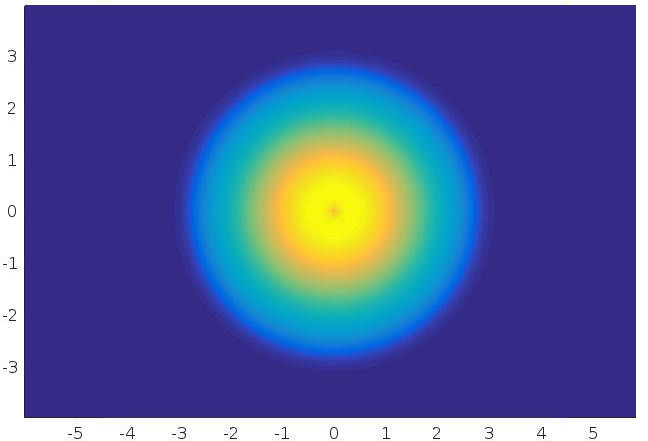}}
  \mbox{
    \rotatebox{90}{\quad$\epsilon=0.0428$}
    \includegraphics[width=\fwidth]{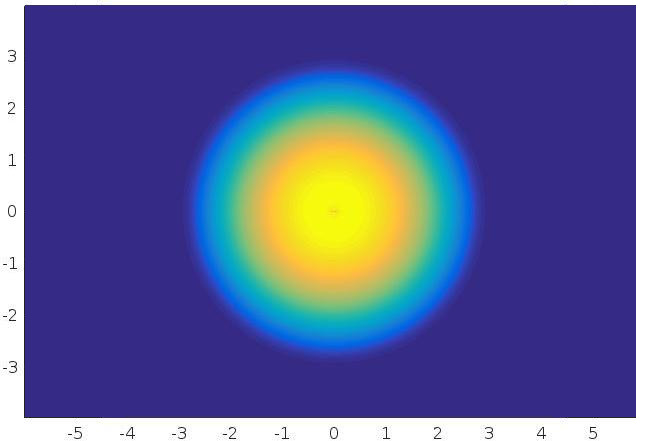}
    \includegraphics[width=\fwidth]{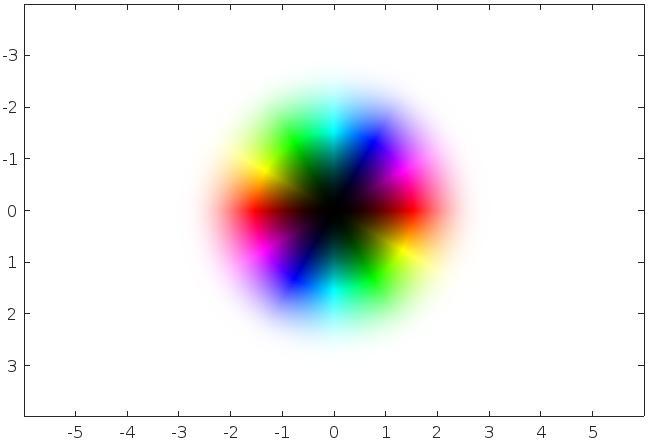}
    \includegraphics[width=\fwidth]{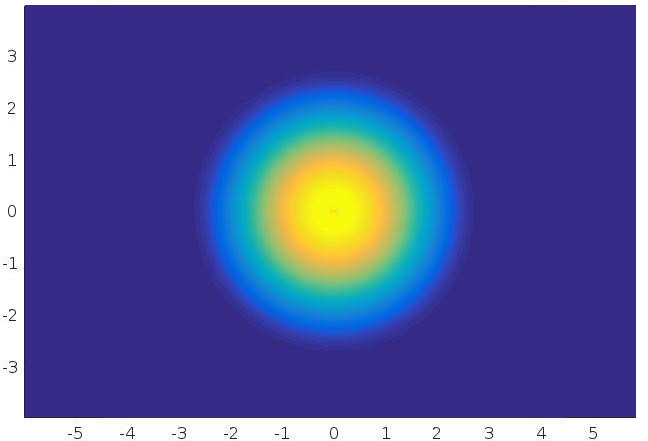}
    \includegraphics[width=\fwidth]{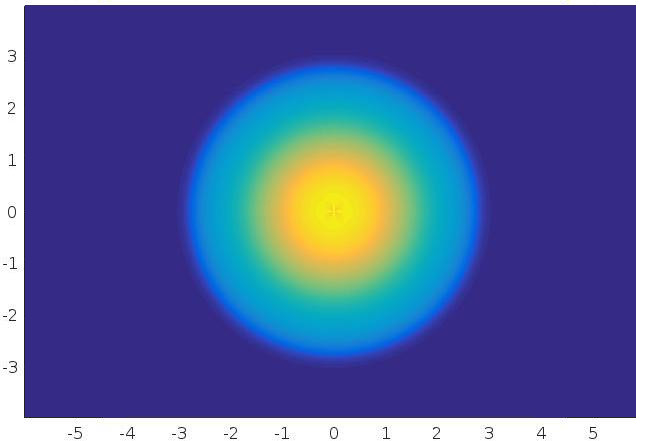}}
  \mbox{
    \rotatebox{90}{\quad$\epsilon=0.0207$}
    \includegraphics[width=\fwidth]{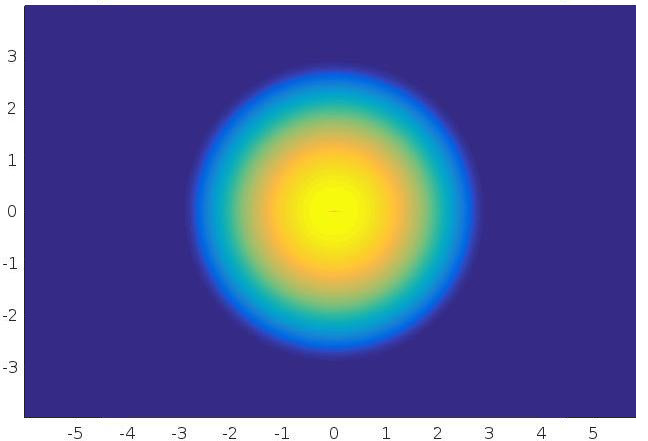}
    \includegraphics[width=\fwidth]{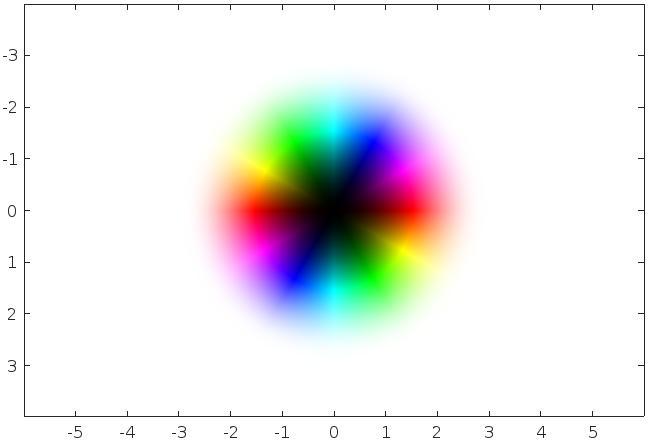}
    \includegraphics[width=\fwidth]{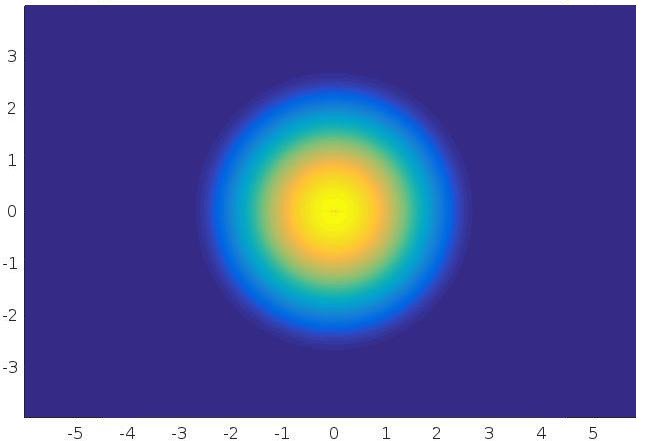}
    \includegraphics[width=\fwidth]{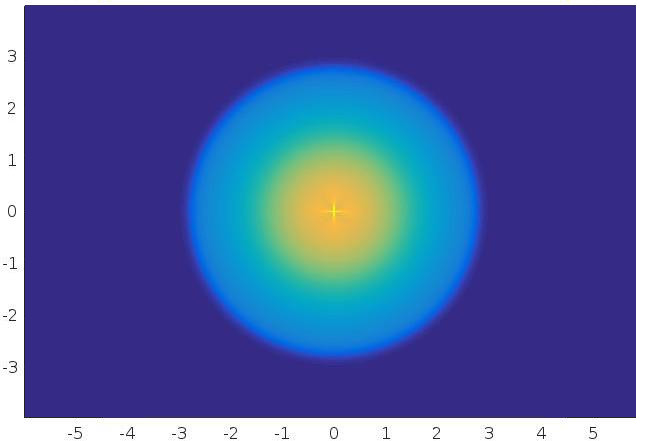}}
  \caption{The $N=2$ baby-Skyrmion solution as a function of
    $\epsilon$, which is the stable $Q=2$ soliton.
  The columns display the topological charge density, the field
  orientation using the color scheme described in the text, the total
  energy and finally the kinetic term $-\epsilon\Lag_2$. }
  \label{fig:N=2}
  \end{center}
\end{figure}

\begin{figure}[!ht]
  \begin{center}
    \mbox{\includegraphics[width=0.49\linewidth]{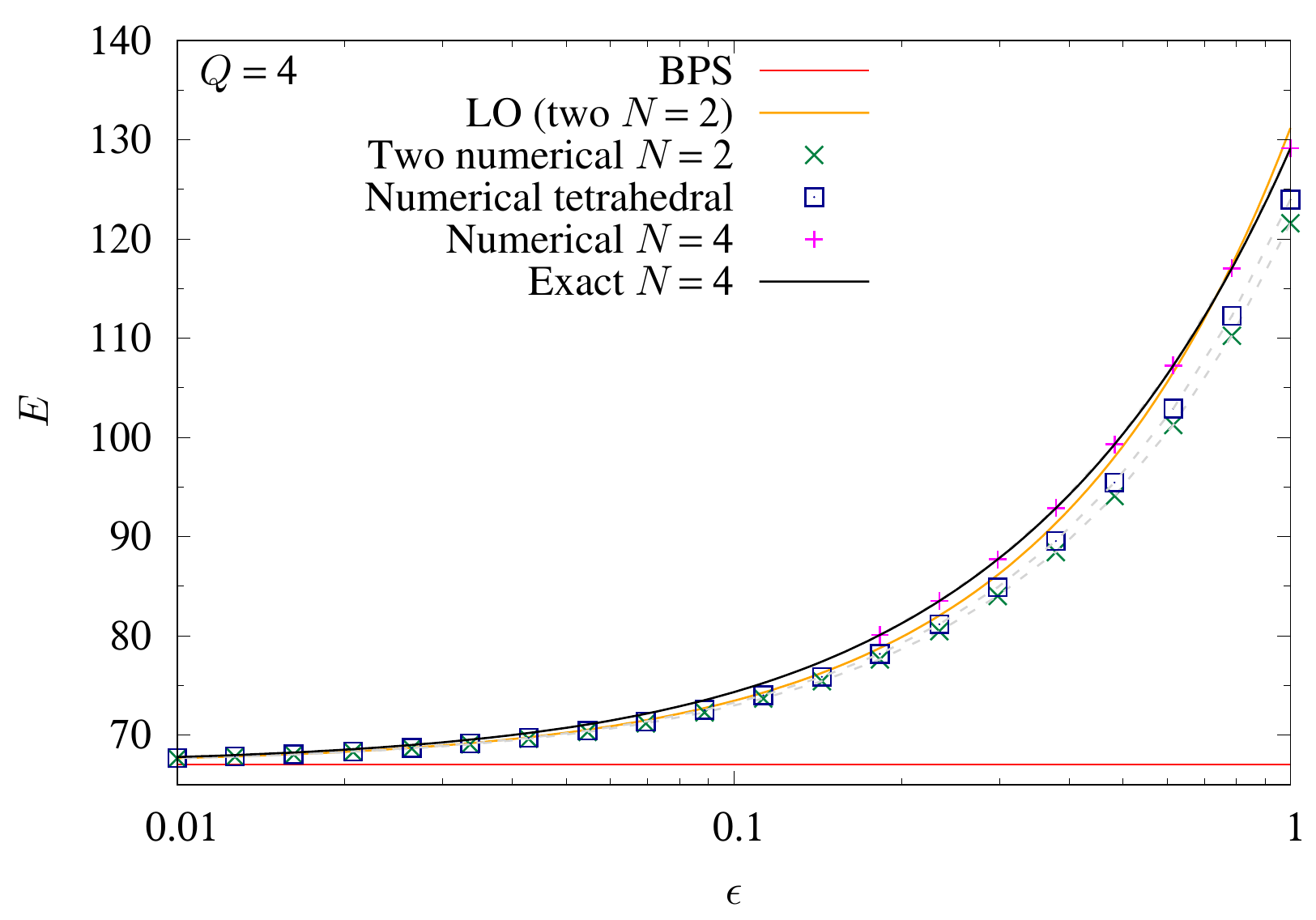}
      \includegraphics[width=0.49\linewidth]{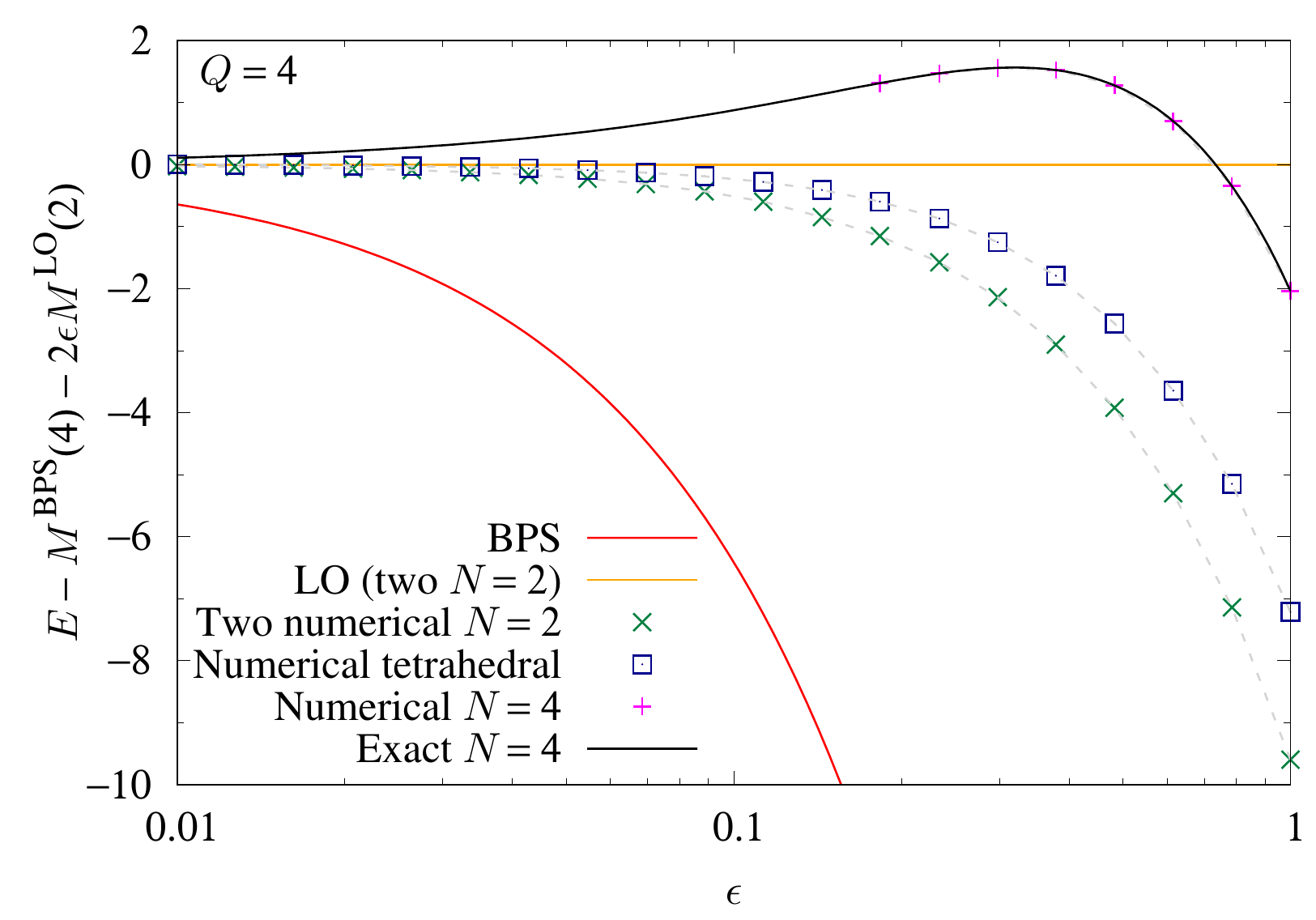}}
    \caption{The total energies of the baby-Skyrmion solutions in the
      $Q=4$ sector. The red line is the BPS bound, the orange curve
      the LO correction (coming from the kinetic term) for two $N=2$
      Skyrmions. The numerical solutions are: the two $N=2$ Skyrmions
      side-by-side (green xs), the tetrahedral Skyrmion (blue squares
      with a dot in the middle), and the axially symmetric $N=4$
      Skyrmion (magenta pluses). Finally, the latter numerical $N=4$
      solution is compared to the exact solution (using the
      1-dimensional ODE).
      For clarity, the right-hand side panel shows the same data
      relative to the orange line, being the LO correction.
      At $\epsilon\lesssim 0.15$ the axially symmetric soliton becomes
      unstable and decays to the tetrahedral one -- which is only
      metastable.
    } 
    \label{fig:NumericalEnQ4}
  \end{center}
\end{figure}

\begin{figure}[!ht]
  \begin{center}
    \mbox{\includegraphics[width=0.49\linewidth]{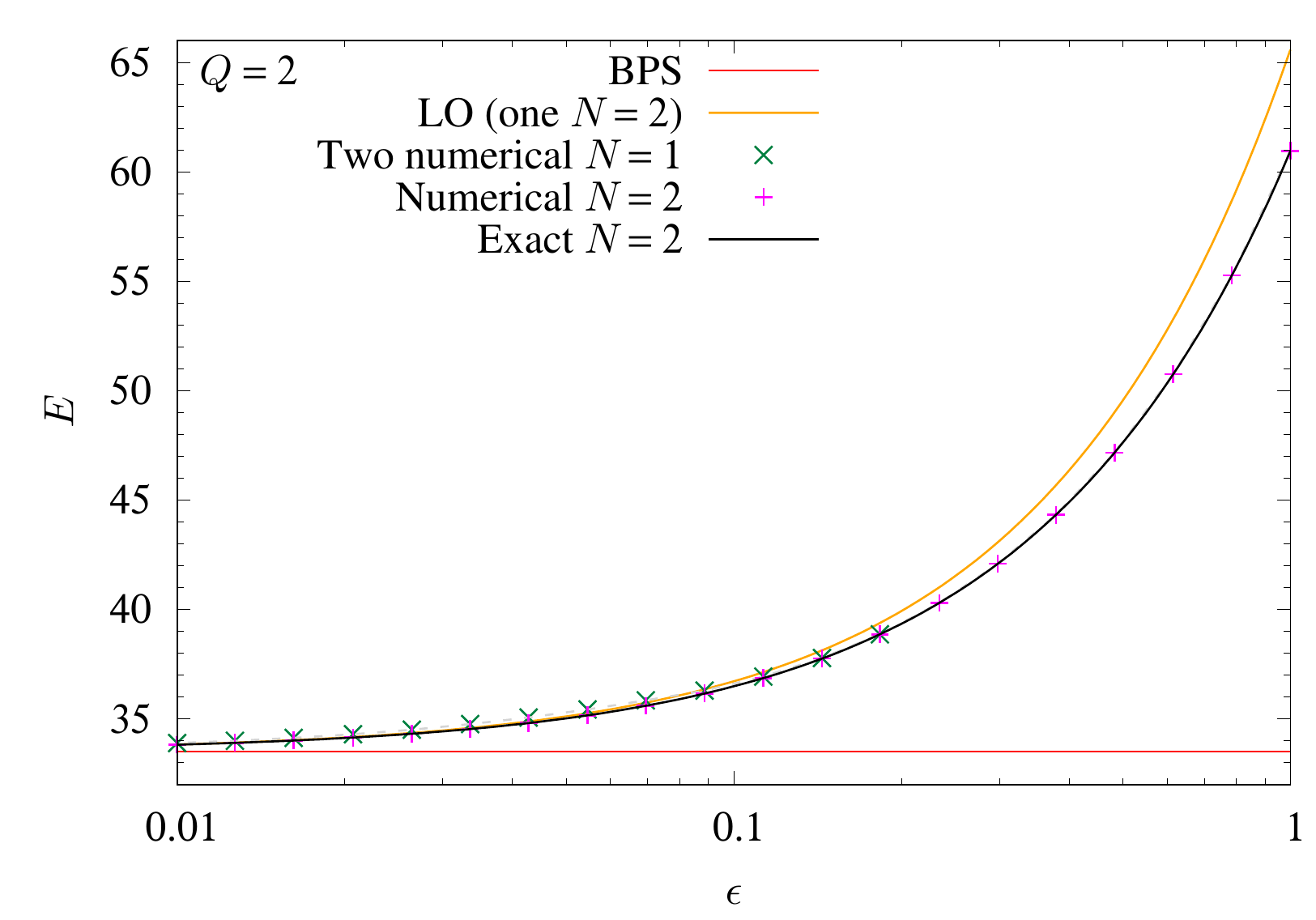}
      \includegraphics[width=0.49\linewidth]{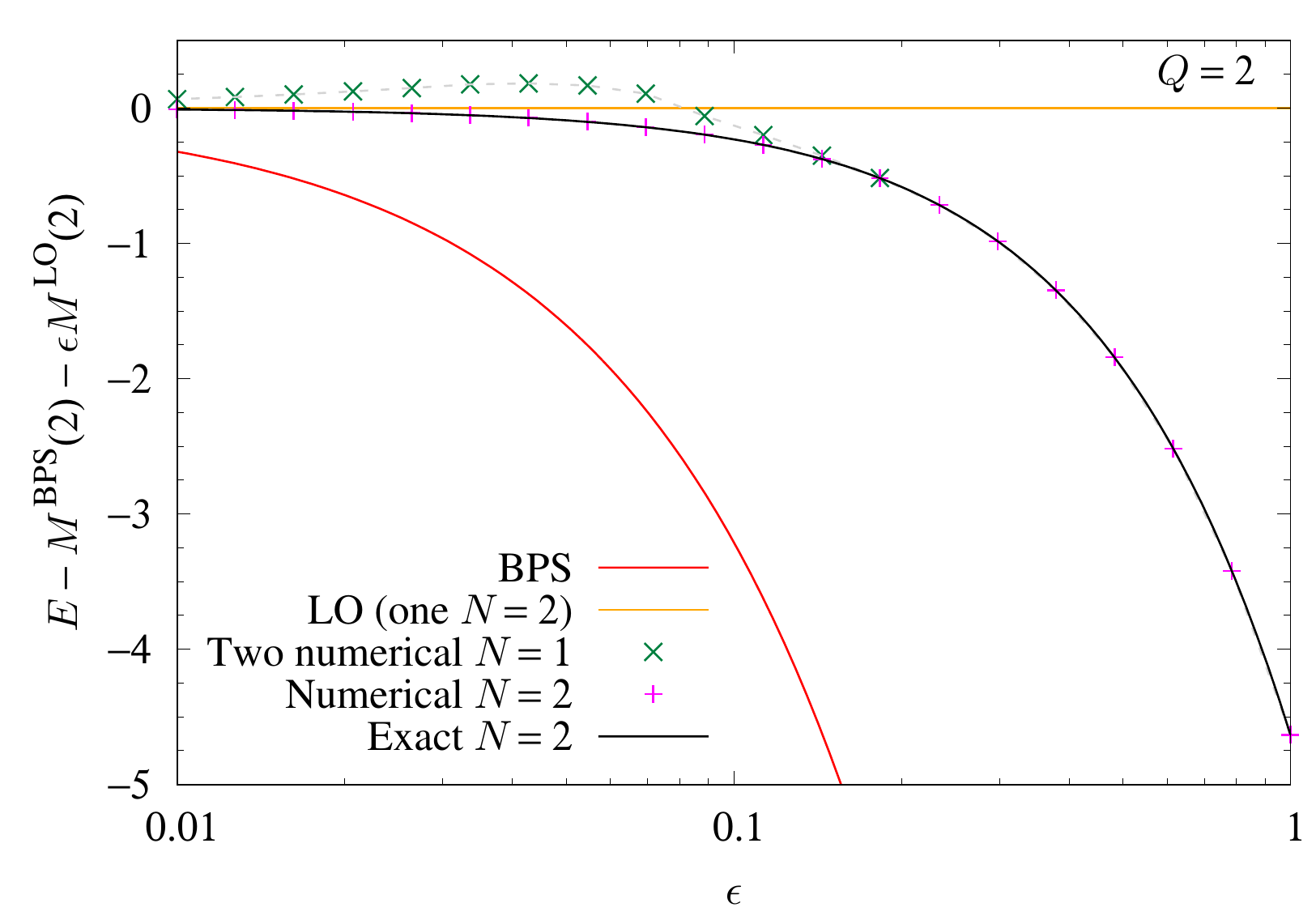}}
    \caption{The total energies of the baby-Skyrmion solutions in the
      $Q=2$ sector. The red line is the BPS bound, the orange curve
      the LO correction (coming from the kinetic term) for a single
      $N=2$ Skyrmion. The numerical
      solutions are: the two $N=1$ Skyrmions side-by-side (green xs),
      and the axially symmetric $N=2$ Skyrmion (magenta
      pluses). Finally, the latter numerical $N=2$ solution is
      compared to the exact solution (using the 1-dimensional ODE).
      For clarity, the right-hand side panel shows the same data
      relative to the orange line, being the LO correction.
      At $\epsilon\gtrsim 0.15$ the two $N=1$'s side-by-side become
      unstable and decay to the axially symmetric $N=2$. }
    \label{fig:NumericalEnQ2}
  \end{center}
\end{figure}

To this end, we start with two $N=2$ baby Skyrmions (hence with a
total topological charge $Q=2+2=4$), situated next to each other.
Some of the numerical results are shown in fig.~\ref{fig:N=2+2}.
The figure is organized into four columns showing the topological
charge density, the field configuration using a color scheme, the
total energy density and finally the energy density of the kinetic
term $(-\epsilon\Lag_2)$.
The colors on each graph is relatively rescaled to the values of the
graph. 
The color scheme for the second column of fig.~\ref{fig:N=2+2}
illustrates the orientation of the fields at each point of the
configuration space $(\mathbb{R}^2)$.
The scheme is defined such that the hue is correlated with the complex
phase $\arg(\phi^1+\i\phi^2)$ and the lightness is determined by the
value of $\phi^3$.
Explicitly, $\phi^3=1$ is white, $\phi^3=-1$ is black,
$\arg(\phi^1+\i\phi^2)=0$ is red, $\arg(\phi^1+\i\phi^2)=2\pi/3$ is
green and $\arg(\phi^1+\i\phi^2)=4\pi/3$ is blue.

In particular, we find that the two $N=2$ baby Skyrmions next to each
other is always stable -- even for $\epsilon=1$.
What happens when $\epsilon$ is decreased, is that the hole in each
$N=2$ soliton is shrunk to a point and asymptotically, the two
Skyrmions become axially symmetric -- almost unaware of each others
presence.
Notice that the orientation of the two baby Skyrmions is always in the
attractive channel, i.e.~such that e.g.~two patches of red touch each
other, see the second column of fig.~\ref{fig:N=2+2}.
At larger values of $\epsilon$, the two $N=2$ solitons are elongated
along the axis joining their centers.
This can be interpreted as an effect due to the strong binding force
between them.
For small $\epsilon$, this effect goes away and the binding energy is
also drastically reduced.
We will discuss the binding energies separately in the next section.

Baby-Skyrmion solitons are free to deform their shape and are able to
split up to the lowest energy state.
Nevertheless, many local minima of the energy functional exist and
they correspond to metastable states.
In order to know which configurations are the stable ones, we
start with many different initial guesses and see what they converge
to by the numerical flow.

We expect from the fact that $N_\star\approx 2$, that the $N=4$
axially symmetric soliton will be only metastable.
As an explicit check, we perform the numerical calculations which are
shown in fig.~\ref{fig:N=4}.
Perhaps surprisingly, it turns from metastable at $\epsilon\approx 1$
into unstable around $\epsilon\approx 0.15$, see the last row of
fig.~\ref{fig:N=4}.
The numerical algorithm finds a lower-energy configuration, which is
composed by four $N=1$ Skyrmions attached to each other in a
tetrahedral arrangement.
This is somewhat surprising, because as we shall see shortly, the
$N=1$ soliton is actually not stable for $\epsilon=0.15$ unless it is
alone. Once it is near another $N=1$ soliton, they will merge into a
deformed $N=2$ soliton at that $\epsilon$.
The high degree of discrete symmetry -- tetrahedral symmetry --
somehow prevents that from happening.

There are two possibilities: Either there is a crossover, so that the
tetrahedral configuration becomes stable and the $N=4$ axially
symmetric soliton becomes unstable around $\epsilon=0.15$ or both the
tetrahedral and axially symmetric solitons exist for
$\epsilon\gtrsim 0.15$ and the axially symmetric solution simply
becomes unstable around $\epsilon=0.15$ such that the decay to the
nearest (in energy) metastable state is the tetrahedral baby
Skyrmion. 
In order to determine the phase diagram of the $Q=4$ sector, we need
to know the energies of the tetrahedral soliton also for larger values
of $\epsilon$.
Thus we use the tetrahedral configuration as a seed (initial guess) in
the calculation and calculate its energy for larger $\epsilon$ all the
way to $\epsilon=1$, see fig.~\ref{fig:N=1+1+1+1_triangle}.
This reveals that the tetrahedral $Q=4$ baby Skyrmion
exists also for large values of $\epsilon$ (even though two $N=1$
baby-Skyrmions next two each other are unstable, see below). 

Then to determine the phase diagram of the $Q=4$, we plot the total
energies of the different solutions in fig.~\ref{fig:NumericalEnQ4}. 
From the figure -- especially the right-hand side panel -- we can see
that the $N=4$ axially symmetric configuration is metastable for
$\epsilon\gtrsim 0.15$ with the highest energy.
The tetrahedral soliton is metastable with an intermediate energy, but
not too far above the lowest-energy state.
Finally, the two $N=2$ Skyrmions side-by-side is the most
energetically favorable solution in the entire range of the $\epsilon$
considered here. 

For $\epsilon=0.0207$ it is rather convincing that the $\epsilon\to 0$
limit turns the solitons into a lattice of axially symmetric solutions
sitting next to each other with the most stable axially symmetric
component being the $N=2$ baby Skyrmion.
The contact to neighboring solitons becomes point-like in the
$\epsilon\to 0$ limit.
This is consistent with the restricted harmonic property that the
near-BPS (baby) Skyrmions should possess.

We will now turn to the $Q=2$ sector. Again, as $N_\star\approx 2$, we
know that the $N=1$ axially symmetric baby Skyrmion is stable only in
isolation.
When exposed to more baby Skyrmions, it has to be either metastable or
unstable; that is, it should be energetically favorable to combine
into $N=2$ Skyrmions.
It turns out that for $\epsilon=1$, two $N=1$ axially symmetric baby
Skyrmions next to each other are unstable and they immediately combine
into a single $N=2$ soliton with axial symmetry \cite{Foster:2009vk}. 
However, for small enough $\epsilon$, the two $N=1$'s side-by-side
become metastable, see fig.~\ref{fig:N=1+1}.
The critical $\epsilon_2$ where the baby Skyrmion departs from axial
symmetry is around $\epsilon\approx 0.2$, whereas the critical
$\epsilon_1$ where the baby Skyrmion is clearly composed of two
individual solitons is around $\epsilon_1\approx 0.1$.
For $\epsilon<0.06$ the two individual solitons are almost only
connected by a ``single point.''

As a check on our numerical calculations, we compute the $N=2$ axially
symmetric baby Skyrmions as well, see fig.~\ref{fig:N=2}.
The figure shows that the hole in the middle of the soliton is shrunk
to a point in the limit of $\epsilon\to 0$.

To check how these solutions fit into our picture, we show their total
energies in fig.~\ref{fig:NumericalEnQ2}.
We can thus confirm that the phase diagram in the $Q=2$ sector is
determined by the LO correction to the energy, which makes the two
$N=1$ baby Skyrmions sitting side-by-side metastable for
$\epsilon\lesssim 0.15$ and unstable above.
We can also confirm by comparing the 2-dimensional PDE calculations
with the 1-dimensional ODE ones (for the $N=2$ case), that our
numerical accuracy is incredibly good for the entire range of
$\epsilon$s considered here.

In order to complete our perturbative scheme, we need to calculate the
binding energies between two Skyrmions, which will be the topic of the
next section.

\section{Binding energies}
\label{cinque}

In this section we will discuss the binding energies of two $N=2$ baby
Skyrmions sitting side-by-side and, in the last part, also the case of
two $N=1$ side-by-side. 
We have performed very high-resolution numerical calculations that
give us the answers we seek for the baby Skyrmion case.
However, for the 3-dimensional Skyrmions, the numerical calculations
become much more difficult -- even with the utilization of adaptive
methods.
Therefore we want to push the perturbative approach to capture the
physics and respective energies of the composite near-BPS solitons. 

Since we have the compacton solutions, the background is analytically
known.
The difficulty is to impose the boundary or cusp condition
\eqref{eq:BCjump} that the derivative of the perturbation fields obey
at the boundary of the compacton, but this can easily be done in polar 
coordinates. 
Furthermore, the composite configuration of two baby Skyrmions
side-by-side will need a gluing condition, which we shall discuss
shortly. 
Finally, the perturbations must go to zero at spatial infinity,
which is the easiest condition to impose.

In section \ref{tre.due} we showed that the restricted harmonic map
for the $Q=2+2$ configuration consists of two separated axially
symmetric compactons with random relative orientation. However, if for
a single baby Skyrmion the problem of the background solution is
solved by the restricted harmonic condition, for the multi-soliton
case further considerations are needed. To understand the reason for
this difference, we must consider the physical meaning of the
zeroth-order background field. This field represents the limit of the
exact solution $\bphi$ of the Lagrangian \eqref{eq:L} when $\epsilon$
tends to zero, i.e. the final configuration that we obtain if we
adiabatically switch off the interaction among the solitons.
The static solution $\bphi$ exists only if the baby Skyrmions interact
with an attractive force that depends on the relative orientation
among them \cite{Piette:1994ug}.
Therefore, we expect that even in the limit of vanishing interaction
($\epsilon\to 0$) the final configuration will conserve a particular
relative orientation. Then, we conclude that for the near-BPS
multi-baby Skyrmion case, the restricted harmonic condition is not
sufficient to identify the correct zeroth-order solution since it does
not yield any restriction on the relative orientation (or distance)
among axially symmetric compactons. 

In order to fix the background configuration among all the restricted
harmonic maps, we propose to identify the right map by looking at the
form of the interaction between two well-separated baby Skyrmions.
The potential $V$ among these solitons has been calculated in
ref.~\cite{Piette:1994ug} for the baby-Skyrme model (the same
Lagrangian \eqref{eq:L} with $\epsilon=1$) and for the $Q=2+2$ case it
takes the form  
\begin{equation}
  V_{2+2}\propto -\cos(\alpha-\beta)\frac{e^{-mR}}{\sqrt{mR}}
  \label{eq:pot}
\end{equation}
where $R$ is the distance between the soliton centers, $\alpha$ and
$\beta$ are the two respective phases and $m$ is the (perturbative)
pion mass. From this expression we recognize that, for fixed distance
$R$, the minimum of the potential is obtained for $\alpha-\beta=0$,
i.e. when the two baby Skyrmions have the same orientation. Therefore,
even if the form of the potential \eqref{eq:pot} holds only at large
distances, we guess that in the limit of vanishing interaction (when
$\epsilon\to 0$) the two solitons keep their relative orientation
unchanged. 

Instead, for what concerns the relative distance, we expect that in
the limit of vanishing interaction, the two baby Skyrmions must flow
to a configuration in which they overlap their tails less and
less. However, since the tail of a near-BPS baby Skyrmion vanishes
rapidly with $\epsilon$ as is clear from eq.~\eqref{eq:tail}, the two
solitons flow to a distance that is very close to the sum of their
radii. 

Summarizing, we guess that in the limit $\epsilon\to 0$, the
background solution for the $Q=2+2$ configuration is given by two
axially symmetric compactons with the same orientation touching each
other at a single point. This guess seems to be confirmed by numerical
calculations in fig.~\ref{fig:N=2+2} and we will thus use it in the
following.

Since the zeroth-order configuration for the two compactons is such
that the other baby Skyrmion is a copy of itself, spatially translated
by a distance of its own diameter (roughly), it is not necessary to
calculate the perturbation over the entire plane, but we can reduce
the problem by identifying appropriate boundary conditions for a field
living in an appropriate half of $\mathbb{R}^2$.

\begin{figure}[!ht]
  \begin{center}
    \includegraphics[width=0.4\linewidth]{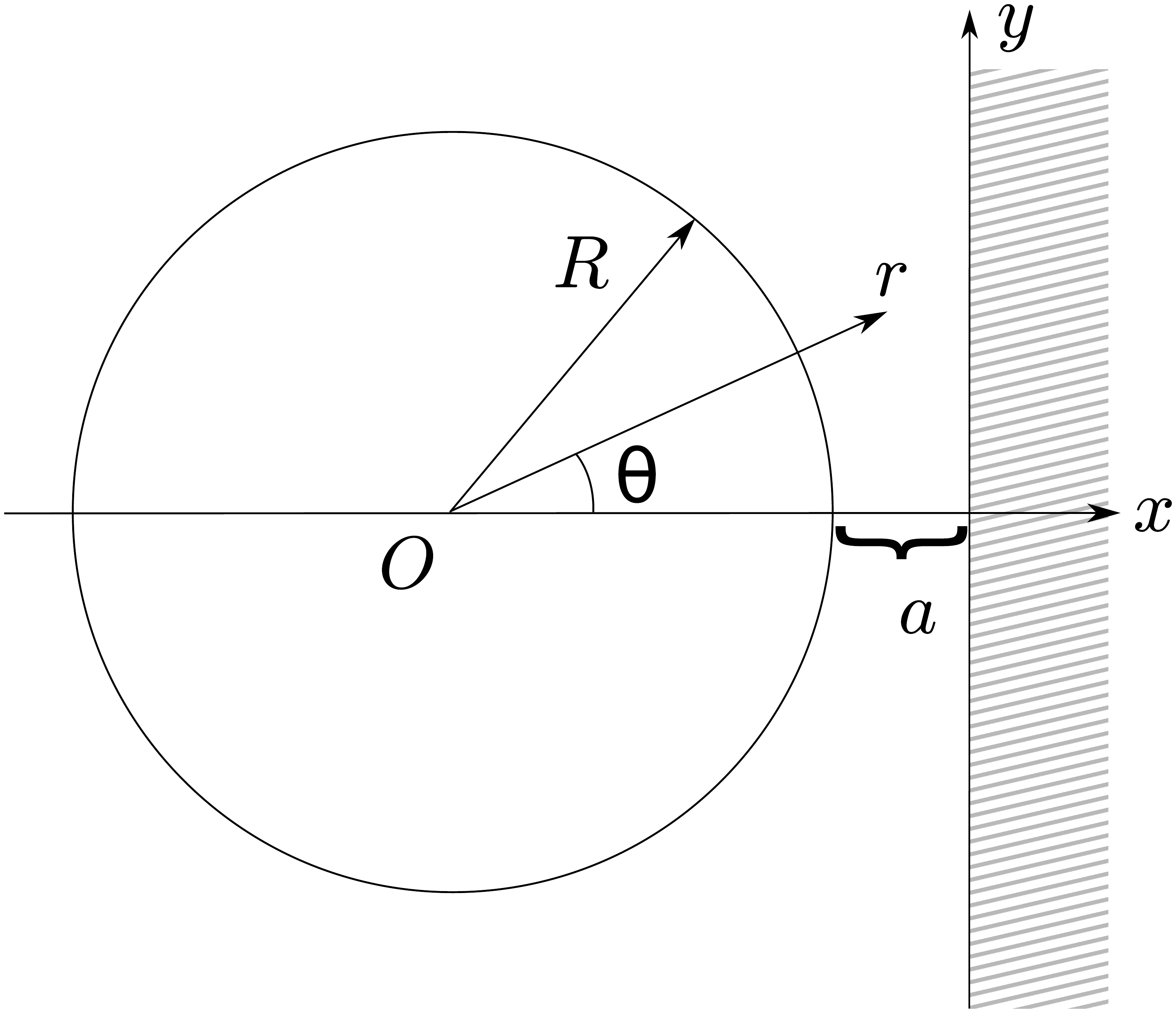}
    \caption{Coordinate system $(r,\theta)$ for perturbation of the
      two $N=2$ baby Skyrmions sitting side-by-side. Boundary
      conditions need to be imposed at $O$ and at the compacton
      boundary $r=R$ as well as at $x=0$ (at $re^{i\theta}=R+a+iy$)
      for attaining the ability of gluing the two solitons together. 
    }
    \label{fig:BC}
  \end{center}
\end{figure}

The boundary conditions are sketched in fig.~\ref{fig:BC}.
In addition to setting the perturbation to zero at $r=0$ and imposing
the condition \eqref{eq:BCjump}, we need to impose the following
boundary (gluing) conditions
\begin{align}
  \p_x\dphi^1(0,y) &= 0,\label{eq:xzeroBC1}\\
  \dphi^2(0,y) &= 0, \label{eq:xzeroBC2}\\
  \p_x\dphi^3(0,y) &= 0,\label{eq:xzeroBC3}
\end{align}
which ensure that the soliton can be glued together with its partner. 
The second boundary condition is ``odd'' because this field is odd
under parity transformations and the only way to connect the
perturbation to a field that is not flipped in $x\to -x$ (because that
would make it an anti-Skyrmion) is to impose the boundary condition as
given in the above equations.
Because the condition \eqref{eq:BCjump} is easier in polar
coordinates, we prefer to impose the $x=0$ gluing conditions on the
equation in polar coordinates.

In terms of the perturbation fields given in
eq.~\eqref{eq:Delta_transverse}, we can write the boundary conditions
as 
\begin{align}
  \df(0,\theta)&=0, &\quad
  \dtheta(0,\theta)&=0, \non
  \df(\infty,\theta)&=0, &\quad
  \dtheta(\infty,\theta)&=0,\non
  \df(r,0)&=\df(r,2\pi), &\quad
  \df_r(r,0)&=\df_r(r,2\pi), \non
  \dtheta(r,0)&=\dtheta(r,2\pi), &\quad
  \dtheta_r(r,0)&=\dtheta_r(r,2\pi),
  \label{eq:PDE_BC}
\end{align}
while at $r=R$ we have
\beq
\df_r(R^-,\theta) - \df_r(R^+,\theta) = \frac{4}{R},
\label{eq:cusp_condition}
\eeq
and finally at $x=0$ -- which is $re^{i\theta}=R+a+iy$ in polar
coordinates -- the boundary conditions
\eqref{eq:xzeroBC1}-\eqref{eq:xzeroBC3} become
\begin{align}
  \dtheta &= -\tan(N\theta)\df,\label{eq:dtheta_gluing_condition}\\
  \p_x(\dtheta^2 + \df^2) &= 0,
\end{align}
and the latter equation can be written as
\begin{equation}
\begin{array}{ll}
  \df = N^{-1}\cot(N\theta)\left(r\cot(\theta)\df_r - \df_\theta\right), \qquad& \theta>0,\label{eq:xzerobccont}\\
  \df_r = \df_\theta = 0, & \theta=0. 
\end{array}
\end{equation}

\begin{figure}[!ht]
  \begin{center}
    \mbox{\includegraphics[width=0.49\linewidth]{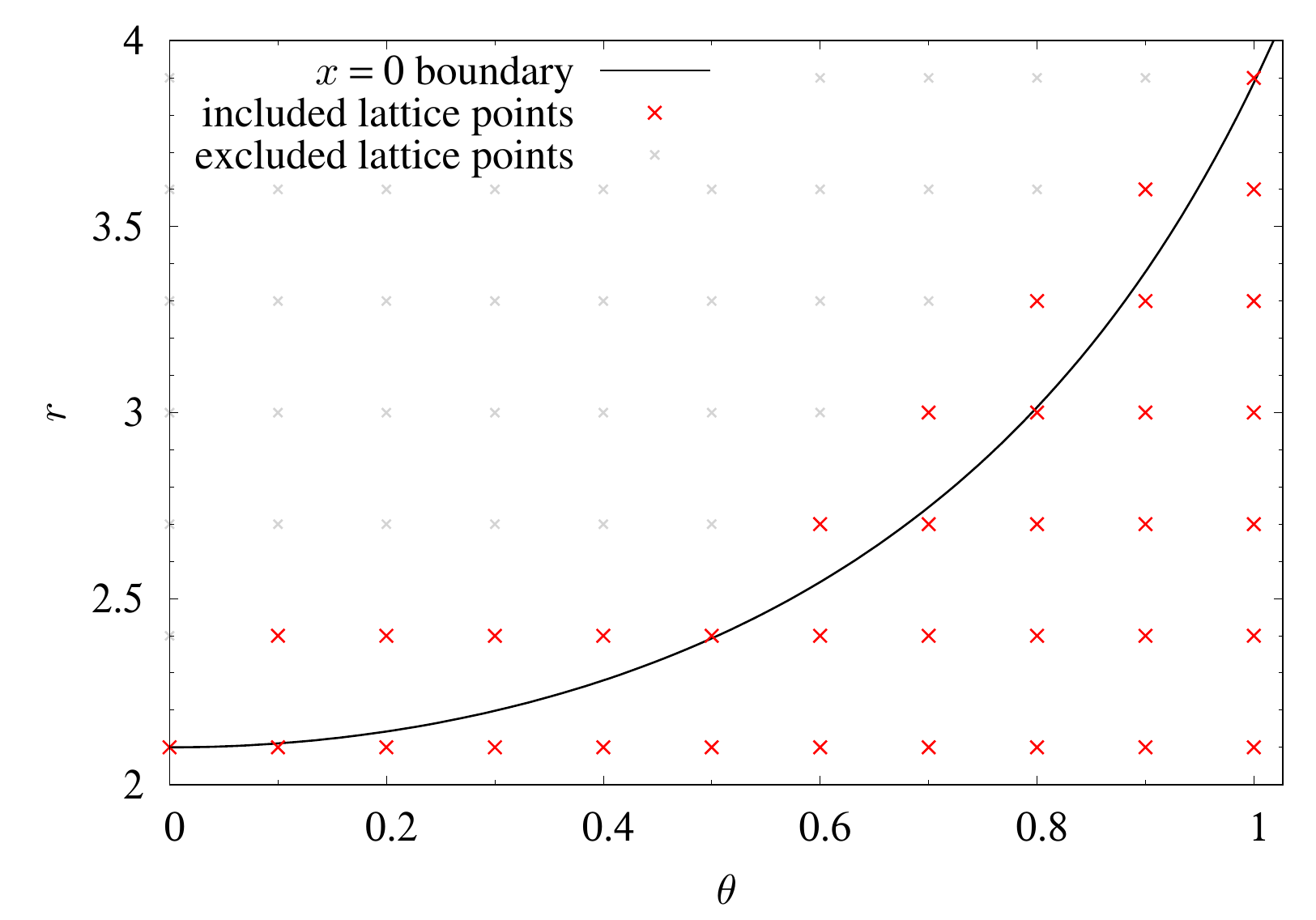}
    \includegraphics[width=0.49\linewidth]{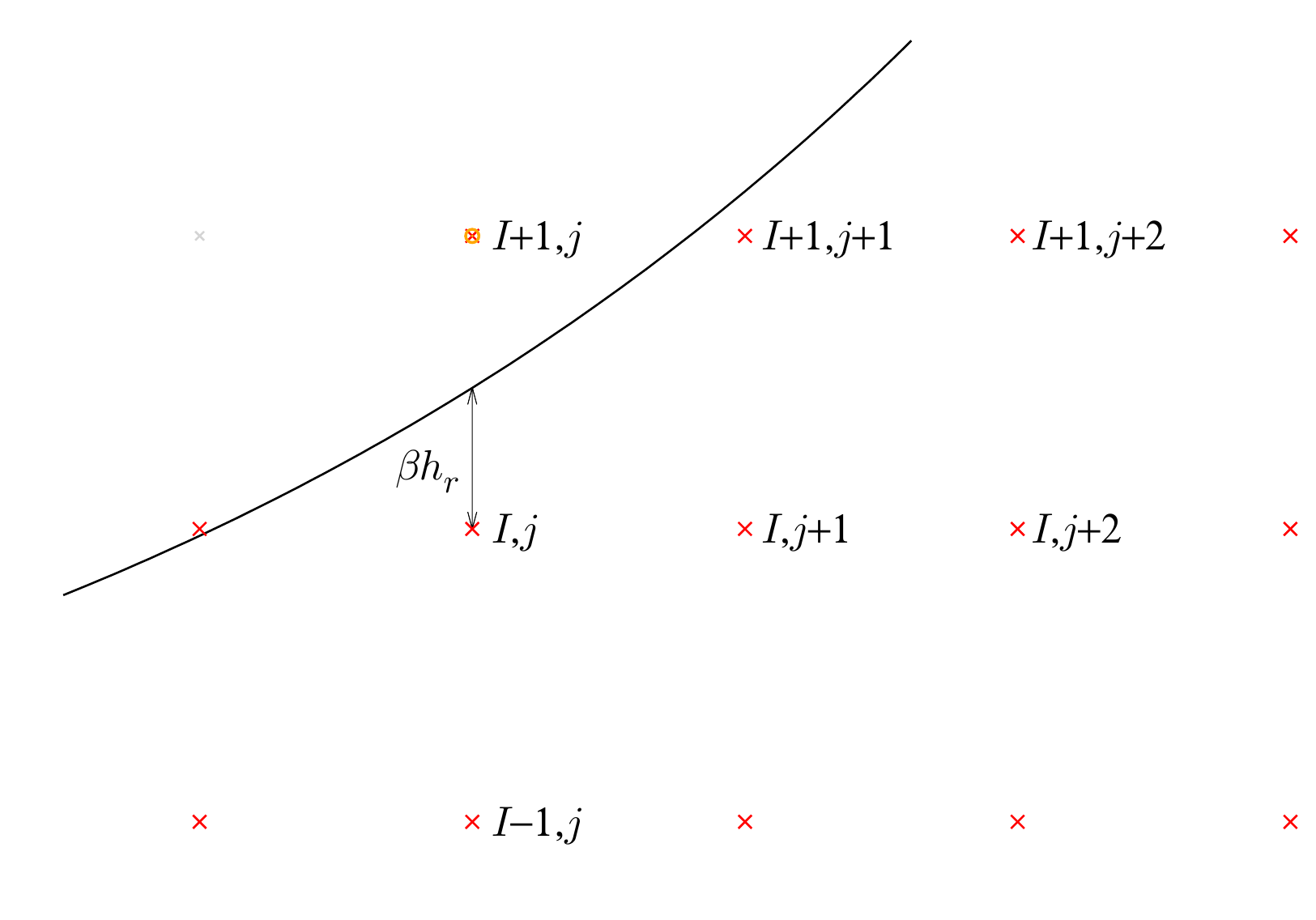}}
    \caption{Boundary (gluing) conditions at $x=0$ for the discretized
      (lattice) problem. This example is shown for $N=1$, $m=1$ and
      $a=0.1$ with an exaggeratedly big step size of $0.3$ chosen for
      illustrative purposes. The right-hand panel shows a zoom with
      the lattice coordinates used in eq.~\eqref{eq:xzerobc}}
    \label{fig:xzerobc}
  \end{center}
\end{figure}

Fig.~\ref{fig:xzerobc} shows the boundary conditions for the lattice
problem (the discretized PDEs) at the $x=0$ boundary.
The lattice point $(I+1,j)$ is determined by the condition
\eqref{eq:xzerobccont} which is written in discretized form as
\begin{align}
\df^{I+1,j}
\left(\beta
-\frac{\cot(N\theta^j)}{N}\left[
  \left(r^I+\beta h_r\right)\cot(\theta^j)\frac{(2\beta+1)}{2h_r}
  + \frac{3\beta}{2h_\theta}\right]\right)\qquad\qquad\non
= (\beta-1)\df^{I,j} + \frac{\cot(N\theta^j)}{N}\bigg[
\left(r^I+\beta h_r\right)\cot(\theta^j) 
\frac{-4\beta\df^{I,j} + (2\beta - 1)\df^{I-1,j}}{2h_r} \qquad\non
-(1-\beta)\frac{-3\df^{I,j} + 4\df^{I,j+1} - \df^{I,j+2}}{2h_\theta} 
-\beta\frac{4\df^{I+1,j+1}-\df^{I+1,j+2}}{2h_\theta}\bigg],
\label{eq:xzerobc}
\end{align}
with $\beta$ the offset of the $x=0$ boundary from the second-last
lattice point in the calculation.

It will be instructive to write the perturbations $\df,\dtheta$ as
follows
\begin{align}
  \bphi &= \bvarphi + \bdphi \non
  &= \bvarphi + \bDelta_\bot\times\bvarphi + \frac12\bDelta_\bot\times(\bDelta_\bot\times\varphi)\non
  &=
  \begin{pmatrix}
    \sin f\cos N\theta\\
    \sin f\sin N\theta\\
    \cos f
  \end{pmatrix}
  +
  \begin{pmatrix}
    \cos f\cos N\theta\\
    \cos f\sin N\theta\\
    -\sin f
  \end{pmatrix}\df
  +
  \begin{pmatrix}
    -\sin N\theta\\
    \cos N\theta\\
    0
  \end{pmatrix}\dtheta
  -
  \begin{pmatrix}
    \sin f\cos N\theta\\
    \sin f\sin N\theta\\
    \cos f
  \end{pmatrix}\frac{\df^2+\dtheta^2}{2}\label{eq:deltaphi_full_form}\\
  &=
  \begin{pmatrix}
    \sin(f + \df)\cos(N\theta + \dtheta)\\
    \sin(f + \df)\sin(N\theta + \dtheta)\\
    \cos(f + \df)
  \end{pmatrix}
  +\mathcal{O}(\dtheta^2,\df\dtheta,\df^3),
\end{align}
where we have used eq.~\eqref{eq:Delta_transverse}.

We now turn to solving the coupled PDEs \eqref{eq:Xpde_polar} with the
boundary conditions \eqref{eq:PDE_BC} and \eqref{eq:xzerobccont}.
As a good starting point, we take the perturbation of the axially
symmetric compacton for given $\epsilon$ and $N$ as the initial
condition. 
What the PDE problem then boils down to, is to implement the gluing
conditions \eqref{eq:xzerobccont} at $x=0$ (i.e.~midway between the
two compactons). 
The gluing conditions break the axial symmetry which provides
nontrivial $\theta$ dependence for $\df$ which in turn acts as a
source for $\dtheta$. 

\def\saelemwidth{0.155}
\def\tabelemwidth{58pt}
\begin{figure}
\begin{center}
\begin{tabular}{p{10pt}p{\tabelemwidth}p{\tabelemwidth}p{\tabelemwidth}p{\tabelemwidth}p{\tabelemwidth}l}
& $\df$ & $\dtheta$ & $\dphi^1$ & $\dphi^2$ & $\dphi^3$ & $\mathcal{E}^{\rm perturb}$
\end{tabular}
\mbox{\rotatebox{90}{\quad$a=0$}
\includegraphics[width=\saelemwidth\linewidth]{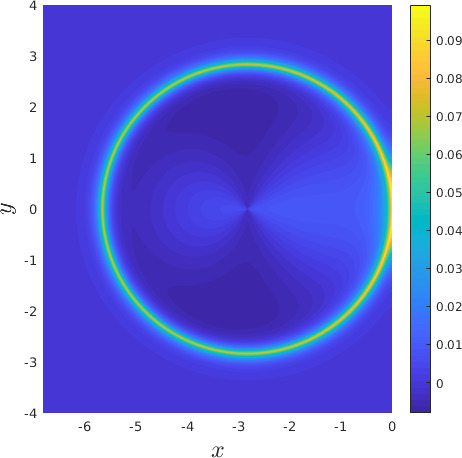}
\includegraphics[width=\saelemwidth\linewidth]{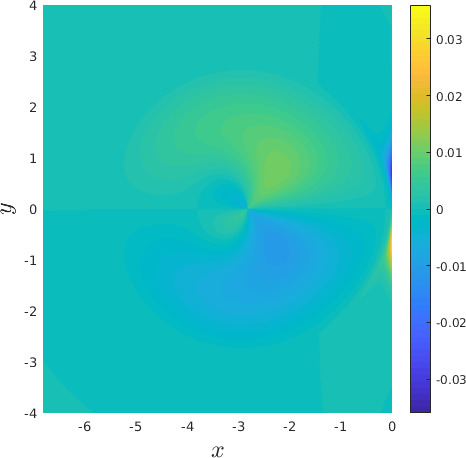}
\includegraphics[width=\saelemwidth\linewidth]{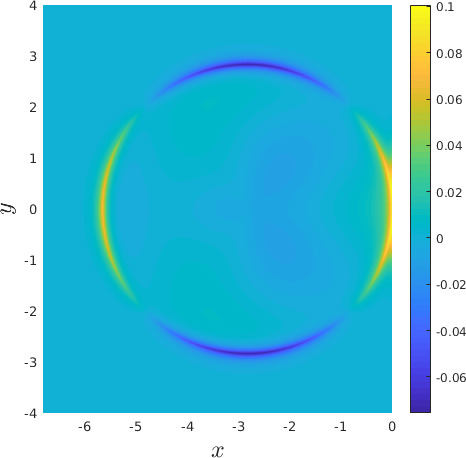}
\includegraphics[width=\saelemwidth\linewidth]{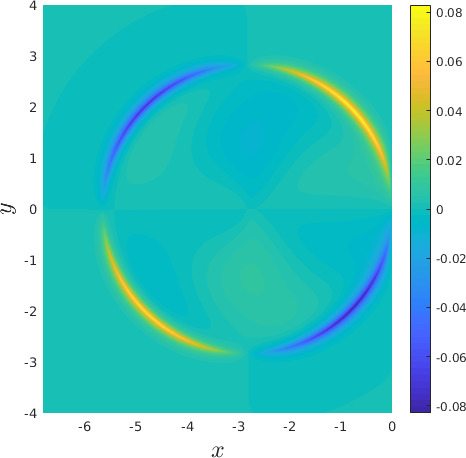}
\includegraphics[width=\saelemwidth\linewidth]{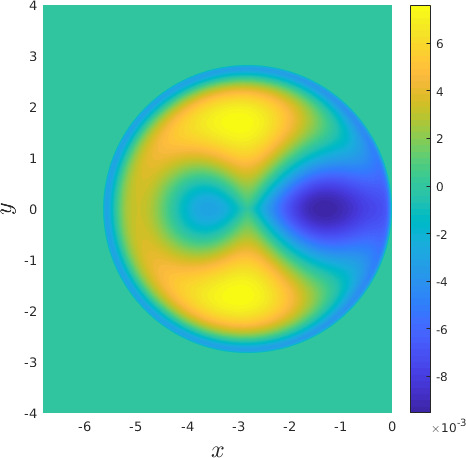}
\includegraphics[width=\saelemwidth\linewidth]{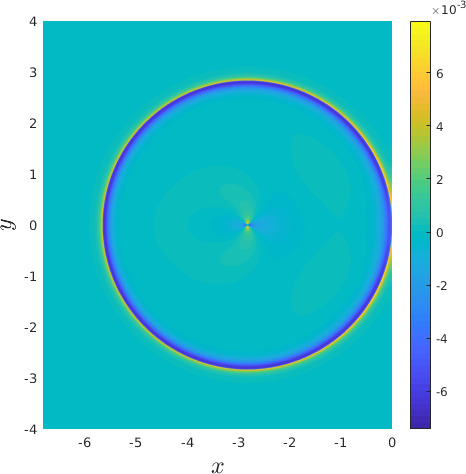}}
\mbox{\rotatebox{90}{\quad$a=0.0221$}
\includegraphics[width=\saelemwidth\linewidth]{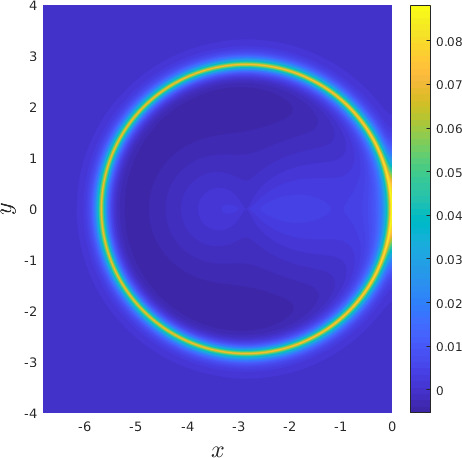}
\includegraphics[width=\saelemwidth\linewidth]{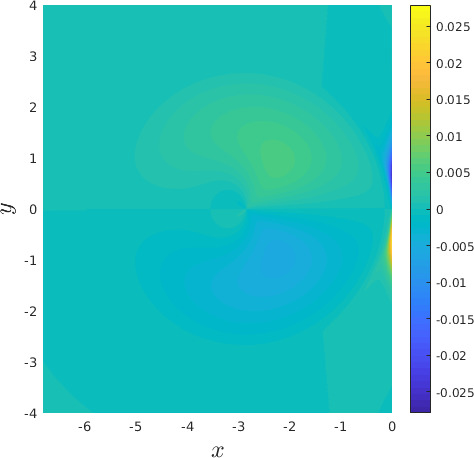}
\includegraphics[width=\saelemwidth\linewidth]{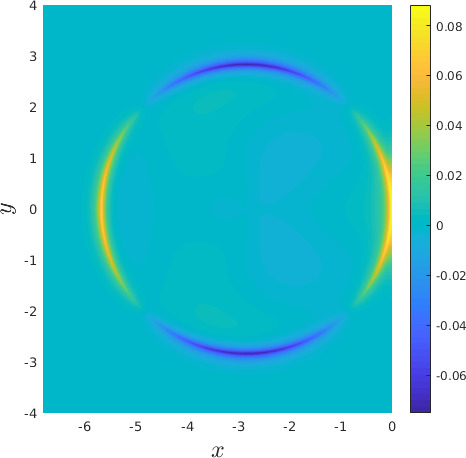}
\includegraphics[width=\saelemwidth\linewidth]{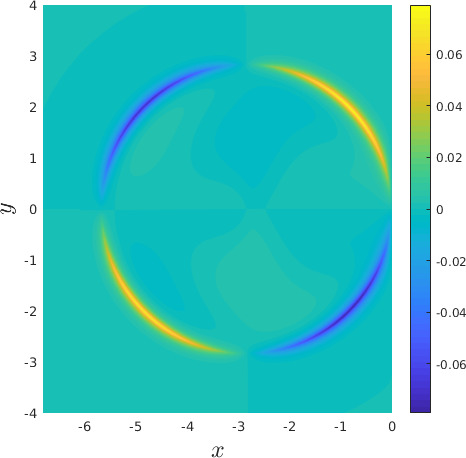}
\includegraphics[width=\saelemwidth\linewidth]{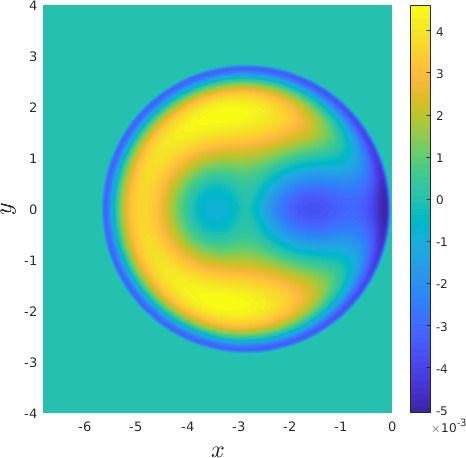}
\includegraphics[width=\saelemwidth\linewidth]{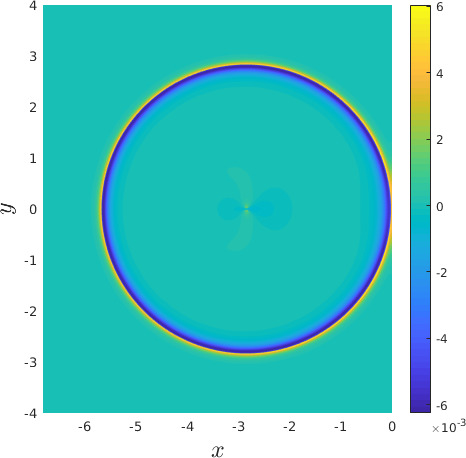}}
\mbox{\rotatebox{90}{\quad$a=0.0664$}
\includegraphics[width=\saelemwidth\linewidth]{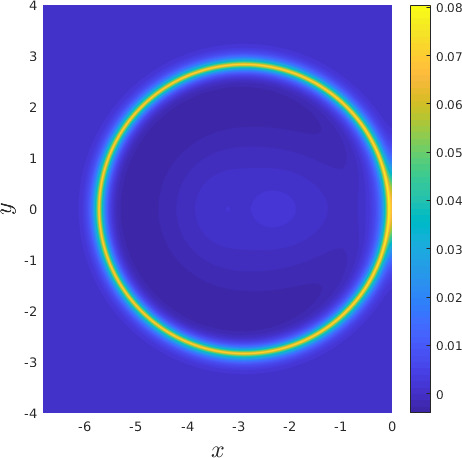}
\includegraphics[width=\saelemwidth\linewidth]{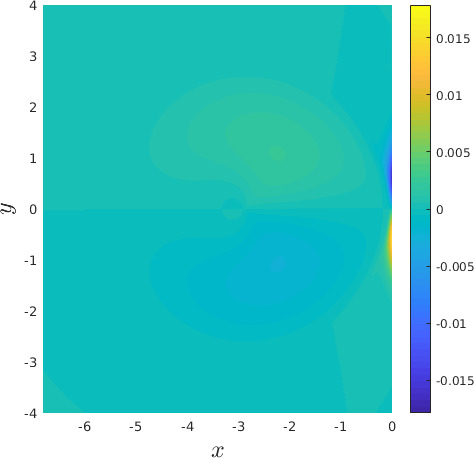}
\includegraphics[width=\saelemwidth\linewidth]{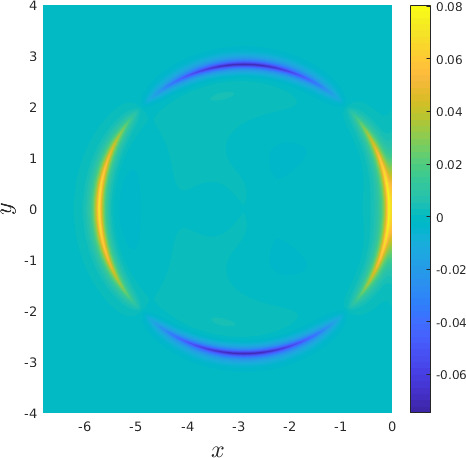}
\includegraphics[width=\saelemwidth\linewidth]{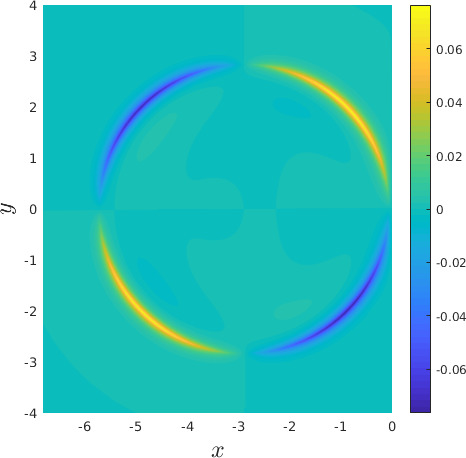}
\includegraphics[width=\saelemwidth\linewidth]{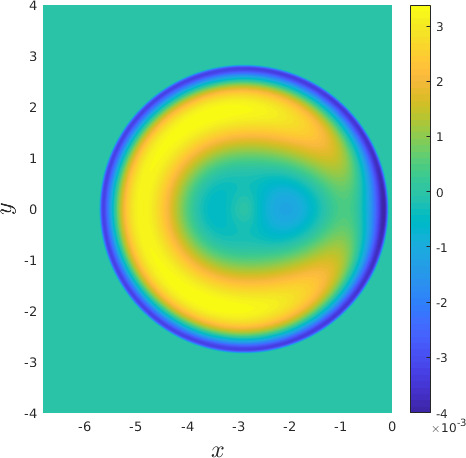}
\includegraphics[width=\saelemwidth\linewidth]{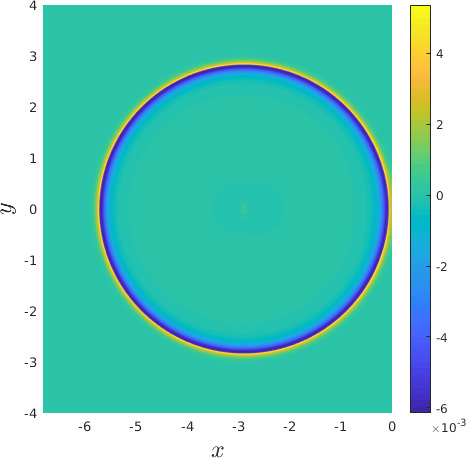}}
\mbox{\rotatebox{90}{\quad$a=0.125$}
\includegraphics[width=\saelemwidth\linewidth]{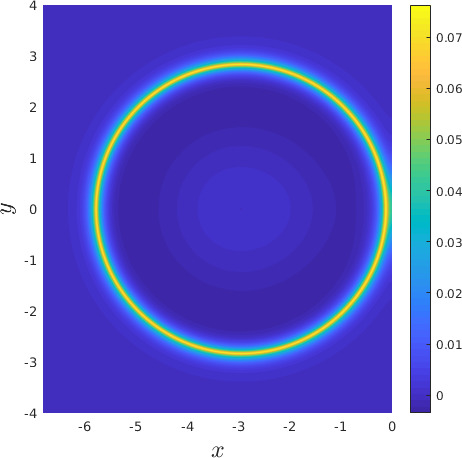}
\includegraphics[width=\saelemwidth\linewidth]{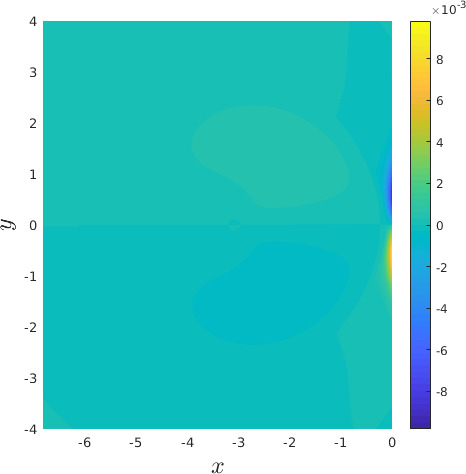}
\includegraphics[width=\saelemwidth\linewidth]{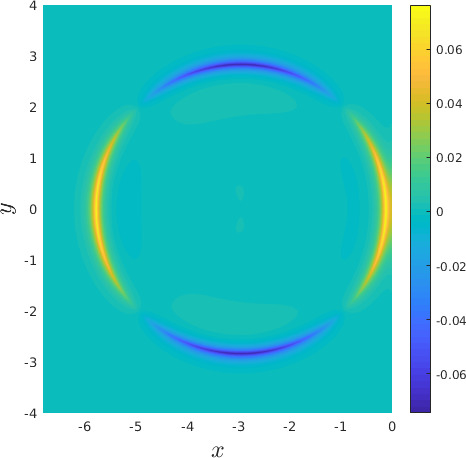}
\includegraphics[width=\saelemwidth\linewidth]{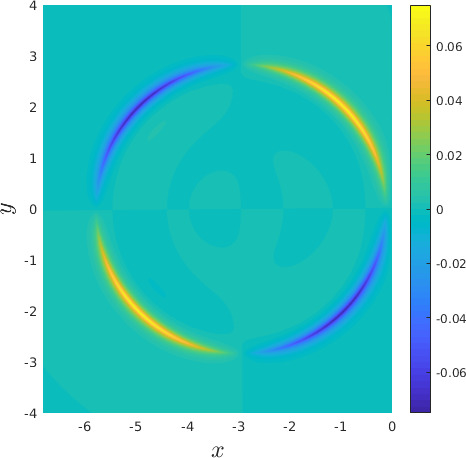}
\includegraphics[width=\saelemwidth\linewidth]{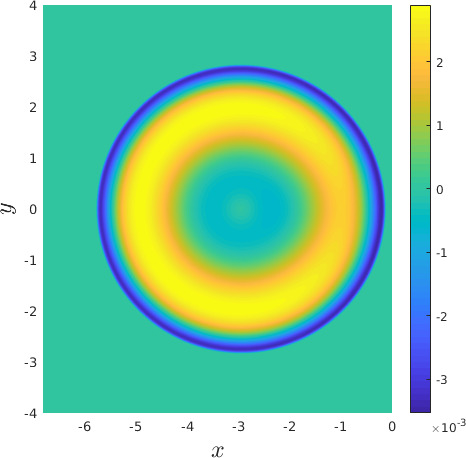}
\includegraphics[width=\saelemwidth\linewidth]{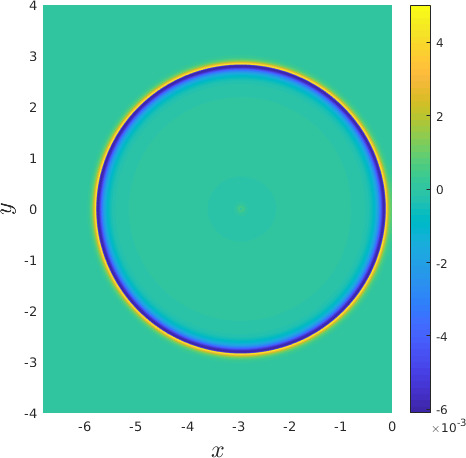}}
\mbox{\rotatebox{90}{\quad$a=0.243$}
\includegraphics[width=\saelemwidth\linewidth]{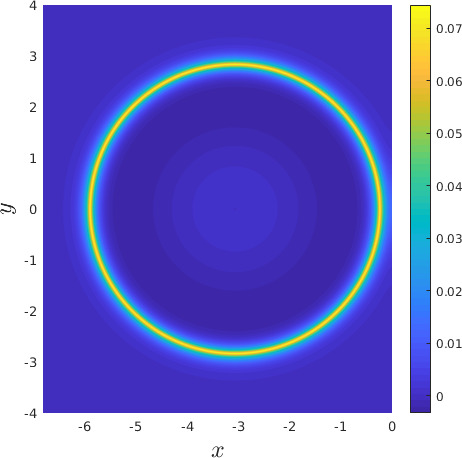}
\includegraphics[width=\saelemwidth\linewidth]{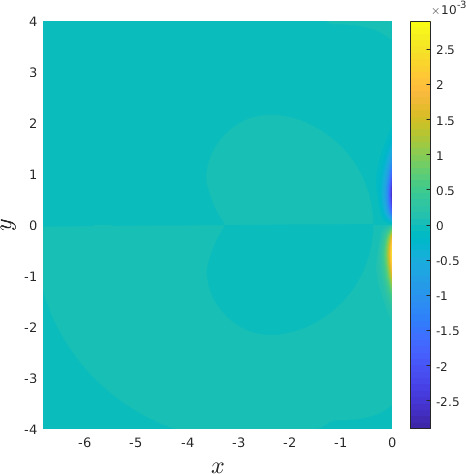}
\includegraphics[width=\saelemwidth\linewidth]{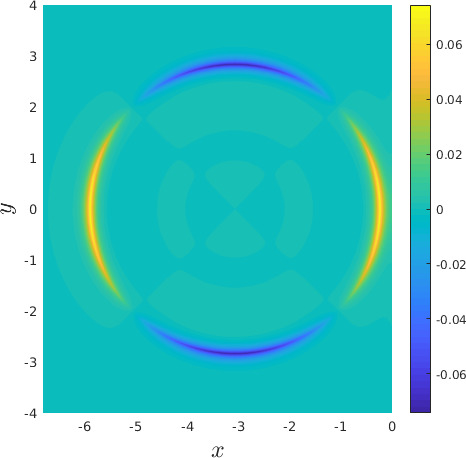}
\includegraphics[width=\saelemwidth\linewidth]{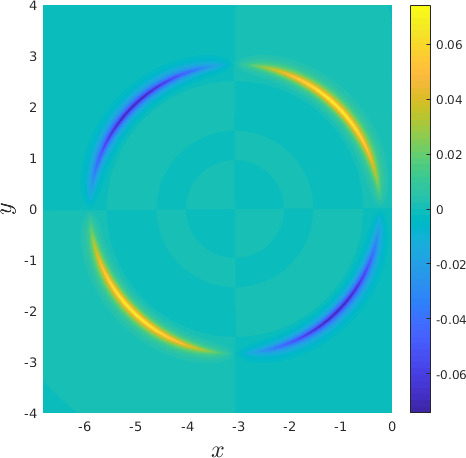}
\includegraphics[width=\saelemwidth\linewidth]{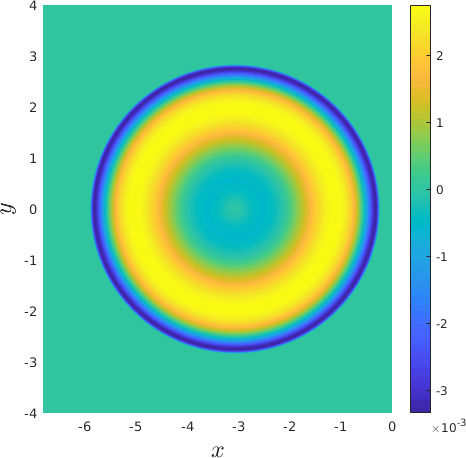}
\includegraphics[width=\saelemwidth\linewidth]{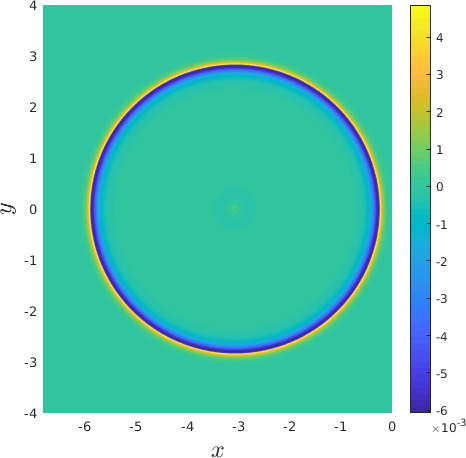}}
\caption{Numerical solution for $\df$ and $\dtheta$ for two $N=2$
  compactons situated next two each other with separation distance
  $2a$ for $\epsilon=0.01$.
  Due to symmetry of the problem, only the left-hand side is
  calculated and shown.
  The gluing conditions \eqref{eq:xzerobccont} are imposed at $x=0$,
  which is located at the right-most part of every panel.
  Each row displays the perturbations $\df$, $\dtheta$, $\dphi^1$,
  $\dphi^2$, $\dphi^3$ and the energy density of the perturbation
  $\mathcal{E}^{\rm perturb}$, each in its column.
  Each panel is cropped so as to render the content as clear as
  possible; the left, top and bottom border do not represent the size
  of the simulations.
}
\label{fig:sapde2c20.01}
\end{center}
\end{figure}

\begin{figure}
\begin{center}
\begin{tabular}{p{10pt}p{\tabelemwidth}p{\tabelemwidth}p{\tabelemwidth}p{\tabelemwidth}p{\tabelemwidth}l}
& $\df$ & $\dtheta$ & $\dphi^1$ & $\dphi^2$ & $\dphi^3$ & $\mathcal{E}^{\rm perturb}$
\end{tabular}
\mbox{\rotatebox{90}{\quad$a=0$}
\includegraphics[width=\saelemwidth\linewidth]{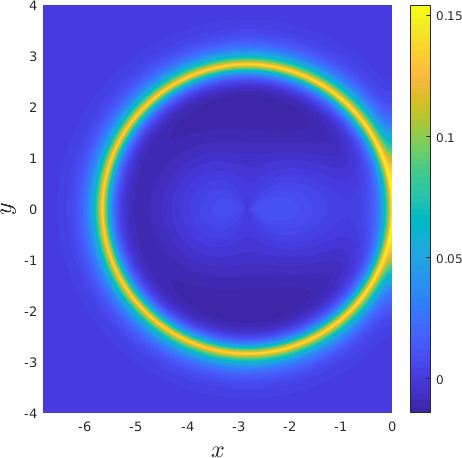}
\includegraphics[width=\saelemwidth\linewidth]{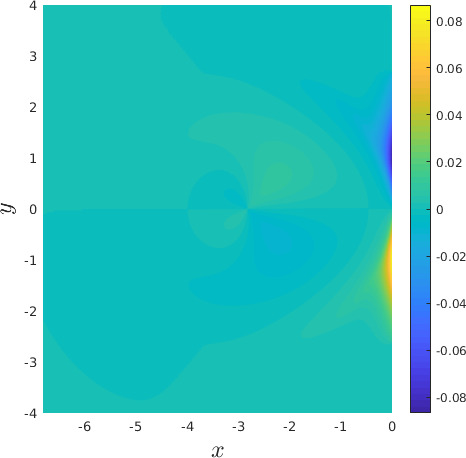}
\includegraphics[width=\saelemwidth\linewidth]{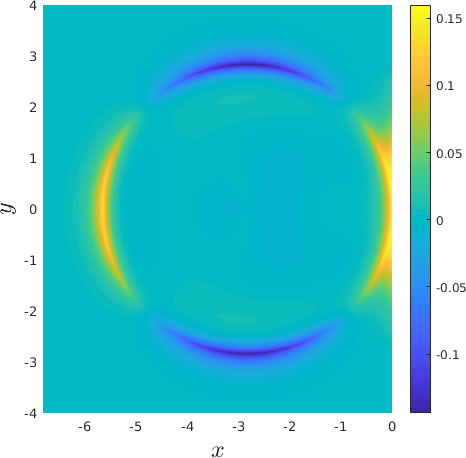}
\includegraphics[width=\saelemwidth\linewidth]{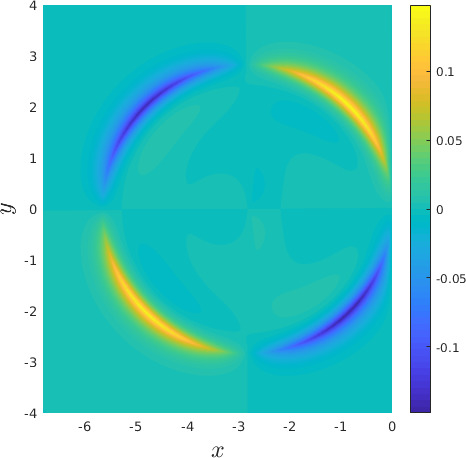}
\includegraphics[width=\saelemwidth\linewidth]{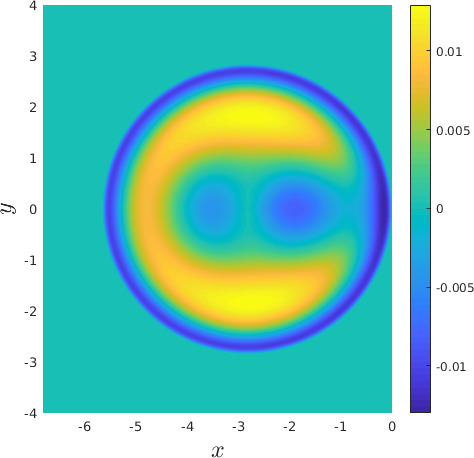}
\includegraphics[width=\saelemwidth\linewidth]{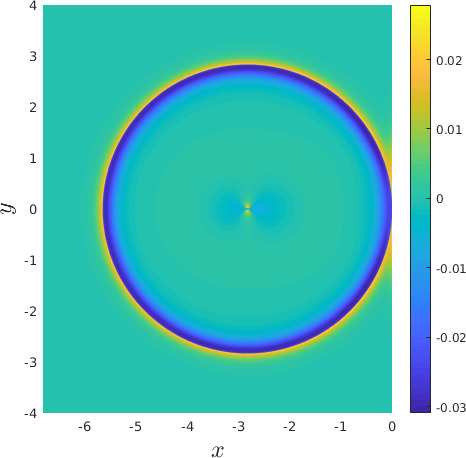}}
\mbox{\rotatebox{90}{\quad$a=0.0221$}
\includegraphics[width=\saelemwidth\linewidth]{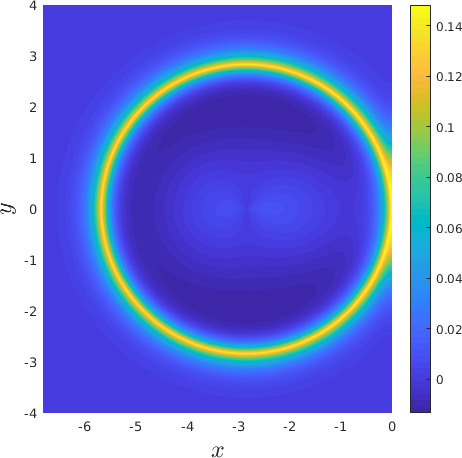}
\includegraphics[width=\saelemwidth\linewidth]{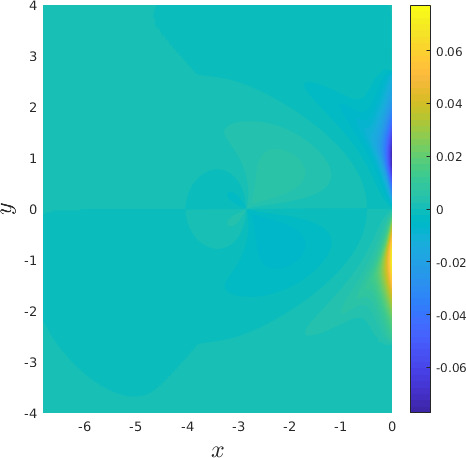}
\includegraphics[width=\saelemwidth\linewidth]{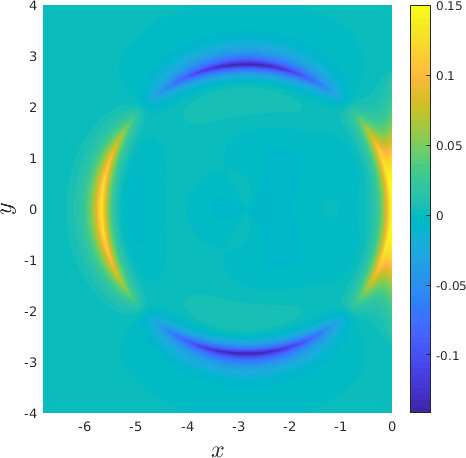}
\includegraphics[width=\saelemwidth\linewidth]{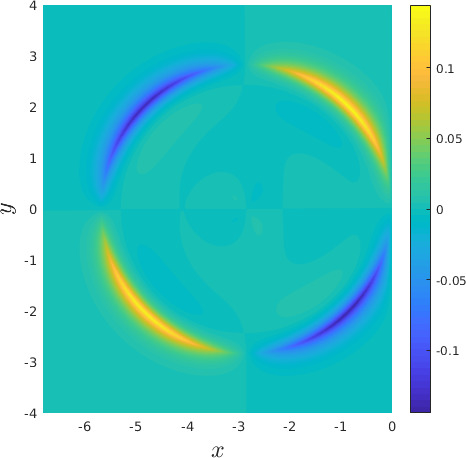}
\includegraphics[width=\saelemwidth\linewidth]{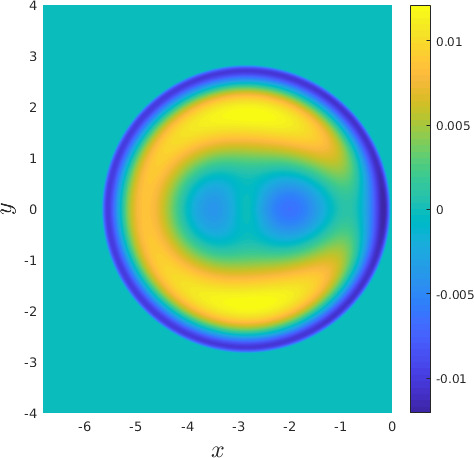}
\includegraphics[width=\saelemwidth\linewidth]{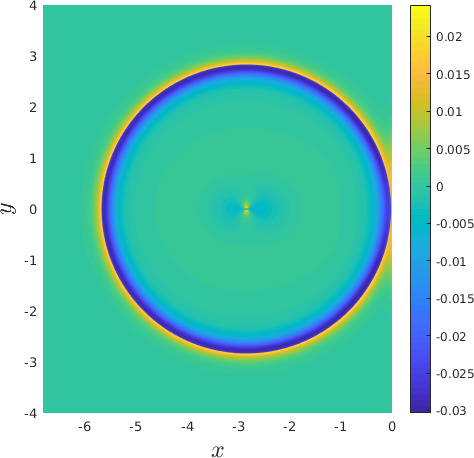}}
\mbox{\rotatebox{90}{\quad$a=0.0664$}
\includegraphics[width=\saelemwidth\linewidth]{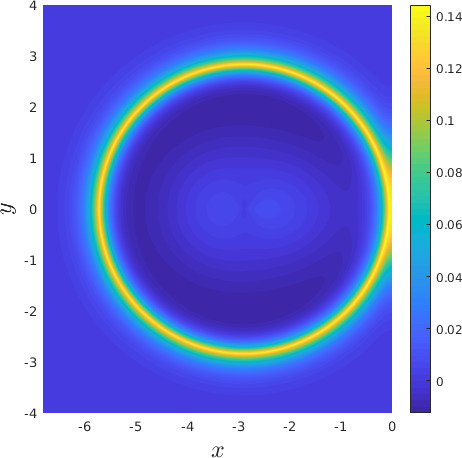}
\includegraphics[width=\saelemwidth\linewidth]{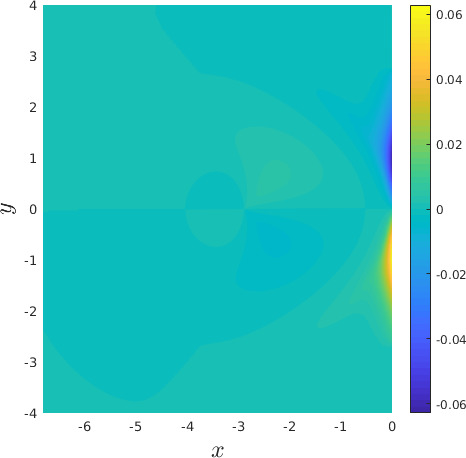}
\includegraphics[width=\saelemwidth\linewidth]{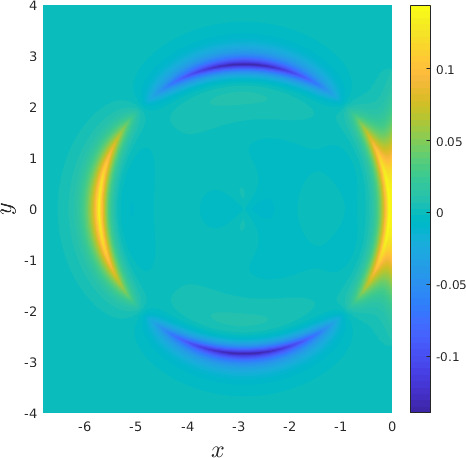}
\includegraphics[width=\saelemwidth\linewidth]{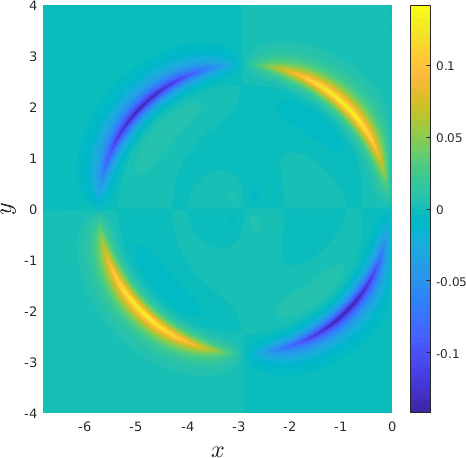}
\includegraphics[width=\saelemwidth\linewidth]{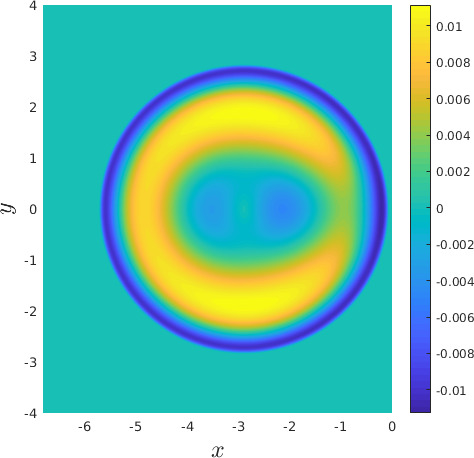}
\includegraphics[width=\saelemwidth\linewidth]{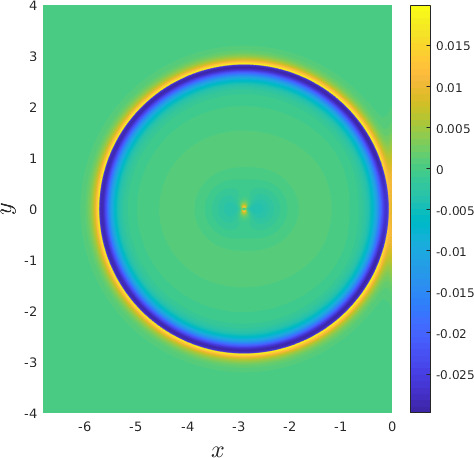}}
\mbox{\rotatebox{90}{\quad$a=0.125$}
\includegraphics[width=\saelemwidth\linewidth]{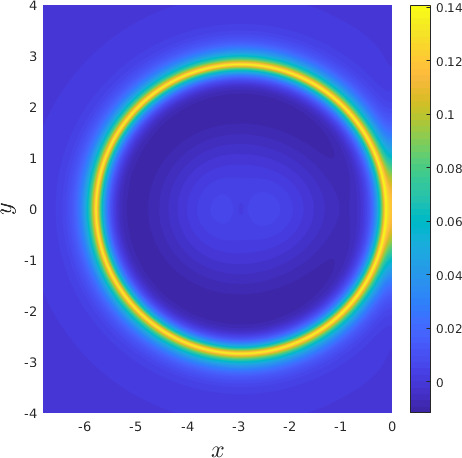}
\includegraphics[width=\saelemwidth\linewidth]{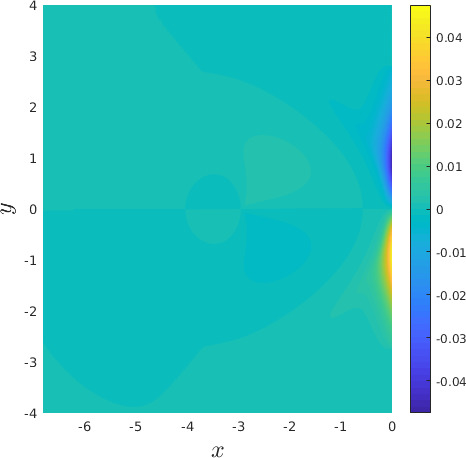}
\includegraphics[width=\saelemwidth\linewidth]{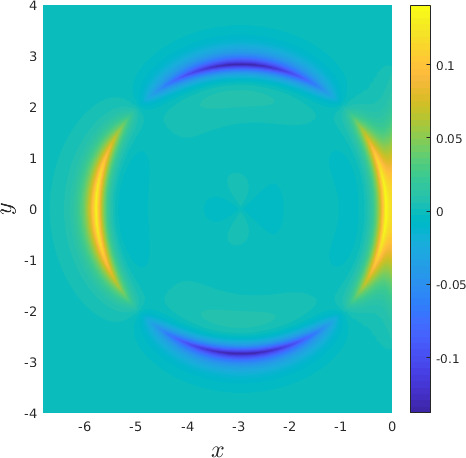}
\includegraphics[width=\saelemwidth\linewidth]{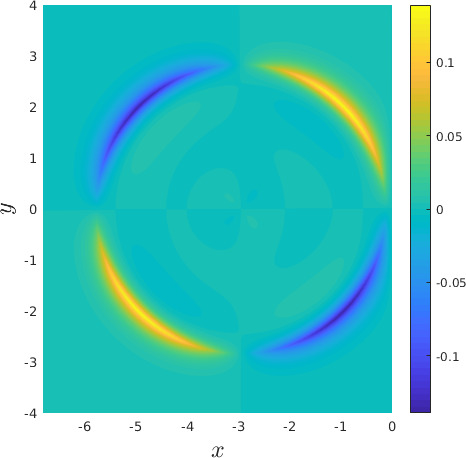}
\includegraphics[width=\saelemwidth\linewidth]{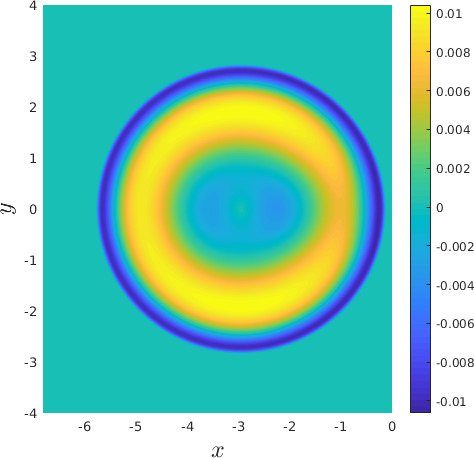}
\includegraphics[width=\saelemwidth\linewidth]{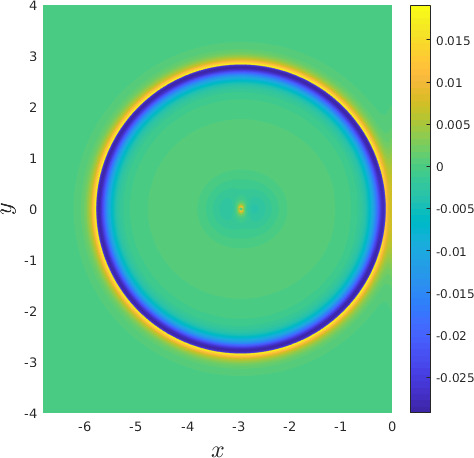}}
\mbox{\rotatebox{90}{\quad$a=0.243$}
\includegraphics[width=\saelemwidth\linewidth]{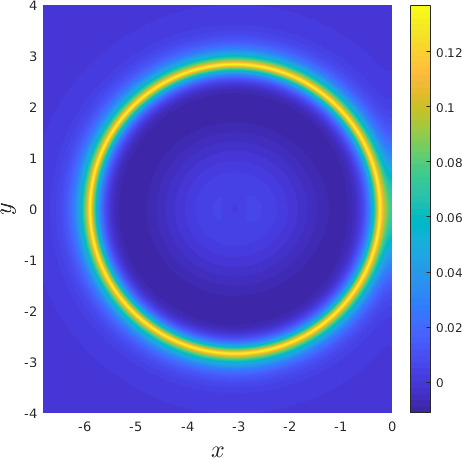}
\includegraphics[width=\saelemwidth\linewidth]{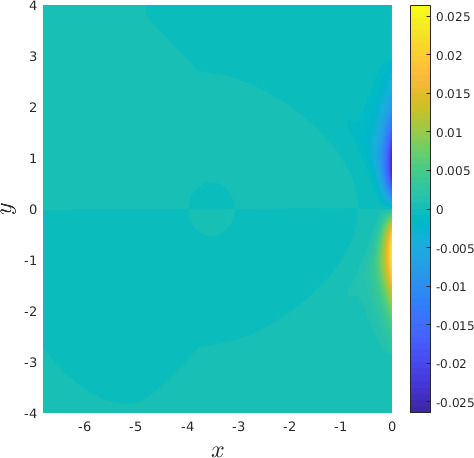}
\includegraphics[width=\saelemwidth\linewidth]{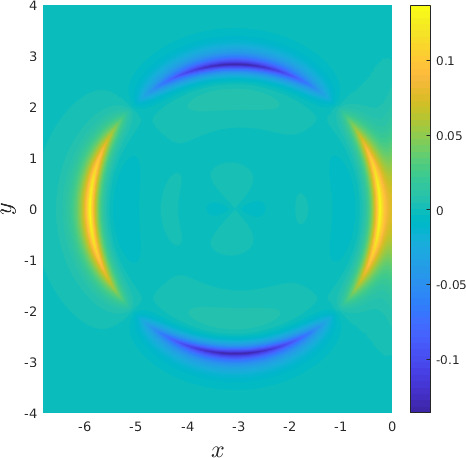}
\includegraphics[width=\saelemwidth\linewidth]{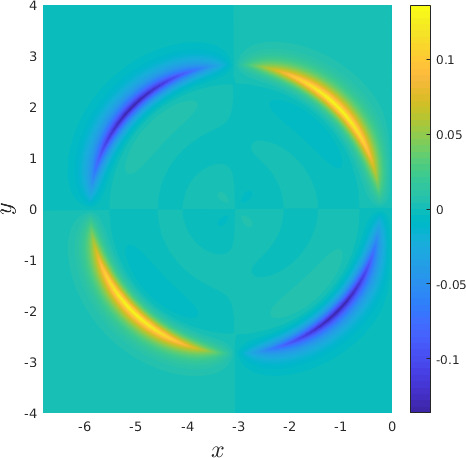}
\includegraphics[width=\saelemwidth\linewidth]{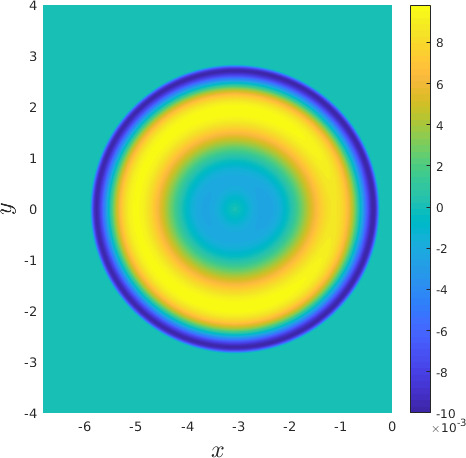}
\includegraphics[width=\saelemwidth\linewidth]{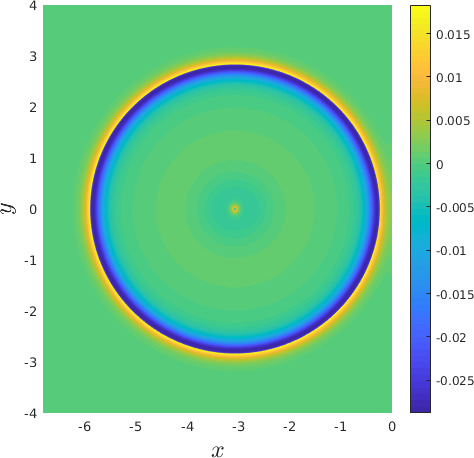}}
\caption{Numerical solution for $\df$ and $\dtheta$ for two $N=2$
  compactons situated next two each other with separation distance
  $2a$ for $\epsilon=0.0428$.
  Due to symmetry of the problem, only the left-hand side is
  calculated and shown.
  The gluing conditions \eqref{eq:xzerobccont} are imposed at $x=0$,
  which is located at the right-most part of every panel.
  Each row displays the perturbations $\df$, $\dtheta$, $\dphi^1$,
  $\dphi^2$, $\dphi^3$ and the energy density of the perturbation
  $\mathcal{E}^{\rm perturb}$, each in its column.
  Each panel is cropped so as to render the content as clear as
  possible; the left, top and bottom border do not represent the size
  of the simulations.
}
\label{fig:sapde2c20.0428}
\end{center}
\end{figure}

Figs.~\ref{fig:sapde2c20.01} and \ref{fig:sapde2c20.0428} display the
solutions for the perturbations $\df$ and $\dtheta$ for $N=2$ and
$\epsilon=0.01,0.0428$, respectively.
Due to the nontrivial relationship between the perturbations
$\df,\dtheta$ and the vector perturbations $\bdphi$ of
eq.~\eqref{eq:deltaphi_full_form}, we show the three vector components
in columns 3 through 5 of the figs.~\ref{fig:sapde2c20.01} and
\ref{fig:sapde2c20.0428}. 
Finally, the last and sixth column of the latter figures shows the
energy density \eqref{eq:Eperturb_df_dtheta} of the perturbation to
N$^2$LO, i.e.~to the third order in $\epsilon$. 
The radial perturbation, $\df$, continues to do the job of obeying the
cusp condition \eqref{eq:cusp_condition}, which is seen in the figures as
a yellow ring (the maximum) at the compacton boundary.
The only $\theta$-dependence is induced by the gluing condition
\eqref{eq:xzerobccont} near the right-hand side boundary of each
panel; the condition on $\df$ is slightly cumbersome as
$\df^2/\cos^2(N\theta)$ must obey the Neumann boundary condition at
$x=0$. The effect is most visible for small values of $a$ and as soon
as $a$ is of order of the thickness of the ``yellow'' ring, which is the
perturbation deformation due to the cusp condition, the
$\theta$-dependence is almost negligible.
This can also be seen in the second column of the figures, where
a nontrivial solution for $\dtheta$ is most pronounced for $a=0$.
The $\theta$-dependence in $\dtheta$ at the $x=0$ boundary is visible
as positive (negative) values for $y$ negative (positive).
This effect persists when $a$ is increased, but the magnitude of the
values of $\dtheta$ at the $x=0$ boundary become exponentially
suppressed in line with the tail of the perturbation around the
compacton does.
The fact that the solutions are $N=2$ baby Skyrmions is clearly visible
in columns 3 and 4, where the cusp in $\df$ switches sign four times
as one goes around the compacton border.
The induced $\theta$-dependence is also very clear for $a=0$ and
$a=0.0221$ in column 5 in the third component of the vector
perturbation, $\dphi^3$. 
The last column in the two figures shows the energy density of the
perturbation and it is not easy to see the result of the integration with
the naked eye. In fact, the largest effect of the gluing is geometric,
meaning that cutting part of the tail of the perturbation at $x=0$,
diminishes the contribution to the energy.
One could anticipate that the gluing of the two compactons would give
a positive contribution counteracting the geometric decrease in the
perturbation energy; although that happens for the field $\df$, the
induction of $\theta$-dependence turns on nontrivial behavior for
$\dtheta$ which lowers the energy slightly.
At the distance $a=0.243$, there is almost no visible effect of the
gluing conditions and the energy is very close to the sum of two
compactons. 
Fig.\ref{fig:sapde2c20.01} shows the case for very small
$\epsilon=0.01$ and in fig.~\ref{fig:sapde2c20.0428} $\epsilon=0.0428$
is slightly larger.
The effect is simply an amplification of perturbations and in
particular, the thickness of the perturbation in $\df$ due to the cusp
condition is far larger for $\epsilon=0.0428$ than for
$\epsilon=0.01$.

\begin{figure}[!htp]
  \begin{center}
    \mbox{\subfloat[]{\includegraphics[width=0.49\linewidth]{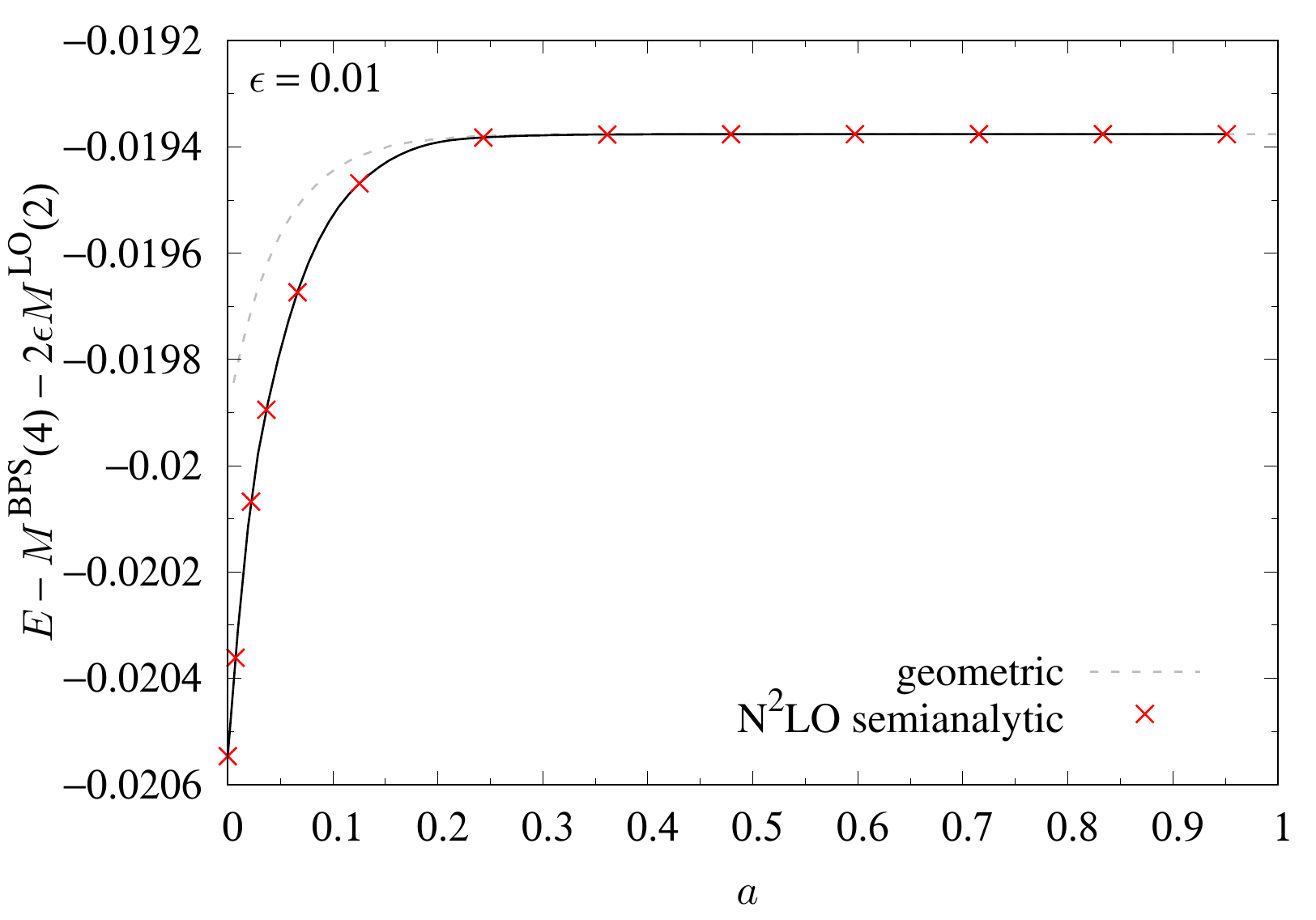}}
      \subfloat[]{\includegraphics[width=0.49\linewidth]{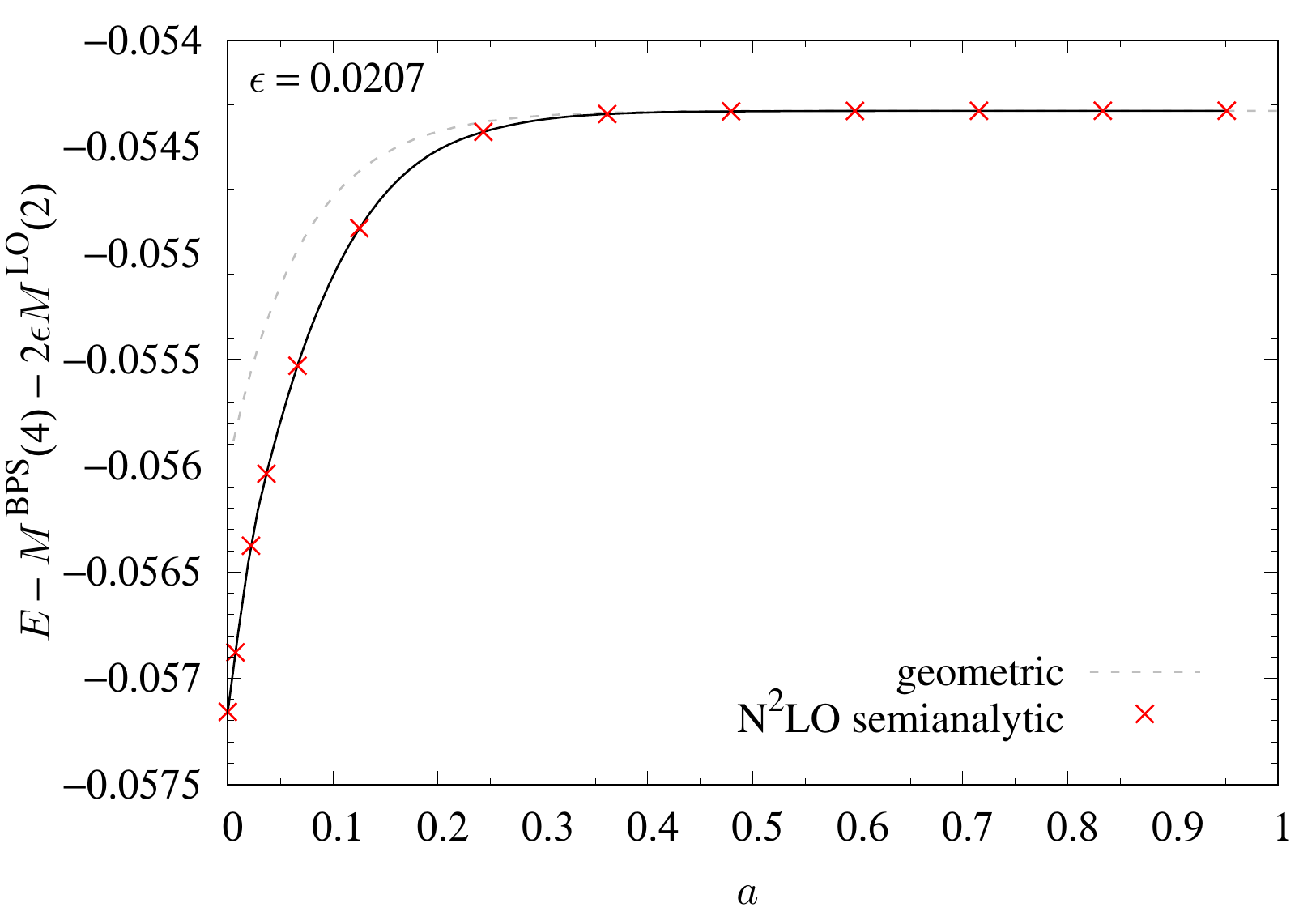}}}
    \mbox{\subfloat[]{\includegraphics[width=0.49\linewidth]{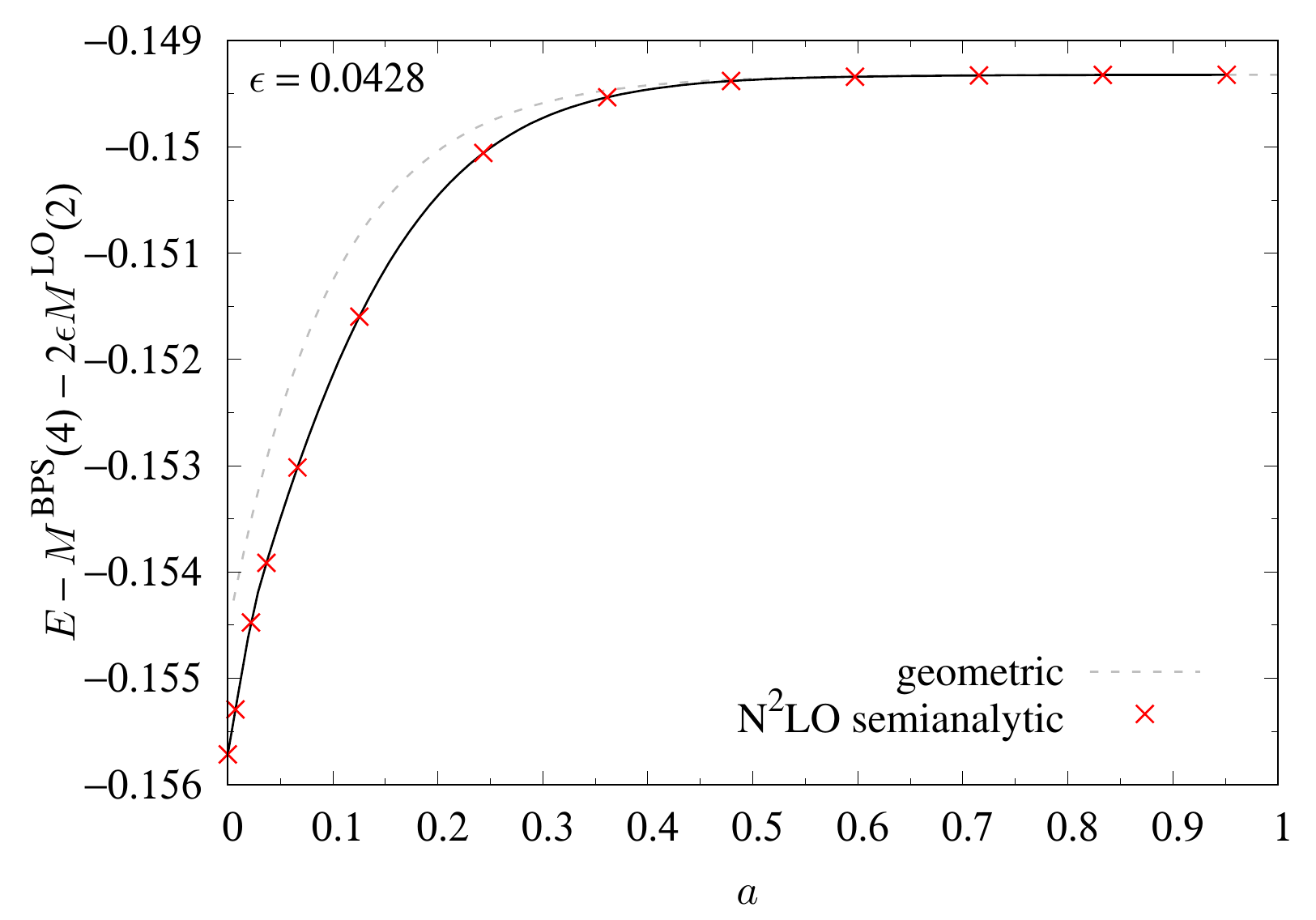}}
      \subfloat[]{\includegraphics[width=0.49\linewidth]{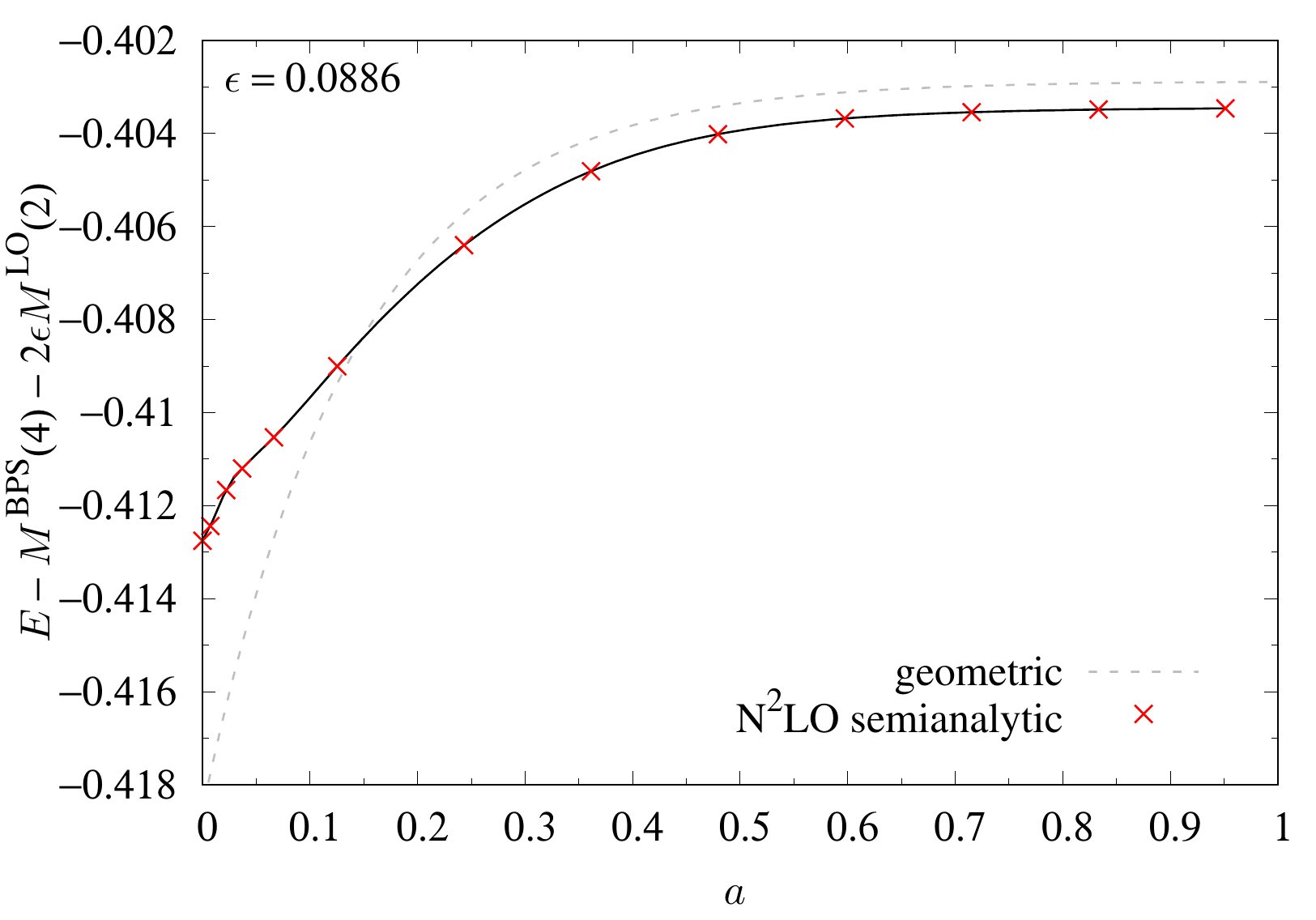}}}
    \caption{The N$^2$LO corrections to the energy for two $N=2$
      baby Skyrmions as a function of the separation distance $2a$, see
      fig.~\ref{fig:xzerobc}. The panels correspond to different
      values of $\epsilon$: (a) $\epsilon=0.01$, (b)
      $\epsilon=0.0207$, (c) $\epsilon=0.0428$ and (d)
      $\epsilon=0.0886$.
      The gray-dashed line is calculated as the geometric cut at $x=0$
      of the spherically symmetric compacton energy.
      The red crosses show the results of the N$^2$LO contribution to
      the energy from the PDE calculation of $\df$ and $\dtheta$.
      The black line is a cubic interpolation between the PDE
      results.
      For $\epsilon=0.0886$ (d) the N$^2$LO perturbation energy
      crosses over the geometric estimate and hence the excess energy
      starts to build up at this ``large'' value of $\epsilon$ for
      $N=2$. 
    }
    \label{fig:ben6}
  \end{center}
\end{figure}

Fig.~\ref{fig:ben6} shows the perturbative contribution to the energy
at next-to-next-to-leading order (N$^2$LO) for two $N=2$ baby
Skyrmions as function of the separation distance $2a$, see
fig.~\ref{fig:xzerobc}.
The four panels show $\epsilon=0.01$, $\epsilon=0.0207$,
$\epsilon=0.0428$ and $\epsilon=0.0886$, respectively and for all
panels, the smallest energy (meaning the largest negative contribution
to the energy) is at $a=0$, which means when the two compactons
exactly touch each other at a point.
For the three smallest values of $\epsilon$
(i.e.~$\epsilon=0.01,0.0207,0.0428$), the perturbation energy turns
out to be lower than the geometric energy for small $a$ (and equal at
large $a$). The geometric energy is simply the perturbation energy of
the axially symmetric compacton cut off at $x=0$. 
However, for $\epsilon=0.0886$ the perturbation energy crosses over
the geometric line at $a\sim 0.12$ and is slightly above the geometric
one at $a=0$, indicating that the gluing condition builds up some
tension or excess energy in the perturbation for this ``large'' value
of $\epsilon$. 
Of course, our perturbation scheme is best trusted at small
$\epsilon$, so this may well be an indication of the approximation
starting to deteriorate.

\begin{figure}[!htp]
  \begin{center}
    \mbox{\subfloat[]{\includegraphics[width=0.49\linewidth]{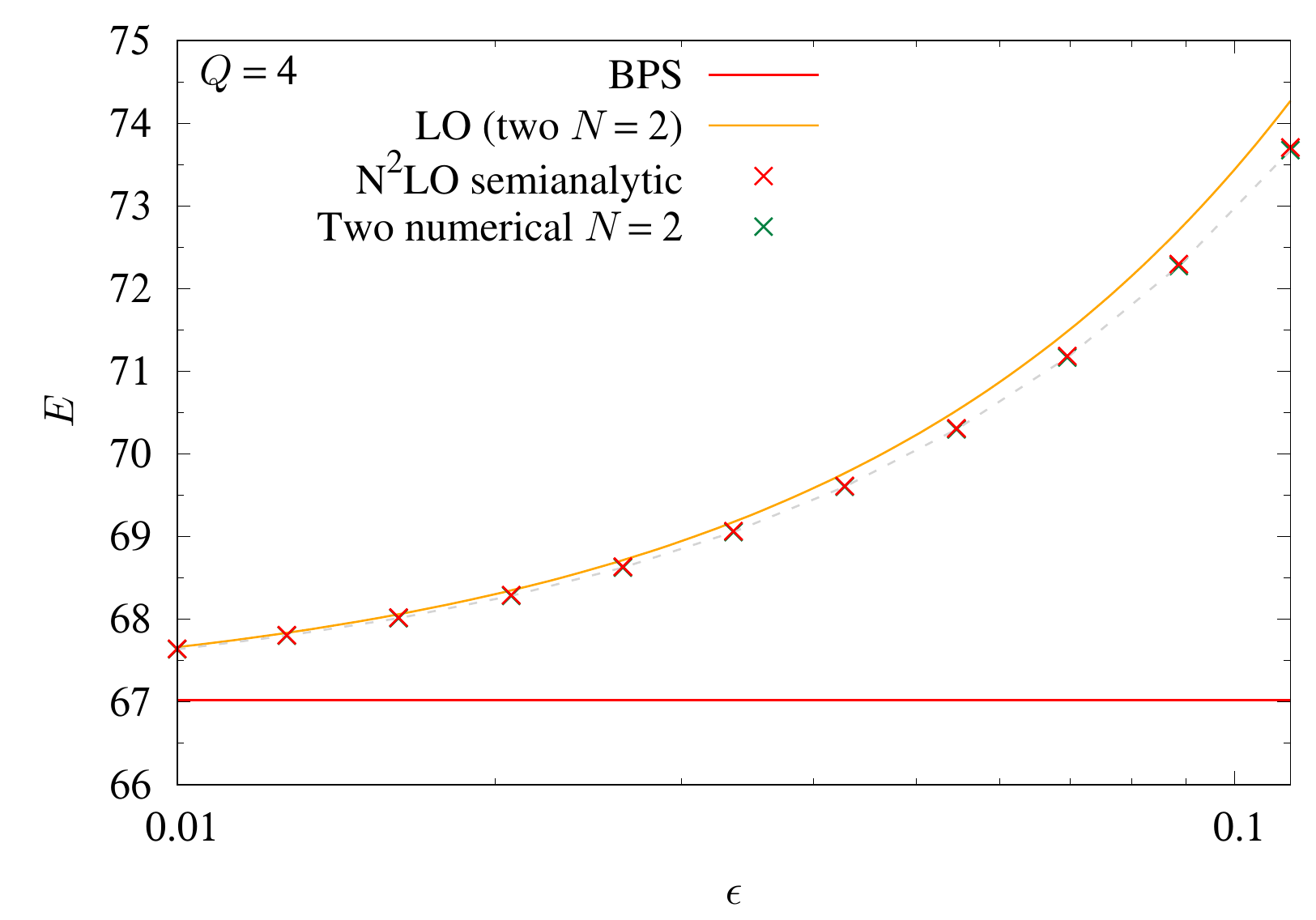}}
      \subfloat[]{\includegraphics[width=0.49\linewidth]{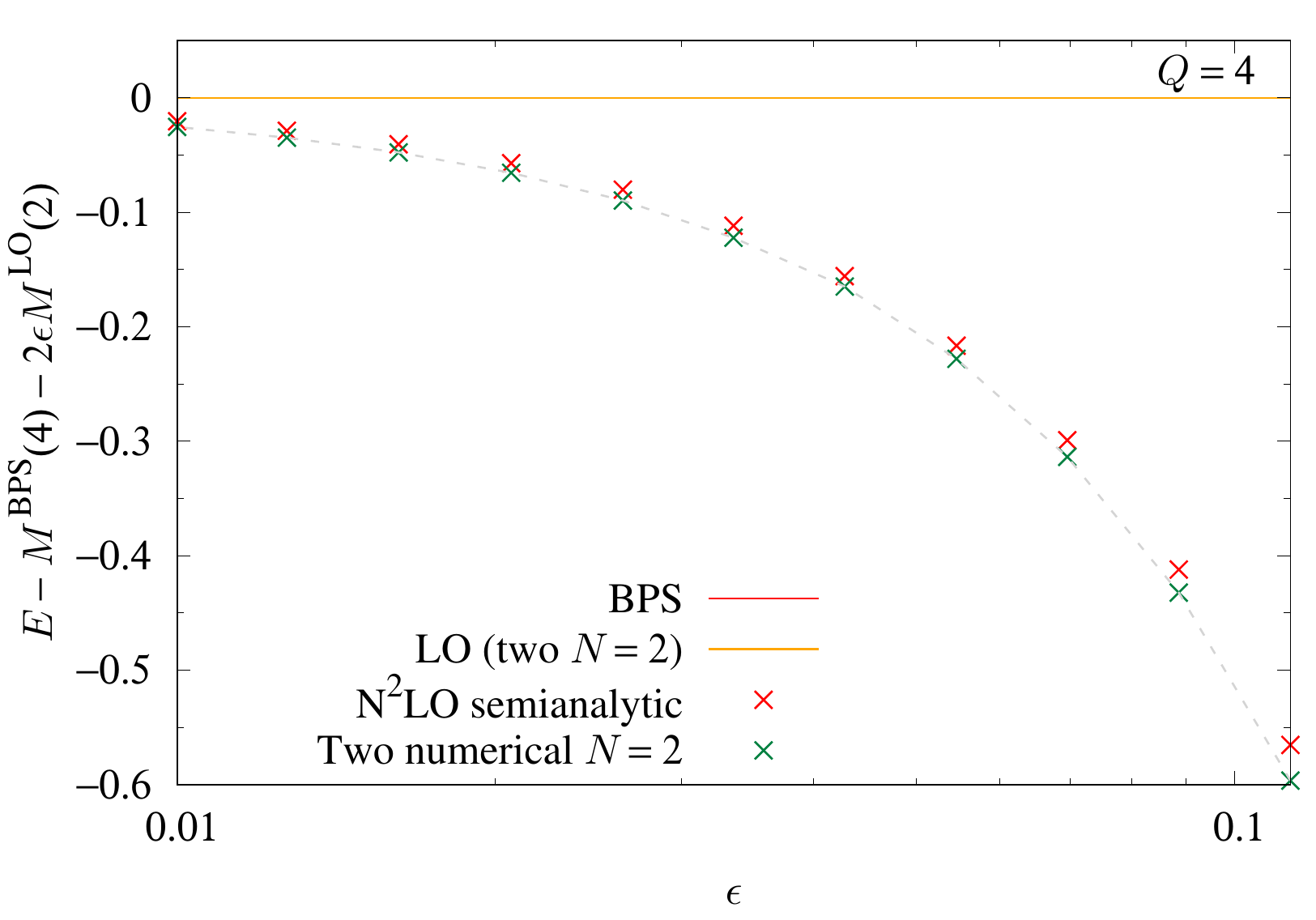}}}
    \caption{The energy (a) and N$^2$LO perturbation energy of the
      lowest energy solutions in the $Q=4$ homotopy class, which is
      made of two $N=2$ compactons with separation distance $2a=0$.
    The total energy (a) shows the BPS energy (red), the LO correction
    from the kinetic term of order $\epsilon$ (orange), the full
    N$^2$LO perturbation energy obtained by PDE calculations (red
    crosses) and finally in comparison with the full PDE calculations
    of the solutions without perturbation theory (green crosses and
    connected with gray-dashed lines), all as functions of
    $\epsilon$. 
    (b) shows the N$^2$LO part of the energy only, so as to better see
    the accuracy of the comparison between the full numerical PDE
    energies (green crosses) and the N$^2$LO perturbation energies
    (red crosses) as functions of $\epsilon$. }
    \label{fig:saen4}
  \end{center}
\end{figure}

We finally, compare the N$^2$LO result of the perturbation theory of
the energy of two $N=2$ baby Skyrmions sitting side-by-side -- taking
into account the binding energy -- with the full numerical
(brute-force) PDE calculations, see fig.~\ref{fig:saen4}.
The two results are incredibly close to each other, in particular for
small values of $\epsilon$. 
Since the binding energy requires a PDE calculation as well, it is not
clear which result is more accurate.
We expect the perturbative result to be most accurate of the two for
small $\epsilon\sim 0.01$, whereas the full numerical PDE calculations are most
accurate for large values of $\epsilon\sim 0.1$.

\begin{figure}[!htp]
\begin{center}
\begin{tabular}{p{10pt}p{\tabelemwidth}p{\tabelemwidth}p{\tabelemwidth}p{\tabelemwidth}p{\tabelemwidth}l}
& $\df$ & $\dtheta$ & $\dphi^1$ & $\dphi^2$ & $\dphi^3$ & $\mathcal{E}^{\rm perturb}$
\end{tabular}
\mbox{\rotatebox{90}{\quad$a=0$}
\includegraphics[width=\saelemwidth\linewidth]{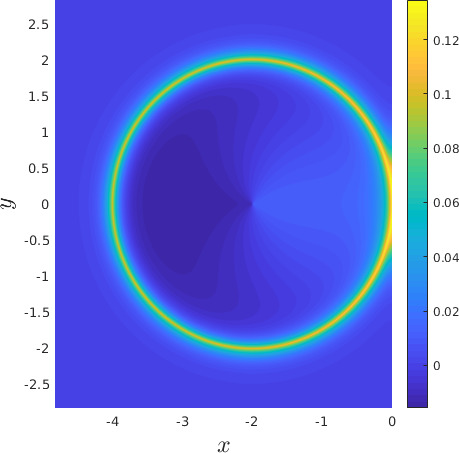}
\includegraphics[width=\saelemwidth\linewidth]{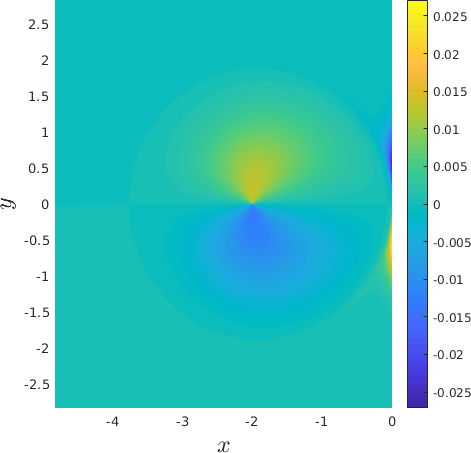}
\includegraphics[width=\saelemwidth\linewidth]{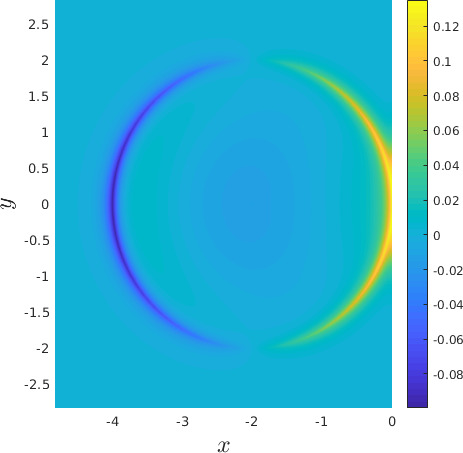}
\includegraphics[width=\saelemwidth\linewidth]{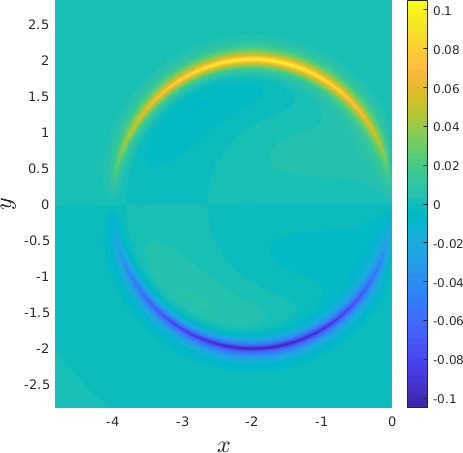}
\includegraphics[width=\saelemwidth\linewidth]{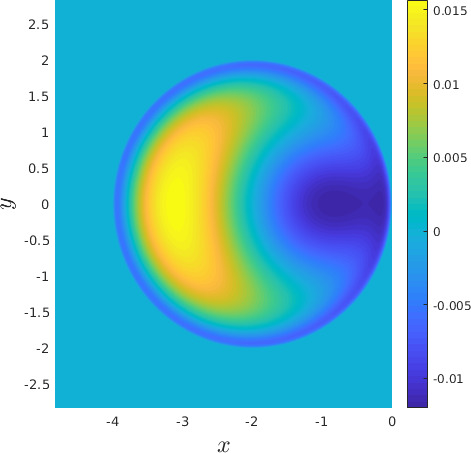}
\includegraphics[width=\saelemwidth\linewidth]{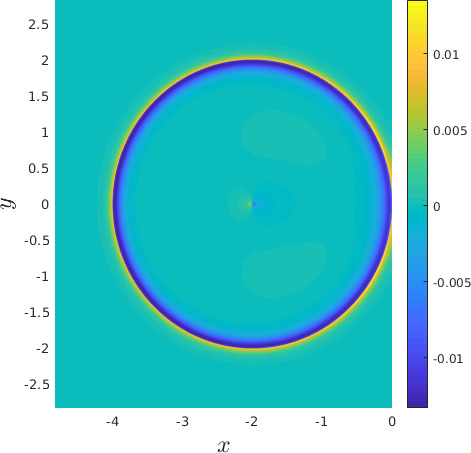}}
\mbox{\rotatebox{90}{\quad$a=0.0156$}
\includegraphics[width=\saelemwidth\linewidth]{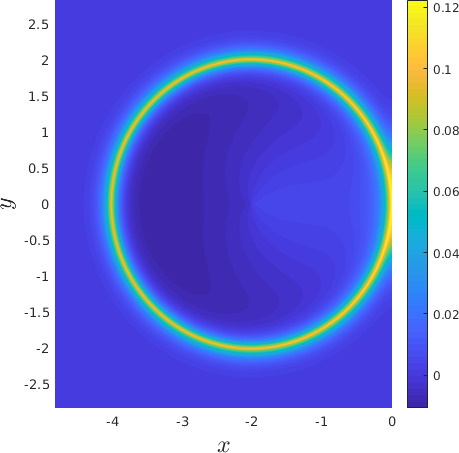}
\includegraphics[width=\saelemwidth\linewidth]{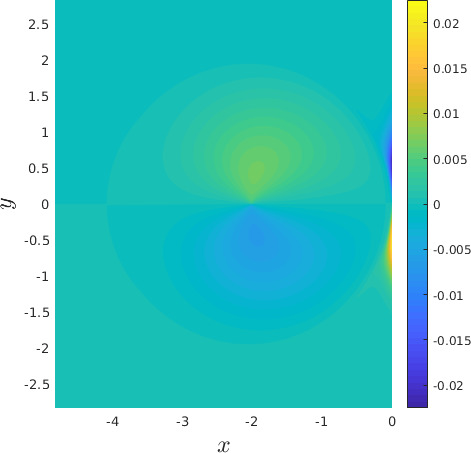}
\includegraphics[width=\saelemwidth\linewidth]{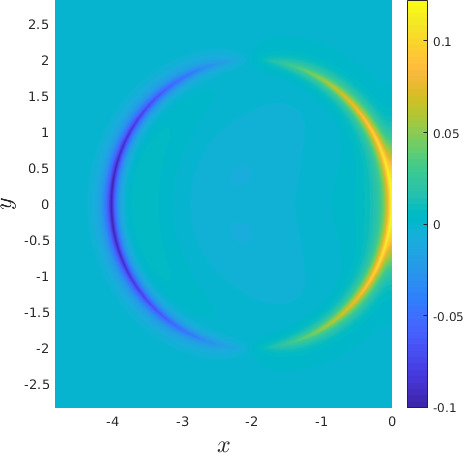}
\includegraphics[width=\saelemwidth\linewidth]{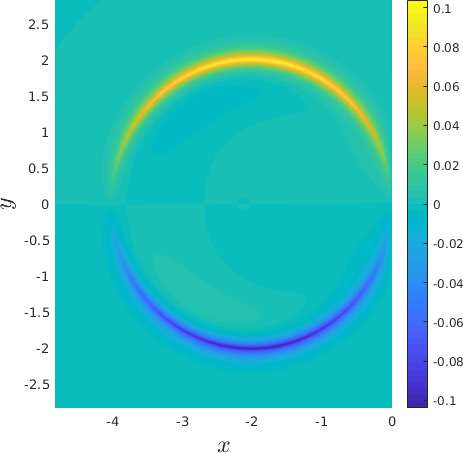}
\includegraphics[width=\saelemwidth\linewidth]{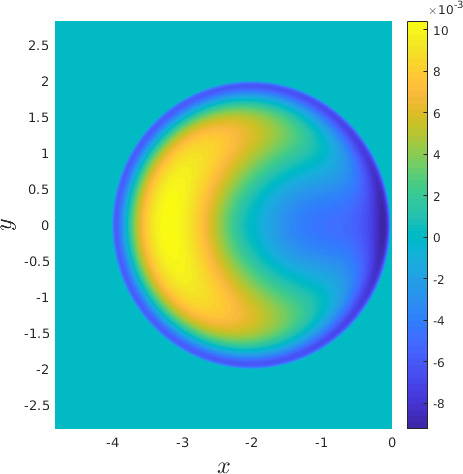}
\includegraphics[width=\saelemwidth\linewidth]{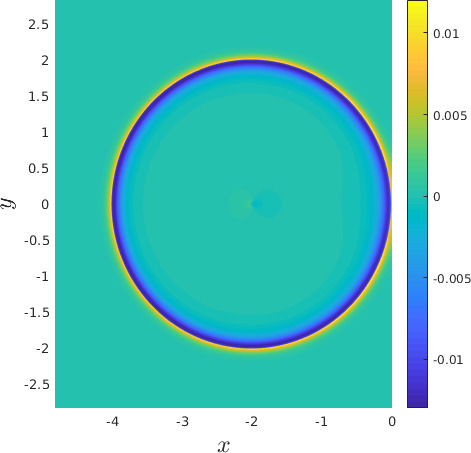}}
\mbox{\rotatebox{90}{\quad$a=0.0469$}
\includegraphics[width=\saelemwidth\linewidth]{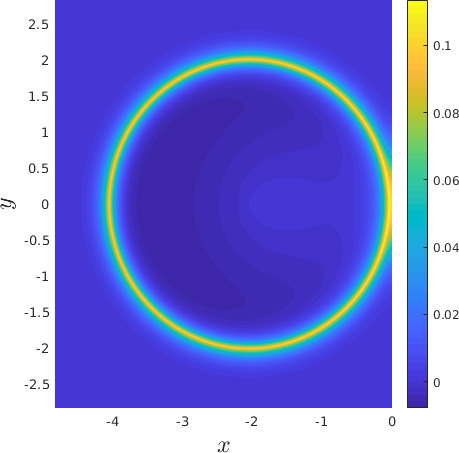}
\includegraphics[width=\saelemwidth\linewidth]{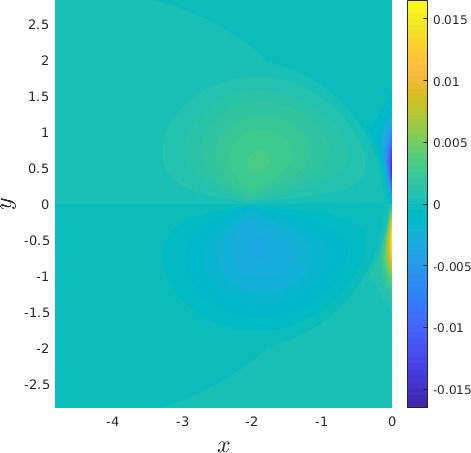}
\includegraphics[width=\saelemwidth\linewidth]{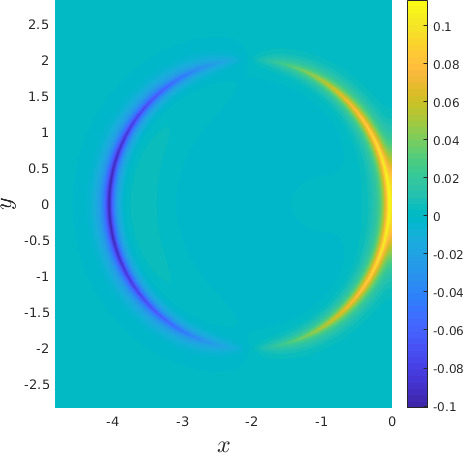}
\includegraphics[width=\saelemwidth\linewidth]{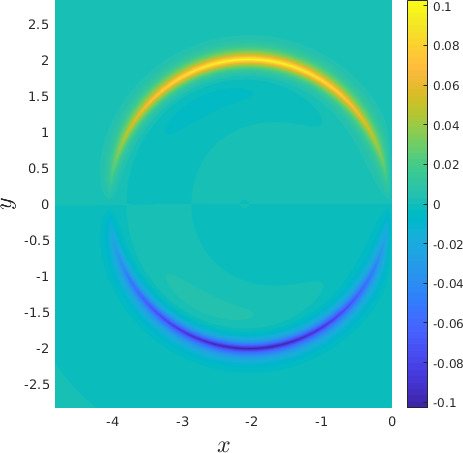}
\includegraphics[width=\saelemwidth\linewidth]{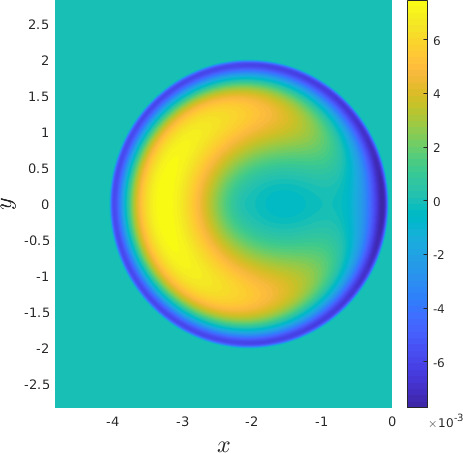}
\includegraphics[width=\saelemwidth\linewidth]{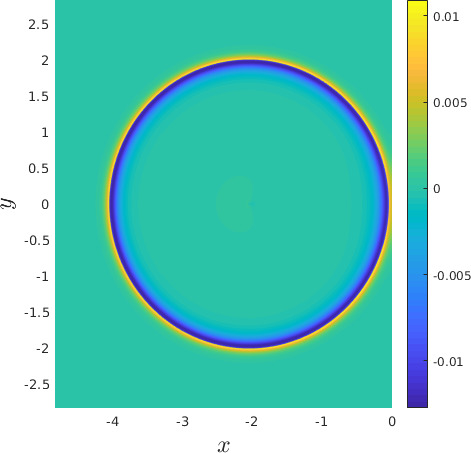}}
\mbox{\rotatebox{90}{\quad$a=0.0887$}
\includegraphics[width=\saelemwidth\linewidth]{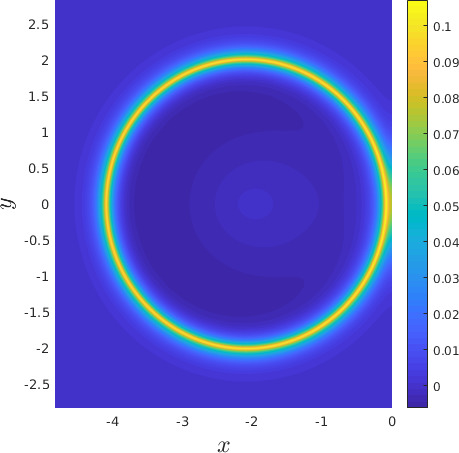}
\includegraphics[width=\saelemwidth\linewidth]{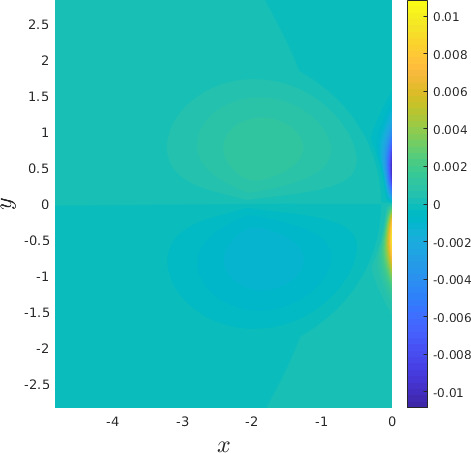}
\includegraphics[width=\saelemwidth\linewidth]{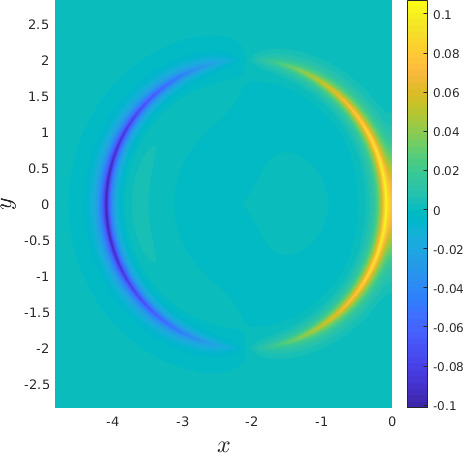}
\includegraphics[width=\saelemwidth\linewidth]{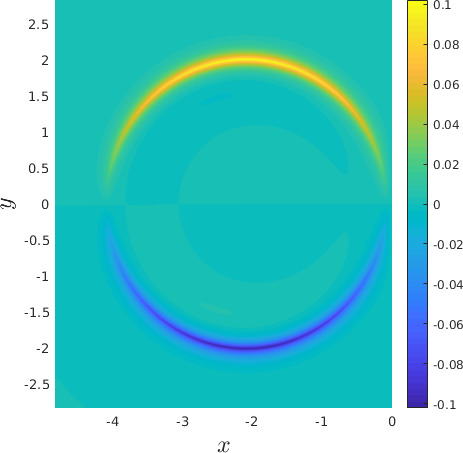}
\includegraphics[width=\saelemwidth\linewidth]{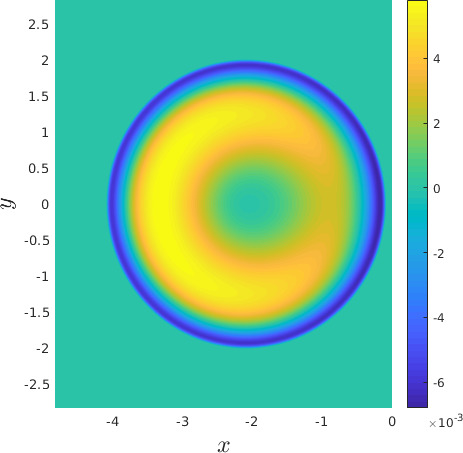}
\includegraphics[width=\saelemwidth\linewidth]{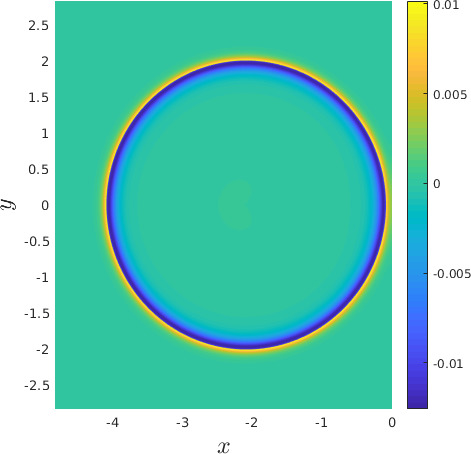}}
\mbox{\rotatebox{90}{\quad$a=0.172$}
\includegraphics[width=\saelemwidth\linewidth]{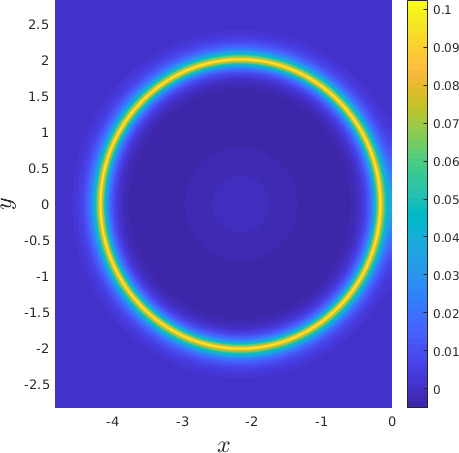}
\includegraphics[width=\saelemwidth\linewidth]{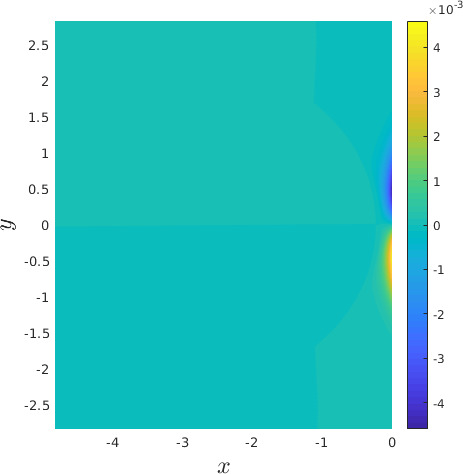}
\includegraphics[width=\saelemwidth\linewidth]{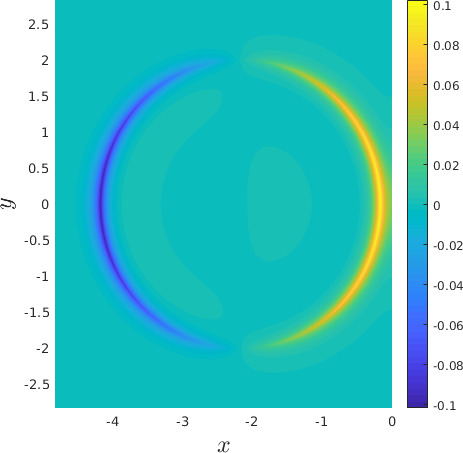}
\includegraphics[width=\saelemwidth\linewidth]{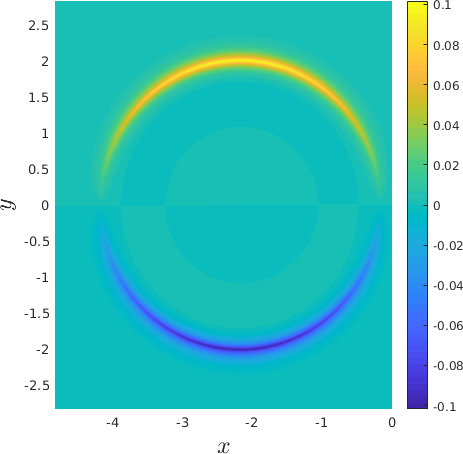}
\includegraphics[width=\saelemwidth\linewidth]{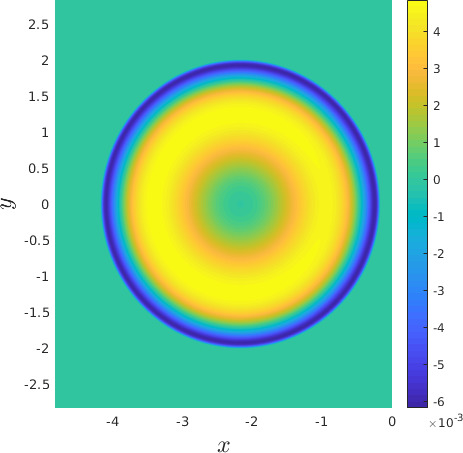}
\includegraphics[width=\saelemwidth\linewidth]{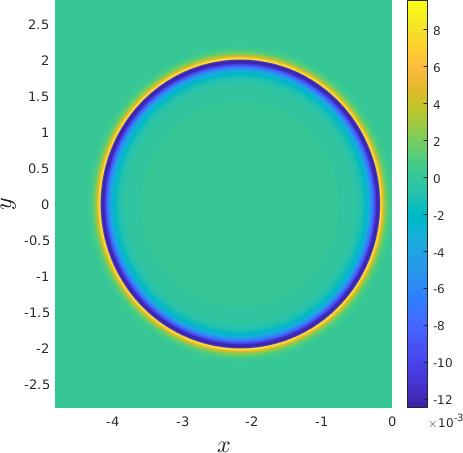}}
\mbox{\rotatebox{90}{\quad$a=0.256$}
\includegraphics[width=\saelemwidth\linewidth]{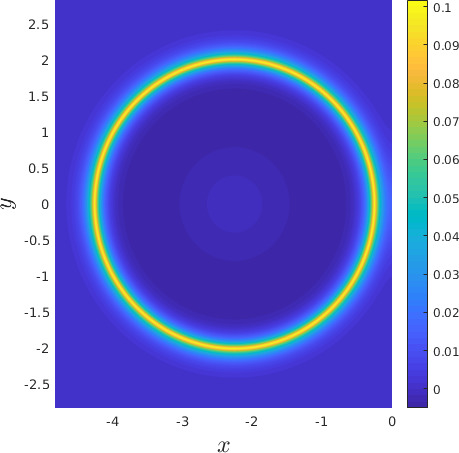}
\includegraphics[width=\saelemwidth\linewidth]{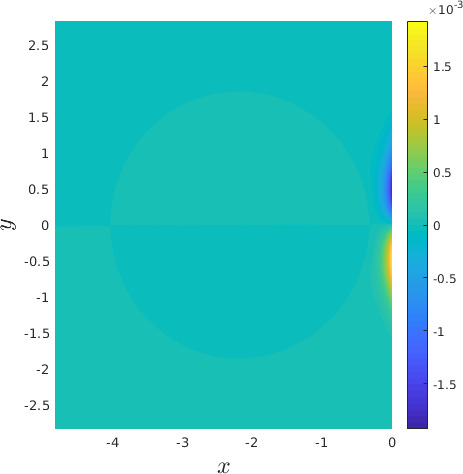}
\includegraphics[width=\saelemwidth\linewidth]{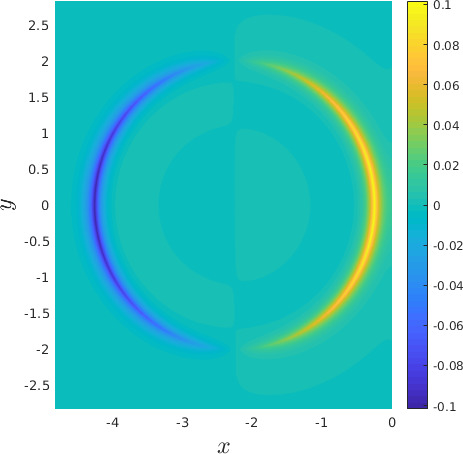}
\includegraphics[width=\saelemwidth\linewidth]{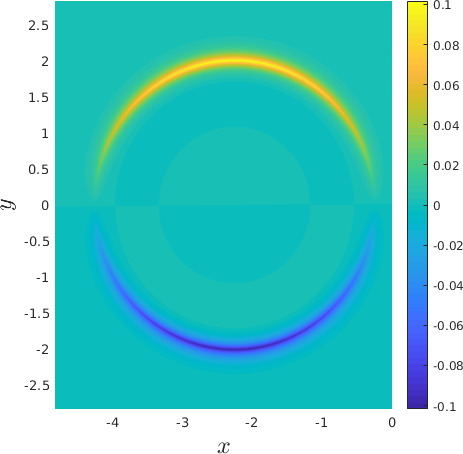}
\includegraphics[width=\saelemwidth\linewidth]{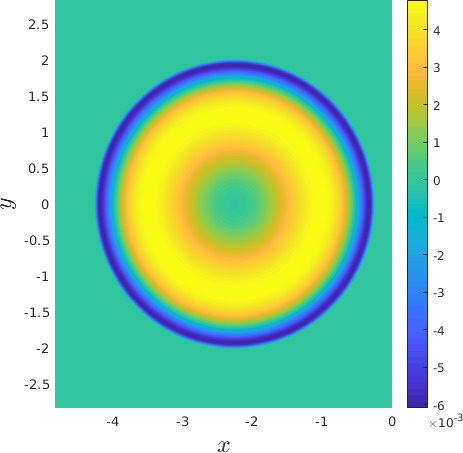}
\includegraphics[width=\saelemwidth\linewidth]{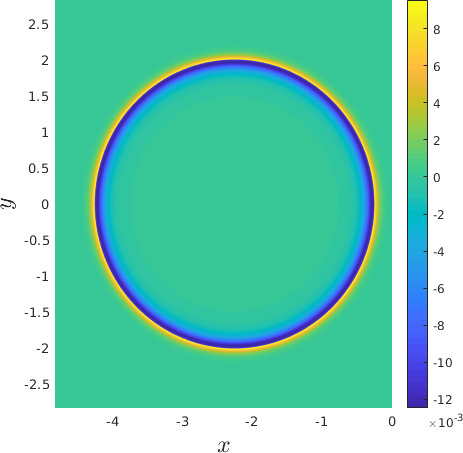}}
\caption{Numerical solution for $\df$ and $\dtheta$ for two $N=1$
  compactons situated next two each other with separation distance
  $2a$ for $\epsilon=0.01$.
  Due to symmetry of the problem, only the left-hand side is
  calculated and shown.
  The gluing conditions \eqref{eq:xzerobccont} are imposed at $x=0$,
  which is located at the right-most part of every panel.
  Each row displays the perturbations $\df$, $\dtheta$, $\dphi^1$,
  $\dphi^2$, $\dphi^3$ and the energy density of the perturbation
  $\mathcal{E}^{\rm perturb}$, each in its column.
  Each panel is cropped so as to render the content as clear as
  possible; the left, top and bottom border do not represent the size
  of the simulations.
}
\label{fig:sapde1c20.01}
\end{center}
\end{figure}

\begin{figure}[!htp]
\begin{center}
\begin{tabular}{p{10pt}p{\tabelemwidth}p{\tabelemwidth}p{\tabelemwidth}p{\tabelemwidth}p{\tabelemwidth}l}
& $\df$ & $\dtheta$ & $\dphi^1$ & $\dphi^2$ & $\dphi^3$ & $\mathcal{E}^{\rm perturb}$
\end{tabular}
\mbox{\rotatebox{90}{\quad$a=0$}
\includegraphics[width=\saelemwidth\linewidth]{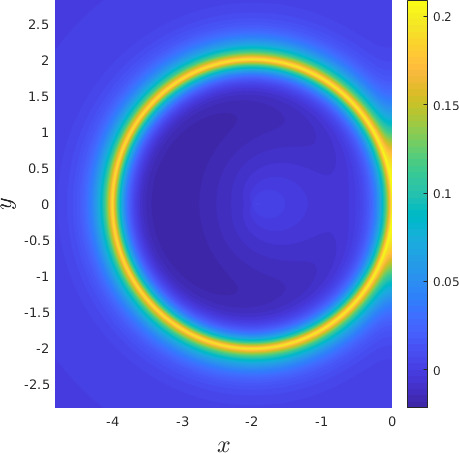}
\includegraphics[width=\saelemwidth\linewidth]{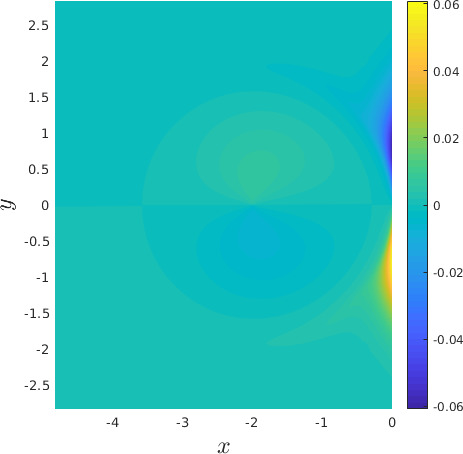}
\includegraphics[width=\saelemwidth\linewidth]{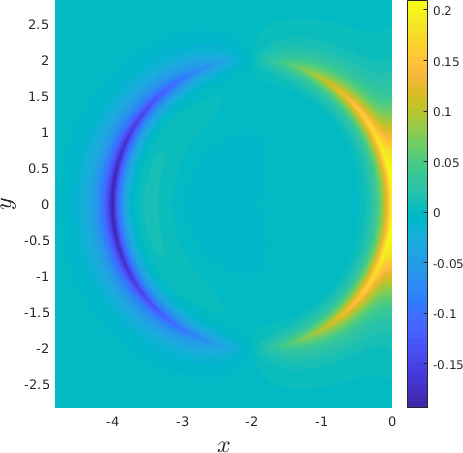}
\includegraphics[width=\saelemwidth\linewidth]{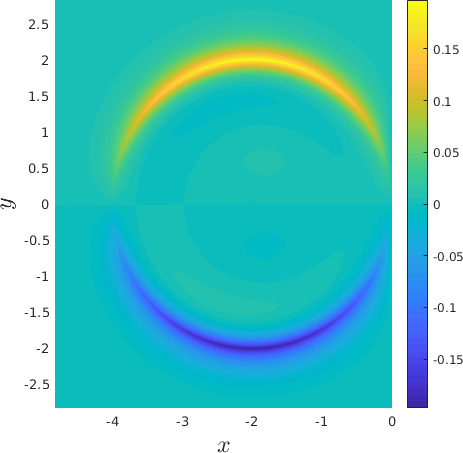}
\includegraphics[width=\saelemwidth\linewidth]{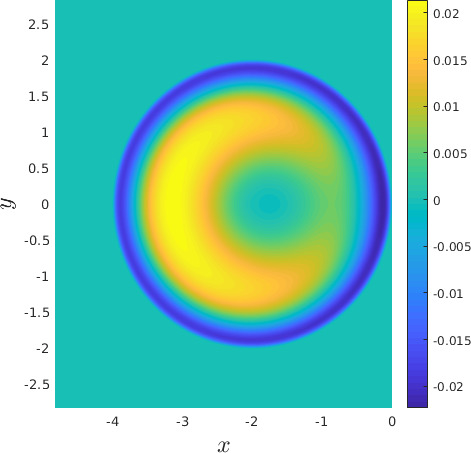}
\includegraphics[width=\saelemwidth\linewidth]{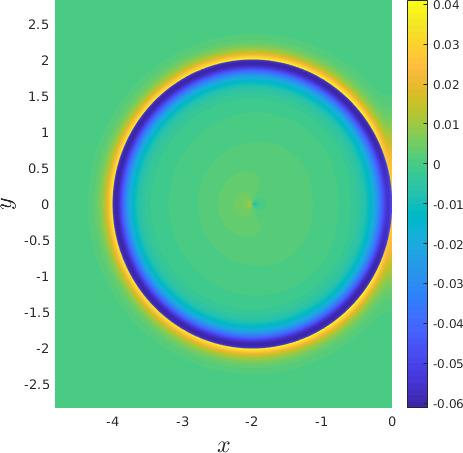}}
\mbox{\rotatebox{90}{\quad$a=0.0156$}
\includegraphics[width=\saelemwidth\linewidth]{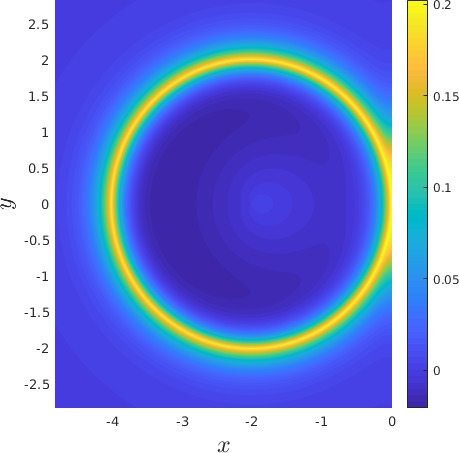}
\includegraphics[width=\saelemwidth\linewidth]{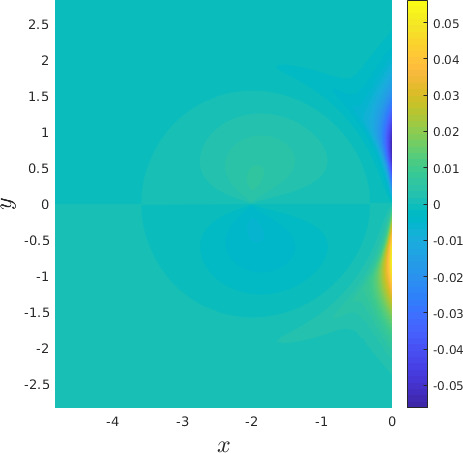}
\includegraphics[width=\saelemwidth\linewidth]{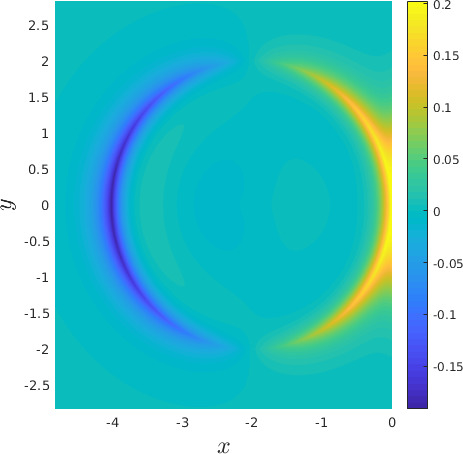}
\includegraphics[width=\saelemwidth\linewidth]{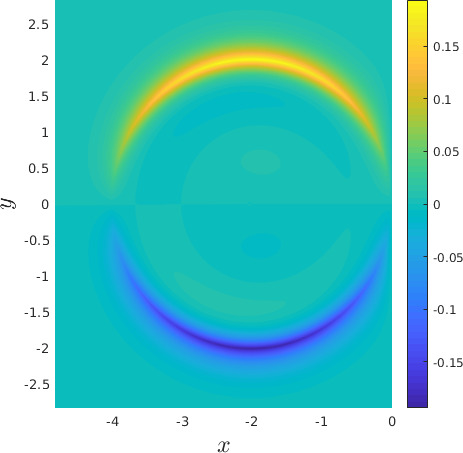}
\includegraphics[width=\saelemwidth\linewidth]{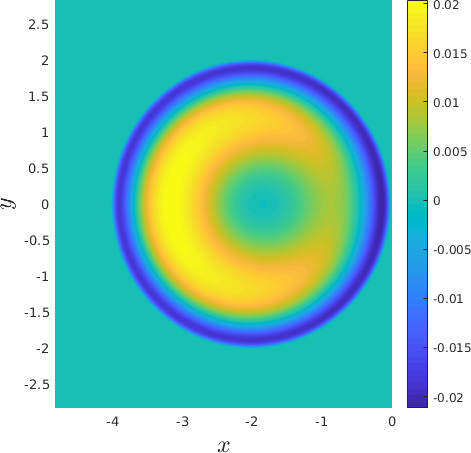}
\includegraphics[width=\saelemwidth\linewidth]{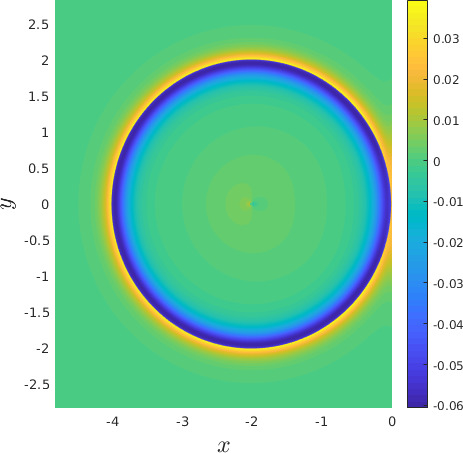}}
\mbox{\rotatebox{90}{\quad$a=0.0469$}
\includegraphics[width=\saelemwidth\linewidth]{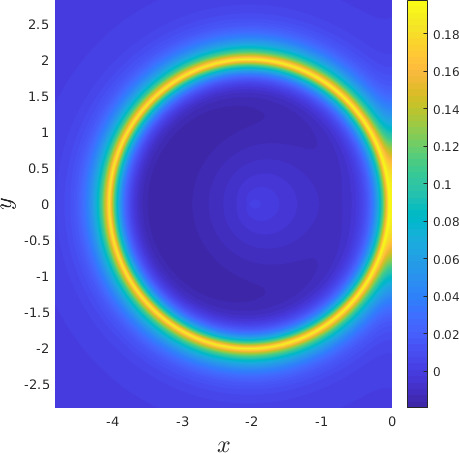}
\includegraphics[width=\saelemwidth\linewidth]{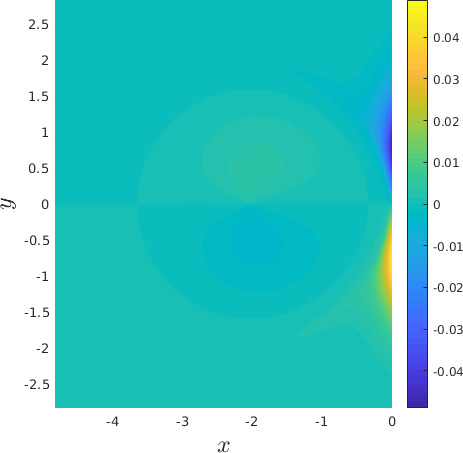}
\includegraphics[width=\saelemwidth\linewidth]{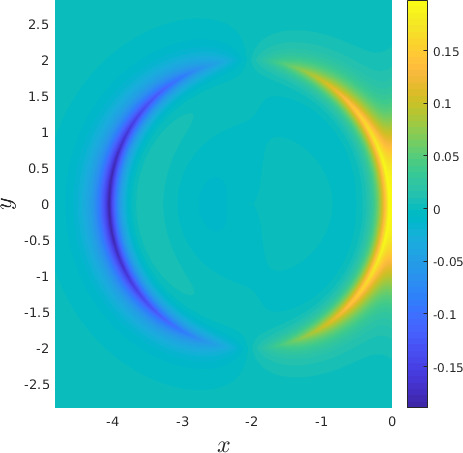}
\includegraphics[width=\saelemwidth\linewidth]{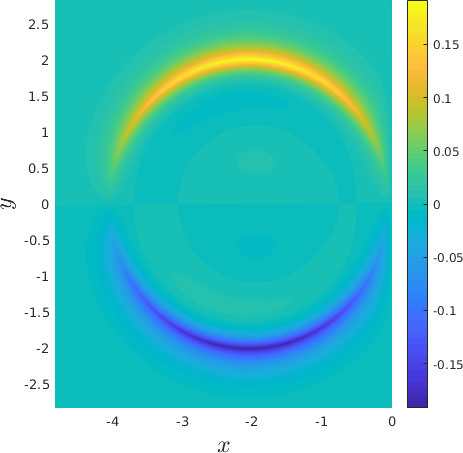}
\includegraphics[width=\saelemwidth\linewidth]{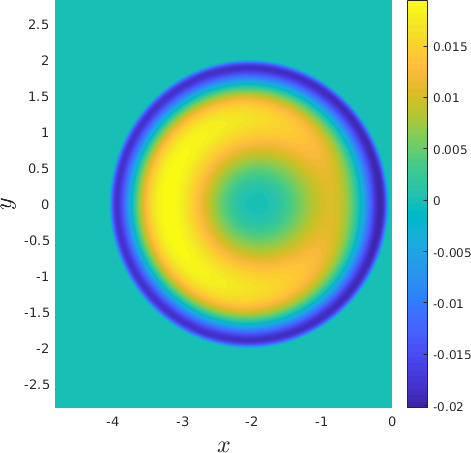}
\includegraphics[width=\saelemwidth\linewidth]{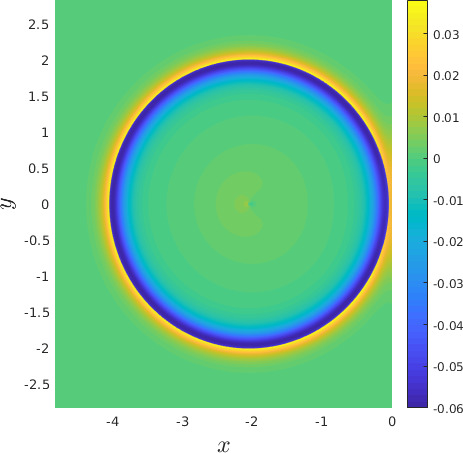}}
\mbox{\rotatebox{90}{\quad$a=0.0887$}
\includegraphics[width=\saelemwidth\linewidth]{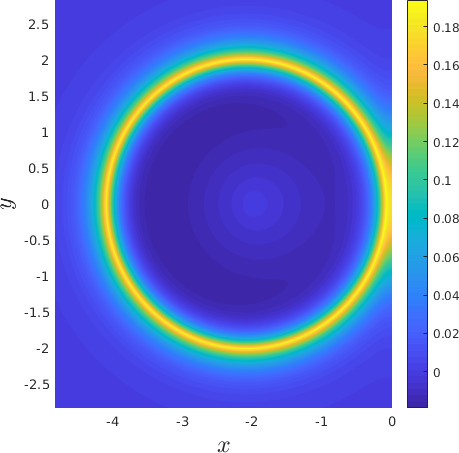}
\includegraphics[width=\saelemwidth\linewidth]{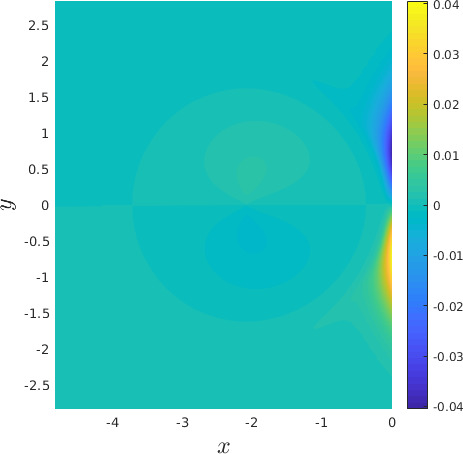}
\includegraphics[width=\saelemwidth\linewidth]{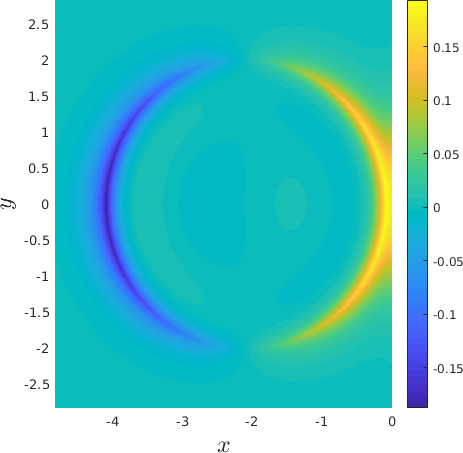}
\includegraphics[width=\saelemwidth\linewidth]{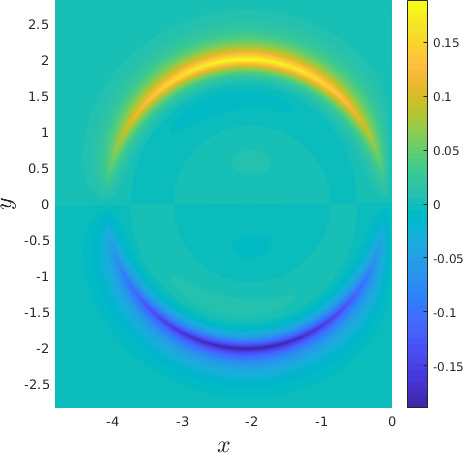}
\includegraphics[width=\saelemwidth\linewidth]{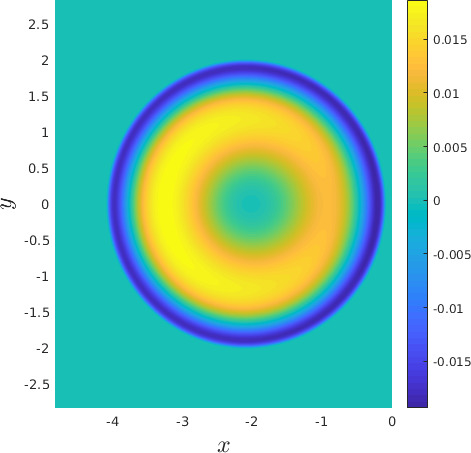}
\includegraphics[width=\saelemwidth\linewidth]{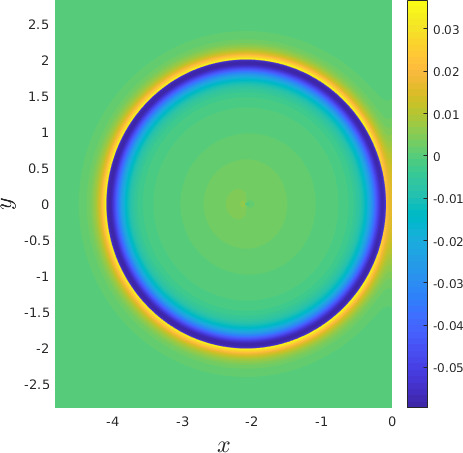}}
\mbox{\rotatebox{90}{\quad$a=0.172$}
\includegraphics[width=\saelemwidth\linewidth]{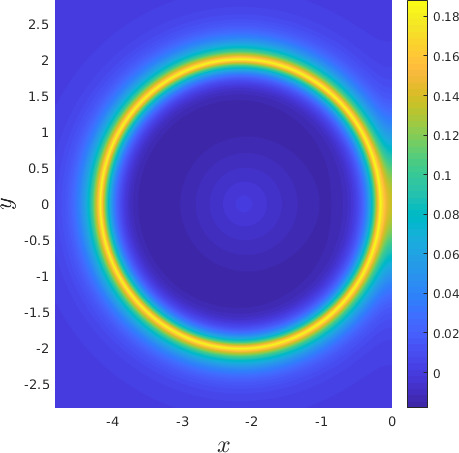}
\includegraphics[width=\saelemwidth\linewidth]{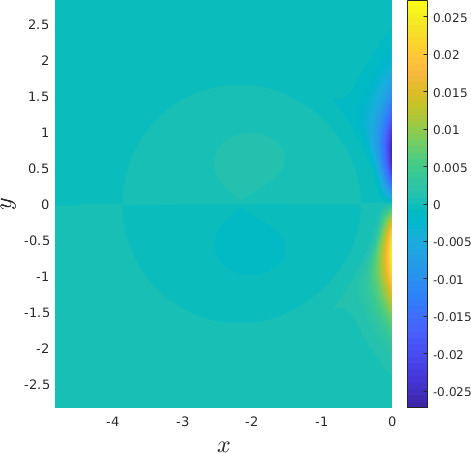}
\includegraphics[width=\saelemwidth\linewidth]{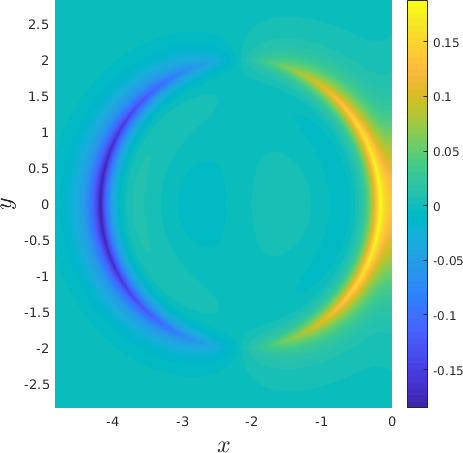}
\includegraphics[width=\saelemwidth\linewidth]{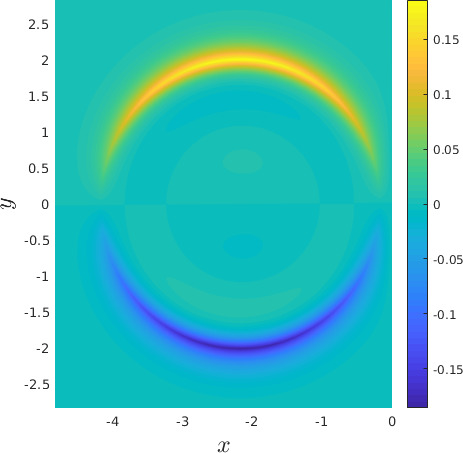}
\includegraphics[width=\saelemwidth\linewidth]{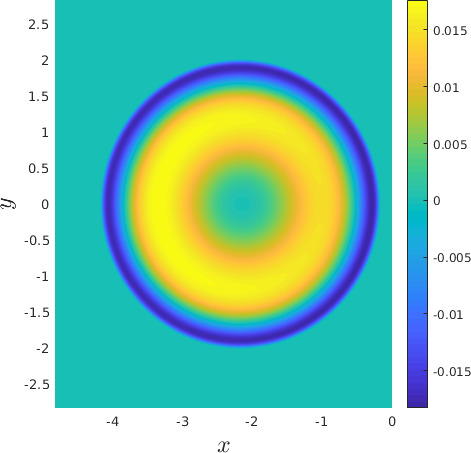}
\includegraphics[width=\saelemwidth\linewidth]{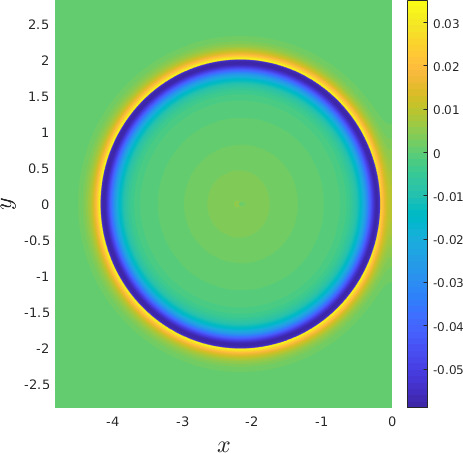}}
\mbox{\rotatebox{90}{\quad$a=0.256$}
\includegraphics[width=\saelemwidth\linewidth]{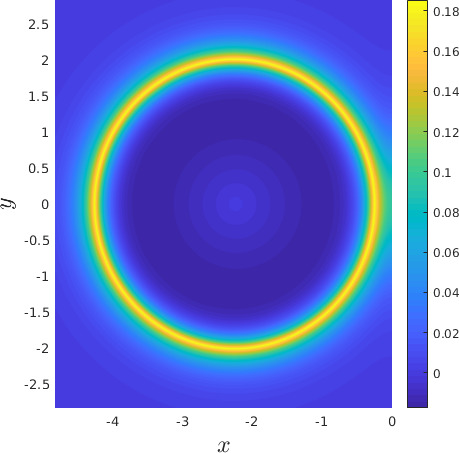}
\includegraphics[width=\saelemwidth\linewidth]{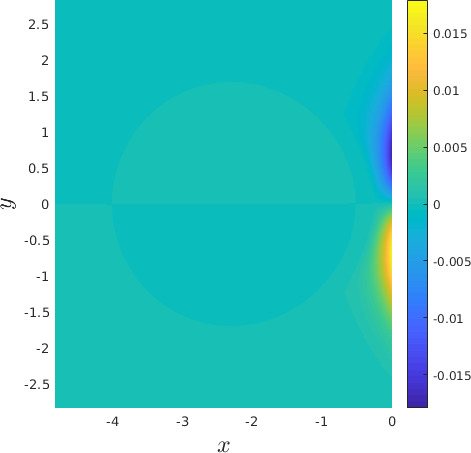}
\includegraphics[width=\saelemwidth\linewidth]{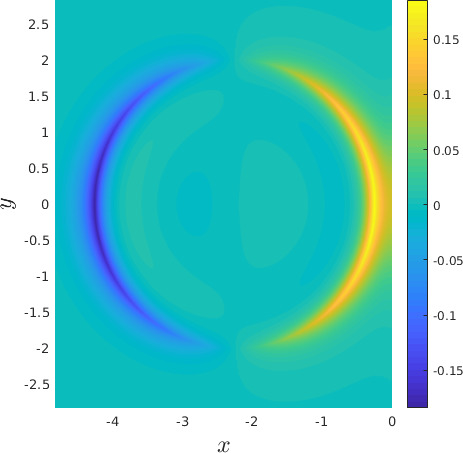}
\includegraphics[width=\saelemwidth\linewidth]{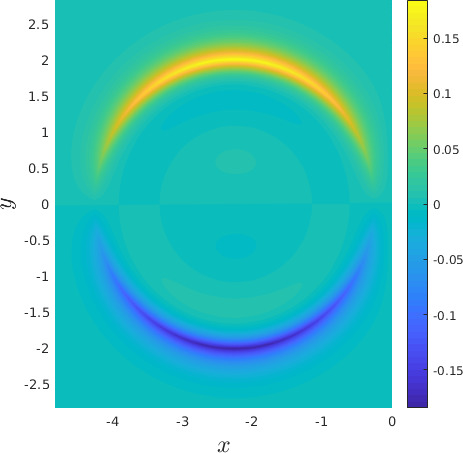}
\includegraphics[width=\saelemwidth\linewidth]{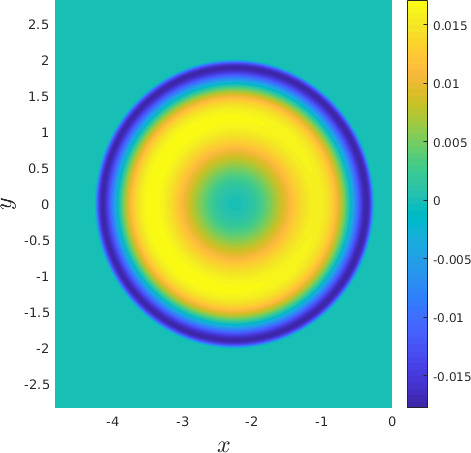}
\includegraphics[width=\saelemwidth\linewidth]{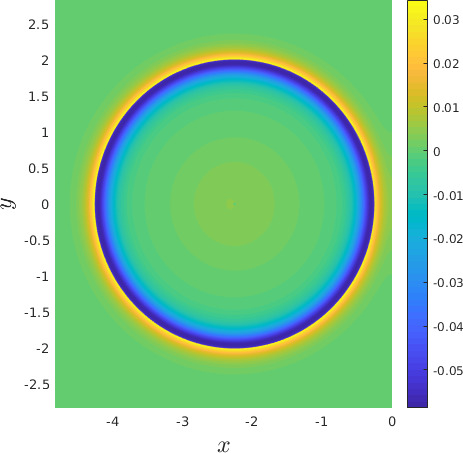}}
\caption{Numerical solution for $\df$ and $\dtheta$ for two $N=1$
  compactons situated next two each other with separation distance
  $2a$ for $\epsilon=0.0428$.
  Due to symmetry of the problem, only the left-hand side is
  calculated and shown.
  The gluing conditions \eqref{eq:xzerobccont} are imposed at $x=0$,
  which is located at the right-most part of every panel.
  Each row displays the perturbations $\df$, $\dtheta$, $\dphi^1$,
  $\dphi^2$, $\dphi^3$ and the energy density of the perturbation
  $\mathcal{E}^{\rm perturb}$, each in its column.
  Each panel is cropped so as to render the content as clear as
  possible; the left, top and bottom border do not represent the size
  of the simulations.
}
\label{fig:sapde1c20.0428}
\end{center}
\end{figure}

Before summarizing the results for the binding energies, we will
repeat the perturbative N$^2$LO calculation of the two baby Skyrmions
side-by-side, but this time for $N=1$ compactons.
This solution is only metastable for small $\epsilon$ while for
$\epsilon\gtrsim0.15$ it is unstable. 
Nevertheless, since we have performed the full PDE computations of
these composite baby Skyrmions in sec.~\ref{quattro}, it is a good
test of the perturbative scheme. 

Similarly to the case of $Q=2+2$, for the case of two $N=1$ baby
Skyrmions side-by-side, we must specify the background solution around
which we calculate the perturbation
\eqref{eq:deltaphi_full_form}. Again the restricted harmonic condition
does not identify uniquely the right compacton configuration that is
needed since it does not specify any condition on the relative
orientation (or relative distance).
Following the discussion of the $Q=2+2$ case, we look at the
long-range potential of two well-separated $N=1$ baby Skyrmions
\cite{Piette:1994ug}:
\begin{equation}
  V_{1+1}\propto \cos(\alpha-\beta)\frac{e^{-mR}}{\sqrt{mR}}
  \label{eq:pot1+1}
\end{equation}
where again $R$ is the relative distance, $\alpha$ and $\beta$ are the
two respective phases and $m$ is the pion mass. The potential
\eqref{eq:pot1+1} generates the maximum attractive force in
case of two baby Skyrmions with opposite orientation,
i.e.~$\alpha=\beta+\pi$. Therefore, in contrast to the $Q=2+2$ case, 
among all the axially symmetric compactons side-by-side, we choose two 
$N=1$ solitons with opposite orientation as the background
solution. Our guess is confirmed by the numerical simulation in
fig.~\ref{fig:N=1+1}. 

Similarly to the case of two $N=2$ compactons side-by-side,
figs.~\ref{fig:sapde1c20.01} and \ref{fig:sapde1c20.0428} which are
for two $N=1$ compactons side-by-side, show the left-hand side of the
composite soliton configuration with the gluing condition
\eqref{eq:xzerobccont} imposed on the right-hand boundary of each
panel for $\epsilon=0.01$ and $\epsilon=0.0428$, respectively.
The other edges of the panels are not the limit of the calculations,
which have merely been cropped so as to render the content as clear
as possible.
The two figs.~\ref{fig:sapde1c20.01} and \ref{fig:sapde1c20.0428}
again display the 6 columns: the perturbations $\df$ and $\dtheta$ as
well as the three components of the vector perturbations $\bdphi$, and
finally the perturbation energy density at N$^2$LO.
Similarly to the $N=2$ case, the largest $\theta$-dependence is
induced for very small separation distances $2a$ and especially for
$a=0$.
It is clearly visible from the third and fourth columns of the figures
that the compacton background solutions are $N=1$ baby Skyrmions,
where cusp condition only changes sign twice around the perimeter of
the compacton.
As before, the width of the perturbation due to the cusp condition
grows with $\epsilon$ and hence is more pronounced in
fig.~\ref{fig:sapde1c20.0428} than in fig.~\ref{fig:sapde1c20.01}.
For both values of $\epsilon$, the effect of the presence of the other
compacton has almost disappeared at the separation distance
$2a=2\times0.256$.
In the third component of the vector perturbations $\dphi^3$, we can
see a dipolar structure emerging in the fifth columns of
figs.~\ref{fig:sapde2c20.01} and \ref{fig:sapde2c20.0428}, which could
be interpreted as the binding force between the two $N=2$ compactons
pulling in the constituent baby Skyrmions inside the $N=2$ compacton.
For larger values of $\epsilon$, this is eventually the effect that
stretches the baby Skyrmions in the $Q=4$ solution made of two
(deformed) $N=2$ baby Skyrmions side-by-side, see
fig.~\ref{fig:N=2+2}.
The effect is clearly absent in figs.~\ref{fig:sapde1c20.01} and
\ref{fig:sapde1c20.0428}, where the compactons are $N=1$ baby
Skyrmions and hence cannot be split up (viz.~fission) into smaller
parts.

\begin{figure}[!htp]
  \begin{center}
    \mbox{\subfloat[]{\includegraphics[width=0.49\linewidth]{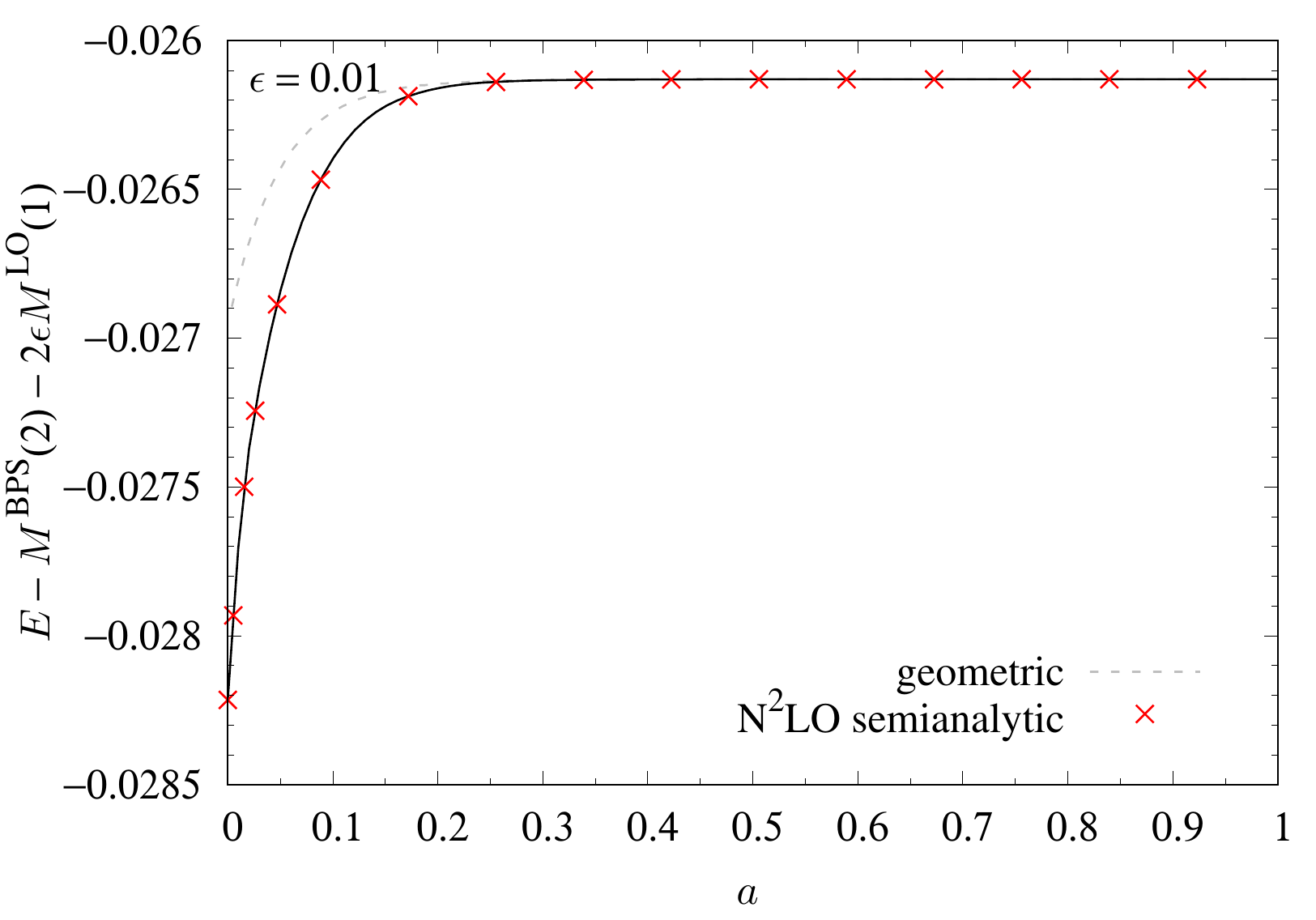}}
      \subfloat[]{\includegraphics[width=0.49\linewidth]{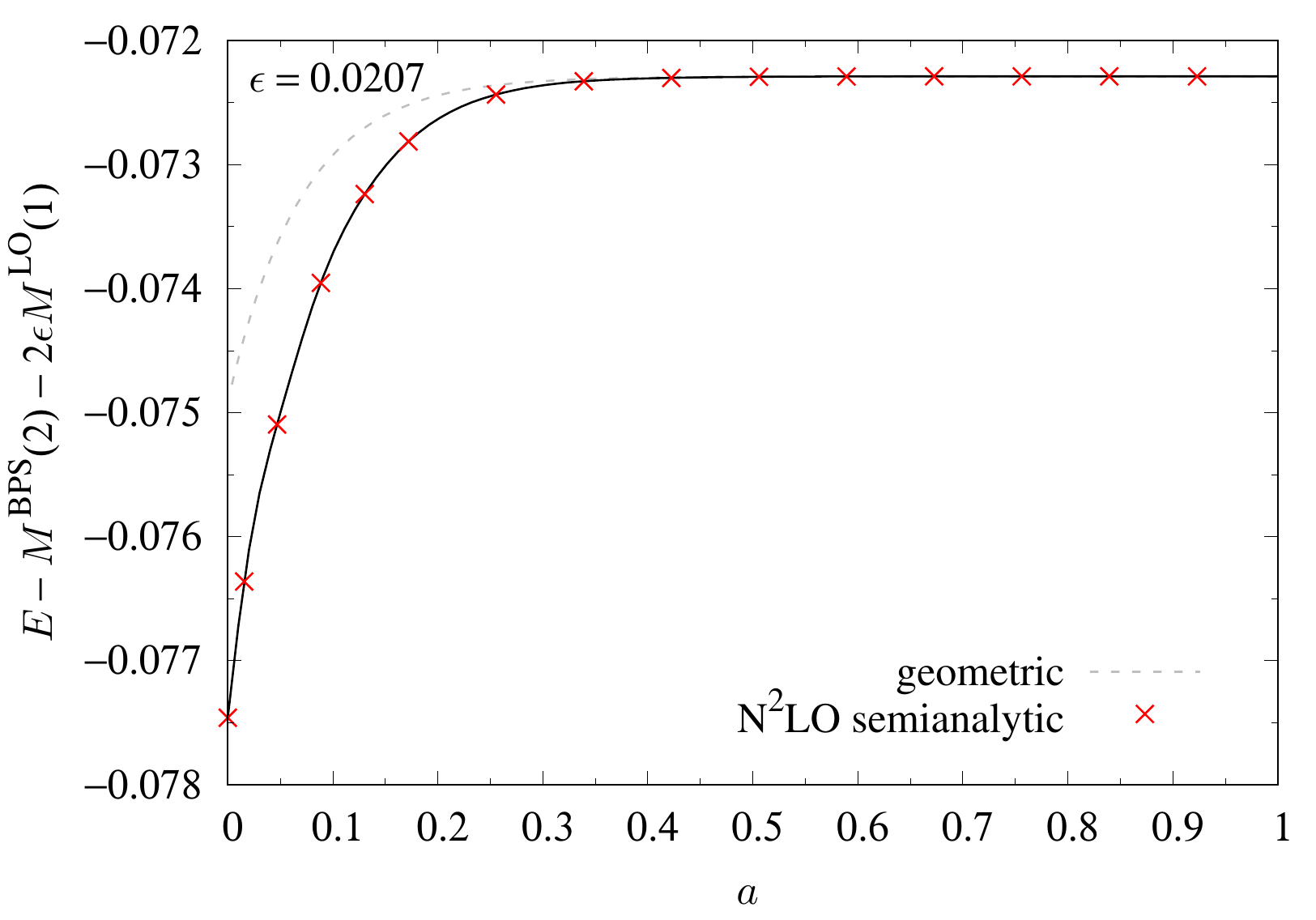}}}
    \mbox{\subfloat[]{\includegraphics[width=0.49\linewidth]{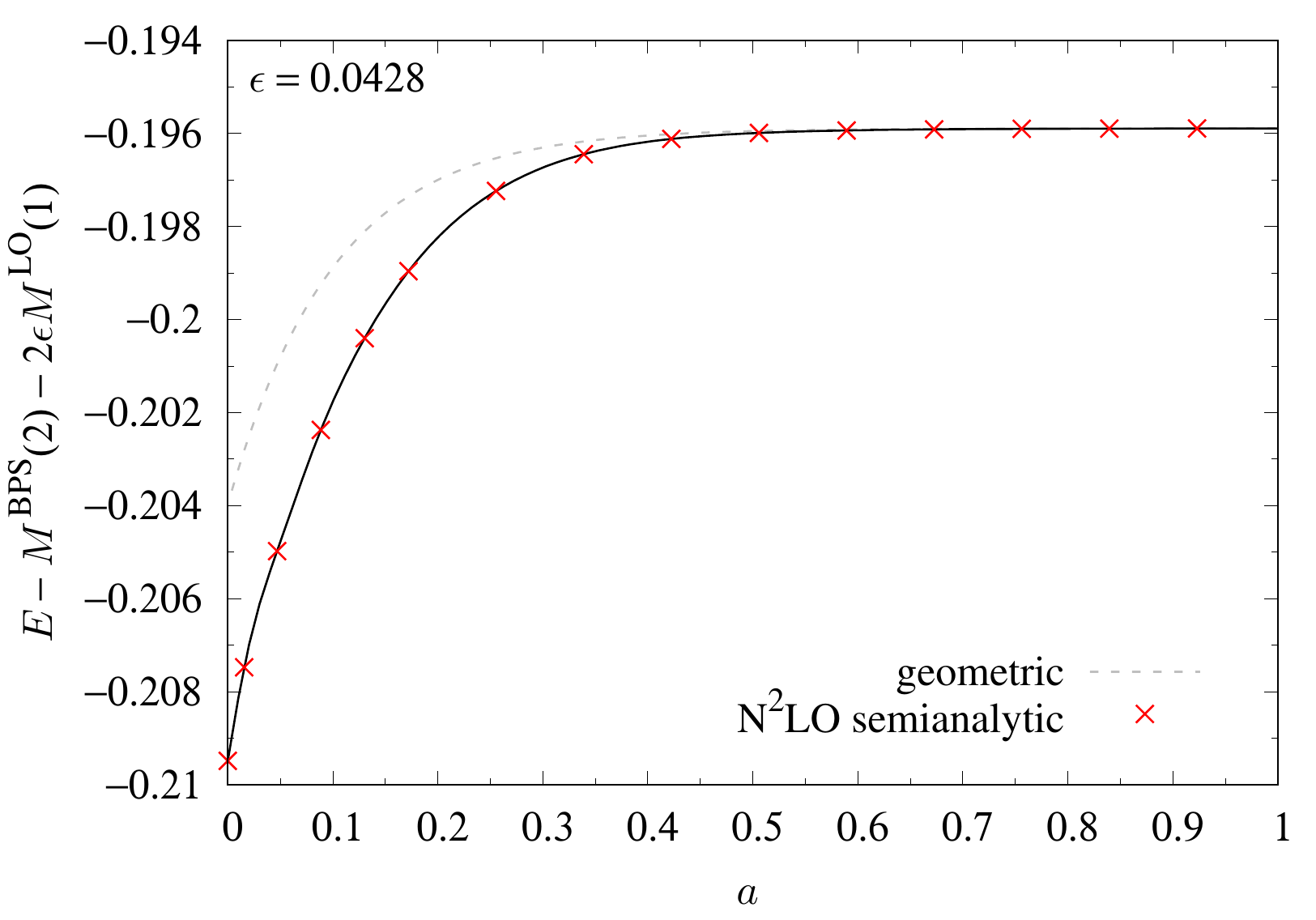}}
      \subfloat[]{\includegraphics[width=0.49\linewidth]{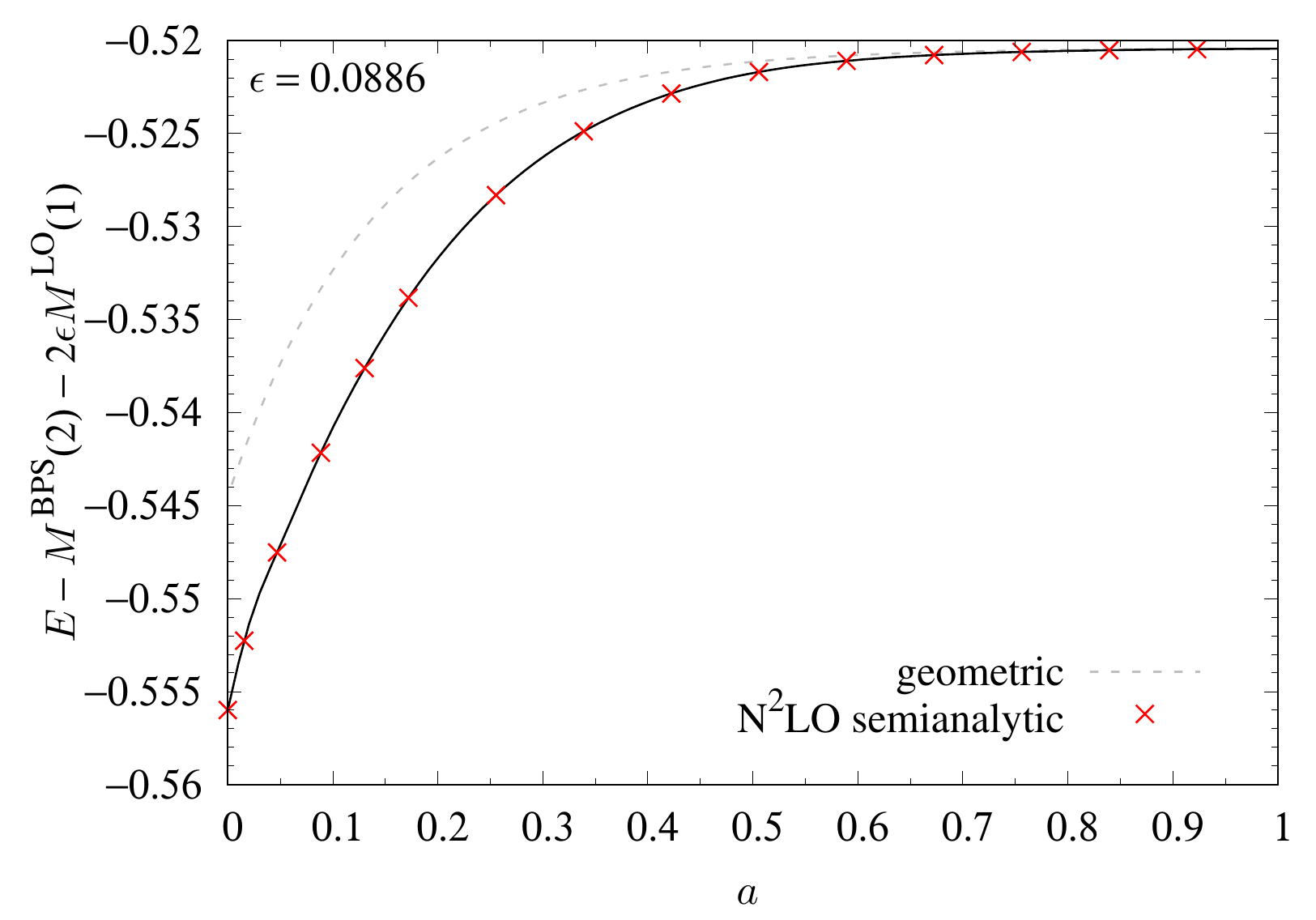}}}
    \caption{The N$^2$LO corrections to the energy for two $N=1$
      baby Skyrmions as a function of the separation distance $2a$, see
      fig.~\ref{fig:xzerobc}. The panels correspond to different
      values of $\epsilon$: (a) $\epsilon=0.01$, (b)
      $\epsilon=0.0207$, (c) $\epsilon=0.0428$ and (d)
      $\epsilon=0.0886$.
      The gray-dashed line is calculated as the geometric cut at $x=0$
      of the spherically symmetric compacton energy.
      The red crosses show the results of the N$^2$LO contribution to
      the energy from the PDE calculation of $\df$ and $\dtheta$.
      The black line is a cubic interpolation between the PDE
      results.
    }
    \label{fig:ben3}
  \end{center}
\end{figure}

Fig.~\ref{fig:ben3} shows the perturbative contribution to the energy
at N$^2$LO for two $N=1$ baby Skyrmions as function of the separation
distance $2a$, see fig.~\ref{fig:xzerobc}.
The four panels show $\epsilon=0.01$, $\epsilon=0.0207$,
$\epsilon=0.0428$ and $\epsilon=0.0886$, respectively and for all
panels, the smallest energy (meaning the largest negative contribution
to the energy) is at $a=0$, which means when the two compactons
exactly touch each other at a point, analogously to the two $N=2$ baby
Skyrmion case.
A difference with respect to the latter case is that for two $N=1$
baby Skyrmions, the N$^2$LO contribution to the energy is always below
the geometric energy, which is simply a geometric cut of the N$^2$LO
energy of the axially symmetric compacton at $x=0$.
In particular, there is no crossing of the N$^2$LO result and the
geometric line, even for $\epsilon=0.0886$.
This is probably because the gluing condition is more relaxed for the
$N=1$ compactons, since the winding of the background solution is less
(i.e.~minimal). In particular, this eases the gluing condition
\eqref{eq:dtheta_gluing_condition} which is more strict for the $N=2$
compacton's $\dphi^2$ component. 

\begin{figure}[!htp]
  \begin{center}
    \mbox{\subfloat[]{\includegraphics[width=0.49\linewidth]{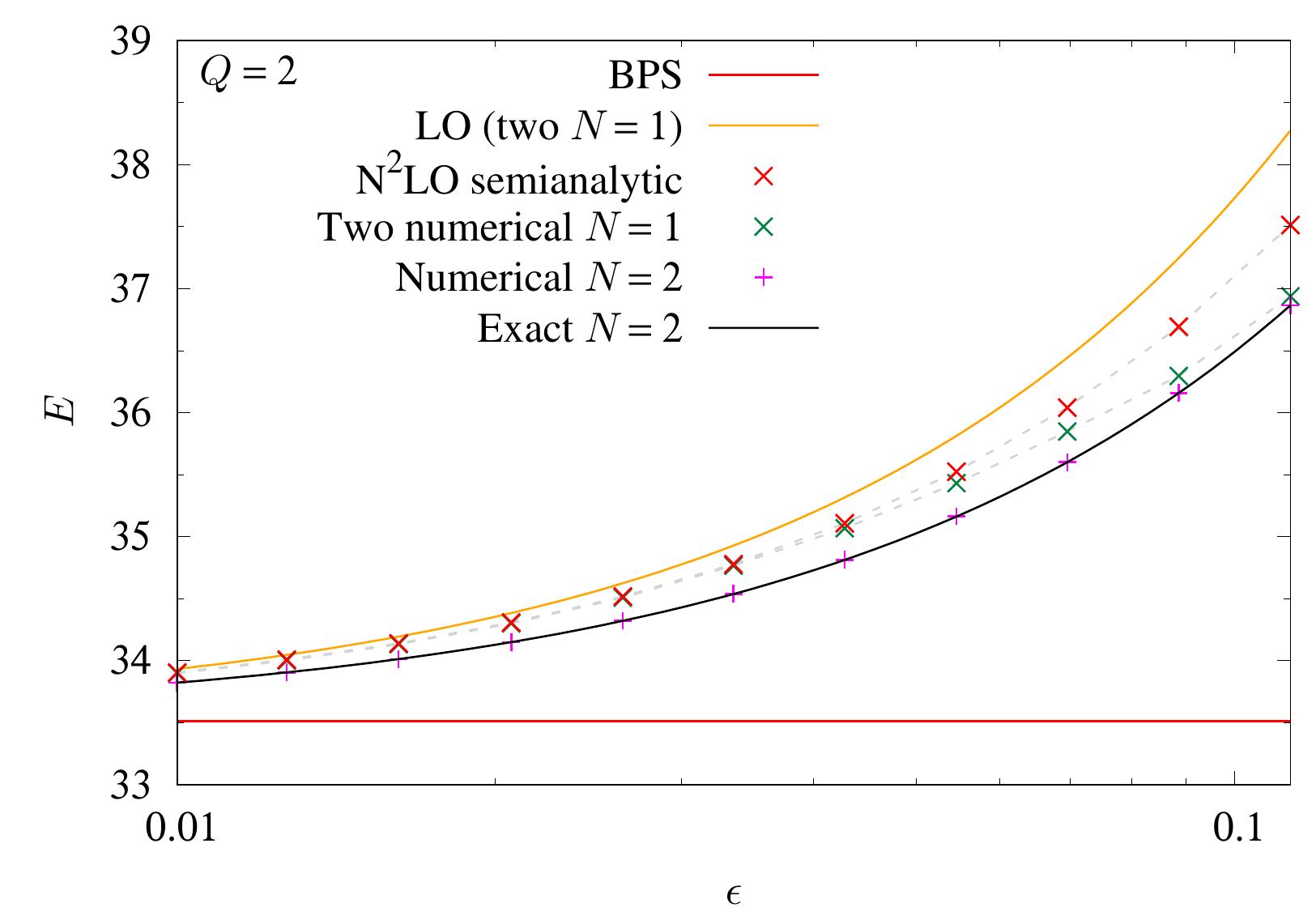}}
      \subfloat[]{\includegraphics[width=0.49\linewidth]{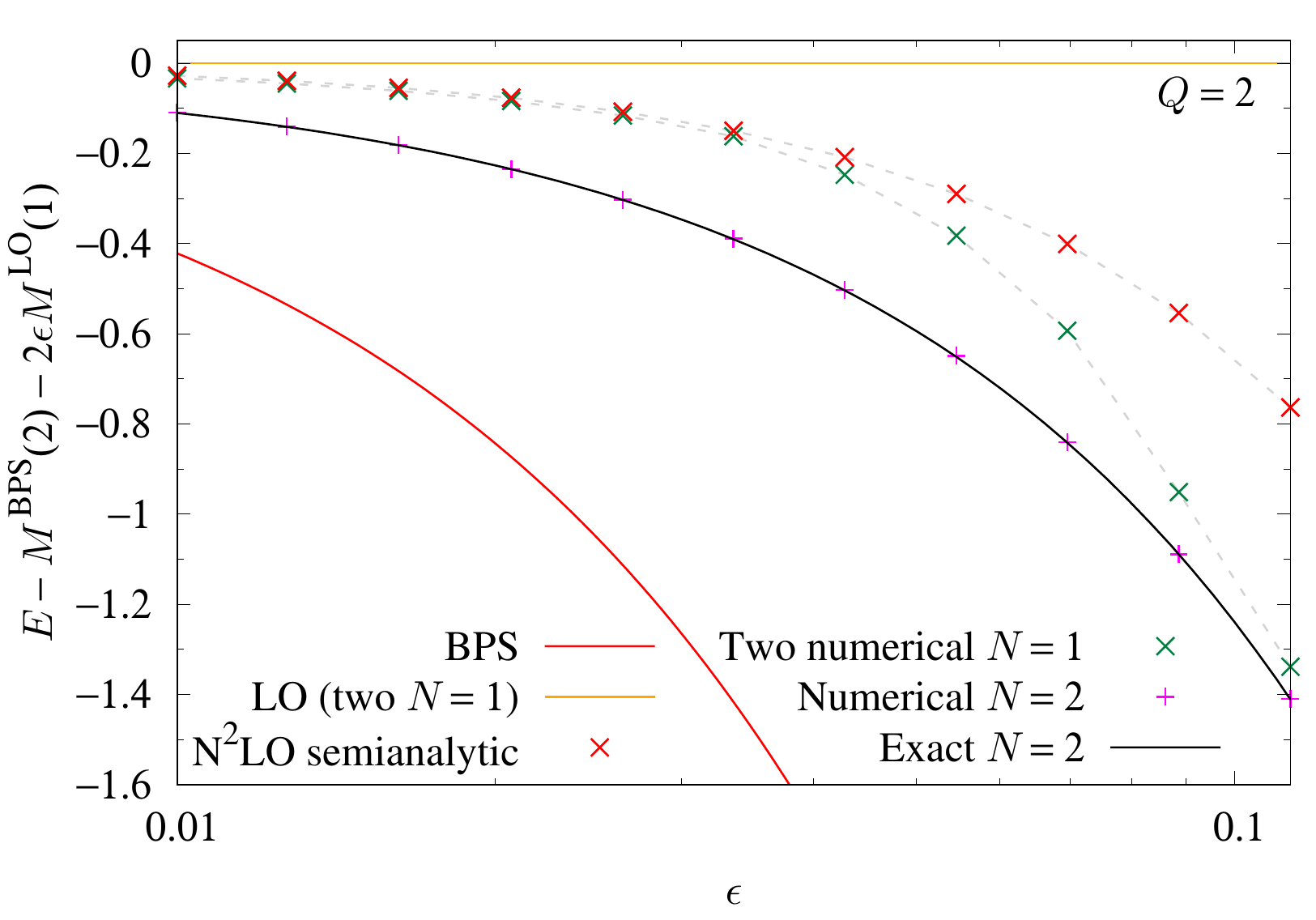}}}
    \caption{The energy (a) and N$^2$LO perturbation energy of the
      \emph{metastable} solution in the $Q=2$ homotopy class, which is
      made of two $N=1$ compactons with separation distance $2a=0$.
    The total energy (a) shows the BPS energy (red), the LO correction
    from the kinetic term of order $\epsilon$ (orange), the full
    N$^2$LO perturbation energy obtained by PDE calculations (red
    crosses) in comparison with the full numerical PDE calculations
    of the solutions without perturbation theory (green crosses), and
    for reference the full numerical PDE solutions of the axially
    symmetric $N=2$ baby Skyrmions (magenta pluses) as well as the
    exact $N=2$ baby Skyrmions obtained by ODEs, all as functions of
    $\epsilon$. 
    (b) shows the N$^2$LO part of the energy only, so as to better see
    the accuracy of the comparison between the full numerical PDE
    energies (green crosses) and the N$^2$LO perturbative energies
    (red crosses) as functions of $\epsilon$. }
    \label{fig:saen2}
  \end{center}
\end{figure}

We will now compare the N$^2$LO result of the perturbative
contribution to the energy of the two $N=1$ baby Skyrmions sitting
side-by-side (taking into account the binding energy between them)
with the full numerical (brute-force) PDE calculation in
fig.~\ref{fig:saen2}.
The result of the N$^2$LO perturbation theory is shown with red
crosses in the figure and should be compared with the green crosses
for the full numerical PDE calculations.
The two $N=1$ baby Skyrmions sitting side-by-side are only
\emph{metastable} and in fact the most stable result in the $Q=2$
sector is the single $N=2$ compacton, for which we do not need to
calculate any binding energy.
Hence this calculation is simply a test of the perturbative scheme on
the two $N=1$ baby Skyrmions sitting side-by-side, even though they
are only metastable.
For large $\epsilon\gtrsim0.15$ we know from the full numerical PDE
computations of sec.~\ref{quattro} that the two $N=1$ baby Skyrmions
side-by-side become unstable and merge into an axially symmetric $N=2$
baby Skyrmion.
This fact is known already from the leading order (LO) result of the
perturbation theory, see eq.~\eqref{eq:Nstar}.
We can thus anticipate that the N$^2$LO perturbative scheme will
become inaccurate for large $\epsilon$, since the instability is due
to the LO result and, in principle, not known to the NLO and N$^2$LO
computations.
For this reason, for reference, we show also the single $N=2$ baby
Skyrmion (with axial symmetry) in fig.~\ref{fig:saen2} as magenta
pluses (full PDE results) and a black solid line (ODE results).
From fig.~\ref{fig:saen2}(b) we can see that the perturbative result
for the N$^2$LO contribution to the energy works impressively well for
$\epsilon\lesssim0.033$ but then deviates and smoothly tends to the
curve for the single axially symmetric $N=2$ solution (black solid
line).

We have now confirmed by comparison with full numerical PDE
computations, that our perturbative scheme at N$^2$LO works very well
for the two $N=2$ baby Skyrmions sitting side-by-side at least for
$\epsilon\lesssim0.1$ and also for the two $N=1$ baby Skyrmions
side-by-side for $\epsilon\lesssim0.033$ (due to the instability of
this solution).
Since we already have a quite precise perturbative formula for the
baby Skyrmion energy for small $N$ and small $\epsilon$ at N$^2$LO,
see eq.~\eqref{eq:E_N2LO_compacton_fit}, it will henceforth be
worthwhile to separate out the binding energy of a ``single bond''
between two baby Skyrmions at N$^2$LO.
This is straightforwardly extracted from the data used in
figs.~\ref{fig:saen4} and \ref{fig:saen2}. 
The result for two baby Skyrmions with total charge $2N$ sitting
side-by-side is hence given by
\begin{align}
E^{\textrm{side-by-side}}(\epsilon,N\oplus N)
&= 2M^{\rm BPS}(N) + 2\epsilon M^{LO}(N) + 2\epsilon^2M^{\rm NLO}(N)
+2\epsilon^3M^{{\rm N}^2{\rm LO}}(N) \non
&\phantom{=\ }
+ 2\epsilon^4M^{\rm residual,4}(N)
+E^{\rm binding}(\epsilon,N),
\end{align}
where the first five terms are given by
eq.~\eqref{eq:E_N2LO_compacton_fit} and the binding energy of a single
bond is given by 
\begin{align}
E^{\rm binding}(\epsilon,N)
=
\begin{cases}
  -8.112\times 10^{-4}
  -15.65\epsilon^2
  +323.8\epsilon^3
  -2882\epsilon^4, &
  N=1,\\
  -5.108\times 10^{-4}
  -8.376\epsilon^2
  +167.8\epsilon^3
  -1097\epsilon^4, &
  N=2,
\end{cases}
\label{eq:Ebinding_fit}
\end{align}
where we have not extracted an $N$-behavior, since for $N=1$ and $N=2$
data only we cannot determine whether the is a quadratic or a linear
behavior (in $N$) in the coefficients.
Obviously, the constant term in both of the above expressions is
expected to vanish.
We have, nevertheless, included it because forcing it to be zero
significantly worsens the fit.
At this stage it is not clear whether the nonvanishing constant term
is numerical error or there is a contact term in the binding energy,
although we did not predict such a term.

\begin{figure}[!htp]
  \begin{center}
    \mbox{\subfloat[]{\includegraphics[width=0.49\linewidth]{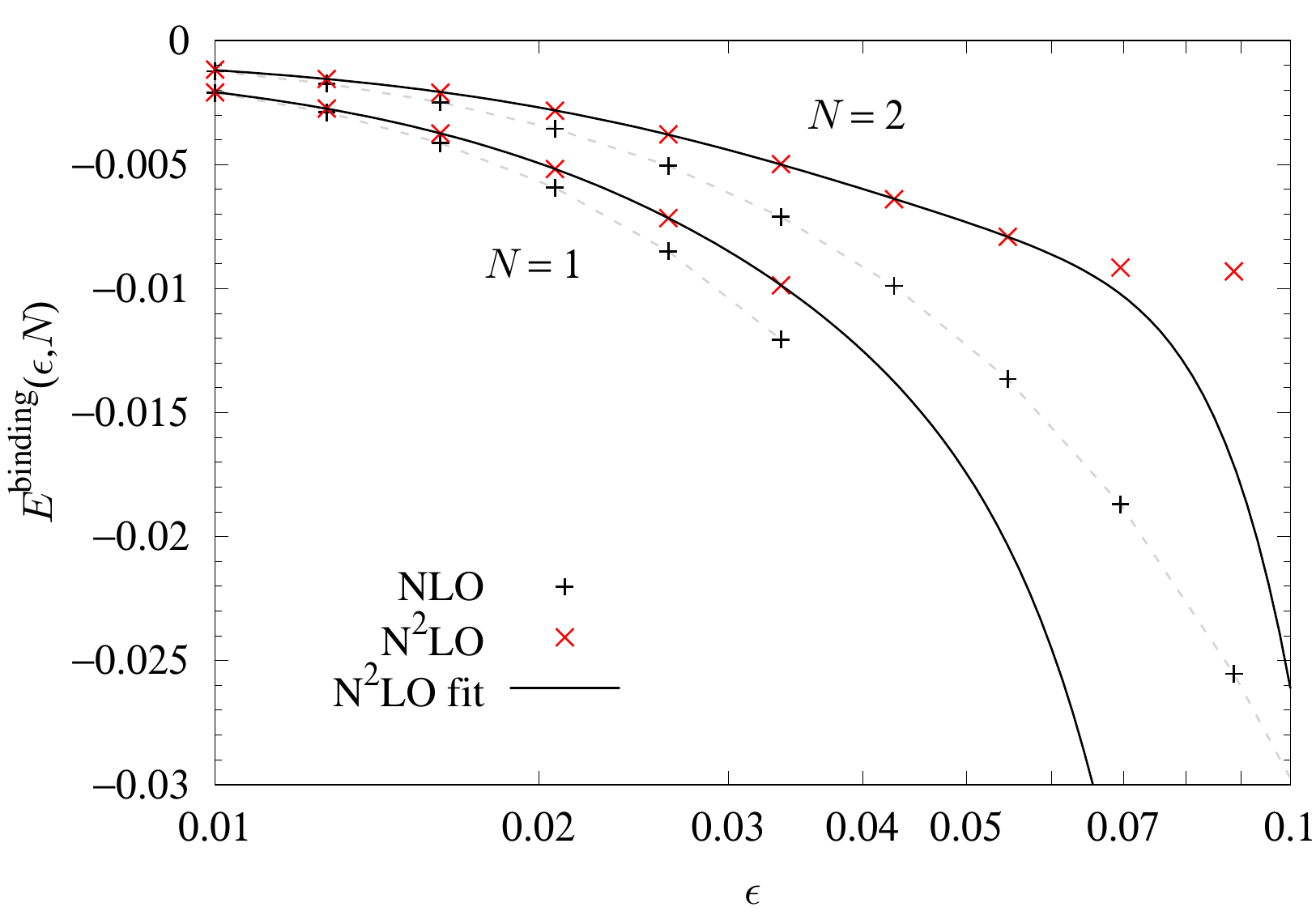}}
      \subfloat[]{\includegraphics[width=0.49\linewidth]{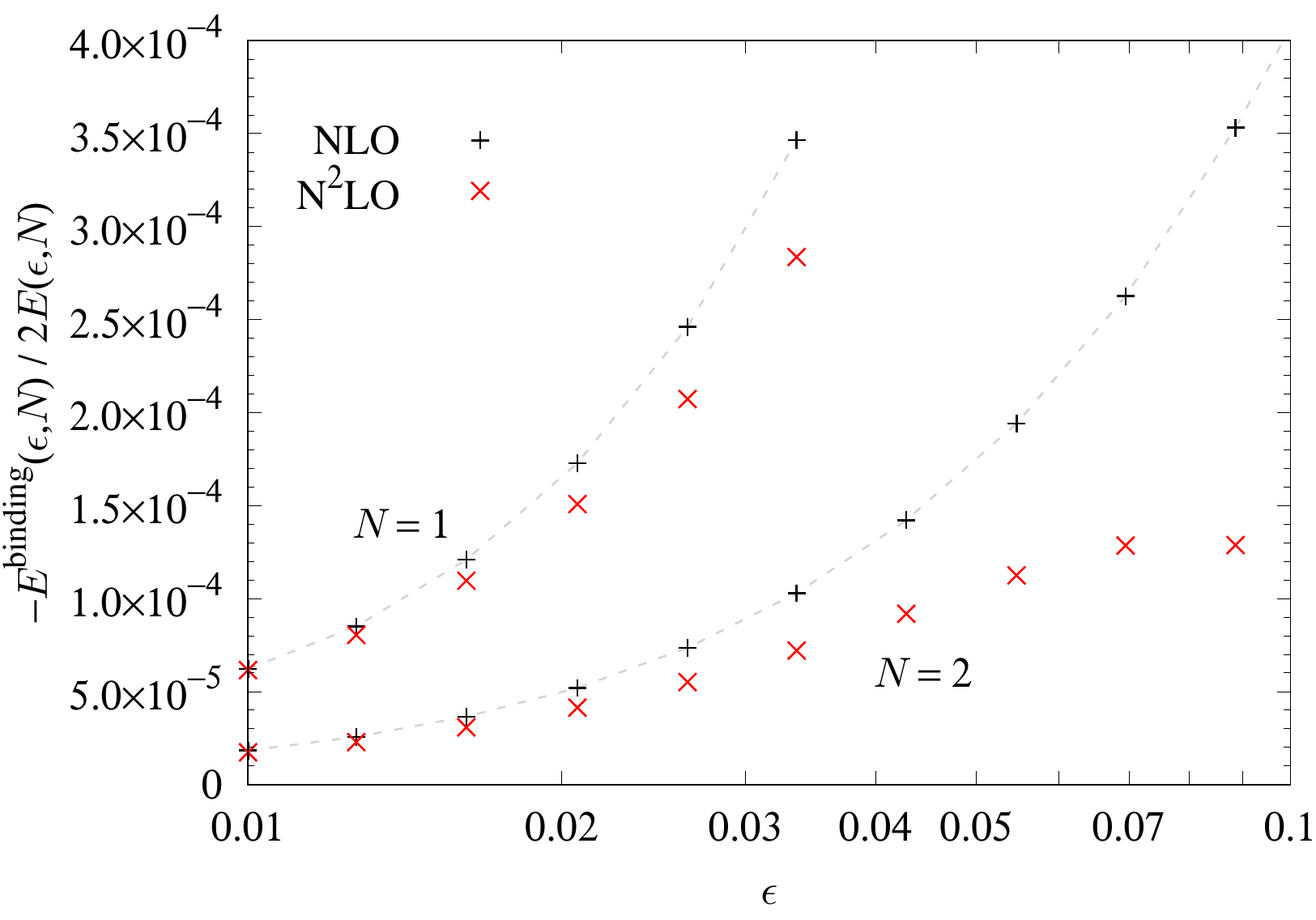}}}
    \caption{(a) Binding energy as a function of $\epsilon$ in form of a
      single ``bond'' between two $N=1,2$ baby Skyrmions.
      The red crosses are the N$^2$LO results, while the black pluses
      are the NLO results shown for reference.
      For $N=1$ the data is truncated at $\epsilon\sim0.033$ since the
      two composite $N=1$ baby Skyrmions side-by-side become unstable
      and will merge.
      The black solid lines are the fits given in
      eq.~\eqref{eq:Ebinding_fit}.
      (b) Binding energy between the two charge-$N$ baby Skyrmions
      divided by the total energy of the two charge-$N$ baby
      Skyrmions.
    }
    \label{fig:bebond}
  \end{center}
\end{figure}

Fig.~\ref{fig:bebond} shows the binding energy extracted from the
N$^2$LO computations of the two baby Skyrmions side-by-side, shown in
figs.~\ref{fig:saen4} and \ref{fig:saen2} for $N=2$ and $N=1$,
respectively.
For reference, we show also the binding energies at NLO.
The fits of eq.~\eqref{eq:Ebinding_fit} are displayed with black solid
lines and are fitted to the region $\epsilon<0.05$.
Fig.~\ref{fig:bebond}(a) shows the actual binding energy in the baby
Skyrme units, which should be compared to the energy of the baby
Skyrmions in baby Skyrme units, see
fig.~\ref{fig:perturb_energy_axial}.
In order to aid the comparison, we show the binding energy per
baby Skyrmion in units of the baby-Skyrmion mass in
fig.~\ref{fig:bebond}(b).

\section{Composite solutions}\label{sei}

One could now in principle construct ``nuclei'' of any $Q$ by rotating
the constituent compactons such that the colors are matching at the
``bond'' and to a first approximation, the total energy can be
calculated from eq.~\eqref{eq:Ebinding_fit} with the addition of
$E^{\rm binding}(\epsilon,N)$ for each ``bond'' binding the nucleus.
The near-BPS property allows one to tune $\epsilon$ until realistic
binding energies are obtained.
Of course, this only provides one with the ground state energies and
further development is needed for calculating the excitational
spectrum.
And of course, the baby Skyrme model is just a toy model.

\begin{figure}[!htp]
  \begin{center}
    \includegraphics[width=0.5\linewidth]{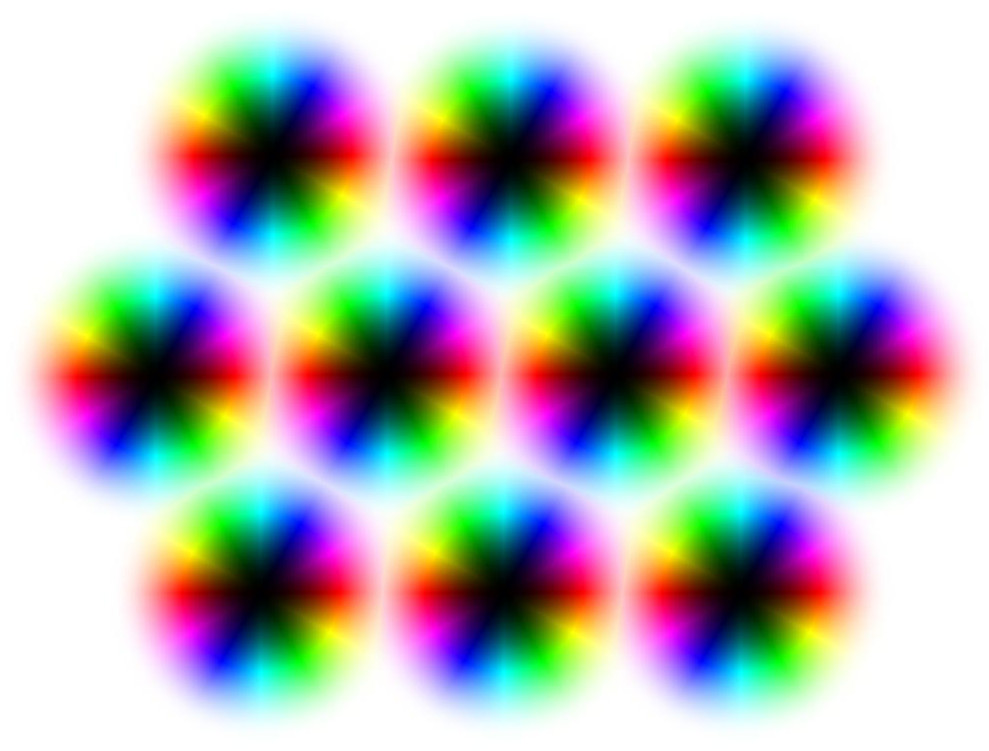}
    \caption{A sketch of a possible lattice for nuclear matter in the
      baby-Skyrme model. Each constituent baby Skyrmion is placed such
    that the colors match between the latter and the neighboring baby
    Skyrmion.}
    \label{fig:lattice}
  \end{center}
\end{figure}

We could further contemplate how the limit of large nuclei or nuclear
matter would look like. In fig.~\ref{fig:lattice} we show a
possibility for an infinite ``crystal'' lattice of which large nuclei
could be cut out from or nuclear matter could be made of.
This particular lattice could be continued indefinitely.
Nevertheless the shown lattice is made of ten baby Skyrmions and the
energy to a first approximation could be calculated for any small
value of $\epsilon$ by taking into account 19 bonds in the formula
\eqref{eq:Ebinding_fit}.

Or course, the lattice phase would be determined by minimizing the
energy and it could be for some $\epsilon$ that the square lattice is
energetically preferred to this triangle lattice, displayed in
fig.~\ref{fig:lattice}. 
Nevertheless, it is expected that for sufficiently large $\epsilon$,
the large-$Q$ solution becomes a chain \cite{Foster:2009vk} and the
lattice would then only exist at finite density.

\section{Conclusion and Discussion}\label{sette}

In this paper, we have studied the near-BPS regime of the baby Skyrme
model with the standard pion mass term as the potential and the
kinetic term as the BPS-breaking perturbation with coefficient
$\epsilon$.
The BPS solutions attain the BPS bound and hence the largest
contribution to the energy is the BPS mass.
The leading order contribution comes from the kinetic term and is of
order $\epsilon$.
To this order, there is no contribution to the binding energy and no
forces between the compactons.
Nevertheless, we find from the leading-order result that the $N=2$
baby Skyrmion is the stable solution for small $\epsilon$. 
In order to get a nonvanishing tail outside the compactons, we must go
to third order in $\epsilon$ and not truncate the solution to a finite
order in $\epsilon$, that is, the solution is of the form
$\exp(-mr/\sqrt{\epsilon})$, which in some sense is an all-order
solution in $\epsilon$.
We test our perturbative scheme on the axially symmetric compacton
solutions and calculate the N$^2$LO corrections to the energy
obtaining impressively good results.
A key to making the perturbation around the compactons capture the
physics of the baby Skyrmion is a delicate cusp condition that must be
imposed on the compacton boundary.
We then turn to a large-scale brute-force computation of the full PDEs
and obtain numerical solutions for the baby Skyrmions for values of
$\epsilon$ in the range $[0.01,1]$ and clearly observe that the baby
Skyrmions tend to almost unperturbed compactons that touch each other
at a single point in the attractive channel.
We also find a new solution, by studying the $N=4$ axially symmetric
baby Skyrmion which for $\epsilon\simeq0.15$ collapses to a new
metastable solution composed of four $N=1$ solutions in a tetrahedral
arrangement -- this solution is to the best of our knowledge new.
The surprise about this solution is that the neighboring $N=1$
solutions should merge into an $N=2$ solution, but this does not
happen for the tetrahedral solution, probably because of the discrete
symmetry yielding a delicate balance of the solution.
We also study the two $N=1$ baby Skyrmions side-by-side and find they
are metastable for small values of $\epsilon\lesssim0.15$.
Finally, we turn to calculating the binding energies of two $N=1,2$
baby Skyrmions sitting side-by-side using our perturbative scheme.
This is possible by imposing a gluing condition on the perturbation
field and the results are in very good agreement with the brute-force
numerical computations for small values of $\epsilon\lesssim0.1$.
Finally, we fit the results we have obtained, yielding an energy
formula as a function of $\epsilon$ and $N$ and the binding energy for
each bond is fitted for the cases of $N=1$ and $N=2$.

In this paper, we considered the case of the pion mass term as the
potential \eqref{masspot}. For this potential, in the near-BPS limit,
the mass of the ``pions'' (the perturbative particles) goes like
$\frac{1}{\sqrt{\epsilon}}$ due to the $\epsilon$ in front of the 
Dirichlet (kinetic) term; thus at a sufficiently small $\epsilon$ they
become even more massive than the baby Skyrmion itself.
Clearly, in QCD terms, this would be phenomenologically unacceptable. 
There are other near-BPS limits that resolve this problem.
For example, by choosing a different type of potential for the main
BPS part of the model, for which the contribution to the pion mass
vanishes, e.g.~that of eq.~\eqref{genpot} with  $p$ bigger than one.
One could then add the massive term in the perturbation together with
the kinetic term (so that it also is of order $\epsilon$ in the
Lagrangian).
These types of near-BPS limits have been discussed in the Skyrme
literature, see for example
refs.~\cite{Beaudoin:2013iza,Bolognesi:2014ova,Gillard:2015eia,Gudnason:2018jia}.
The main new observation we wish to make here is the following.
We showed that the binding energy for the compacton case goes like
$\epsilon^2$ and not as $\epsilon$ as would be
expected from the restricted harmonic argument. Thus a considerably
small binding energy could possibly be obtained without pushing the
pion mass too high.

Both the analytical guess and the numerical type of work and in
particular the semianalytic numerics we performed in the present paper
are tailored to the compacton case, so the non-compacton case will
require a different approach or at least crucial modifications.  
So far we know of several types of near-BPS solitonic models which can 
produce small classical binding energies.
There are big differences among them, and it is not clear which one is
most suitable for phenomenological applications to nuclear physics.
For example there are cases in which the small binding energy is
achieved by nuclei with size much smaller than the separation, and a
relative position fixed by a potential which can be computed in the
linear approximation \cite{Gillard:2015eia,Baldino:2017mqq},
i.e.~quite different from the type of bound state studied in the
present paper. 
Sometimes it is possible to interpolate between these two regimes by 
dialing a parameter, for example in holographic QCD with the 't
Hooft coupling \cite{Bartolini:2017sxi}.
It may be that by studying different types of potentials of the baby
Skyrme model even more diverse near-BPS behaviors will be discovered.

A crucial point of the perturbative scheme deployed in this paper is
that the leading order (LO) energy is finite. In
app.~\ref{app:potentials}, we have extended the BPS solutions and the
leading-order mass correction to the case of various generalized
potentials.
For the straightforward generalization of the pion mass term, which
is simply the latter to the power $s$ (i.e.~$V\sim (1-\phi^3)^s$), the
BPS solution yields a finite integral for the kinetic energy for all
the values of $s$ studied in the appendix.
Changing the potential to the modified pion mass, which possesses
domain walls, (i.e.~$V\sim (1-(\phi^3)^2)^s$, the BPS solutions can be
found only indirectly, except for $s=1$.
But yet worse, for the solution with $s=1$, the leading order mass
correction diverges.
This demonstrates that the problem of perturbation around the
compacton is not due to the cusp, that one might naively expect, but
is due to the behavior of the BPS solution at the origin (i.e.~at
$r=0$).
For the BPS Skyrme model in 3+1 dimensions, a similar issue with the
leading-order mass correction is known, for example for the solutions
in ref.~\cite{Adam:2010fg}.
As illustrated in the app.~\ref{app:potentials}, it may well be that
one must find an appropriate class of potentials for the BPS Skyrme
model, in order to be able to port the perturbative scheme to the 3+1
dimensional model.

\section*{Acknowledgments}

We thank Calum Ross, Martin Speight, Andrzej Wereszczynski and Wojtek
Zakrzewski for useful discussions. 
The work of M.B.~and S.B.~is supported by the INFN special project
grant ``GAST (Gauge and String Theory)''.   
S.B.G.~thanks the Outstanding Talent Program of Henan University for
partial support.
The work of S.B.G.~is supported by the National Natural Science
Foundation of China (Grant No.~11675223 and 12071111).

\appendix

\section{The leading order energy for various potentials}\label{app:potentials}

In this appendix, will extend the leading order calculation, in
$\epsilon$, of the energy of the baby Skyrmion to the cases of
different potentials than the traditional mass term.

Although the leading order (LO) contribution to the energy comes from
the kinetic term, it depends on the potential via the solution which
depends strongly on the choice of potential.
The BPS equation \eqref{eq:BPSeq_zeta} for a generic potential takes
the form 
\beq
\frac{\p_r\zeta}{r} = \pm \frac{(1+\zeta^2)^2}{2\sqrt{2}N\zeta}\sqrt{V(\zeta)}.
\eeq
We will consider a few cases in turn in the next subsections.

\subsection{\texorpdfstring{$(1-\phi^3)^s$}{(1-phi3)**s}}

Starting with the simplest generalization of the standard pion mass
term, we consider the latter to the power $s$:
\begin{align}
  V &= \frac{m^2}{s}(1 - \phi^3)^s \non
  &= \frac{m^2}{s}\left(\frac{2\zeta^2}{1+\zeta^2}\right)^s,
  \label{eq:generalized_pion_mass}
\end{align}
which yields the BPS equation
\beq
\frac{\p_r\zeta}{r}
= \pm \frac{2^{\frac{s-3}{2}}m}{N\sqrt{s}}
(1 + \zeta^2)^{2-\frac{s}{2}} \zeta^{s-1},
\eeq
which we can integrate as
\beq
\int\frac{\d\zeta}{\zeta^{s-1}(1+\zeta^2)^{2-\frac{s}{2}}}
= \pm \frac{2^{\frac{s-5}{2}}m}{N\sqrt{s}} r^2
\mp \frac{\xi_0^2}{2-s}, \qquad
0 < s < 2.
\eeq
Defining the compacton radius
\beq
R\equiv \sqrt{\frac{2^{\frac{5-s}{2}}\sqrt{s}N}{m}}, \qquad
0 < s < 2,
\eeq
we can write the equation with the lower sign as
\beq
\left(\frac{\zeta}{\sqrt{1+\zeta^2}}\right)^{2-s}
= \xi_0^2 - \xi^2, \qquad
0 < s < 2,
\eeq
which can be inverted to give the solution
\beq
\zeta = \frac{(\xi_0^2 - \xi^2)^{\frac{1}{2-s}}}{\sqrt{1 - (\xi_0^2 - \xi^2)^{\frac{2}{2-s}}}}, \qquad
0 < s < 2.
\eeq
Finally, we move the singularity to $\xi=0$ by setting $\xi_0=1$:
\beq
\zeta = \frac{(1 - \xi^2)^{\frac{1}{2-s}}}{\sqrt{1 - (1 - \xi^2)^{\frac{2}{2-s}}}}, \qquad
0 < s < 2.
\eeq
The BPS bound is still attained, but now we can calculate the LO
contribution to the energy using the above solution
\begin{align}
&\epsilon M^{\rm LO}(N) = 4\pi\epsilon
\int_0^1 \d\xi\;\frac{\xi\zeta_\xi^2 +
  \frac{N^2}{\xi}\zeta^2}{(1+\zeta^2)^2}\non
&\qquad= -4\pi\epsilon\int_0^1\d\xi\;
\frac{(1-\zeta^2)^{\frac{4}{s-2}}\left[
  4\zeta^4(1-\zeta^2)^{-2-\frac{4}{s-2}}
  +N^2(2-s)^2\left(1 - (1-\zeta^2)^{\frac{2}{s-2}}\right)^2\right]}
     {(s-2)^2\zeta\left(1 - (1-\zeta^2)^{\frac{2}{s-2}}\right)},
\end{align}
with $0<s<2$.
Unfortunately, we do not know how to perform this integral for arbitrary
values of $s$ in the given interval, but it can be carried out for
certain specific values
\begin{align}
&\epsilon M^{\rm LO}(N) = 4\pi\epsilon F(N)\non
  &F(N)=
  \begin{cases}
  \left(\log2 - \frac{5}{12}\right)N^2
  +\frac18(\sqrt{3}\pi+9\log3), & s=\frac{2}{3},\\
  \frac{7}{24}N^2 + 2\log 2, & s=1,\\
  \frac{533N^2}{1680} - \pi + 6\log 2, & s=\frac{3}{2},\\
  \frac{37N^2}{120} - \frac14\sqrt{3}\pi + \frac{9}{4}\log3, &
  s=\frac{4}{3},\\
  \frac{18107N^2}{55440} - \frac32\sqrt{3}\pi + \frac92\log3 + 6\log2,
  & s=\frac{5}{3},\\
  \frac{95549N^2}{288288}
  -2(1+\sqrt{2})\pi
  +4\sqrt{2}\arccoth(\sqrt{2})
  +16\log2, & s=\frac{7}{4},\\
  \frac{3602044091N^2}{10708457760}
  -3(2+\sqrt{3})\pi
  +9\log12
  -\frac34\sqrt{3}\log(18817 - 10864\sqrt{3}), & s=\frac{11}{6}.
\end{cases}
\end{align}

\begin{figure}[!htp]
  \begin{center}
    \includegraphics[width=0.5\linewidth]{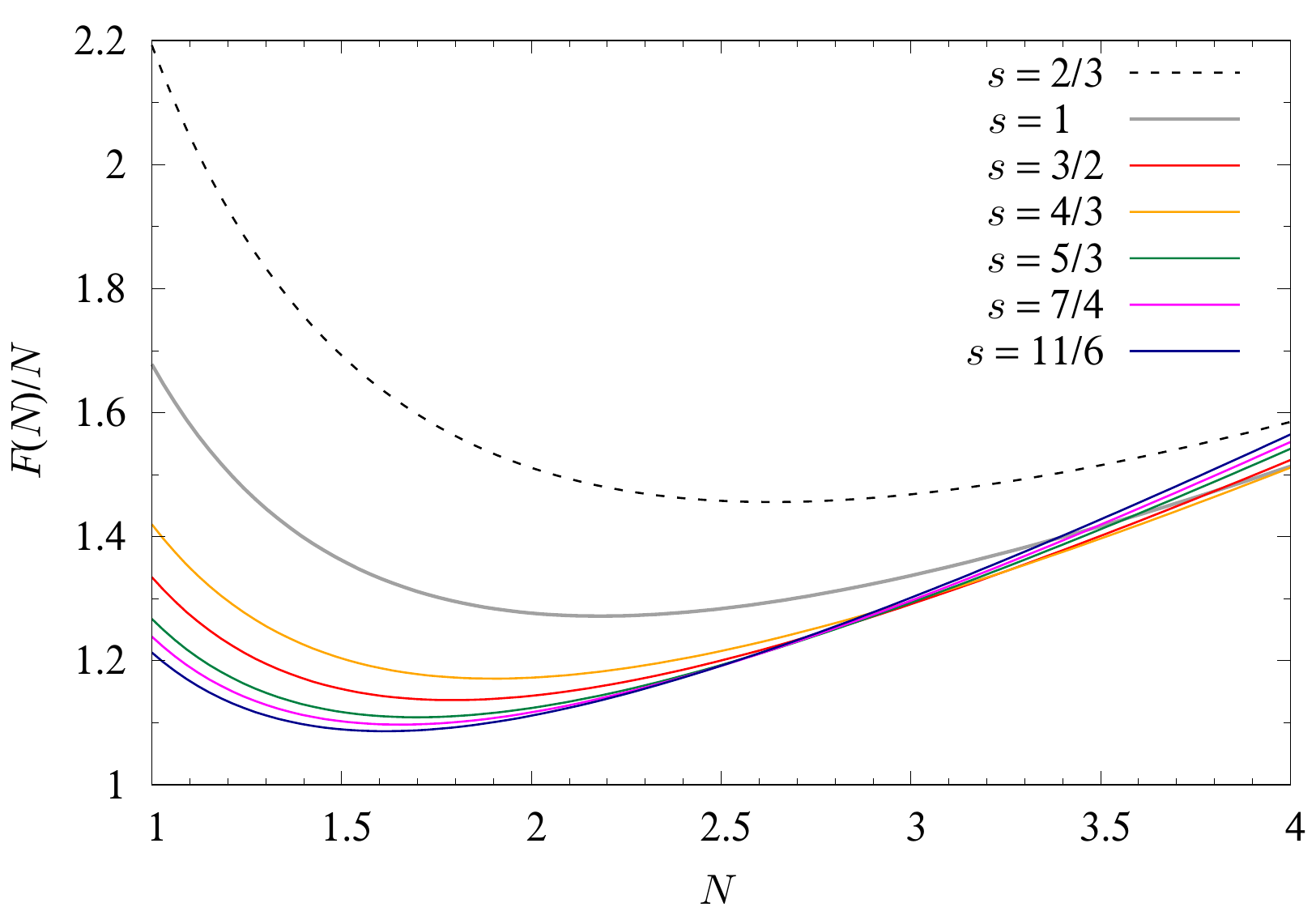}
    \caption{The leading-order mass correction function
      $F(N)=M^{\rm LO}(N)/4\pi$ as function of $N$ for various values
      of $s$ of the potential \eqref{eq:generalized_pion_mass}.
      Clearly, in all cases, the energy per $N$ is smaller for $N=2$
      than for $N=1$. 
    }
    \label{fig:FN}
  \end{center}
\end{figure}

Fig.~\ref{fig:FN} shows the function $F(N)$ for various values of $s$.

Since the BPS mass does not contribute to the determination of
$N_\star$, to leading order (LO), $N_\star$ is determined by
\beq
\frac{\d}{\d N}\left(\frac{E(\epsilon,N)}{N}\right)
= \frac{\d}{\d N}\left(\frac{F(N)}{N}\right) = 0,
\eeq
yielding
\begin{equation}
N_\star =
\begin{cases}
  \sqrt{\frac{3(\sqrt{3}\pi + 9\log3)}{24\log2-10}}, & s=\frac23,\\
  4\sqrt{\frac{3\log 2}{7}}, & s=1,\\
  4\sqrt{\frac{105}{533}(6\log2-\pi)}, & s=\frac32,\\
  \sqrt{\frac{30}{37}(9\log3-\sqrt{3}\pi)}, & s=\frac43,\\
  6\sqrt{\frac{2310}{18107}(\log432-\sqrt{3}\pi)}, & s=\frac53,\\
  12\sqrt{\frac{2002}{95549}\left(-2(1+\sqrt{2})\pi+4\sqrt{2}\arccoth(\sqrt{2})+16\log2\right)}, & s=\frac74,\\
  6\sqrt{\frac{223092870}{3602044091}\left(-4(2+\sqrt{2})\pi+12\log12-\sqrt{3}\log(18817-10864\sqrt{3})\right)}, & s=\frac{11}{6},
\end{cases}
\end{equation}
or in approximate numbers
\beq
N_\star \simeq
\begin{cases}
  2.633, & s=\frac23,\\
  2.180, & s=1,\\
  1.791, & s=\frac32,\\
  1.899, & s=\frac43,\\
  1.697, & s=\frac53,\\
  1.654, & s=\frac74,\\
  1.614, & s=\frac{11}{6}.
\end{cases}
\eeq

\begin{table}[!htp]
  \begin{center}
    \begin{tabular}{c||llll}
      $s$ & $N=1$ & $N=2$ & $N=3$ & $N=4$\\
      \hline\hline
      2/3  & 2.193 & 1.511        & 1.468$^\star$ & 1.585\\
      1    & 1.678 & 1.276$^\star$ & 1.337        & 1.513\\
      3/2  & 1.335 & 1.143$^\star$ & 1.291        & 1.523\\
      4/3  & 1.420 & 1.172$^\star$ & 1.296        & 1.511\\
      5/3  & 1.267 & 1.123$^\star$ & 1.293        & 1.542\\
      7/4  & 1.239 & 1.116$^\star$ & 1.297        & 1.553\\
      11/6 & 1.213 & 1.111$^\star$ & 1.301        & 1.565
    \end{tabular}
    \caption{The function $F(N)/N$ determining which solution has the
      smallest energy per $N$ as function of $N$ at LO.
      The smallest number is marked with a ${}^\star$.
      Except for $s=\frac{2}{3}$ for which $N=3$ has the smallest
      energy at LO, all the cases have the smallest energy at LO for
      $N=2$. 
    }
    \label{tab:FN}
  \end{center}
\end{table}

When $N_\star$ is not close to an integer, it is less clear which $N$
actually provides the minimal energy at LO.
Thus we provide the data for $F(N)/N$ in tab.~\ref{tab:FN}

\subsection{\texorpdfstring{$(1-(\phi^3)^2)^s$}{(1-phi3**2)**s}}

A straightforward generalization of the so-called modified pion mass
term can be written as
\begin{align}
  V &= \frac{m^2}{2s}\left(1 - (\phi^3)^2\right)^s \non
  &= \frac{m^2}{2s}\left(\frac{2\zeta}{1+\zeta^2}\right)^{2s},
\end{align}
which yields the BPS equation
\beq
\frac{\p_r\zeta}{r} = \pm \frac{2^{s-2}m}{N\sqrt{s}}(1+\zeta^2)^{2-s}\zeta^{s-1},
\eeq
which we can integrate as
\beq
\int\frac{\d\zeta}{\zeta^{s-1}(1+\zeta^2)^{2-s}}
= \pm \frac{2^{s-3}m}{N\sqrt{s}} r^2 \mp \xi_0^2, \qquad
0 < s < 2.
\eeq
Defining the compacton radius
\beq
R \equiv \sqrt{\frac{2^{3-s}\sqrt{s}N}{m}}, 
\eeq
we can write the equation with the lower sign as
\beq
\frac{\zeta^{2-s}}{2-s} {}_2F_1\left[2-s,1-\frac{s}{2};2-\frac{s}{2};
  -\zeta^2\right]
= \xi_0^2 - \xi^2, \qquad
0 < s < 2,
\eeq
where ${}_2F_1$ is Gauss' hypergeometric function. 
Unfortunately, we do not know how to invert this equation to get a
closed expression for $\zeta$, which makes it difficult to evaluate
the LO energy analytically.

However, if we restrict to the case of $s=1$, we get the equation
\beq
\arctan\zeta = \xi_0^2 - \xi^2,
\eeq
which can readily be inverted as
\beq
\zeta = \tan(1 - \xi^2),
\eeq
where we have set $\xi_0=1$.
This solution, however, does not yield a finite result for the
leading-order correction to the mass.

\end{document}